\documentclass[aps,prd,a4paper,twocolumn]{revtex4}

\usepackage{graphicx}
\usepackage{bm}
\usepackage{amsfonts}
\usepackage{amsmath}

\usepackage[russian,ngerman,english]{babel}

\newcommand{\muas}[0]{\hbox{\rm $\mu$as}}

\newcommand{\ve}[1]{\mbox{\boldmath$#1$}}
\newcommand*\Laplace{\mathop{}\!\mathbin\bigtriangleup}

\let\oldbibitem\bibitem
\renewcommand\bibitem[2][]{\oldbibitem{#2}}

\begin{document}

\title{Light propagation in the 2PN approximation in the monopole and quadrupole field of a body at rest: Boundary value problem}

\author{Sven Zschocke}

\affiliation{Lohrmann Observatory, TUD Dresden University of Technology, Helmholtzstrasse 10, D-01069 Dresden, Germany}


\begin{abstract}
In a recent investigation, the initial value problem of light propagation in the gravitational field of a  
body at rest with monopole and quadrupole structure has been determined in the second post-Newtonian (2PN) 
approximation. In reality, the light source as well as the observer are located at finite distances  
from the solar system bodies. This fact requires solving the boundary value problem of light propagation. 
In this investigation, the solution of the boundary value problem is deduced from the initial value problem 
of light propagation in 2PN approximation. These results are a basic requirement for subsequent investigations 
aiming at ultra-highly precise tests of light deflection and time delay in the solar system. 
\end{abstract}

\pacs{95.10.Jk, 95.10.Ce, 95.30.Sf, 04.25.Nx, 04.80.Cc}

\maketitle

\begin{center} 


\end{center}

\section{Introduction}\label{Introduction}\label{Section0}

The precision in astrometric angular measurements has made a giant step from the milli-arcsecond (mas) level, as achieved by the astrometry 
mission Hipparcos \cite{Hipparcos} to the micro-arcsecond (\muas) scale of accuracy, as achieved by the astrometry mission Gaia \cite{Gaia}.  
These astrometry missions of the European Space Agency (ESA) were the first space missions developed for highly precise astrometric measurements 
of positions, distances, and proper motions of celestial objects. Meanwhile, there are several development proposals, both space-based astrometry 
missions \cite{Gaia_NIR,Theia,Astrod1,Astrod2,Lator1,Lator2,Odyssey,Sagas,TIPO} and ground-based facilities \cite{nas_telescopes}, which are aiming 
at the sub-micro-arcsecond (sub-\muas) and even at the nano-arcsecond (nas) scale of accuracy. The science cases of sub-\muas{} astrometry   
are overwhelming, like detection of Earth-like exoplanets, measurements of dark matter distributions, tests of general relativity (GR) in the solar 
system, progress in determining natural constants, considerable extension of a model-independent cosmic distance ladder, and even detection of gravitational waves by 
astrometric measurements \cite{Sub_Micro_1,Sub_Micro_2,Sub_Micro_3,Sub_Micro_4,Sub_Micro_5}. 

A fundamental problem in relativistic astrometry concerns the precise determination of the trajectory of a light signal, emitted by some celestial 
light source and propagating through the curved space-time of the solar system towards the observer. The importance of this fact has also been emphasized by the 
ESA-Senior-Survey-Committee (SSC) in response of the selection of near-future space missions. In particular, the SSC has recommended a further development of the 
GR framework to model photon trajectories to the required accuracy of high-precision astrometry of the next generation \cite{SSC}.

There are several difficult and serious issues regarding a general-relativistic modeling of light propagation aiming at the sub-\muas${}$ level of accuracy. 
Let us consider three of these aspects: 

{\bf (a)} Solar system bodies can be of complicated shape and inner structure, which can be described by the multipole expansion of the metric tensor of these 
bodies \cite{Thorne,Blanchet_Damour1,Multipole_Damour_2}. The light trajectories in the gravitational field of such arbitrarily shaped bodies at rest have been determined 
in the first orders of the post-Newtonian (PN) scheme: namely in the 1PN approximation (terms of order ${\cal O}(c^{-2})$) and in the 1.5PN approximation 
(terms of order ${\cal O}(c^{-3})$), both for stationary multipoles \cite{Kopeikin1997} as well as for time-dependent multipoles \cite{KopeikinKorobkovPolnarev2006}. 

{\bf (b)} The second problem concerns the motion of solar system bodies along their complicated world lines. Several investigations have demonstrated that 
on the \muas{} level it is sufficient to determine the light trajectories in the gravitational fields of bodies at rest, if one implements their retarded positions 
in the GR model \cite{KopeikinSchaefer1999,KopeikinMashhoon2002,KopeikinMakarov2007,KlionerPeip2003,Klioner_LT}. In order to investigate this problem further, 
the light trajectories in the gravitational field of slowly moving solar system bodies with full multipole structure have been determined in the 1PN and 1.5PN 
approximation \cite{Zschocke_1PN,Zschocke_15PN} by means of the approach developed in \cite{Kopeikin1997,KopeikinKorobkovPolnarev2006}. 

{\bf (c)} Higher orders of the post-Newtonian (PN) expansion of metric tensor of the solar system need to be taken into account, both for defining highly 
accurate reference systems as well as for highly precise GR modeling of light trajectories in the solar system. Regarding this problem, it can be stated 
that the metric tensor for solar system bodies at rest is, in principle, well-known up to the second post-Newtonian approximation (2PN)  
\cite{Thorne,Blanchet_Damour1,Multipole_Damour_2,Blanchet2,Zschocke_2PM_Metric,2PN_Metric1}. However, such an understanding has by far not been achieved for the 
light propagation in 2PN approximation. In fact, 2PN light trajectories have only been solved for gravitational fields generated by monopoles, that means spherically 
symmetric bodies at rest \cite{Brumberg1987,Brumberg1991}. This problem has later been reconsidered under several aspects in a series of subsequent investigations 
\cite{KlionerKopeikin1992,Klioner_Zschocke,Ashby_Bertotti,Teyssandier,Deng_Xie,2PN_Light_PropagationA,Deng_2015,Xie_Huang,Minazzoli2,Hees_Bertone_Poncin_Lafitte_2014b,Bruegmann2005,Zschocke3,Zschocke4,Zschocke5}. The next term in the multipole decomposition of the metric tensor of solar system bodies is the quadrupole term. Accordingly,  
the impact of the quadrupole structure of solar system bodies on light trajectories is the most significant effect beyond the monopole. Such a rigorous 2PN solution 
for light trajectories has been obtained only recently in our investigation \cite{Zschocke_Quadrupole_1}. In that investigation it has also been found that the 
2PN quadrupole light deflection amounts up to $0.95$ \muas${}$ and $0.29$ \muas${}$ for grazing light rays at the giant planets Jupiter and Saturn, respectively. 

Another aspect of astrometric science concerns the progress in ultra-highly precise time measurements, both by ground-based facilities as well as space-based atomic 
clocks. In our recent investigation in \cite{Zschocke_Time_Delay_2PN}, the impact of the quadrupole structure of solar system bodies on time delay has been determined 
in the 2PN approximation, which amounts up to $0.14$ and $0.04$ pico-seconds (ps) for grazing rays at the giant planets Jupiter and Saturn. 

Recent development in the atomic clock and time transfer techniques, like the atomic clocks NIST-F1 and NIST-F2 at the National Institute of
Standard and Technology (NIST) \cite{NIST}, the optical atomic clock facility at NIST \cite{Atomic_Clock1}, as well as the atomic clocks onboard a 
satellite like the Deep Space Atomic Clock (DSAC) by National Aeronautics and Space Administration (NASA) \cite{DSAC1} and the Atomic Clock Ensemble
in Space (ACES) of ESA \cite{ACES}, fosters the hope that in the foreseeable future one could measure the travel time of the electromagnetic signals
in the solar system to a precision on the sub-pico-second level or perhaps even better.

Thus, in view of advancements of the precision in angular observations as well as in time measurements, it is clear that in near-future the quadrupole effects in 
2PN approximation become relevant for high-precision tests of relativity in the solar system. As mentioned above, the 2PN solution of the initial value problem 
for the light trajectory in the quadrupole field of solar system bodies has been achieved in our recent investigation in \cite{Zschocke_Quadrupole_1}. 
The initial value problem (Cauchy problem) is characterized by two given initial values: the initial direction of the light ray and the spatial position 
of the light source. However, for high-precision tests of relativity in the solar system it is necessary to determine the 2PN light ray solution 
of the boundary value problem. The boundary value problem is characterized by two given boundary values: the spatial position of the light source  
and the spatial position of the observer. The solution of the boundary value problem is considerably more involved than the initial value problem. 
Some parts of the boundary value problem have been considered in \cite{Zschocke_Time_Delay_2PN}. A fundamental and important step 
towards a comprehensive solution of the boundary value problem will be given in this investigation, where the light trajectory is written in a new form, which 
contains the spatial positions of the light source and of the observer.  

The manuscript is organized as follows: The 2PN solution of the initial value problem of light propagation in the monopole and quadrupole field of a body at rest 
is summarized in Sec.~\ref{Section1}. The problem and the procedure of how to rewrite the initial value solution of light propagation into a new form, which is 
required to solve the boundary value problem, is described in Sections~\ref{Section2a} and \ref{Section2b}. The 2PN solution of the initial value problem, 
which is appropriate to implement the spatial positions of source and observer, is given in Sec.~\ref{Section3}. The solution of the boundary value problem of 
light propagation, that means the coordinate velocity and trajectory of the light signal in terms of the spatial positions of source and observer, is presented in 
Sec.~\ref{Section4}. A summary and outlook is given in Sec.~\ref{Section_Summary}. The notations, some details of the calculations, the tensorial coefficients 
as well as the scalar functions are shifted into a set of several appendices.

\section{Initial value problem of light propagation in the old form} \label{Section1}

We consider the gravitational field generated by a massive solar system body. The curved space-time is assumed to be covered by harmonic four-coordinates 
$\left(x^0,x^1,x^2,x^3\right)$, in line with the resolutions of the International Astronomical Union \cite{IAU_Resolutions}. They are treated 
like Cartesian coordinates \cite{Thorne,Brumberg1991}; see also Section III in \cite{Zschocke_2PM_Metric}. 
The origin of the spatial axes is assumed to be located at the barycenter of this body. In case of weak gravitational fields and slow motions of the bodies, 
the metric tensor can be series expanded in inverse powers of the speed of light. This PN expansion reads in the 2PN approximation
\begin{eqnarray}
        g_{\alpha\beta} &=& \eta_{\alpha\beta} + h_{\alpha\beta}^{\left(2\right)} + h_{\alpha\beta}^{\left(3\right)} + h_{\alpha\beta}^{\left(4\right)} 
	+ {\cal O}\left(c^{-5}\right), 
        \label{PN_Expansion_1}
\end{eqnarray}

\noindent
where $h_{\alpha\beta}^{\left(n\right)} = {\cal O}(c^{-n})$. In the general case, the solar system body can be of arbitrary shape and inner structure and can 
also be in arbitrary rotational motions. In order to describe the gravitational field generated by such bodies, the metric tensor is decomposed in terms of 
six symmetric trace-free (STF) source-multipoles $\{I_L,J_L,W_L,X_L,Y_L,Z_L\}$, which are integrals over the energy-momentum tensor of the body 
\cite{Thorne,Blanchet_Damour1,Multipole_Damour_2,Blanchet2}. In the so-called canonical harmonic gauge the metric tensor can finally be expressed in terms of only two STF 
multipoles: mass-multipoles $M_L = I_L + {\cal O}\left(c^{-5}\right)$ and spin-multipoles $S_L = J_L + {\cal O}\left(c^{-5}\right)$ \cite{Thorne,Blanchet_Damour1,Blanchet2}. 
The mass-multipoles describe the shape and inner structure of the body, while the spin-multipoles account for the rotational motions and inner currents 
of the body. In the stationary case these multipoles are time-independent and then they are given by 
\begin{eqnarray}
	M_L &=& \int d^3 x \; x_{< L >}\;\frac{T^{00} + T^{kk}}{c^2}\;, 
\label{M_L}
\\
\nonumber\\ 
S_L &=& \int d^3 x \; x_{< L-1}\;\epsilon_{i_l >\,j k}\;x^j\;\frac{T^{0k}}{c}\;,
\label{S_L}
\end{eqnarray}

\noindent 
where $T^{\alpha\beta}$ is the stress-energy tensor of the body and $x_{< L >}$ is the symmetric and trace-free part of ${x}_L$ with respect to the spatial indices, 
and $L = {i_1 i_2 \dots i_l}$ is a multi-index of these spatial indices. A solar system body, described by these multipoles in (\ref{M_L}) and (\ref{S_L}), 
can still be of arbitrary shape, inner structure, and can also be in rotational motions, but the body cannot oscillate and the rotational motions and 
inner currents have to be time-independent. Then, the post-Newtonian expansion of the metric tensor in (\ref{PN_Expansion_1}) reads
\begin{eqnarray}
        g_{\alpha\beta} &=& \eta_{\alpha\beta} + h_{\alpha\beta}^{\left(2\right)}\left(M_L\right) + h_{\alpha\beta}^{\left(3\right)}\left(S_L\right) 
        + h_{\alpha\beta}^{\left(4\right)}\left(M_L\right) 
        \label{PN_Expansion_2}
\end{eqnarray}

\noindent
up to terms of the order ${\cal O}\left(c^{-5}\right)$. 
The metric perturbations in (\ref{PN_Expansion_2}) have been deduced in \cite{Zschocke_2PM_Metric} from the metric density achieved in the basic 
investigations in \cite{Thorne,Blanchet_Damour1,2PN_Metric1}. 
In our investigation \cite{Zschocke_Quadrupole_1} we have considered the problem of light propagation in 2PN approximation 
in the gravitational field of a body at rest and have taken into account the mass-monopole, $M$, and mass-quadrupole terms, $M_{ab}$, 
\begin{eqnarray}
M &=& \int d^3 x \,\frac{T^{00} + T^{kk}}{c^2}\,, 
\label{Mass}
\\
M_{ab} &=& \int d^3 x \,{x}_{< ab > }\,\frac{T^{00} + T^{kk}}{c^2}\,, 
\label{Quadrupole}
\end{eqnarray}

\noindent
where the integrals run over the three-dimensional volume of the body, and $x_{< ab >} = x_a x_b - \frac{1}{3}\,|\ve{x}|^2\,\delta_{ab}$.
The mass-dipole terms vanish, $M_i = 0$, because the origin of the spatial axes is located the center of mass of the source.
Then, the metric (\ref{PN_Expansion_2}) simplifies to the form
\begin{eqnarray}
        g_{\alpha\beta} &=& \eta_{\alpha\beta} + h_{\alpha\beta}^{\left(2\right)}\left(M,M_{ab}\right)
        + h_{\alpha\beta}^{\left(4\right)}\left(M,M_{ab}\right)
        \label{PN_Expansion_3}
\end{eqnarray}

\noindent
up to terms of the order ${\cal O}\left(c^{-6}\right)$. 
The metric perturbations in (\ref{PN_Expansion_3}) were explicitly given in \cite{Zschocke_2PM_Metric,Frutos_Alfaro_Soffel}.

Light trajectories are null-geodesics and they are governed by the geodesic equation, which in terms of coordinate time reads \cite{MTW,Brumberg1991,Kopeikin_Efroimsky_Kaplan}:
\begin{eqnarray}
	\frac{\ddot{x}^{i}\left(t\right)}{c^2}  
	+ \Gamma^{i}_{\mu\nu} \frac{\dot{x}^{\mu}\left(t\right)}{c} \frac{\dot{x}^{\nu}\left(t\right)}{c}   
	&=& \Gamma^{0}_{\mu\nu} \frac{\dot{x}^{\mu}\left(t\right)}{c} \frac{\dot{x}^{\nu}\left(t\right)}{c} \frac{\dot{x}^{i}\left(t\right)}{c}\,, 
\label{Geodetic_Equation}
\end{eqnarray}

\noindent
where $\Gamma^{\alpha}_{\mu\nu} = g^{\alpha\beta} \left(g_{\beta\mu,\nu} + g_{\beta\nu,\mu} - g_{\mu\nu,\beta}\right)/2$ 
are the Christoffel symbols, which are functions of the metric tensor $g_{\alpha\beta}$.  
The geodesic equation is a differential equation of second order, thus a unique solution requires two initial conditions,
\begin{eqnarray}
        \ve{\sigma} &=& \frac{\dot{\ve{x}}\left(t\right)}{c}\bigg|_{t = - \infty} \;,
	\label{Initional_Condition_1}
	\\
	\ve{x}_0 &=& \ve{x}\left(t\right)\bigg|_{t=t_0}\;,
        \label{Initional_Condition_2}
\end{eqnarray}

\noindent
with $\ve{\sigma}$ and $\ve{x}_0$ being the unit-direction of the light ray at past infinity and the spatial position of the light source at the moment of 
emission of the light signal. The first integration of the geodesic equation yields the coordinate velocity of the light signal, and the second integration 
of the geodesic equation yields the trajectory of the light signal. 
By inserting the metric tensor (\ref{PN_Expansion_2}) into the geodesic equation (\ref{Geodetic_Equation}) one arrives at the geodesic equation in 2PN approximation 
for the light propagation, as given, for instance, in the Refs.~\cite{KlionerKopeikin1992,Bruegmann2005,Zschocke_Quadrupole_1}. 

\begin{figure}[!ht]
\begin{center}
\includegraphics[scale=0.15]{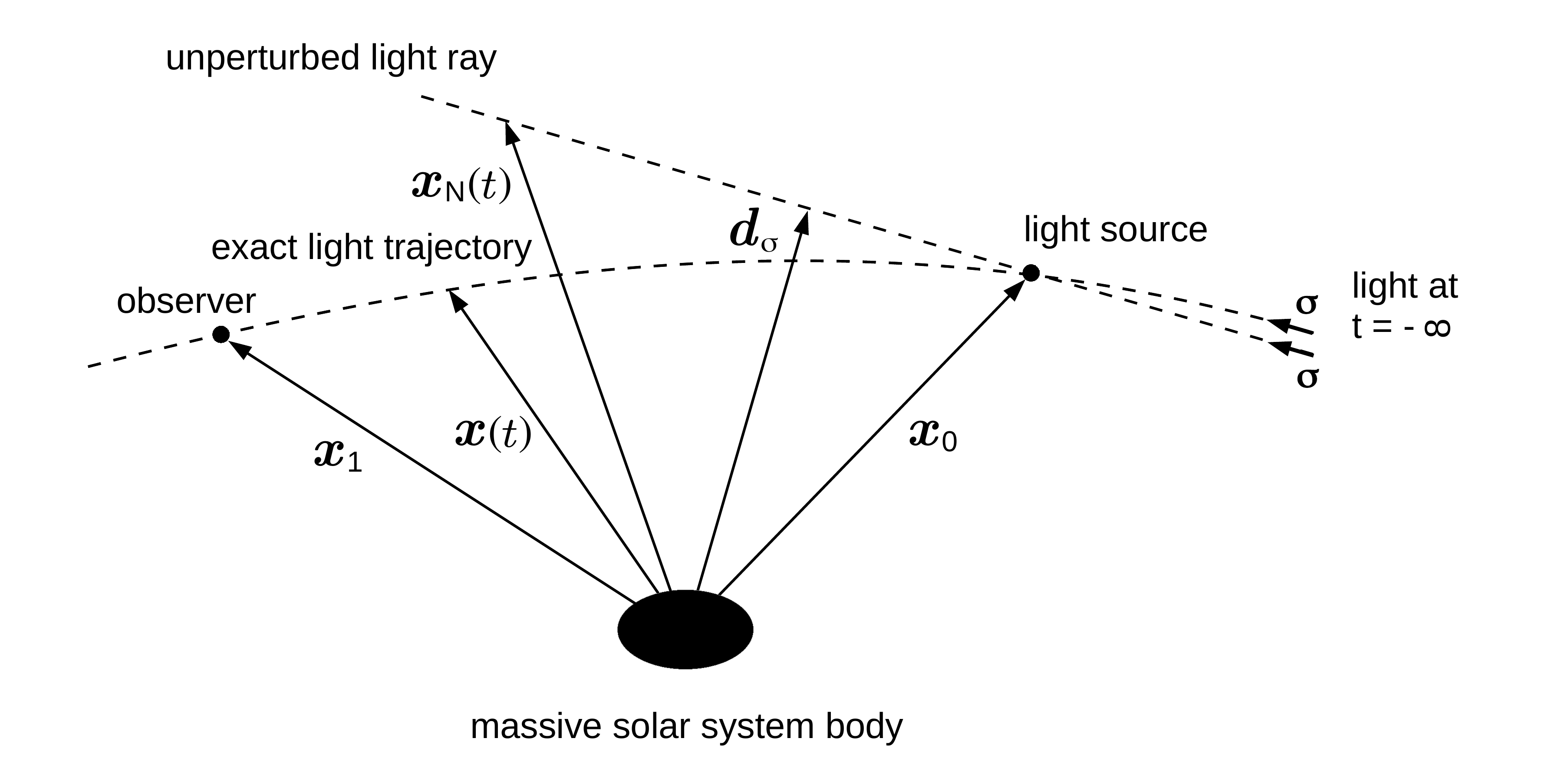}
\end{center}
        \caption{A geometrical representation of the propagation of a light signal through the gravitational field of a massive solar system body
        at rest. The light signal is emitted by the light source at $\ve{x}_0$ and propagates along the exact light trajectory $\ve{x}\left(t\right)$.
        The three-vector $\ve{x}_1$ points from the origin of the coordinate system towards the spatial position of the observer.
        The unit tangent vector along the light trajectory at past infinity is $\ve{\sigma}$.
        The unperturbed light ray $\ve{x}_{\rm N}\left(t\right)$ is given by Eq.~(\ref{Second_Integration_N_old}) and propagates in the direction
        of $\ve{\sigma}$ along a straight line through the position of the light source at $\ve{x}_0$. The impact vector $\ve{d}_{\sigma}$ of the
        unperturbed light ray is given by Eq.~(\ref{impact_vector_N}).} 
\label{Diagram}
\end{figure}

\noindent 
The solutions of the first and second integration of geodesic equation in 2PN approximation have recently been determined for light rays propagating 
in the gravitational field of a body at rest by means of an iterative approach \cite{Zschocke_Quadrupole_1}, where the monopole and quadrupole structure 
of the body have been taken into account. These iterative solutions are given in the following form:  
\begin{eqnarray}
\frac{\dot{\ve{x}}_{\rm N}}{c} &=& \ve{\sigma}\;,
\label{First_Integration_N_old}
\\
\ve{x}_{\rm N} &=& \ve{x}_0 + c \left(t - t_0\right) \ve{\sigma}\;,
\label{Second_Integration_N_old}
\\
\frac{\dot{\ve{x}}_{\rm 1PN}}{c} &=& \ve{\sigma} + \frac{\Delta \dot{\ve{x}}_{\rm 1PN}\left(\ve{x}_{\rm N}\right)}{c}\,,
\label{First_Integration_1PN_old}
\\
\ve{x}_{\rm 1PN} &=& \ve{x}_{\rm N} + \Delta \ve{x}_{\rm 1PN}\left(\ve{x}_{\rm N}\right) - \Delta \ve{x}_{\rm 1PN}\left(\ve{x}_0\right),
\label{Second_Integration_1PN_old}
\\
\frac{\dot{\ve{x}}_{\rm 2PN}}{c} &=& \ve{\sigma} + \frac{\Delta\dot{\ve{x}}_{\rm 1PN}\left(\ve{x}_{\rm N}\right)}{c}
+ \frac{\Delta \dot{\ve{x}}_{\rm 2PN}\left(\ve{x}_{\rm N}\right)}{c}\;,
\label{First_Integration_2PN_old}
\\
\ve{x}_{\rm 2PN} &=& \ve{x}_{\rm N} + \Delta \ve{x}_{\rm 1PN}\left(\ve{x}_{\rm N}\right)
- \Delta \ve{x}_{\rm 1PN}\left(\ve{x}_0\right)
\nonumber\\ 
&& \quad\;\; +\, \Delta \ve{x}_{\rm 2PN}\left(\ve{x}_{\rm N}\right) - \Delta \ve{x}_{\rm 2PN}\left(\ve{x}_0\right), 
\label{Second_Integration_2PN_old}
\end{eqnarray}

\noindent
where the time-arguments have been omitted. 
The first two equations, (\ref{First_Integration_N_old}) and (\ref{Second_Integration_N_old}), represent the homogeneous solution of geodesic equation 
(vanishing Christoffel symbols), which yields the unperturbed light ray propagating along a straight line. The 1PN perturbations $\Delta\dot{\ve{x}}_{\rm 1PN}$ 
and $\Delta\ve{x}_{\rm 1PN}$ are terms of the order ${\cal O}(c^{-2})$ which have been determined long time ago in \cite{Klioner1991}, while the 
2PN perturbations $\Delta\dot{\ve{x}}_{\rm 2PN}$ and $\Delta\ve{x}_{\rm 2PN}$ are terms of the order ${\cal O}(c^{-4})$, which were the primary 
results of our recent investigation \cite{Zschocke_Quadrupole_1}.

The effort to integrate the geodesic equation can significantly be simplified, if one separates the time-dependent scalar functions from the time-independent tensorial coefficients. 
This procedure has been applied in \cite{Zschocke_Quadrupole_1} and has lead to eight master integrals, which can be solved in closed form. However, this approach 
leads to tensorial coefficients, which contain linearly dependent tensors. For the subsequent investigations, it would be more appropriate, to rearrange these tensorial 
terms into such a form, that the tensorial coefficients contain only tensors which are linearly independent of each other. In this way, we achieve the following expressions 
for the perturbations: the spatial components of 1PN terms in (\ref{First_Integration_1PN_old}) - (\ref{Second_Integration_2PN_old}) are given by 
\begin{eqnarray} 
	\frac{\Delta \dot{x}^i_{\rm 1PN}\left(\ve{x}_{\rm N}\right)}{c} &=& 
        \frac{G M}{c^2} \sum\limits_{n=1}^{2} U^i_{(n)}\left(\ve{x}_{\rm N}\right)\, \dot{F}_{(n)}\left(\ve{x}_{\rm N}\right)
	\nonumber\\ 
	&& \hspace{-2.0cm} + \frac{G M_{ab}}{c^2} \sum\limits_{n=1}^{8} V^{ab\,i}_{(n)}\left(\ve{x}_{\rm N}\right)\, \dot{G}_{(n)}\left(\ve{x}_{\rm N}\right),  
        \label{First_Integration_1PN_Terms_old} 
        \\
        \Delta x^i_{\rm 1PN}\left(\ve{x}_{\rm N}\right) &=& 
        \frac{G M}{c^2} \sum\limits_{n=1}^{2} U^i_{(n)}\left(\ve{x}_{\rm N}\right)\,F_{(n)}\left(\ve{x}_{\rm N}\right)
	\nonumber\\ 
	&& \hspace{-2.0cm} + \frac{G M_{ab}}{c^2} \sum\limits_{n=1}^{8} V^{ab\,i}_{(n)}\left(\ve{x}_{\rm N}\right)\,G_{(n)}\left(\ve{x}_{\rm N}\right), 
\label{Second_Integration_1PN_Terms_old}
\end{eqnarray}

\noindent 
and the spatial components of 2PN terms are given by
\begin{eqnarray}
        \frac{\Delta \dot{x}^i_{\rm 2PN}\left(\ve{x}_{\rm N}\right)}{c} &=& 
        \frac{G M}{c^2}\,\frac{G M}{c^2} \sum\limits_{n=1}^{2} U^i_{(n)}\left(\ve{x}_{\rm N}\right)\, \dot{A}_{(n)}\left(\ve{x}_{\rm N}\right)
	\nonumber\\ 
	&& \hspace{-2.0cm} + \frac{G M}{c^2}\,\frac{G M_{ab}}{c^2} \sum\limits_{n=1}^{8} V^{ab\,i}_{(n)}\left(\ve{x}_{\rm N}\right)\, \dot{B}_{(n)}\left(\ve{x}_{\rm N}\right)
        \nonumber\\ 
        && \hspace{-2.0cm} + \frac{G M_{ab}}{c^2}\,\frac{G M_{cd}}{c^2} \sum\limits_{n=1}^{28} W^{abcd\,i}_{(n)}\left(\ve{x}_{\rm N}\right)\, \dot{C}_{(n)}\left(\ve{x}_{\rm N}\right),
        \label{First_Integration_2PN_Terms_old}
\\ 
       \Delta x^i_{\rm 2PN}\left(\ve{x}_{\rm N}\right) &=&  
       \frac{G M}{c^2}\,\frac{G M}{c^2} \sum\limits_{n=1}^{2} U^i_{(n)}\left(\ve{x}_{\rm N}\right)\, A_{(n)}\left(\ve{x}_{\rm N}\right) 
	\nonumber\\ 
       && \hspace{-2.0cm} + \frac{G M}{c^2}\,\frac{G M_{ab}}{c^2} \sum\limits_{n=1}^{8} V^{ab\,i}_{(n)}\left(\ve{x}_{\rm N}\right)\,B_{(n)}\left(\ve{x}_{\rm N}\right)
       \nonumber\\ 
       && \hspace{-2.0cm} + \frac{G M_{ab}}{c^2}\,\frac{G M_{cd}}{c^2} \sum\limits_{n=1}^{28} W^{abcd\,i}_{(n)}\left(\ve{x}_{\rm N}\right)\,C_{(n)}\left(\ve{x}_{\rm N}\right).
       \label{Second_Integration_2PN_Terms_old}
\end{eqnarray}

\noindent 
The terms $\Delta\ve{x}_{\rm 1PN}\left(\ve{x}_0\right)$ and $\Delta\ve{x}_{\rm 2PN}\left(\ve{x}_0\right)$ in Eqs.~(\ref{Second_Integration_1PN_old}) and 
(\ref{Second_Integration_2PN_old}) are obtained from Eqs.~(\ref{Second_Integration_1PN_Terms_old}) and (\ref{Second_Integration_2PN_Terms_old}) by replacing 
the arguments $\ve{x}_{\rm N}$ by $\ve{x}_0$. The tensorial coefficients, $U_{(n)}^i, V_{(n)}^{ab\,i}, W_{(n)}^{abcd\,i}$, are given in Appendix~\ref{Appendix1}. 
The scalar functions in (\ref{First_Integration_1PN_Terms_old}) - (\ref{Second_Integration_1PN_Terms_old}) are given in 
Appendix~\ref{Appendix2}. The scalar functions in Eqs.~(\ref{First_Integration_2PN_Terms_old}) - (\ref{Second_Integration_2PN_Terms_old}) can straightforwardly be deduced 
from \cite{Zschocke_Quadrupole_1}, namely by comparison of the tensorial coefficients in Eqs.~(83) - (85) and Eqs.~(89) - (91) in \cite{Zschocke_Quadrupole_1} with 
Eqs.~(\ref{First_Integration_2PN_Terms_old}) and (\ref{Second_Integration_2PN_Terms_old}), respectively. These scalar functions are of simple but extensive algebraic 
structure. Therefore, in favor of a simpler representation, only the functions $A_{(n)}$ and $B_{(1)},B_{(8)}$ as well as $C_{(1)},C_{(28)}$ and their time derivatives 
are given in Appendix~\ref{Appendix3}, while the full set of these functions is represented as supplementary material \cite{supplementary_script}. 

The tensorial coefficients contain the impact vector $\ve{d}_{\sigma}$ of the unperturbed light ray,
\begin{eqnarray}
\ve{d}_{\sigma} &=& \ve{\sigma} \times \left(\ve{x}_{\rm N} \times \ve{\sigma}\right) = \ve{\sigma} \times \left(\ve{x}_0 \times \ve{\sigma}\right), 
\label{impact_vector_N}
\end{eqnarray}

\noindent
and its absolute value $d_{\sigma} = \left|\ve{d}_{\sigma}\right|$ which is called impact parameter $d_{\sigma}$. The impact vector in (\ref{impact_vector_N}) is 
time-independent and points from the origin of the coordinate system towards the unperturbed light ray at the moment of its closest encounter; see also Fig.~\ref{Diagram}.

\section{Statement of the problem}\label{Section2a}

The solution of the initial value problem, defined by Eqs.~(\ref{Initional_Condition_1}) and (\ref{Initional_Condition_2}), can only be the first step,  
because in reality the light source and the observer are located at finite distances from the gravitating body. Therefore, real astrometric measurements 
require the solution of the boundary value problem, that means solving the geodesic equation for light rays in terms of the boundary values, 
\begin{eqnarray}
        \ve{x}_0 &=& \ve{x}\left(t\right)\,\,\bigg|_{t = t_0}\,, 
        \label{boundary0}
	\\
	\ve{x}_1 &=& \ve{x}\left(t\right)\,\,\bigg|_{t = t_1}\,, 
        \label{boundary1}
\end{eqnarray}

\noindent
where $\ve{x}_0$ is the spatial position of the light source at the moment of emission of the light signal $t_0$, while $\ve{x}_1$ is the spatial position 
of the observer at the moment of reception of the light signal $t_1$. These equations state, that the spatial position of the exact trajectory of the light 
signal, $\ve{x}\left(t\right)$, is in coincidence with the spatial positions of light source and observer at the moment of emission and the moment of reception, 
respectively; see also Fig.~\ref{Diagram}. 

The solution of the boundary value problem can uniquely be deduced from the solution of the initial value problem derived in our investigation \cite{Zschocke_Quadrupole_1} 
and represented by Eqs.~(\ref{First_Integration_N_old}) - (\ref{Second_Integration_2PN_old}). 
The iterative approach in \cite{Zschocke_Quadrupole_1} implies, that coordinate velocity and trajectory of the light signal are given in terms of the unperturbed light 
ray $\ve{x}_{\rm N}$. It is essential to realize, that the 1PN perturbations in Eqs.~(\ref{First_Integration_2PN_old}) and (\ref{Second_Integration_2PN_old}) 
are terms of the order ${\cal O}(c^{-2})$. Therefore, in the arguments of these 1PN perturbations in Eqs.~(\ref{First_Integration_2PN_old}) and 
(\ref{Second_Integration_2PN_old}) one cannot replace the unperturbed light ray at the moment of reception by the spatial coordinate of the observer, because such a replacement  
\begin{eqnarray}
	\ve{x}_1 &=& \ve{x}_{\rm N}\left(t_1\right) + {\cal O}\left(c^{-2}\right) 
        \label{statement_replacement_N}
\end{eqnarray}

\noindent
would cause terms of the order ${\cal O}(c^{-4})$, that means, such a replacement would spoil the 2PN approximation. 
Therefore, one has to rewrite the 2PN solution in Eqs.~(\ref{First_Integration_2PN_old}) and (\ref{Second_Integration_2PN_old}) into such a form, 
where the arguments of these 1PN perturbations are the spatial positions of the 1PN light ray, $\ve{x}_{\rm 1PN}$, while the arguments of the 
2PN perturbations are the spatial position of the unperturbed light ray $\ve{x}_{\rm N}$. Afterwards, one may replace the arguments of the 
1PN terms according to the following relation, 
\begin{eqnarray}
	\ve{x}_1 &=& \ve{x}_{\rm 1PN}\left(t_1\right) + {\cal O}\left(c^{-4}\right).  
        \label{statement_replacement_1PN}
\end{eqnarray}

\noindent
Such a replacement in these 1PN perturbations would cause terms of the order ${\cal O}(c^{-6})$ which are neglected, in line with 
the 2PN approximation. The procedure of how to arrive at this new form of the 2PN solution of the initial value problem is described 
in the subsequent section.

\section{Description of the procedure}\label{Section2b}

The procedure, to rewrite the initial value problem into such a new form as described in the previous section, is relevant for 
calculations in 2PN approximation and calculations of higher order, but is not necessary in the 1PN or 1.5PN approximation. Furthermore, if one takes into 
account only the monopole structure of a massive solar system body, then the performance of this technique in 2PN calculations is more or less straightforward. 
That might be the reason, that this treatment has almost not been discussed explicitly in the literature. It might well be, that the comments in the 
text below Eq.~(3.2.42) in \cite{Brumberg1991}, in the text below Eq.~(6.8) in \cite{KlionerKopeikin1992}, as well as in the text below Eq.~(52) 
in \cite{Klioner_Zschocke} are just three of only a very few explicit hints in the literature about this specific issue. 
However, if one accounts for the quadrupole structure of a 
massive solar system body, then the calculations become rather involved. This section is devoted to representing the method of how to treat  
that specific problem of 2PN calculations. The procedure is subdivided into three steps.

\noindent
{\bf First step}: One changes the arguments of the 1PN perturbations in Eqs.~(\ref{First_Integration_2PN_old}) and (\ref{Second_Integration_2PN_old}) from the
unperturbed light ray, $\ve{x}_{\rm N}$, to the light ray in 1PN approximation, $\ve{x}_{\rm 1PN}$. Such a replacement generates additional terms of 2PN order.
In order to identify these additional terms, one has to consider the difference between these 1PN perturbations, formally given by,
\begin{eqnarray}
        \frac{\delta \dot{\ve{x}}_{\rm 2PN}\left(\ve{x}_{\rm N}\right)}{c}
        &=& \frac{\Delta\dot{\ve{x}}_{\rm 1PN}\left(\ve{x}_{\rm N}\right)}{c} - \frac{\Delta\dot{\ve{x}}_{\rm 1PN}\left(\ve{x}_{\rm 1PN}\right)}{c}\,,
\label{Relation_1}
\\
        \delta \ve{x}_{\rm 2PN}\left(\ve{x}_{\rm N}\right)
        &=& \Delta\ve{x}_{\rm 1PN}\left(\ve{x}_{\rm N}\right) - \Delta\ve{x}_{\rm 1PN}\left(\ve{x}_{\rm 1PN}\right),
\label{Relation_2}
\end{eqnarray}

\noindent
where label 2PN on the left-hand side in (\ref{Relation_1}) and (\ref{Relation_2}) indicates, that these differences are terms of second post-Newtonian order. The first terms 
on the right-hand side in (\ref{Relation_1}) and (\ref{Relation_2}) are given by Eqs.~(\ref{First_Integration_1PN_Terms_old}) and (\ref{Second_Integration_1PN_Terms_old}), 
while the second terms on the right-hand side in (\ref{Relation_1}) and (\ref{Relation_2}) read 
\begin{eqnarray}
        \frac{\Delta \dot{x}^i_{\rm 1PN}\left(\ve{x}_{\rm 1PN}\right)}{c} &=&
        \frac{G M}{c^2} \sum\limits_{n=1}^{2} U^i_{(n)}\left(\ve{x}_{\rm 1PN}\right)\,\dot{F}_{(n)}\left(\ve{x}_{\rm 1PN}\right) 
	\nonumber\\ 
	&& \hspace{-2.0cm} + \frac{G M_{ab}}{c^2} \sum\limits_{n=1}^{8} V^{ab\,i}_{(n)}\left(\ve{x}_{\rm 1PN}\right)\,\dot{G}_{(n)}\left(\ve{x}_{\rm 1PN}\right),  
\label{New_First_Integration_1PN}
\\
        \Delta x^i_{\rm 1PN}\left(\ve{x}_{\rm 1PN}\right) &=&
        \frac{G M}{c^2} \sum\limits_{n=1}^{2} U^i_{(n)}\left(\ve{x}_{\rm 1PN}\right)\,F_{(n)}\left(\ve{x}_{\rm 1PN}\right)
	\nonumber\\ 
	&& \hspace{-2.0cm} + \frac{G M_{ab}}{c^2} \sum\limits_{n=1}^{8} V^{ab\,i}_{(n)}\left(\ve{x}_{\rm 1PN}\right)\,G_{(n)}\left(\ve{x}_{\rm 1PN}\right), 
\label{New_Second_Integration_1PN}
\end{eqnarray}

\noindent
where the tensorial coefficients are given in Appendix~\ref{Appendix4a} and the scalar functions are given in Appendix~\ref{Appendix4b}. 

\noindent
{\bf Second step}: In order to obtain the terms $\delta\dot{\ve{x}}_{\rm 2PN}$ in (\ref{Relation_1}) and $\delta\ve{x}_{\rm 2PN}$ in (\ref{Relation_2}),
one has to perform a series expansion of $\Delta\dot{\ve{x}}_{\rm 1PN}\left(\ve{x}_{\rm 1PN}\right)$ and
$\Delta\ve{x}_{\rm 1PN}\left(\ve{x}_{\rm 1PN}\right)$ in (\ref{New_First_Integration_1PN}) and (\ref{New_Second_Integration_1PN}), respectively.
The tensorial coefficients and the scalar functions in (\ref{New_First_Integration_1PN}) and (\ref{New_Second_Integration_1PN}) contain the 
light ray in 1PN approximation, $\ve{x}_{\rm 1PN}$, and its absolute value, $x_{\rm 1PN}$. Therefore, for performing that series expansion, one needs the 
following relations, which are valid up to terms of the order ${\cal O}(c^{-4})$, 
\begin{eqnarray}
        \ve{x}_{\rm 1PN} &=& \ve{x}_{\rm N} + \Delta \ve{x}_{\rm 1PN}\left(\ve{x}_{\rm N}\right) - \Delta \ve{x}_{\rm 1PN}\left(\ve{x}_0\right)\;,
        \label{Appendix_x_A}
        \\
        \frac{1}{\left(x_{\rm 1PN}\right)^n} &=& \frac{1}{\left(x_{\rm N}\right)^n}
        - n\,
        \frac{\ve{x}_{\rm N} \cdot
        \left(\Delta \ve{x}_{\rm 1PN}\left(\ve{x}_{\rm N}\right) - \Delta \ve{x}_{\rm 1PN}\left(\ve{x}_0\right)\right)}{\left(x_{\rm N}\right)^{n+2}}\;,
	\nonumber\\ 
        \label{Appendix_x_C1}
\end{eqnarray}

\noindent
where $n$ is an arbitrary integer. It is noticed, that relation (\ref{Appendix_x_C1}) can also be written in the form
\begin{eqnarray}
        \frac{1}{\left(x_{\rm 1PN}\right)^n} &=& \frac{1}{\left(x_{\rm N}\right)^n}
        - n\,
        \frac{\ve{d}_{\sigma} \cdot
        \left(\Delta \ve{x}_{\rm 1PN}\left(\ve{x}_{\rm N}\right) - \Delta \ve{x}_{\rm 1PN}\left(\ve{x}_0\right)\right)}{\left(x_{\rm N}\right)^{n+2}}
	\nonumber\\ 
	&& \hspace{-1.0cm} - n \left(\ve{\sigma} \cdot \ve{x}_{\rm N}\right)
        \frac{\ve{\sigma} \cdot
        \left(\Delta \ve{x}_{\rm 1PN}\left(\ve{x}_{\rm N}\right) - \Delta \ve{x}_{\rm 1PN}\left(\ve{x}_0\right)\right)}{\left(x_{\rm N}\right)^{n+2}}\;,
        \label{Appendix_x_C2}
\end{eqnarray}

\noindent
where the unperturbed light ray (\ref{Second_Integration_N_old}) has been used in the form 
$\ve{x}_{\rm N} = \ve{d}_{\sigma} + \left(\ve{\sigma} \cdot \ve{x}_{\rm N}\right) \, \ve{\sigma}$, which follows from (\ref{impact_vector_N}). 

The 1PN perturbations (\ref{New_First_Integration_1PN}) and (\ref{New_Second_Integration_1PN}) do also contain the impact vector with respect to the 
light ray in 1PN approximation (cf. Eq.~(J4) in \cite{Zschocke_Quadrupole_1}),
\begin{eqnarray}
	\widehat{\ve{d}_{\sigma}} &=& \ve{\sigma} \times \left(\ve{x}_{\rm 1PN} \times \ve{\sigma} \right),
        \label{impact_vector_x_1PN}
\end{eqnarray}

\noindent
as well as its absolute value $\widehat{d}_{\sigma} = |\widehat{\ve{d}}_{\sigma}|$. 
This impact vector, like the impact vector with respect to the unperturbed light ray (\ref{impact_vector_N}), is perpendicular to three-vector $\ve{\sigma}$.
By inserting (\ref{Appendix_x_A}) into (\ref{impact_vector_x_1PN}), one finds that these impact vectors in (\ref{impact_vector_N}) and (\ref{impact_vector_x_1PN})
and their absolute values are related to each other as follows (cf. Eqs.~(J5) and (J7) in \cite{Zschocke_Quadrupole_1}),
\begin{eqnarray}
        \widehat{\ve{d}_{\sigma}} &=&
        \ve{d}_{\sigma} + \ve{\sigma}\times 
        \bigg[\left(\Delta \ve{x}_{\rm 1PN}\left(\ve{x}_{\rm N}\right) - \Delta \ve{x}_{\rm 1PN}\left(\ve{x}_0\right)\right) \times \ve{\sigma}\bigg],
	\nonumber\\ 
        \label{Appendix_x_D}
        \\
        \frac{1}{(\widehat{d_{\sigma}})^n} &=& \frac{1}{\left(d_{\sigma}\right)^n} 
	\nonumber\\ 
	&-& \frac{n}{\left(d_{\sigma}\right)^n} \,
        \frac{\ve{d}_{\sigma} \cdot 
        \left(\Delta \ve{x}_{\rm 1PN}\left(\ve{x}_{\rm N}\right) - \Delta \ve{x}_{\rm 1PN}\left(\ve{x}_0\right)\right)}{\left(d_{\sigma}\right)^2}\;,
        \label{Appendix_x_F}
\end{eqnarray}

\noindent
up to terms of the order ${\cal O}(c^{-4})$; $n$ is an arbitrary integer. The relation (\ref{Appendix_x_F})
represents the first term of an infinite series expansion (cf. text below Eq.~(J7) in \cite{Zschocke_Quadrupole_1}). 
This series expansion has a convergence radius determined by the condition $n \left|\ve{d}_{\sigma} \cdot \Delta \ve{x}_{\rm 1PN}\right| \le \left(d_{\sigma}\right)^2$.
Using the same arguments as given in text below Eq.~(J7) in \cite{Zschocke_Quadrupole_1}, one finds that this convergence condition is satisfied for any realistic 
observer, which is naturally assumed to be located in the solar system. Nevertheless, this convergence condition is even satisfied for observers which are 
located at far distances of more than a few hundred astronomical units from the solar system. The impact vector in (\ref{impact_vector_x_1PN}) is only needed as an 
intermediate step, because later the replacement in (\ref{statement_replacement_1PN}) will be performed, where the intermediate impact vector in (\ref{impact_vector_x_1PN}) 
becomes literally the impact vector of the boundary value problem in Eq.~(\ref{impact_vector_1}), up to terms of the order ${\cal O}(c^{-4})$.

\noindent
{\bf Third step}: The terms in (\ref{Relation_1}) and (\ref{Relation_2}) are 2PN terms, hence they can be decomposed into terms of the 
same set of linearly independent tensors like the 2PN terms in (\ref{First_Integration_2PN_Terms_old}) and (\ref{Second_Integration_2PN_Terms_old}), that means 
\begin{eqnarray}
        \frac{\delta \dot{x}^i_{\rm 2PN}\left(\ve{x}_{\rm N}\right)}{c}
	&=& \frac{G M}{c^2}\,\frac{G M}{c^2} \sum\limits_{n=1}^{2} U^i_{(n)}\left(\ve{x}_{\rm N}\right)\, \dot{\widetilde{A}}_{(n)}\left(\ve{x}_{\rm N}\right)
        \nonumber\\ 
	&& \hspace{-2.0cm} 
	+ \frac{G M}{c^2}\,\frac{G M_{ab}}{c^2} \sum\limits_{n=1}^{8} V^{ab\,i}_{(n)}\left(\ve{x}_{\rm N}\right)\,\dot{\widetilde{B}}_{(n)}\left(\ve{x}_{\rm N}\right)
        \nonumber\\ 
	&& \hspace{-2.0cm} 
	+ \frac{G M_{ab}}{c^2}\,\frac{G M_{cd}}{c^2} \sum\limits_{n=1}^{28} W^{abcd\,i}_{(n)}\left(\ve{x}_{\rm N}\right)\,\dot{\widetilde{C}}_{(n)}\left(\ve{x}_{\rm N}\right),
\label{Relation_1_Decomposition} 
       \end{eqnarray}

\noindent 
and 
\begin{eqnarray} 
	\delta {x}^i_{\rm 2PN}\left(\ve{x}_{\rm N}\right)
	&=& \frac{G M}{c^2}\,\frac{G M}{c^2} \sum\limits_{n=1}^{2} U^i_{(n)}\left(\ve{x}_{\rm N}\right)\, \widetilde{A}_{(n)}\left(\ve{x}_{\rm N}\right) 
       \nonumber\\  
       && \hspace{-2.0cm} + \frac{G M}{c^2}\,\frac{G M_{ab}}{c^2} \sum\limits_{n=1}^{8} V^{ab\,i}_{(n)}\left(\ve{x}_{\rm N}\right)\,\widetilde{B}_{(n)}\left(\ve{x}_{\rm N}\right)
       \nonumber\\  
       && \hspace{-2.0cm} 
	+ \frac{G M_{ab}}{c^2}\,\frac{G M_{cd}}{c^2} \sum\limits_{n=1}^{28} W^{abcd\,i}_{(n)}\left(\ve{x}_{\rm N}\right)\,\widetilde{C}_{(n)}\left(\ve{x}_{\rm N}\right), 
\label{Relation_2_Decomposition}
\end{eqnarray}

\noindent 
where the tensorial coefficients are given in Appendix~\ref{Appendix1}. 
The calculations in order to get these terms in (\ref{Relation_1_Decomposition}) and (\ref{Relation_2_Decomposition}) are not complicated but lengthy, and the scalar functions 
$\dot{\widetilde{A}}_{(n)}$, $\dot{\widetilde{B}}_{(n)}$, $\dot{\widetilde{C}}_{(n)}$, as well as $\widetilde{A}_{(n)}$, $\widetilde{B}_{(n)}$, $\widetilde{C}_{(n)}$ are of  
extensive algebraic structure. In view of this fact and because these functions are considered as an intermediate step, they will not be presented here in their explicit form. 

As final step, these expressions in (\ref{Relation_1_Decomposition}) and (\ref{Relation_2_Decomposition}) have to be added to the 2PN terms 
in (\ref{First_Integration_2PN_Terms_old}) and (\ref{Second_Integration_2PN_Terms_old}), respectively, which leads to the following expressions: 
\begin{eqnarray}
	\frac{\Laplace \dot{\ve{x}}_{\rm 2PN}\left(\ve{x}_{\rm N}\right)}{c} &=& \frac{\Delta \dot{\ve{x}}_{\rm 2PN}\left(\ve{x}_{\rm N}\right)}{c} 
	+ \frac{\delta \dot{\ve{x}}_{\rm 2PN}\left(\ve{x}_{\rm N}\right)}{c}\,,
        \label{First_Integration_2PN_Laplace}
\\ 
	\Laplace \ve{x}_{\rm 2PN}\left(\ve{x}_{\rm N}\right) &=& \Delta \ve{x}_{\rm 2PN}\left(\ve{x}_{\rm N}\right) + \delta \ve{x}_{\rm 2PN}\left(\ve{x}_{\rm N}\right).
       \label{Second_Integration_2PN_Laplace}
\end{eqnarray}

\noindent
The symbol $\Laplace$ (Laplace) instead of $\Delta$ (Delta) on the left-hand side in (\ref{First_Integration_2PN_Laplace}) and (\ref{Second_Integration_2PN_Laplace}) indicates, 
that these functions have carefully to be distinguished from the 2PN functions in (\ref{First_Integration_2PN_Terms_old}) and (\ref{Second_Integration_2PN_Terms_old}).

\section{Initial value problem of light propagation in the new form}\label{Section3} 

By performing this procedure, which has been described in the previous section, one arrives at a new representation of the same iterative solution 
as given by Eqs.~(\ref{First_Integration_N_old}) - (\ref{Second_Integration_2PN_old}), but in the following form: 
\begin{eqnarray}
\frac{\dot{\ve{x}}_{\rm N}}{c} &=& \ve{\sigma}\;,
\label{First_Integration_N_new}
\\
\ve{x}_{\rm N} &=& \ve{x}_0 + c \left(t - t_0\right) \ve{\sigma}\;,
\label{Second_Integration_N_new}
\\
\frac{\dot{\ve{x}}_{\rm 1PN}}{c} &=& \ve{\sigma} + \frac{\Delta \dot{\ve{x}}_{\rm 1PN}\left(\ve{x}_{\rm N}\right)}{c}\,,
\label{First_Integration_1PN_new}
\\
\ve{x}_{\rm 1PN} &=& \ve{x}_{\rm N} + \Delta \ve{x}_{\rm 1PN}\left(\ve{x}_{\rm N}\right) - \Delta \ve{x}_{\rm 1PN}\left(\ve{x}_0\right),
\label{Second_Integration_1PN_new}
\\ 
\frac{\dot{\ve{x}}_{\rm 2PN}}{c} &=& \ve{\sigma} + \frac{\Delta\dot{\ve{x}}_{\rm 1PN}\left(\ve{x}_{\rm 1PN}\right)}{c}
+ \frac{\Laplace \dot{\ve{x}}_{\rm 2PN}\left(\ve{x}_{\rm N}\right)}{c}\;,
\label{First_Integration_2PN_new}
\\
\ve{x}_{\rm 2PN} &=& \ve{x}_{\rm N} + \Delta \ve{x}_{\rm 1PN}\left(\ve{x}_{\rm 1PN}\right)
- \Delta \ve{x}_{\rm 1PN}\left(\ve{x}_0\right)
\nonumber\\ 
&& \quad\;\; +\, \Laplace \ve{x}_{\rm 2PN}\left(\ve{x}_{\rm N}\right) - \Laplace \ve{x}_{\rm 2PN}\left(\ve{x}_0\right), 
\label{Second_Integration_2PN_new}
\end{eqnarray}

\noindent
where the time-arguments have been omitted. 
The only differences between Eqs.~(\ref{First_Integration_N_old}) - (\ref{Second_Integration_2PN_old}) and Eqs.~(\ref{First_Integration_N_new}) - (\ref{Second_Integration_2PN_new})
are the arguments $\ve{x}_{\rm 1PN}$ in the 1PN terms in (\ref{First_Integration_2PN_new}) and (\ref{Second_Integration_2PN_new}) and the new scalar functions
$\Laplace \dot{\ve{x}}_{\rm 2PN}$ and $\Laplace \ve{x}_{\rm 2PN}$ in (\ref{First_Integration_2PN_new}) and (\ref{Second_Integration_2PN_new}). It is, however, essential to
realize that the iterative solution in Eqs.~(\ref{First_Integration_N_new}) - (\ref{Second_Integration_2PN_new}) is identical to the iterative solution given above by
Eqs.~(\ref{First_Integration_N_old}) - (\ref{Second_Integration_2PN_old}), up to terms beyond the 2PN approximation.

These Eqs.~(\ref{First_Integration_1PN_new}) - (\ref{Second_Integration_2PN_new}) represent a generalization of Eqs.~(3.2.35) - (3.2.38) in \cite{Brumberg1991}, 
which are valid for the 2PN light propagation in the monopole field of a body, while the equations presented here are valid for the 2PN light propagation in the monopole 
and quadrupole field of a body. 

The 1PN terms in (\ref{First_Integration_1PN_new}) and (\ref{Second_Integration_1PN_new}) are given by Eqs.~(\ref{First_Integration_1PN_Terms_old}) and 
(\ref{Second_Integration_1PN_Terms_old}). The 1PN terms in (\ref{First_Integration_2PN_new}) and (\ref{Second_Integration_2PN_new}) are given by Eqs.~(\ref{New_First_Integration_1PN}) 
and (\ref{New_Second_Integration_1PN}). The 2PN perturbations in (\ref{First_Integration_2PN_new}) - (\ref{Second_Integration_2PN_new}) are given by
\begin{eqnarray}
        \frac{\Laplace \dot{x}^i_{\rm 2PN}\left(\ve{x}_{\rm N}\right)}{c} &=&
        \frac{G M}{c^2}\,\frac{G M}{c^2} \sum\limits_{n=1}^{2} U^i_{(n)}\left(\ve{x}_{\rm N}\right)\,\dot{X}_{(n)}\left(\ve{x}_{\rm N}\right)
	\nonumber\\ 
	&& \hspace{-2.0cm} + \frac{G M}{c^2}\,\frac{G M_{ab}}{c^2} \sum\limits_{n=1}^{8} V^{ab\,i}_{(n)}\left(\ve{x}_{\rm N}\right)\,\dot{Y}_{(n)}\left(\ve{x}_{\rm N}\right)
        \nonumber\\
        && \hspace{-2.0cm} + \frac{G M_{ab}}{c^2}\,\frac{G M_{cd}}{c^2} \sum\limits_{n=1}^{28} W^{abcd\,i}_{(n)}\left(\ve{x}_{\rm N}\right)\,\dot{Z}_{(n)}\left(\ve{x}_{\rm N}\right),
\label{New_First_Integration_2PN}
\end{eqnarray}

\noindent 
and 
\begin{eqnarray}
       \Laplace x^i_{\rm 2PN}\left(\ve{x}_{\rm N}\right) &=&
       \frac{G M}{c^2}\,\frac{G M}{c^2} \sum\limits_{n=1}^{2} U^i_{(n)}\left(\ve{x}_{\rm N}\right)\,X_{(n)}\left(\ve{x}_{\rm N}\right)
	\nonumber\\ 
	&& \hspace{-2.0cm} + \frac{G M}{c^2}\,\frac{G M_{ab}}{c^2} \sum\limits_{n=1}^{8} V^{ab\,i}_{(n)}\left(\ve{x}_{\rm N}\right)\,Y_{(n)}\left(\ve{x}_{\rm N}\right)
        \nonumber\\
	&& \hspace{-2.0cm} + \frac{G M_{ab}}{c^2}\,\frac{G M_{cd}}{c^2} \sum\limits_{n=1}^{28} W^{abcd\,i}_{(n)}\left(\ve{x}_{\rm N}\right)\,Z_{(n)}\left(\ve{x}_{\rm N}\right).
\label{New_Second_Integration_2PN}
\end{eqnarray}

\noindent
It is noticed here, that the 2PN monopole terms in (\ref{New_First_Integration_2PN}) and (\ref{New_Second_Integration_2PN}) are in agreement with the results obtained 
in \cite{Brumberg1991,KlionerKopeikin1992,Klioner_Zschocke}. The tensorial coefficients in (\ref{New_First_Integration_2PN}) and (\ref{New_Second_Integration_2PN}) are 
the same as in Eqs.~(\ref{First_Integration_2PN_Terms_old}) and (\ref{Second_Integration_2PN_Terms_old}) and they are given in Appendix~\ref{Appendix1}. The scalar 
functions are presented in Appendix~\ref{Appendix6} (arguments $\ve{x}_1$ in Appendix~\ref{Appendix6} substituted with the arguments $\ve{x}_{\rm N}$). 
The 2PN terms $\Delta\ve{x}_{\rm 2PN}\left(\ve{x}_0\right)$ in (\ref{Second_Integration_2PN_new}) are obtained from (\ref{New_Second_Integration_2PN}) by the 
replacements $\ve{x}_{\rm N}$ by $\ve{x}_0$. A final comment is in order here about the dot over the scalar functions in Eq.~(\ref{New_First_Integration_2PN}). 
This dot does not mean that these functions are time derivatives of the scalar functions in Eq.~(\ref{New_Second_Integration_2PN}). Instead, this dot means only, 
that these functions belong to the coordinate velocity of the photon.
 
The final step of this procedure concerns the replacement of the arguments in Eqs.~(\ref{First_Integration_N_new}) - (\ref{Second_Integration_2PN_new}) 
by the spatial positions of source and observer. This issue will be the subject of the subsequent section.

\section{Boundary value problem of light propagation}\label{Section4}

From Eqs.~(\ref{Second_Integration_N_new}), (\ref{Second_Integration_1PN_new}), and (\ref{Second_Integration_2PN_new}) follows that
the position of the light source is given by the position of the light ray at emission time $t_0$, in any order of the post-Newtonian expansion, 
that means 
\begin{eqnarray}
        \ve{x}_0 &=& \ve{x}_{\rm N}\left(t_0\right) = \ve{x}_{\rm 1PN}\left(t_0\right) = \ve{x}_{\rm 2PN}\left(t_0\right), 
        \label{x0_1PN}
\end{eqnarray}

\noindent
represent exact relations. 

On the other side, the spatial position of the observer, $\ve{x}_1$, coincides with the spatial position of the light signal, propagating along the exact light trajectory, 
at the moment of reception, $\ve{x}\left(t_1\right)$, which is determined up to the given order in the
post-Newtonian expansion. Therefore, the position of the observer can be obtained by replacing the spatial coordinate of the light signal at the moment
of observation by accounting for the correct order, that means according to Eqs.~(\ref{statement_replacement_N}) and (\ref{statement_replacement_1PN}); 
see also \cite{Klioner_Zschocke,Zschocke_Quadrupole_1,Zschocke_Time_Delay_2PN}. 
Only that new representation of the initial value problem, as represented by Eqs.~(\ref{First_Integration_N_new}) - (\ref{Second_Integration_2PN_new}),
allows one to insert in these relations (\ref{statement_replacement_N}) and (\ref{statement_replacement_1PN}). 
This procedure leads finally to the 1PN and 2PN expressions, which are given in their explicit form in the following two subsections.

\subsection{The 1PN terms}

The coordinate velocity and spatial position of the light signal in the 1PN approximation are given by Eqs.~(\ref{First_Integration_1PN_new}) 
and (\ref{Second_Integration_1PN_new}). According to relation (\ref{statement_replacement_N}), one may replace the arguments at the moment of 
reception of the light signal by the spatial position of the observer. In this way, one obtains for the coordinate velocity and spatial position 
of the light ray in the 1PN approximation: 
\begin{eqnarray}
\frac{\dot{\ve{x}}_{\rm 1PN}\left(t_1\right)}{c} &=& \ve{\sigma} + \frac{\Delta\dot{\ve{x}}_{\rm 1PN}\left(\ve{x}_1\right)}{c}\,,
\label{First_Integration_1PN_boundary}
\\
\ve{x}_{\rm 1PN}\left(t_1\right) &=& \ve{x}_0 + c \left(t_1 - t_0\right) \ve{\sigma} 
\nonumber\\ 
&+& \Delta\ve{x}_{\rm 1PN}\left(\ve{x}_1\right) - \Delta\ve{x}_{\rm 1PN}\left(\ve{x}_0\right),
\label{Second_Integration_1PN_boundary}
\end{eqnarray}

\noindent
where the 1PN perturbation terms in (\ref{First_Integration_1PN_boundary}) and (\ref{Second_Integration_1PN_boundary}) are given by
\begin{eqnarray}
        \frac{\Delta\dot{x}^i_{\rm 1PN}\left(\ve{x}_1\right)}{c} &=&
        \frac{G M}{c^2} \sum\limits_{n=1}^{2} U^i_{(n)}\left(\ve{x}_1\right)\,\dot{F}_{(n)}\left(\ve{x}_1\right)
	\nonumber\\ 
	&+& \frac{G M_{ab}}{c^2} \sum\limits_{n=1}^{8} V^{ab\,i}_{(n)}\left(\ve{x}_1\right)\,\dot{G}_{(n)}\left(\ve{x}_1\right),
\label{First_Integration_1PN_Final}
\\
        \Delta x^i_{\rm 1PN}\left(\ve{x}_1\right) &=&
        \frac{G M}{c^2} \sum\limits_{n=1}^{2} U^i_{(n)}\left(\ve{x}_1\right)\,F_{(n)}\left(\ve{x}_1\right)
	\nonumber\\ 
	&+& \frac{G M_{ab}}{c^2} \sum\limits_{n=1}^{8} V^{ab\,i}_{(n)}\left(\ve{x}_1\right)\, G_{(n)}\left(\ve{x}_1\right). 
\label{Second_Integration_1PN_Final}
\end{eqnarray}

\noindent
The tensorial coefficients and scalar functions are given in Appendices~\ref{Tensorial_Coefficients_1PN_Final} and \ref{Appendix5}. Clearly, 
the term $\Delta \ve{x}_{\rm 1PN}\left(\ve{x}_0\right)$ in (\ref{Second_Integration_1PN_boundary}) is obtained from (\ref{Second_Integration_1PN_Final}) 
by substituting the argument $\ve{x}_1$ with the argument $\ve{x}_0$.

\subsection{The 2PN terms} 

The coordinate velocity and spatial position of the light signal in the 2PN approximation are given by Eqs.~(\ref{First_Integration_2PN_new}) 
and (\ref{Second_Integration_2PN_new}). One may replace the arguments at the moment of reception of the light signal by the spatial position 
of the observer according to relations (\ref{statement_replacement_N}) and (\ref{statement_replacement_1PN}). In this way, one obtains for the 
coordinate velocity and spatial position of the light ray in the 2PN approximation:
\begin{eqnarray} 
\frac{\dot{\ve{x}}_{\rm 2PN}\left(t_1\right)}{c} &=& \ve{\sigma} + \frac{\Delta\dot{\ve{x}}_{\rm 1PN}\left(\ve{x}_1\right)}{c} 
+ \frac{\Laplace \dot{\ve{x}}_{\rm 2PN}\left(\ve{x}_1\right)}{c}\;,
\label{First_Integration_2PN_boundary} 
\\
\ve{x}_{\rm 2PN}\left(t_1\right) &=& \ve{x}_0 + c \left(t_1 - t_0\right) \ve{\sigma} 
\nonumber\\ 
&& + \Delta \ve{x}_{\rm 1PN}\left(\ve{x}_1\right) - \Delta \ve{x}_{\rm 1PN}\left(\ve{x}_0\right) 
\nonumber\\ 
&& + \Laplace \ve{x}_{\rm 2PN}\left(\ve{x}_1\right) - \Laplace \ve{x}_{\rm 2PN}\left(\ve{x}_0\right), 
\label{Second_Integration_2PN_boundary}
\end{eqnarray}

\noindent 
where the 2PN perturbation terms in (\ref{First_Integration_2PN_boundary}) and (\ref{Second_Integration_2PN_boundary}) are given by
\begin{eqnarray}
        \frac{\Laplace \dot{x}^i_{\rm 2PN}\left(\ve{x}_1\right)}{c} &=& 
        \frac{G M}{c^2}\,\frac{G M}{c^2} \sum\limits_{n=1}^{2} U^i_{(n)}\left(\ve{x}_1\right)\,\dot{X}_{(n)}\left(\ve{x}_1\right)
	\nonumber\\ 
	&& \hspace{-2.0cm} + \frac{G M}{c^2}\,\frac{G M_{ab}}{c^2} \sum\limits_{n=1}^{8} V^{ab\,i}_{(n)}\left(\ve{x}_1\right)\,\dot{Y}_{(n)}\left(\ve{x}_1\right)
        \nonumber\\
        && \hspace{-2.0cm} + \frac{G M_{ab}}{c^2}\,\frac{G M_{cd}}{c^2} \sum\limits_{n=1}^{28} W^{abcd\,i}_{(n)}\left(\ve{x}_1\right)\,\dot{Z}_{(n)}\left(\ve{x}_1\right),
\label{First_Integration_2PN_Final}
\end{eqnarray}

\noindent 
and
\begin{eqnarray}
       \Laplace x^i_{\rm 2PN}\left(\ve{x}_1\right) &=&
       \frac{G M}{c^2}\,\frac{G M}{c^2} \sum\limits_{n=1}^{2} U^i_{(n)}\left(\ve{x}_1\right)\,X_{(n)}\left(\ve{x}_1\right)
       \nonumber\\
       && \hspace{-2.0cm} + \frac{G M}{c^2}\,\frac{G M_{ab}}{c^2} \sum\limits_{n=1}^{8} V^{ab\,i}_{(n)}\left(\ve{x}_1\right)\,Y_{(n)}\left(\ve{x}_1\right)
       \nonumber\\
       && \hspace{-2.0cm} + \frac{G M_{ab}}{c^2}\,\frac{G M_{cd}}{c^2} \sum\limits_{n=1}^{28} W^{abcd\,i}_{(n)}\left(\ve{x}_1\right)\,Z_{(n)}\left(\ve{x}_1\right). 
\label{Second_Integration_2PN_Final}
\end{eqnarray}

\noindent
The tensorial coefficients are given in the Appendices~\ref{Tensorial_Coefficients_1PN_Final} and \ref{Tensorial_Coefficients_2PN_Final}, while the scalar functions are 
presented in Appendix~\ref{Appendix6}. Clearly, the term $\Laplace \ve{x}_{\rm 2PN}\left(\ve{x}_0\right)$ in (\ref{Second_Integration_2PN_boundary}) is obtained 
from (\ref{Second_Integration_2PN_Final}) by substituting the argument $\ve{x}_1$ with the argument $\ve{x}_0$. 

Finally, it should be noticed that the replacements (\ref{statement_replacement_N}) and (\ref{x0_1PN}) into the impact vector (\ref{impact_vector_N}), 
as well as the replacements in (\ref{statement_replacement_1PN}) and (\ref{x0_1PN}) into the impact vector (\ref{impact_vector_x_1PN}) imply the occurrence 
of two new impact vectors,  
\begin{eqnarray}
	\ve{d}^{\,0}_{\sigma} &=& \ve{\sigma} \times \left(\ve{x}_0 \times \ve{\sigma}\right), 
	\label{impact_vector_0}
	\\
        \ve{d}^{\,1}_{\sigma} &=& \ve{\sigma} \times \left(\ve{x}_1 \times \ve{\sigma}\right), 
        \label{impact_vector_1}
\end{eqnarray}

\noindent
where their absolute values are $d^{\,0}_{\sigma} = |\ve{d}^{\,0}_{\sigma}|$ and $d^{\,1}_{\sigma} = |\ve{d}^{\,1}_{\sigma}|$. These two impact vectors appear in 
a natural way if one considers the boundary value problem. It is not surprising, that the impact vector in (\ref{impact_vector_0}) is actually identical 
to the impact vector in (\ref{impact_vector_N}). The reason, that this impact vector appears both in the boundary value problem as well as in the initial value problem, 
is based in the fact that the initial condition (\ref{Initional_Condition_2}) of the initial value problem is identical to the boundary condition (\ref{boundary0}) of 
the boundary value problem.

\section{Summary and Outlook}\label{Section_Summary}

In our recent investigation \cite{Zschocke_Quadrupole_1} the coordinate velocity and the trajectory of a light signal in the gravitational field of a body 
at rest has been determined in the 2PN approximation, where the monopole and quadrupole terms of the gravitational field have been taken into account. 
The unique solution of the geodesic equation has been determined in the scheme of the initial value problem. In reality, however, the light source as well as 
the observer are located at finite distances $\ve{x}_0$ and $\ve{x}_1$ from the gravitating solar system body. This fact requires solving the boundary value problem  
of the geodesic equation. The solution of 2PN light propagation in terms of these boundary values $\ve{x}_0$ and $\ve{x}_1$ is represented in Sec.~\ref{Section4}. 
Notably the 2PN terms in Eqs.~(\ref{First_Integration_2PN_boundary}) and (\ref{Second_Integration_2PN_boundary}) with 
Eqs.~(\ref{First_Integration_2PN_Final}) - (\ref{Second_Integration_2PN_Final}) are the primary results of this investigation. These results are a basic requirement 
for highly precise measurements of light deflection on the sub-\muas{} scale and time delay on sub-pico-second level in the solar system. 

The final ambition of the boundary value problem is the determination of three fundamental transformations: 
$\ve{k} \rightarrow \ve{\sigma}$, $\ve{\sigma} \rightarrow \ve{n}$, $\ve{k} \rightarrow \ve{n}$, where $\ve{k}$ is the unit direction from the source towards 
the observer, $\ve{\sigma}$ is the unit tangent vector of light trajectory at minus infinity, and $\ve{n}$ is the unit tangent vector of the light ray at the spatial 
position of the observer. These transformations represent the basis of the Gaia relativistic model \cite{Klioner2003}, 
which has later been refined by our investigations in \cite{Klioner_Zschocke} and \cite{Zschocke_Klioner}. The determination of these transformations will be represented 
in a subsequent investigation. These transformations would also be implemented in relativistic models for data reduction of possible future space astrometry missions, like 
the Gaia successor GaiaNIR \cite{Gaia_NIR} or Theia \cite{Theia}, planned to be launched in an optimistic scenario in $2045$ as medium-sized missions of ESA \cite{Eric_Hoeg}.

\section*{Acknowledgments}

This work was funded by the German Research Foundation (Deutsche Forschungsgemeinschaft DFG) under Grant No. 447922800. Sincere gratitude is expressed to
Professor Sergei A. Klioner for continual support and inspiring discussions about astrometry and general theory of relativity. Professor Michael H. Soffel, 
Professor Ralf Sch\"utzhold, Professor William G. Unruh, Priv.-Doz. G\"unter Plunien, Dr. Alexey Butkevich, Dipl-Inf. Robin Geyer, 
Professor Burkhard K\"ampfer, and Professor Laszlo Csernai, are greatly acknowledged for interesting discussions about general theory of relativity.

\newpage

\appendix

\section{Notation}\label{Appendix0}

\noindent 
Throughout the investigation the following notation is in use:

\begin{itemize}
\item $G$ is the Newtonian constant of gravitation.
\item $c$ is the vacuum speed of light in Minkowskian space-time.
\item $M$ is the rest mass of the body.
\item $M_{ab}$ is the symmetric trace-free quadrupole moment of the body.
\item Lower case Latin indices $i$, $j$, \dots take values $1,2,3$.
\item $\dot{f}$ denotes total derivative of $f$ with respect to global coordinate time.
\item $\delta_{ij} = \delta^{ij} = {\rm diag} \left(+1,+1,+1\right)$ is Kronecker delta.
\item Three-vectors are in boldface: e.g. $\ve{a}$, $\ve{b}$, $\ve{\sigma}$, $\ve{x}$.
\item Contravariant components of three-vectors: $a^{i} = \left(a^{\,1},a^2,a^3\right)$.
\item Scalar product of three-vectors: $\ve{a}\,\cdot\,\ve{b}=\delta_{ij}\,a^i\,b^j$.
\item Absolute value of three-vector: $|\ve{a}| = \sqrt{\delta_{ij}\,a^i\,a^j}$.
\item Levi-Civita symbol: $\varepsilon_{ijk} = \varepsilon^{ijk}$ with $\varepsilon_{123} = + 1$.
\item Cross-product of three-vectors: $\left(\ve{a}\times\ve{b}\right)^i\!=\!\varepsilon_{ijk} a^j b^k$.
\item Lower case Greek indices take values 0,1,2,3.
\item $\eta_{\alpha\beta} = \eta^{\alpha \beta} = {\rm diag}\left(-1,+1,+1,+1\right)$ is the metric tensor of flat space-time.
\item $g_{\alpha\beta}$ and $g^{\alpha\beta}$ are the covariant and contravariant components of the metric tensor.
\item Contravariant components of four-vectors: $a^{\mu} = \left(a^{\,0},a^{\,1},a^2,a^3\right)$.
\item milli-arcsecond (mas): $1\,{\rm mas} = \pi/(180 \times 60 \times 60)\,\times 10^{-3}\,{\rm rad}$.
\item micro-arcsecond (\muas): $1\,\muas = \pi/(180 \times 60 \times 60)\,\times 10^{-6}\,{\rm rad}$.
\item nano-arcsecond (nas): $1\,{\rm nas} = \pi/(180 \times 60 \times 60)\,\times 10^{-9}\,{\rm rad}$.
\item pico-second (ps): $1\,{\rm ps} = 10^{-12}\,{\rm second}$. 
\item repeated indices are implicitly summed over ({\it Einstein's} sum convention).
\end{itemize}

\section{Tensorial coefficients of 1PN and 2PN solution in (\ref{First_Integration_1PN_Terms_old}) - (\ref{Second_Integration_2PN_Terms_old})}\label{Appendix1}

The tensorial coefficients of the 1PN perturbation terms in (\ref{First_Integration_1PN_Terms_old}) and (\ref{Second_Integration_1PN_Terms_old}) are given by
\begin{eqnarray}
        U^{i}_{\left(1\right)}\left(\ve{x}_{\rm N}\right) &=& \sigma^i\;,
        \label{coefficient_U1_N}
        \\
        U^{i}_{\left(2\right)}\left(\ve{x}_{\rm N}\right) &=& d_{\sigma}^i \;, 
        \label{coefficient_U2_N}
        \\
        V^{ab\,i}_{\left(1\right)}\left(\ve{x}_{\rm N}\right) &=& \sigma^a \delta^{bi}\;, 
        \label{coefficient_V1_N} 
        \\
        V^{ab\,i}_{\left(2\right)}\left(\ve{x}_{\rm N}\right) &=& d_{\sigma}^a \delta^{bi}\;, 
        \label{coefficient_V2_N}
        \\
        V^{ab\,i}_{\left(3\right)}\left(\ve{x}_{\rm N}\right) &=& \sigma^a \sigma^b \sigma^i\;, 
        \label{coefficient_V3_N} 
        \\
        V^{ab\,i}_{\left(4\right)}\left(\ve{x}_{\rm N}\right) &=& \sigma^a d_{\sigma}^b \sigma^i\;, 
        \label{coefficient_V4_N} 
        \\
        V^{ab\,i}_{\left(5\right)}\left(\ve{x}_{\rm N}\right) &=& d_{\sigma}^a d_{\sigma}^b \sigma^i\;, 
        \label{coefficient_V5_N} 
        \\
        V^{ab\,i}_{\left(6\right)}\left(\ve{x}_{\rm N}\right) &=& d_{\sigma}^a d_{\sigma}^b d_{\sigma}^i\;, 
        \label{coefficient_V6_N} 
        \\
        V^{ab\,i}_{\left(7\right)}\left(\ve{x}_{\rm N}\right) &=& \sigma^a \sigma^b d_{\sigma}^i\;, 
        \label{coefficient_V7_N} 
        \\
        V^{ab\,i}_{\left(8\right)}\left(\ve{x}_{\rm N}\right) &=& \sigma^a d_{\sigma}^b d_{\sigma}^i\;.
        \label{coefficient_V8_N} 
\end{eqnarray}

\noindent 
The coefficients in (\ref{coefficient_V1_N}) - (\ref{coefficient_V8_N}) represent a complete set of linearly independent tensors with three spatial indices, which can be 
constructed from two independent three-vectors, $\sigma^a$ and $d_{\sigma}^b$, and the Kronecker symbol (there are no tensors which contain $\delta_{ab}$
because the quadrupole tensor is trace-free). Note, that a permutation of the spatial indices 
$\left(a \leftrightarrow b\right)$ is of no relevance, because of the symmetry of the quadrupole tensor. For instance, there is no need to distinguish between the tensors 
$\sigma^a \delta^{bi}$ and $\sigma^b \delta^{ai}$, because they yield same result when contracted with the quadrupole tensor: 
$\sigma^a \delta^{bi}\,M_{ab} = \sigma^b \delta^{ai}\,M_{ab}$. 

The tensorial coefficients of the 2PN perturbation terms in (\ref{First_Integration_2PN_Terms_old}) and (\ref{Second_Integration_2PN_Terms_old}) are given by
\begin{eqnarray}
        W^{abcd\,i}_{\left(1\right)}\left(\ve{x}_{\rm N}\right) &=& \delta^{ac} \sigma^b \delta^{di} \;, 
        \label{coefficient_W1}
        \\
        W^{abcd\,i}_{\left(2\right)}\left(\ve{x}_{\rm N}\right) &=& \delta^{ac} d_{\sigma}^b \delta^{di} \;,
        \label{coefficient_W2}
        \\
        W^{abcd\,i}_{\left(3\right)}\left(\ve{x}_{\rm N}\right) &=& \sigma^a \sigma^b \sigma^c \delta^{di} \;, 
        \label{coefficient_W3}
        \\
        W^{abcd\,i}_{\left(4\right)}\left(\ve{x}_{\rm N}\right) &=& \sigma^a \sigma^b d_{\sigma}^c \delta^{di} \;,
        \label{coefficient_W4}
        \\
        W^{abcd\,i}_{\left(5\right)}\left(\ve{x}_{\rm N}\right) &=& \sigma^a d_{\sigma}^b \sigma^c \delta^{di} \;, 
        \label{coefficient_W5}
        \\
        W^{abcd\,i}_{\left(6\right)}\left(\ve{x}_{\rm N}\right) &=& \sigma^a d_{\sigma}^b d_{\sigma}^c \delta^{di} \;, 
        \label{coefficient_W6}
        \\
        W^{abcd\,i}_{\left(7\right)}\left(\ve{x}_{\rm N}\right) &=& d_{\sigma}^a d_{\sigma}^b \sigma^c \delta^{di} \;,
        \label{coefficient_W7}
        \\
        W^{abcd\,i}_{\left(8\right)}\left(\ve{x}_{\rm N}\right) &=& d_{\sigma}^a d_{\sigma}^b d_{\sigma}^c \delta^{di} \;, 
        \label{coefficient_W8}
        \\
        W^{abcd\,i}_{\left(9\right)}\left(\ve{x}_{\rm N}\right) &=& \delta^{ac} \delta^{bd} \sigma^i \;, 
        \label{coefficient_W9} 
        \\
        W^{abcd\,i}_{\left(10\right)}\left(\ve{x}_{\rm N}\right) &=& \delta^{ac} \sigma^b \sigma^d \sigma^i \;,
        \label{coefficient_W10} 
        \\
        W^{abcd\,i}_{\left(11\right)}\left(\ve{x}_{\rm N}\right) &=& \delta^{ac} \sigma^b d_{\sigma}^d \sigma^i \;,
        \label{coefficient_W11}
        \\
        W^{abcd\,i}_{\left(12\right)}\left(\ve{x}_{\rm N}\right) &=& \delta^{ac} d_{\sigma}^b d_{\sigma}^d \sigma^i \;,
        \label{coefficient_W12}
        \\
        W^{abcd\,i}_{\left(13\right)}\left(\ve{x}_{\rm N}\right) &=& \sigma^a \sigma^b \sigma^c \sigma^d \sigma^i \;, 
        \label{coefficient_W13}
        \\
        W^{abcd\,i}_{\left(14\right)}\left(\ve{x}_{\rm N}\right) &=& \sigma^a \sigma^b \sigma^c d_{\sigma}^d \sigma^i \;,
        \label{coefficient_W14}
        \\ 
        W^{abcd\,i}_{\left(15\right)}\left(\ve{x}_{\rm N}\right) &=& \sigma^a \sigma^b d_{\sigma}^c d_{\sigma}^d \sigma^i \;,
        \label{coefficient_W15}
        \\
        W^{abcd\,i}_{\left(16\right)}\left(\ve{x}_{\rm N}\right) &=& \sigma^a d_{\sigma}^b \sigma^c d_{\sigma}^d \sigma^i \;,
        \label{coefficient_W16}
        \\ 
        W^{abcd\,i}_{\left(17\right)}\left(\ve{x}_{\rm N}\right) &=& \sigma^a d_{\sigma}^b d_{\sigma}^c d_{\sigma}^d \sigma^i \;,
        \label{coefficient_W17}
        \\
        W^{abcd\,i}_{\left(18\right)}\left(\ve{x}_{\rm N}\right) &=& d_{\sigma}^a d_{\sigma}^b d_{\sigma}^c d_{\sigma}^d \sigma^i \;,
        \label{coefficient_W18}
        \\ 
        W^{abcd\,i}_{\left(19\right)}\left(\ve{x}_{\rm N}\right) &=& \delta^{ac} \delta^{bd} d_{\sigma}^i \;,
        \label{coefficient_W19} 
        \\
        W^{abcd\,i}_{\left(20\right)}\left(\ve{x}_{\rm N}\right) &=& \delta^{ac} \sigma^b \sigma^d d_{\sigma}^i \;,
        \label{coefficient_W20} 
        \\
        W^{abcd\,i}_{\left(21\right)}\left(\ve{x}_{\rm N}\right) &=& \delta^{ac} \sigma^b d_{\sigma}^d d_{\sigma}^i \;,
        \label{coefficient_W21}
        \\
        W^{abcd\,i}_{\left(22\right)}\left(\ve{x}_{\rm N}\right) &=& \delta^{ac} d_{\sigma}^b d_{\sigma}^d d_{\sigma}^i \;,
        \label{coefficient_W22}
        \\
        W^{abcd\,i}_{\left(23\right)}\left(\ve{x}_{\rm N}\right) &=& \sigma^a \sigma^b \sigma^c \sigma^d d_{\sigma}^i \;,
        \label{coefficient_W23}
        \\
        W^{abcd\,i}_{\left(24\right)}\left(\ve{x}_{\rm N}\right) &=& \sigma^a \sigma^b \sigma^c d_{\sigma}^d d_{\sigma}^i \;,
        \label{coefficient_W24}
        \\
        W^{abcd\,i}_{\left(25\right)}\left(\ve{x}_{\rm N}\right) &=& \sigma^a \sigma^b d_{\sigma}^c d_{\sigma}^d d_{\sigma}^i \;,
        \label{coefficient_W25}
        \\
        W^{abcd\,i}_{\left(26\right)}\left(\ve{x}_{\rm N}\right) &=& \sigma^a d_{\sigma}^b \sigma^c d_{\sigma}^d d_{\sigma}^i \;,
        \label{coefficient_W26}
        \\
        W^{abcd\,i}_{\left(27\right)}\left(\ve{x}_{\rm N}\right) &=& \sigma^a d_{\sigma}^b d_{\sigma}^c d_{\sigma}^d d_{\sigma}^i \;,
        \label{coefficient_W27}
        \\
        W^{abcd\,i}_{\left(28\right)}\left(\ve{x}_{\rm N}\right) &=& d_{\sigma}^a d_{\sigma}^b d_{\sigma}^c d_{\sigma}^d d_{\sigma}^i \;.
        \label{coefficient_W28}
\end{eqnarray}

\noindent
The coefficients in (\ref{coefficient_W1}) - (\ref{coefficient_W28}) represent a complete set of tensors
with five spatial indices, which can be constructed from two independent three-vectors, $\sigma^a$ and $d_{\sigma}^b$,
and the Kronecker symbol (there are no tensors which contain $\delta_{ab}$ or $\delta_{cd}$ because the quadrupole tensor is trace-free), 
and where the symmetry of the quadrupole tensor is accounted for. That means, the permutations of the spatial indices  
$\left(a \leftrightarrow b\, \land \,c \leftrightarrow d\right)$, $\left(a \leftrightarrow b \, \lor \, c \leftrightarrow d\right)$, 
$\left(a \leftrightarrow c\,\land \,b \leftrightarrow d\right)$, $\left(a \leftrightarrow d \, \land \, b \leftrightarrow c\right)$ have no relevance; 
cf. text below Eq.~(\ref{coefficient_V8_N}).

\section{Scalar functions of 1PN and 2PN solution in (\ref{First_Integration_1PN_Terms_old}) - (\ref{Second_Integration_1PN_Terms_old})}\label{Appendix2}

\subsection{Scalar functions of the 1PN monopole term in (\ref{First_Integration_1PN_Terms_old}) and (\ref{Second_Integration_1PN_Terms_old})}

By comparing with Eqs.~(62) and (66) in \cite{Zschocke_Quadrupole_1} one obtains the scalar functions of the 1PN monopole terms in Eqs.~(\ref{First_Integration_1PN_Terms_old}) 
and (\ref{Second_Integration_1PN_Terms_old}): 
\begin{eqnarray}
	\dot{F}_{\left(1\right)}\left(\ve{x}_{\rm N}\right) &=& + 2\,\dot{\cal W}_{\left(3\right)}\left(\ve{x}_{\rm N}\right)\,,
        \label{scalar_dot_F1}
        \\
	\dot{F}_{\left(2\right)}\left(\ve{x}_{\rm N}\right) &=& - 2\,\dot{\cal X}_{\left(3\right)}\left(\ve{x}_{\rm N}\right)\,,
        \label{scalar_dot_F2}
\end{eqnarray}

\noindent
with 
\begin{eqnarray}
	\dot{\cal W}_{\left(3\right)}\left(\ve{x}_{\rm N}\right) &=& - \frac{1}{x_{\rm N}}\,,
        \label{scalar_dot_W3}
	\\
	\dot{\cal X}_{\left(3\right)}\left(\ve{x}_{\rm N}\right) &=& + \frac{1}{\left(d_{\sigma}\right)^2} \left(1 + \frac{\ve{\sigma} \cdot \ve{x}_{\rm N}}{x_{\rm N}}\right), 
        \label{scalar_dot_X3}
\end{eqnarray}

\noindent 
and
\begin{eqnarray}
        F_{\left(1\right)}\left(\ve{x}_{\rm N}\right) &=&  + 2\,{\cal W}_{\left(3\right)}\left(\ve{x}_{\rm N}\right)\,,
        \label{scalar_F1}
        \\
        F_{\left(2\right)}\left(\ve{x}_{\rm N}\right) &=& - 2\,{\cal X}_{\left(3\right)}\left(\ve{x}_{\rm N}\right)\,, 
        \label{scalar_F2}
\end{eqnarray}

\noindent 
with
\begin{eqnarray}
	{\cal W}_{\left(3\right)}\left(\ve{x}_{\rm N}\right) &=& + \ln \left(x_{\rm N} - \ve{\sigma} \cdot \ve{x}_{\rm N}\right),
	\label{scalar_W3}
        \\
	{\cal X}_{\left(3\right)}\left(\ve{x}_{\rm N}\right) &=& +  \frac{1}{\left(d_{\sigma}\right)^2} \left(x_{\rm N} + \ve{\sigma} \cdot \ve{x}_{\rm N}\right), 
        \label{scalar_X3}
\end{eqnarray}

\noindent
where $\ve{x}_{\rm N} = \ve{x}_{\rm N}\left(t\right)$. 

\subsection{Scalar functions of the 1PN Quadrupole term in (\ref{First_Integration_1PN_Terms_old}) and (\ref{Second_Integration_1PN_Terms_old})}

By comparing with Eqs.~(63) and (67) in \cite{Zschocke_Quadrupole_1} one obtains the scalar functions of the 1PN quadrupole terms 
in Eqs.~(\ref{First_Integration_1PN_Terms_old}) and (\ref{Second_Integration_1PN_Terms_old}): 
\begin{eqnarray}
	\dot{G}_{\left(1\right)}\left(\ve{x}_{\rm N}\right) &=& + 6\,\dot{\cal W}_{\left(5\right)}\left(\ve{x}_{\rm N}\right), 
        \label{scalar_dot_G1}
        \\
	\dot{G}_{\left(2\right)}\left(\ve{x}_{\rm N}\right) &=& + 6\,\dot{\cal X}_{\left(5\right)}\left(\ve{x}_{\rm N}\right),
        \label{scalar_dot_G2}
        \\
	\dot{G}_{\left(3\right)}\left(\ve{x}_{\rm N}\right) &=& + 3\,\dot{\cal W}_{\left(5\right)}\left(\ve{x}_{\rm N}\right) 
	- 15 \left(d_{\sigma}\right)^2 \dot{\cal W}_{\left(7\right)}\left(\ve{x}_{\rm N}\right),
	\nonumber\\ 
        \label{scalar_dot_G3}
        \\
        \dot{G}_{\left(4\right)}\left(\ve{x}_{\rm N}\right) &=& + 18\,\dot{\cal X}_{\left(5\right)}\left(\ve{x}_{\rm N}\right) 
	- 30 \left(d_{\sigma}\right)^2 \dot{\cal X}_{\left(7\right)}\left(\ve{x}_{\rm N}\right),
        \label{scalar_dot_G4}
	\nonumber\\ 
        \\
        \dot{G}_{\left(5\right)}\left(\ve{x}_{\rm N}\right) &=& + 15\,\dot{\cal W}_{\left(7\right)}\left(\ve{x}_{\rm N}\right), 
        \label{scalar_dot_G5}
        \\
        \dot{G}_{\left(6\right)}\left(\ve{x}_{\rm N}\right) &=& - 15\,\dot{\cal X}_{\left(7\right)}\left(\ve{x}_{\rm N}\right), 
       \label{scalar_dot_G6}
        \\
        \dot{G}_{\left(7\right)}\left(\ve{x}_{\rm N}\right) &=& - 15\,\dot{\cal X}_{\left(5\right)}\left(\ve{x}_{\rm N}\right) 
	+ 15 \left(d_{\sigma}\right)^2 \dot{\cal X}_{\left(7\right)}\left(\ve{x}_{\rm N}\right),
	\nonumber\\ 
        \label{scalar_dot_G7}
        \\
        \dot{G}_{\left(8\right)}\left(\ve{x}_{\rm N}\right) &=& - 30\,\dot{\cal W}_{\left(7\right)}\left(\ve{x}_{\rm N}\right), 
        \label{scalar_dot_G8}
\end{eqnarray}

\noindent
with  
\begin{eqnarray}
	\dot{\cal W}_{\left(5\right)}\left(\ve{x}_{\rm N}\right) &=& - \frac{1}{3}\,\frac{1}{\left(x_{\rm N}\right)^3}\,,
        \label{scalar_dot_W5}
        \\
        \dot{\cal W}_{\left(7\right)}\left(\ve{x}_{\rm N}\right) &=& - \frac{1}{5}\,\frac{1}{\left(x_{\rm N}\right)^5}\,,  
        \label{scalar_dot_W7}
	\\
	\dot{\cal X}_{\left(5\right)}\left(\ve{x}_{\rm N}\right) &=& + \frac{2}{3}\,\frac{1}{\left(d_{\sigma}\right)^2} 
	\nonumber\\ 
	&& \hspace{-1.5cm} \times \left(\frac{1}{\left(d_{\sigma}\right)^2} + \frac{1}{\left(d_{\sigma}\right)^2}\,\frac{\ve{\sigma} \cdot \ve{x}_{\rm N}}{x_{\rm N}} 
	+ \frac{1}{2}\,\frac{\ve{\sigma} \cdot \ve{x}_{\rm N}}{\left(x_{\rm N}\right)^3}\right), 
        \label{scalar_dot_X5}
	\\
	\dot{\cal X}_{\left(7\right)}\left(\ve{x}_{\rm N}\right) &=& + \frac{8}{15}\,\frac{1}{\left(d_{\sigma}\right)^2} 
	\Bigg(\frac{1}{\left(d_{\sigma}\right)^4} + \frac{1}{\left(d_{\sigma}\right)^4}\,\frac{\ve{\sigma} \cdot \ve{x}_{\rm N}}{x_{\rm N}} 
	\nonumber\\ 
	&& \hspace{-1.5cm} 
        + \frac{1}{2}\,\frac{1}{\left(d_{\sigma}\right)^2}\,\frac{\ve{\sigma} \cdot \ve{x}_{\rm N}}{\left(x_{\rm N}\right)^3} 
	+ \frac{3}{8}\,\frac{\ve{\sigma} \cdot \ve{x}_{\rm N}}{\left(x_{\rm N}\right)^5}\Bigg),
        \label{scalar_dot_X7}
\end{eqnarray}

\noindent 
and 
\begin{eqnarray}
	G_{\left(1\right)}\left(\ve{x}_{\rm N}\right) &=& + 6\,{\cal W}_{\left(5\right)}\left(\ve{x}_{\rm N}\right), 
        \label{scalar_G1}
        \\
        G_{\left(2\right)}\left(\ve{x}_{\rm N}\right) &=& + 6\,{\cal X}_{\left(5\right)}\left(\ve{x}_{\rm N}\right), 
        \label{scalar_G2}
        \\
	G_{\left(3\right)}\left(\ve{x}_{\rm N}\right) &=& + 3\,{\cal W}_{\left(5\right)}\left(\ve{x}_{\rm N}\right) 
	- 15 \left(d_{\sigma}\right)^2 {\cal W}_{\left(7\right)}\left(\ve{x}_{\rm N}\right),
	\nonumber\\ 
        \label{scalar_G3}
        \\
        G_{\left(4\right)}\left(\ve{x}_{\rm N}\right) &=& + 18\,{\cal X}_{\left(5\right)}\left(\ve{x}_{\rm N}\right) 
	- 30 \left(d_{\sigma}\right)^2 {\cal X}_{\left(7\right)}\left(\ve{x}_{\rm N}\right),
	\nonumber\\ 
        \label{scalar_G4}
        \\
        G_{\left(5\right)}\left(\ve{x}_{\rm N}\right) &=& + 15\,{\cal W}_{\left(7\right)}\left(\ve{x}_{\rm N}\right),
        \label{scalar_G5}
        \\
        G_{\left(6\right)}\left(\ve{x}_{\rm N}\right) &=& - 15\,{\cal X}_{\left(7\right)}\left(\ve{x}_{\rm N}\right), 
        \label{scalar_G6}
        \\
        G_{\left(7\right)}\left(\ve{x}_{\rm N}\right) &=& - 15\,{\cal X}_{\left(5\right)}\left(\ve{x}_{\rm N}\right) 
	+ 15 \left(d_{\sigma}\right)^2 {\cal X}_{\left(7\right)}\left(\ve{x}_{\rm N}\right),
	\nonumber\\ 
        \label{scalar_G7}
        \\
        G_{\left(8\right)}\left(\ve{x}_{\rm N}\right) &=& - 30\,{\cal W}_{\left(7\right)}\left(\ve{x}_{\rm N}\right), 
        \label{scalar_G8}
\end{eqnarray}

\noindent 
with
\begin{eqnarray}
	{\cal W}_{\left(5\right)}\left(\ve{x}_{\rm N}\right) &=& - \frac{1}{3}\,\frac{1}{\left(d_{\sigma}\right)^2}\,\frac{\ve{\sigma} \cdot \ve{x}_{\rm N}}{x_{\rm N}}\,,
        \label{scalar_W5}
        \\
        {\cal W}_{\left(7\right)}\left(\ve{x}_{\rm N}\right) &=& - \frac{2}{15}\,\frac{1}{\left(d_{\sigma}\right)^2} 
	\left(\frac{1}{\left(d_{\sigma}\right)^2}\frac{\ve{\sigma} \cdot \ve{x}_{\rm N}}{x_{\rm N}} + \frac{1}{2}\,\frac{\ve{\sigma} \cdot \ve{x}_{\rm N}}{\left(x_{\rm N}\right)^3}\right),
	\nonumber\\ 
        \label{scalar_W7}
        \\
	{\cal X}_{\left(5\right)}\left(\ve{x}_{\rm N}\right) &=& + \frac{2}{3}\,\frac{1}{\left(d_{\sigma}\right)^2} 
	\left(\frac{x_{\rm N} + \ve{\sigma} \cdot \ve{x}_{\rm N}}{\left(d_{\sigma}\right)^2} - \frac{1}{2}\,\frac{1}{x_{\rm N}}\right), 
        \label{scalar_X5}
        \\
	{\cal X}_{\left(7\right)}\left(\ve{x}_{\rm N}\right) &=& + \frac{8}{15}\,\frac{1}{\left(d_{\sigma}\right)^2}
	\nonumber\\ 
	&& \hspace{-1.5cm} 
	\times \left(\frac{x_{\rm N} + \ve{\sigma} \cdot \ve{x}_{\rm N}}{\left(d_{\sigma}\right)^4} - \frac{1}{2}\,\frac{1}{\left(d_{\sigma}\right)^2}\,\frac{1}{x_{\rm N}} 
	- \frac{1}{8}\,\frac{1}{\left(x_{\rm N}\right)^3}\right), 
        \label{scalar_X7}
\end{eqnarray}

\noindent 
where $\ve{x}_{\rm N} = \ve{x}_{\rm N}\left(t\right)$.

\clearpage 

\begin{widetext}

\section{The scalar functions in (\ref{First_Integration_2PN_Terms_old}) - (\ref{Second_Integration_2PN_Terms_old})}\label{Appendix3}

\subsection{Scalar functions of the 2PN monopole-monopole term in (\ref{First_Integration_2PN_Terms_old}) and (\ref{Second_Integration_2PN_Terms_old})} 

By comparing with Eqs.~(83) and (89) in \cite{Zschocke_Quadrupole_1} one obtains the scalar functions for the 2PN monopole-monopole terms 
in Eqs.~(\ref{First_Integration_2PN_Terms_old}) and (\ref{Second_Integration_2PN_Terms_old}):  
\begin{eqnarray}
        \dot{A}_{\left(1\right)}\left(\ve{x}_{\rm N}\right) &=& - 4\,\dot{\cal W}_{\left(4\right)}
        - 12 \left(x_0 + \ve{\sigma} \cdot \ve{x}_0\right) \dot{\cal W}_{\left(5\right)}
        - 2 \left(d_{\sigma}\right)^2 \dot{\cal W}_{\left(6\right)}
        + 4\,\dot{\cal X}_{\left(3\right)}
        - 12 \left(d_{\sigma}\right)^2 \dot{\cal X}_{\left(5\right)}
        - 8\,\dot{\cal Z}_{\left(3\right)}
        + 12 \left(d_{\sigma}\right)^2 \dot{\cal Z}_{\left(5\right)}\,,  
        \label{scalar_function_dot_A1_2PN}
        \\
        \dot{A}_{\left(2\right)}\left(\ve{x}_{\rm N}\right) &=& - \frac{4}{\left(d_{\sigma}\right)^2}\,\dot{\cal W}_{\left(3\right)}
	- 12\,\dot{\cal W}_{\left(5\right)} - \frac{4}{\left(d_{\sigma}\right)^2}\,\dot{\cal X}_{\left(2\right)} 
	- \frac{4}{\left(d_{\sigma}\right)^2} \left(x_0 + \ve{\sigma} \cdot \ve{x}_0\right) \dot{\cal X}_{\left(3\right)}  
	+ 6\,\dot{\cal X}_{\left(4\right)} + 12 \left(x_0 + \ve{\sigma} \cdot \ve{x}_0\right) \dot{\cal X}_{\left(5\right)} 
	\nonumber\\
	&& - 2 \left(d_{\sigma}\right)^2 \dot{\cal X}_{\left(6\right)}  
        + 12 \,\dot{\cal Y}_{\left(5\right)}\,,
        \label{scalar_function_dot_A2_2PN}
\end{eqnarray}

\noindent
and 
\begin{eqnarray}
        A_{\left(1\right)}\left(\ve{x}_{\rm N}\right) &=& - 4\,{\cal W}_{\left(4\right)}
        - 12 \left(x_0 + \ve{\sigma} \cdot \ve{x}_0\right) {\cal W}_{\left(5\right)}
        - 2 \left(d_{\sigma}\right)^2 {\cal W}_{\left(6\right)}
        + 4\,{\cal X}_{\left(3\right)}
        - 12 \left(d_{\sigma}\right)^2 {\cal X}_{\left(5\right)}
        - 8\,{\cal Z}_{\left(3\right)}
        + 12 \left(d_{\sigma}\right)^2 {\cal Z}_{\left(5\right)}\,,
        \label{scalar_function_A1_2PN}
        \\
	A_{\left(2\right)}\left(\ve{x}_{\rm N}\right) &=& - \frac{4}{\left(d_{\sigma}\right)^2}\,{\cal W}_{\left(3\right)}
        - 12\,{\cal W}_{\left(5\right)} - \frac{4}{\left(d_{\sigma}\right)^2}\,{\cal X}_{\left(2\right)}
        - \frac{4}{\left(d_{\sigma}\right)^2} \left(x_0 + \ve{\sigma} \cdot \ve{x}_0\right) {\cal X}_{\left(3\right)}
        + 6\,{\cal X}_{\left(4\right)} + 12 \left(x_0 + \ve{\sigma} \cdot \ve{x}_0\right) {\cal X}_{\left(5\right)}
        \nonumber\\
        && - 2 \left(d_{\sigma}\right)^2 {\cal X}_{\left(6\right)}
        + 12 \,{\cal Y}_{\left(5\right)}\,,
        \label{scalar_function_A2_2PN}
\end{eqnarray}

\noindent 
where the scalar functions $\dot{\cal W}_{\left(n\right)}$, $\dot{\cal X}_{\left(n\right)}$, $\dot{\cal Y}_{\left(n\right)}$, $\dot{\cal Z}_{\left(n\right)}$ and 
${\cal W}_{\left(n\right)}$, ${\cal X}_{\left(n\right)}$, ${\cal Y}_{\left(n\right)}$, ${\cal Z}_{\left(n\right)}$
are given in \cite{Zschocke_Quadrupole_1}; the argument $\ve{x}_{\rm N}\left(t\right)$ of these functions has been omitted here.

\subsection{Scalar functions of the 2PN monopole-quadrupole term in (\ref{First_Integration_2PN_Terms_old}) and (\ref{Second_Integration_2PN_Terms_old})} 

By comparing with Eqs.~(84) and (90) in \cite{Zschocke_Quadrupole_1} one obtains the scalar functions for the 2PN monopole-quadrupole terms 
in Eqs.~(\ref{First_Integration_2PN_Terms_old}) and (\ref{Second_Integration_2PN_Terms_old}): 
\begin{eqnarray}
        \dot{B}_{\left(1\right)}\left(\ve{x}_{\rm N}\right) &=& + \frac{4}{\left(d_{\sigma}\right)^2}\,\dot{\cal W}_{\left(4\right)} 
	+ 22\,\dot{\cal W}_{\left(6\right)} - 60 \left(x_0 + \ve{\sigma} \cdot \ve{x}_0\right) \dot{\cal W}_{\left(7\right)}  
	+ \frac{21}{2} \left(d_{\sigma}\right)^2 \dot{\cal W}_{\left(8\right)} 
	- \frac{4}{\left(d_{\sigma}\right)^2}\,\frac{\ve{\sigma} \cdot \ve{x}_0}{x_0}\,\dot{\cal X}_{\left(3\right)} 
	+ 60\,\dot{\cal X}_{\left(5\right)} 
	\nonumber\\ 
	&& - 60 \left(d_{\sigma}\right)^2 \dot{\cal X}_{\left(7\right)} 
	- 48\,\dot{\cal Z}_{\left(5\right)} + 60 \left(d_{\sigma}\right)^2 \dot{\cal Z}_{\left(7\right)} \,, 
        \label{scalar_function_dot_B1_2PN}
        \\
	\vdots 
        \\
        \dot{B}_{\left(8\right)}\left(\ve{x}_{\rm N}\right) &=& - \frac{8}{\left(d_{\sigma}\right)^4}\,\dot{\cal W}_{\left(4\right)}
        - \frac{12}{\left(x_0\right)^3}\,\dot{\cal W}_{\left(5\right)}
        + \frac{32}{\left(d_{\sigma}\right)^2}\,\dot{\cal W}_{\left(6\right)} 
        - \frac{120}{\left(d_{\sigma}\right)^2} \left(x_0 + \ve{\sigma} \cdot \ve{x}_0\right) \dot{\cal W}_{\left(7\right)}
        - 63\,\dot{\cal W}_{\left(8\right)}
        \nonumber\\ 
        && + 420 \left(x_0 + \ve{\sigma} \cdot \ve{x}_0\right) \dot{\cal W}_{\left(9\right)}
        - 60 \left(d_{\sigma}\right)^2 \dot{\cal W}_{\left(10\right)}
        + \frac{8}{\left(d_{\sigma}\right)^4}\,\frac{\ve{\sigma} \cdot \ve{x}_0}{x_0}\,\dot{\cal X}_{\left(3\right)}
        + \frac{4}{\left(d_{\sigma}\right)^2}\,\frac{\ve{\sigma} \cdot \ve{x}_0}{\left(x_0\right)^3}\,\dot{\cal X}_{\left(3\right)}
        + \frac{24}{\left(d_{\sigma}\right)^2}\,\dot{\cal X}_{\left(5\right)}
        \nonumber\\
        && - \frac{12}{\left(d_{\sigma}\right)^2}\,\frac{\ve{\sigma} \cdot \ve{x}_0}{x_0}\,\dot{\cal X}_{\left(5\right)}
        - 12\,\frac{\ve{\sigma} \cdot \ve{x}_0}{\left(x_0\right)^3}\,\dot{\cal X}_{\left(5\right)}
        - 420\,\dot{\cal X}_{\left(7\right)} + 420 \left(d_{\sigma}\right)^2 \dot{\cal X}_{\left(9\right)}
        + 360 \,\dot{\cal Z}_{\left(7\right)} - 420 \left(d_{\sigma}\right)^2 \dot{\cal Z}_{\left(9\right)}\,. 
        \label{scalar_function_dot_B8_2PN}
\end{eqnarray}

\noindent 
and 
\begin{eqnarray}
        B_{\left(1\right)}\left(\ve{x}_{\rm N}\right) &=& + \frac{4}{\left(d_{\sigma}\right)^2}\,{\cal W}_{\left(4\right)} 
        + 22\,{\cal W}_{\left(6\right)} - 60 \left(x_0 + \ve{\sigma} \cdot \ve{x}_0\right) {\cal W}_{\left(7\right)}  
        + \frac{21}{2} \left(d_{\sigma}\right)^2 {\cal W}_{\left(8\right)} 
        - \frac{4}{\left(d_{\sigma}\right)^2}\,\frac{\ve{\sigma} \cdot \ve{x}_0}{x_0}\,{\cal X}_{\left(3\right)} 
        + 60\,{\cal X}_{\left(5\right)} 
        \nonumber\\ 
        && - 60 \left(d_{\sigma}\right)^2 {\cal X}_{\left(7\right)} 
        - 48\,{\cal Z}_{\left(5\right)} + 60 \left(d_{\sigma}\right)^2 {\cal Z}_{\left(7\right)} \,, 
        \label{scalar_function_B1_2PN}
        \\
        \vdots 
        \\
        B_{\left(8\right)}\left(\ve{x}_{\rm N}\right) &=& - \frac{8}{\left(d_{\sigma}\right)^4}\,{\cal W}_{\left(4\right)}
        - \frac{12}{\left(x_0\right)^3}\,{\cal W}_{\left(5\right)}
        + \frac{32}{\left(d_{\sigma}\right)^2}\,{\cal W}_{\left(6\right)}
        - \frac{120}{\left(d_{\sigma}\right)^2} \left(x_0 + \ve{\sigma} \cdot \ve{x}_0\right) {\cal W}_{\left(7\right)}
        - 63\,{\cal W}_{\left(8\right)}
        \nonumber\\
        && + 420 \left(x_0 + \ve{\sigma} \cdot \ve{x}_0\right) {\cal W}_{\left(9\right)}
        - 60 \left(d_{\sigma}\right)^2 {\cal W}_{\left(10\right)}
        + \frac{8}{\left(d_{\sigma}\right)^4}\,\frac{\ve{\sigma} \cdot \ve{x}_0}{x_0}\,{\cal X}_{\left(3\right)}
        + \frac{4}{\left(d_{\sigma}\right)^2}\,\frac{\ve{\sigma} \cdot \ve{x}_0}{\left(x_0\right)^3}\,{\cal X}_{\left(3\right)}
        + \frac{24}{\left(d_{\sigma}\right)^2}\,{\cal X}_{\left(5\right)}
        \nonumber\\
        && - \frac{12}{\left(d_{\sigma}\right)^2}\,\frac{\ve{\sigma} \cdot \ve{x}_0}{x_0}\,{\cal X}_{\left(5\right)}
        - 12\,\frac{\ve{\sigma} \cdot \ve{x}_0}{\left(x_0\right)^3}\,{\cal X}_{\left(5\right)}
        - 420\,{\cal X}_{\left(7\right)} + 420 \left(d_{\sigma}\right)^2 {\cal X}_{\left(9\right)}
        + 360 \,{\cal Z}_{\left(7\right)} - 420 \left(d_{\sigma}\right)^2 {\cal Z}_{\left(9\right)}\,.
        \label{scalar_function_B8_2PN}
\end{eqnarray}

\noindent 
where the scalar functions $\dot{\cal W}_{\left(n\right)}$, $\dot{\cal X}_{\left(n\right)}$, $\dot{\cal Y}_{\left(n\right)}$, $\dot{\cal Z}_{\left(n\right)}$ and 
${\cal W}_{\left(n\right)}$, ${\cal X}_{\left(n\right)}$, ${\cal Y}_{\left(n\right)}$, ${\cal Z}_{\left(n\right)}$
are given in \cite{Zschocke_Quadrupole_1}; the argument $\ve{x}_{\rm N}\left(t\right)$ of these functions has been omitted here.

\subsection{Scalar functions of the 2PN Quadrupole-Quadrupole term in (\ref{First_Integration_2PN_Terms_old}) and (\ref{Second_Integration_2PN_Terms_old})} 

By comparing with Eqs.~(85) and (91) in \cite{Zschocke_Quadrupole_1} one obtains the scalar functions for the 2PN quadrupole-quadrupole terms 
in (\ref{First_Integration_2PN_Terms_old}) and (\ref{Second_Integration_2PN_Terms_old}): 
\begin{eqnarray}
        \dot{C}_{\left(1\right)}\left(\ve{x}_{\rm N}\right) &=& - \frac{12}{\left(d_{\sigma}\right)^2}\,\dot{\cal W}_{\left(6\right)} 
	+ 9\,\dot{\cal W}_{\left(8\right)} - 13 \left(d_{\sigma}\right)^2 \dot{\cal W}_{\left(10\right)} 
	+ \frac{12}{\left(d_{\sigma}\right)^2}\,\frac{\ve{\sigma} \cdot \ve{x}_0}{x_0}\,\dot{\cal X}_{\left(5\right)}\,, 
        \label{scalar_function_dot_C1_2PN}
	\\
	\vdots
	\\ 
        \dot{C}_{\left(28\right)}\left(\ve{x}_{\rm N}\right) &=& + \frac{120}{\left(d_{\sigma}\right)^6}\,\dot{\cal W}_{\left(7\right)} 
	- \frac{420}{\left(d_{\sigma}\right)^4}\,\dot{\cal W}_{\left(9\right)}
	+ \frac{210}{\left(d_{\sigma}\right)^4}\,\frac{\ve{\sigma} \cdot \ve{x}_0}{x_0}\,\dot{\cal W}_{\left(9\right)}
	+ \frac{105}{\left(d_{\sigma}\right)^2}\,\frac{\ve{\sigma} \cdot \ve{x}_0}{\left(x_0\right)^3}\,\dot{\cal W}_{\left(9\right)}
	+ \frac{120}{\left(d_{\sigma}\right)^6}\,\dot{\cal X}_{\left(6\right)} 
	+ \frac{180}{\left(d_{\sigma}\right)^4}\,\frac{\dot{\cal X}_{\left(7\right)}}{x_0}
	\nonumber\\ 
	&& + \frac{45}{\left(d_{\sigma}\right)^2}\,\frac{\dot{\cal X}_{\left(7\right)}}{\left(x_0\right)^3}
	- \frac{360}{\left(d_{\sigma}\right)^6} \left(x_0 + \ve{\sigma} \cdot \ve{x}_0\right) \dot{\cal X}_{\left(7\right)} 
	- \frac{690}{\left(d_{\sigma}\right)^4}\,\dot{\cal X}_{\left(8\right)}
	+ \frac{300}{\left(d_{\sigma}\right)^2}\,\dot{\cal X}_{\left(10\right)}
	+ \frac{1005}{2}\,\dot{\cal X}_{\left(12\right)} 
	- \frac{225}{2} \left(d_{\sigma}\right)^2 \dot{\cal X}_{\left(14\right)}\,,
	\nonumber\\ 
        \label{scalar_function_dot_C28_2PN}
\end{eqnarray}

\noindent 
and 
\begin{eqnarray}
        C_{\left(1\right)}\left(\ve{x}_{\rm N}\right) &=& - \frac{12}{\left(d_{\sigma}\right)^2}\,{\cal W}_{\left(6\right)} 
        + 9\,{\cal W}_{\left(8\right)} - 13 \left(d_{\sigma}\right)^2 {\cal W}_{\left(10\right)} 
	+ \frac{12}{\left(d_{\sigma}\right)^2}\,\frac{\ve{\sigma} \cdot \ve{x}_0}{x_0}\,{\cal X}_{\left(5\right)}\,, 
        \label{scalar_function_C1_2PN}
        \\
	\vdots
	\\
        C_{\left(28\right)}\left(\ve{x}_{\rm N}\right) &=& + \frac{120}{\left(d_{\sigma}\right)^6}\,{\cal W}_{\left(7\right)} 
        - \frac{420}{\left(d_{\sigma}\right)^4}\,{\cal W}_{\left(9\right)}
        + \frac{210}{\left(d_{\sigma}\right)^4}\,\frac{\ve{\sigma} \cdot \ve{x}_0}{x_0}\,{\cal W}_{\left(9\right)}
        + \frac{105}{\left(d_{\sigma}\right)^2}\,\frac{\ve{\sigma} \cdot \ve{x}_0}{\left(x_0\right)^3}\,{\cal W}_{\left(9\right)}
        + \frac{120}{\left(d_{\sigma}\right)^6}\,{\cal X}_{\left(6\right)} 
        + \frac{180}{\left(d_{\sigma}\right)^4}\,\frac{{\cal X}_{\left(7\right)}}{x_0} 
        \nonumber\\ 
        && + \frac{45}{\left(d_{\sigma}\right)^2}\,\frac{{\cal X}_{\left(7\right)}}{\left(x_0\right)^3}
        - \frac{360}{\left(d_{\sigma}\right)^6} \left(x_0 + \ve{\sigma} \cdot \ve{x}_0\right) {\cal X}_{\left(7\right)} 
        - \frac{690}{\left(d_{\sigma}\right)^4}\,{\cal X}_{\left(8\right)}
        + \frac{300}{\left(d_{\sigma}\right)^2}\,{\cal X}_{\left(10\right)}
        + \frac{1005}{2}\,{\cal X}_{\left(12\right)} 
        - \frac{225}{2} \left(d_{\sigma}\right)^2 {\cal X}_{\left(14\right)}\,,
	\nonumber\\
        \label{scalar_function_C28_2PN}
\end{eqnarray}

\noindent 
where the scalar functions $\dot{\cal W}_{\left(n\right)}$, $\dot{\cal X}_{\left(n\right)}$, $\dot{\cal Y}_{\left(n\right)}$, $\dot{\cal Z}_{\left(n\right)}$ and 
${\cal W}_{\left(n\right)}$, ${\cal X}_{\left(n\right)}$, ${\cal Y}_{\left(n\right)}$, ${\cal Z}_{\left(n\right)}$
are given in \cite{Zschocke_Quadrupole_1}; the argument $\ve{x}_{\rm N}\left(t\right)$ of these functions has been omitted here.

\clearpage 

\end{widetext}

\section{Tensorial coefficients in (\ref{New_First_Integration_1PN}) - (\ref{New_Second_Integration_1PN})}\label{Appendix4a} 

The tensorial coefficients in (\ref{New_First_Integration_1PN}) - (\ref{New_Second_Integration_1PN}) are given by 
\begin{eqnarray}
        U^{i}_{\left(1\right)}\left(\ve{x}_{\rm 1PN}\right) &=& \sigma^i\;,
        \label{coefficient_U1_1PN}
        \\
	U^{i}_{\left(2\right)}\left(\ve{x}_{\rm 1PN}\right) &=& \widehat{d_{\sigma}}^i \;.
        \label{coefficient_U2_1PN}
	\\
        V^{ab\,i}_{\left(1\right)}\left(\ve{x}_{\rm 1PN}\right) &=& \sigma^a \delta^{bi}\;, 
        \label{coefficient_V1_1PN} 
        \\
	V^{ab\,i}_{\left(2\right)}\left(\ve{x}_{\rm 1PN}\right) &=& \widehat{d_{\sigma}}^a \delta^{bi}\;, 
        \label{coefficient_V2_1PN}
        \\
        V^{ab\,i}_{\left(3\right)}\left(\ve{x}_{\rm 1PN}\right) &=& \sigma^a \sigma^b \sigma^i\;, 
        \label{coefficient_V3_1PN} 
        \\
	V^{ab\,i}_{\left(4\right)}\left(\ve{x}_{\rm 1PN}\right) &=& \sigma^a \widehat{d_{\sigma}}^b \sigma^i\;, 
        \label{coefficient_V4_1PN} 
        \\
	V^{ab\,i}_{\left(5\right)}\left(\ve{x}_{\rm 1PN}\right) &=& \widehat{d_{\sigma}}^a \;\widehat{d_{\sigma}}^b \;\sigma^i\;, 
        \label{coefficient_V5_1PN} 
        \\
	V^{ab\,i}_{\left(6\right)}\left(\ve{x}_{\rm 1PN}\right) &=& \widehat{d_{\sigma}}^a \;\widehat{d_{\sigma}}^b \;\widehat{d_{\sigma}}^i\;, 
        \label{coefficient_V6_1PN} 
        \\
	V^{ab\,i}_{\left(7\right)}\left(\ve{x}_{\rm 1PN}\right) &=& \sigma^a \sigma^b \;\widehat{d_{\sigma}}^i\;, 
        \label{coefficient_V7_1PN} 
        \\
	V^{ab\,i}_{\left(8\right)}\left(\ve{x}_{\rm 1PN}\right) &=& \sigma^a \;\widehat{d_{\sigma}}^b \;\widehat{d_{\sigma}}^i\;. 
        \label{coefficient_V8_1PN} 
\end{eqnarray}

\section{Scalar functions in (\ref{New_First_Integration_1PN}) - (\ref{New_Second_Integration_1PN})}\label{Appendix4b} 

The scalar functions for the monopole term in Eqs.~(\ref{New_First_Integration_1PN}) - (\ref{New_Second_Integration_1PN}) are given by 
\begin{eqnarray}
        \dot{F}_{\left(1\right)}\left(\ve{x}_{\rm 1PN}\right) &=& + 2\,\dot{\cal W}_{\left(3\right)}\left(\ve{x}_{\rm 1PN}\right)\,,
        \label{scalar_dot_F1_1PN}
        \\
        \dot{F}_{\left(2\right)}\left(\ve{x}_{\rm 1PN}\right) &=& - 2\,\dot{\cal X}_{\left(3\right)}\left(\ve{x}_{\rm 1PN}\right)\,,
        \label{scalar_dot_F2_1PN}
\end{eqnarray}

\noindent
with
\begin{eqnarray}
        \dot{\cal W}_{\left(3\right)}\left(\ve{x}_{\rm 1PN}\right) &=& - \frac{1}{x_{\rm 1PN}}\,,
        \label{scalar_dot_W3_1PN}
        \\
        \dot{\cal X}_{\left(3\right)}\left(\ve{x}_{\rm 1PN}\right) &=& 
	+ \frac{1}{(\widehat{d}_{\sigma})^2} \left(1 + \frac{\ve{\sigma} \cdot \ve{x}_{\rm 1PN}}{x_{\rm 1PN}}\right), 
        \label{scalar_dot_X3_1PN}
\end{eqnarray}

\noindent
and
\begin{eqnarray}
        F_{\left(1\right)}\left(\ve{x}_{\rm 1PN}\right) &=& + 2\,{\cal W}_{\left(3\right)}\left(\ve{x}_{\rm 1PN}\right)\,,
        \label{scalar_F1_1PN}
        \\
        F_{\left(2\right)}\left(\ve{x}_{\rm 1PN}\right) &=& - 2\,{\cal X}_{\left(3\right)}\left(\ve{x}_{\rm 1PN}\right)\,, 
        \label{scalar_F2_1PN}
\end{eqnarray}

\noindent
with
\begin{eqnarray}
        {\cal W}_{\left(3\right)}\left(\ve{x}_{\rm 1PN}\right) &=& + \ln \left(x_{\rm 1PN} - \ve{\sigma} \cdot \ve{x}_{\rm 1PN}\right),
        \label{scalar_W3_1PN}
        \\
        {\cal X}_{\left(3\right)}\left(\ve{x}_{\rm 1PN}\right) &=& 
	+ \frac{1}{(\widehat{d}_{\sigma})^2} \left(x_{\rm 1PN} + \ve{\sigma} \cdot \ve{x}_{\rm 1PN}\right), 
        \label{scalar_X3_1PN}
\end{eqnarray}

\noindent
where $\ve{x}_{\rm 1PN} = \ve{x}_{\rm 1PN}\left(t\right)$. 

The scalar functions for the quadrupole term in Eqs.~(\ref{New_First_Integration_1PN}) - (\ref{New_Second_Integration_1PN}) are given by
\begin{eqnarray}
        \dot{G}_{\left(1\right)}\left(\ve{x}_{\rm 1PN}\right) &=& + 6\,\dot{\cal W}_{\left(5\right)}\left(\ve{x}_{\rm 1PN}\right)\,, 
        \label{scalar_dot_G1_1PN}
        \\
        \dot{G}_{\left(2\right)}\left(\ve{x}_{\rm 1PN}\right) &=& + 6\,\dot{\cal X}_{\left(5\right)}\left(\ve{x}_{\rm 1PN}\right)\,,
        \label{scalar_dot_G2_1PN}
        \\
        \dot{G}_{\left(3\right)}\left(\ve{x}_{\rm 1PN}\right) &=& + 3\,\dot{\cal W}_{\left(5\right)}\left(\ve{x}_{\rm 1PN}\right) 
        - 15 (\widehat{d}_{\sigma})^2\, \dot{\cal W}_{\left(7\right)}\left(\ve{x}_{\rm 1PN}\right)\,,
	\nonumber\\ 
        \label{scalar_dot_G3_1PN}
        \\
        \dot{G}_{\left(4\right)}\left(\ve{x}_{\rm 1PN}\right) &=& + 18\,\dot{\cal X}_{\left(5\right)}\left(\ve{x}_{\rm 1PN}\right) 
        - 30 (\widehat{d}_{\sigma})^2 \, \dot{\cal X}_{\left(7\right)}\left(\ve{x}_{\rm 1PN}\right)\,,
	\nonumber\\ 
        \label{scalar_dot_G4_1PN}
        \\
        \dot{G}_{\left(5\right)}\left(\ve{x}_{\rm 1PN}\right) &=& + 15\,\dot{\cal W}_{\left(7\right)}\left(\ve{x}_{\rm 1PN}\right)\,, 
        \label{scalar_dot_G5_1PN}
        \\
        \dot{G}_{\left(6\right)}\left(\ve{x}_{\rm 1PN}\right) &=& - 15\,\dot{\cal X}_{\left(7\right)}\left(\ve{x}_{\rm 1PN}\right)\,, 
       \label{scalar_dot_G6_1PN}
        \\
        \dot{G}_{\left(7\right)}\left(\ve{x}_{\rm 1PN}\right) &=& - 15\,\dot{\cal X}_{\left(5\right)}\left(\ve{x}_{\rm 1PN}\right) 
        + 15 (\widehat{d}_{\sigma})^2 \,\dot{\cal X}_{\left(7\right)}\left(\ve{x}_{\rm 1PN}\right)\,,
	\nonumber\\ 
        \label{scalar_dot_G7_1PN}
        \\
        \dot{G}_{\left(8\right)}\left(\ve{x}_{\rm 1PN}\right) &=& - 30\,\dot{\cal W}_{\left(7\right)}\left(\ve{x}_{\rm 1PN}\right)\,, 
        \label{scalar_dot_G8_1PN}
\end{eqnarray}

\noindent
with  
\begin{eqnarray}
        \dot{\cal W}_{\left(5\right)}\left(\ve{x}_{\rm 1PN}\right) &=& - \frac{1}{3}\,\frac{1}{\left(x_{\rm 1PN}\right)^3}\,,
        \label{scalar_dot_W5_1PN}
        \\
        \dot{\cal W}_{\left(7\right)}\left(\ve{x}_{\rm 1PN}\right) &=& - \frac{1}{5}\,\frac{1}{\left(x_{\rm 1PN}\right)^5}\,,  
        \label{scalar_dot_W7_1PN}
        \\
	\dot{\cal X}_{\left(5\right)}\left(\ve{x}_{\rm 1PN}\right) &=& + \frac{2}{3}\,\frac{1}{(\widehat{d}_{\sigma})^2}
	\nonumber\\ 
	&& \hspace{-2.0cm} \times \left(\frac{1}{(\widehat{d}_{\sigma})^2} + \frac{1}{(\widehat{d}_{\sigma})^2}\,\frac{\ve{\sigma} \cdot \ve{x}_{\rm 1PN}}{x_{\rm 1PN}} 
        + \frac{1}{2}\,\frac{\ve{\sigma} \cdot \ve{x}_{\rm 1PN}}{\left(x_{\rm 1PN}\right)^3}\right), 
        \label{scalar_dot_X5_1PN}
        \\
	\dot{\cal X}_{\left(7\right)}\left(\ve{x}_{\rm 1PN}\right) &=& + \frac{8}{15}\,\frac{1}{(\widehat{d}_{\sigma})^2} 
	\Bigg(\frac{1}{(\widehat{d}_{\sigma})^4} + \frac{1}{(\widehat{d}_{\sigma})^4}\,\frac{\ve{\sigma} \cdot \ve{x}_{\rm 1PN}}{x_{\rm 1PN}}
	\nonumber\\ 
	&& \hspace{-2.0cm} + \frac{1}{2}\,\frac{1}{(\widehat{d}_{\sigma})^2}\,\frac{\ve{\sigma} \cdot \ve{x}_{\rm 1PN}}{\left(x_{\rm 1PN}\right)^3} 
        + \frac{3}{8}\,\frac{\ve{\sigma} \cdot \ve{x}_{\rm 1PN}}{\left(x_{\rm 1PN}\right)^5}\Bigg), 
        \label{scalar_dot_X7_1PN}
\end{eqnarray}

\noindent
and
\begin{eqnarray}
        G_{\left(1\right)}\left(\ve{x}_{\rm 1PN}\right) &=& + 6\,{\cal W}_{\left(5\right)}\left(\ve{x}_{\rm 1PN}\right)\,, 
        \label{scalar_G1_1PN}
        \\
        G_{\left(2\right)}\left(\ve{x}_{\rm 1PN}\right) &=& + 6\,{\cal X}_{\left(5\right)}\left(\ve{x}_{\rm 1PN}\right)\,, 
        \label{scalar_G2_1PN}
        \\
        G_{\left(3\right)}\left(\ve{x}_{\rm 1PN}\right) &=& + 3\,{\cal W}_{\left(5\right)}\left(\ve{x}_{\rm 1PN}\right) 
        - 15 (\widehat{d}_{\sigma})^2 \,{\cal W}_{\left(7\right)}\left(\ve{x}_{\rm 1PN}\right)\,, 
	\nonumber\\ 
        \label{scalar_G3_1PN}
        \\
        G_{\left(4\right)}\left(\ve{x}_{\rm 1PN}\right) &=& + 18\,{\cal X}_{\left(5\right)}\left(\ve{x}_{\rm 1PN}\right) 
        - 30 (\widehat{d}_{\sigma})^2 \, {\cal X}_{\left(7\right)}\left(\ve{x}_{\rm 1PN}\right)\,,
	\nonumber\\ 
        \label{scalar_G4_1PN}
        \\
        G_{\left(5\right)}\left(\ve{x}_{\rm 1PN}\right) &=& + 15\,{\cal W}_{\left(7\right)}\left(\ve{x}_{\rm 1PN}\right)\,,
        \label{scalar_G5_1PN}
        \\
        G_{\left(6\right)}\left(\ve{x}_{\rm 1PN}\right) &=& - 15\,{\cal X}_{\left(7\right)}\left(\ve{x}_{\rm 1PN}\right)\,, 
        \label{scalar_G6_1PN}
        \\
        G_{\left(7\right)}\left(\ve{x}_{\rm 1PN}\right) &=& - 15\,{\cal X}_{\left(5\right)}\left(\ve{x}_{\rm 1PN}\right) 
        + 15 (\widehat{d}_{\sigma})^2\, {\cal X}_{\left(7\right)}\left(\ve{x}_{\rm 1PN}\right)\,,
	\nonumber\\ 
        \label{scalar_G7_1PN}
        \\
        G_{\left(8\right)}\left(\ve{x}_{\rm 1PN}\right) &=& - 30\,{\cal W}_{\left(7\right)}\left(\ve{x}_{\rm 1PN}\right)\,, 
        \label{scalar_G8_1PN}
\end{eqnarray}

\noindent
with
\begin{eqnarray}
	{\cal W}_{\left(5\right)}\left(\ve{x}_{\rm 1PN}\right) &=& - \frac{1}{3}\,\frac{1}{(\widehat{d}_{\sigma})^2}\,\frac{\ve{\sigma} \cdot \ve{x}_{\rm 1PN}}{x_{\rm 1PN}}\,,
        \label{scalar_W5_1PN}
        \\
	{\cal W}_{\left(7\right)}\left(\ve{x}_{\rm 1PN}\right) &=& - \frac{2}{15}\,\frac{1}{(\widehat{d}_{\sigma})^2} 
	\left(\frac{1}{(\widehat{d}_{\sigma})^2}\frac{\ve{\sigma} \cdot \ve{x}_{\rm 1PN}}{x_{\rm 1PN}} 
	+ \frac{1}{2}\,\frac{\ve{\sigma} \cdot \ve{x}_{\rm 1PN}}{\left(x_{\rm 1PN}\right)^3}\right),
	\nonumber\\ 
        \label{scalar_W7_1PN}
        \\
	{\cal X}_{\left(5\right)}\left(\ve{x}_{\rm 1PN}\right) &=& + \frac{2}{3}\,\frac{1}{(\widehat{d}_{\sigma})^2}
	\nonumber\\ 
	&& \hspace{-2.0cm} \times \left(\frac{x_{\rm 1PN} + \ve{\sigma} \cdot \ve{x}_{\rm 1PN}}{(\widehat{d}_{\sigma})^2} - \frac{1}{2}\,\frac{1}{x_{\rm 1PN}}\right), 
        \label{scalar_X5_1PN}
        \\
	{\cal X}_{\left(7\right)}\left(\ve{x}_{\rm 1PN}\right) &=& + \frac{8}{15}\,\frac{1}{(\widehat{d}_{\sigma})^2} 
	\nonumber\\ 
	&& \hspace{-2.0cm} 
	\times \Bigg(\frac{x_{\rm 1PN} + \ve{\sigma} \cdot \ve{x}_{\rm 1PN}}{(\widehat{d}_{\sigma})^4} - \frac{1}{2}\,\frac{1}{(\widehat{d}_{\sigma})^2}\,\frac{1}{x_{\rm 1PN}} 
        - \frac{1}{8}\,\frac{1}{\left(x_{\rm 1PN}\right)^3}\Bigg), 
        \label{scalar_X7_1PN}
\end{eqnarray}

\noindent 
where $\ve{x}_{\rm 1PN} = \ve{x}_{\rm 1PN} \left(t\right)$ is given by Eq.~(\ref{Appendix_x_A}) and 
the impact vector $\widehat{\ve{d}}_{\sigma}$ has been defined by Eq.~(\ref{impact_vector_x_1PN}).

\section{Tensorial coefficients in (\ref{First_Integration_1PN_Final}) and (\ref{Second_Integration_1PN_Final})}\label{Tensorial_Coefficients_1PN_Final}

The tensorial coefficients of the 1PN perturbation terms in (\ref{First_Integration_1PN_Final}) and (\ref{Second_Integration_1PN_Final}) are given by
\begin{eqnarray}
        U^{i}_{\left(1\right)}\left(\ve{x}_1\right) &=& \sigma^i\;,
        \label{coefficient_U1_Final}
        \\
	U^{i}_{\left(2\right)}\left(\ve{x}_1\right) &=& d_{\sigma}^{\,1\,i}\;,
        \label{coefficient_U2_Final}
        \\
        V^{ab\,i}_{\left(1\right)}\left(\ve{x}_1\right) &=& \sigma^a \delta^{bi}\;,
        \label{coefficient_V1_Final}
        \\
	V^{ab\,i}_{\left(2\right)}\left(\ve{x}_1\right) &=& d_{\sigma}^{\,1\,a} \delta^{bi}\;,
        \label{coefficient_V2_Final}
        \\
        V^{ab\,i}_{\left(3\right)}\left(\ve{x}_1\right) &=& \sigma^a \sigma^b \sigma^i\;,
        \label{coefficient_V3_Final}
        \\
	V^{ab\,i}_{\left(4\right)}\left(\ve{x}_1\right) &=& \sigma^a d_{\sigma}^{\,1\,b} \sigma^i\;,
        \label{coefficient_V4_Final}
        \\
	V^{ab\,i}_{\left(5\right)}\left(\ve{x}_1\right) &=& d_{\sigma}^{\,1\,a} d_{\sigma}^{\,1\,b} \sigma^i\;,
        \label{coefficient_V5_Final}
        \\
	V^{ab\,i}_{\left(6\right)}\left(\ve{x}_1\right) &=& d_{\sigma}^{\,1\,a} d_{\sigma}^{\,1\,b} d_{\sigma}^{\,1\,i}\;,
        \label{coefficient_V6_Final}
        \\
	V^{ab\,i}_{\left(7\right)}\left(\ve{x}_1\right) &=& \sigma^a \sigma^b d_{\sigma}^{\,1\,i}\;,
        \label{coefficient_V7_Final}
        \\
	V^{ab\,i}_{\left(8\right)}\left(\ve{x}_1\right) &=& \sigma^a d_{\sigma}^{\,1\,b} d_{\sigma}^{\,1\,i}\;,
        \label{coefficient_V8_Final}
\end{eqnarray}

\noindent 
where the impact vector $\ve{d}^{\,1}_{\sigma}$ is defined by Eq.~(\ref{impact_vector_1}).

\section{Tensorial coefficients in (\ref{First_Integration_2PN_Final}) and (\ref{Second_Integration_2PN_Final})}\label{Tensorial_Coefficients_2PN_Final}

The tensorial coefficients of the 2PN perturbation terms in (\ref{First_Integration_2PN_Final}) and (\ref{Second_Integration_2PN_Final}) are given by
\begin{eqnarray}
        W^{abcd\,i}_{\left(1\right)}\left(\ve{x}_1\right) &=& \delta^{ac} \sigma^b \delta^{di} \;, 
        \label{coefficient_W1_Final}
        \\
	W^{abcd\,i}_{\left(2\right)}\left(\ve{x}_1\right) &=& \delta^{ac} d_{\sigma}^{\,1\,b} \delta^{di} \;,
        \label{coefficient_W2_Final}
        \\
        W^{abcd\,i}_{\left(3\right)}\left(\ve{x}_1\right) &=& \sigma^a \sigma^b \sigma^c \delta^{di} \;, 
        \label{coefficient_W3_Final}
        \\
        W^{abcd\,i}_{\left(4\right)}\left(\ve{x}_1\right) &=& \sigma^a \sigma^b d_{\sigma}^{\,1\,c} \delta^{di} \;,
        \label{coefficient_W4_Final}
        \\
	W^{abcd\,i}_{\left(5\right)}\left(\ve{x}_1\right) &=& \sigma^a d_{\sigma}^{\,1\,b} \sigma^c \delta^{di} \;, 
        \label{coefficient_W5_Final}
        \\
        W^{abcd\,i}_{\left(6\right)}\left(\ve{x}_1\right) &=& \sigma^a d_{\sigma}^{\,1\,b} d_{\sigma}^{\,1\,c} \delta^{di} \;, 
        \label{coefficient_W6_Final}
        \\
        W^{abcd\,i}_{\left(7\right)}\left(\ve{x}_1\right) &=& d_{\sigma}^{\,1\,a} d_{\sigma}^{\,1\,b} \sigma^c \delta^{di} \;,
        \label{coefficient_W7_Final}
        \\
        W^{abcd\,i}_{\left(8\right)}\left(\ve{x}_1\right) &=& d_{\sigma}^{\,1\,a} d_{\sigma}^{\,1\,b} d_{\sigma}^{\,1\,c} \delta^{di} \;, 
        \label{coefficient_W8_Final}
        \\
        W^{abcd\,i}_{\left(9\right)}\left(\ve{x}_1\right) &=& \delta^{ac} \delta^{bd} \sigma^i \;, 
        \label{coefficient_W9_Final}
        \\
        W^{abcd\,i}_{\left(10\right)}\left(\ve{x}_1\right) &=& \delta^{ac} \sigma^b \sigma^d \sigma^i \;,
        \label{coefficient_W10_Final} 
        \\
        W^{abcd\,i}_{\left(11\right)}\left(\ve{x}_1\right) &=& \delta^{ac} \sigma^b d_{\sigma}^{\,1\,d} \sigma^i \;,
        \label{coefficient_W11_Final}
        \\
        W^{abcd\,i}_{\left(12\right)}\left(\ve{x}_1\right) &=& \delta^{ac} d_{\sigma}^{\,1\,b} d_{\sigma}^{\,1\,d} \sigma^i \;,
        \label{coefficient_W12_Final}
        \\
        W^{abcd\,i}_{\left(13\right)}\left(\ve{x}_1\right) &=& \sigma^a \sigma^b \sigma^c \sigma^d \sigma^i \;, 
        \label{coefficient_W13_Final}
        \\
        W^{abcd\,i}_{\left(14\right)}\left(\ve{x}_1\right) &=& \sigma^a \sigma^b \sigma^c d_{\sigma}^{\,1\,d} \sigma^i \;,
        \label{coefficient_W14_Final}
        \\ 
        W^{abcd\,i}_{\left(15\right)}\left(\ve{x}_1\right) &=& \sigma^a \sigma^b d_{\sigma}^{\,1\,c} d_{\sigma}^{\,1\,d} \sigma^i \;,
        \label{coefficient_W15_Final}
        \\
        W^{abcd\,i}_{\left(16\right)}\left(\ve{x}_1\right) &=& \sigma^a d_{\sigma}^{\,1\,b} \sigma^c d_{\sigma}^{\,1\,d} \sigma^i \;,
        \label{coefficient_W16_Final}
        \\ 
        W^{abcd\,i}_{\left(17\right)}\left(\ve{x}_1\right) &=& \sigma^a d_{\sigma}^{\,1\,b} d_{\sigma}^{\,1\,c} d_{\sigma}^{\,1\,d} \sigma^i \;,
        \label{coefficient_W17_Final}
        \\
        W^{abcd\,i}_{\left(18\right)}\left(\ve{x}_1\right) &=& d_{\sigma}^{\,1\,a} d_{\sigma}^{\,1\,b} d_{\sigma}^{\,1\,c} d_{\sigma}^{\,1\,d} \sigma^i \;,
        \label{coefficient_W18_Final}
        \\ 
        W^{abcd\,i}_{\left(19\right)}\left(\ve{x}_1\right) &=& \delta^{ac} \delta^{bd} d_{\sigma}^{\,1\,i} \;,
        \label{coefficient_W19_Final} 
        \\
        W^{abcd\,i}_{\left(20\right)}\left(\ve{x}_1\right) &=& \delta^{ac} \sigma^b \sigma^d d_{\sigma}^{\,1\,i} \;,
        \label{coefficient_W20_Final} 
        \\
        W^{abcd\,i}_{\left(21\right)}\left(\ve{x}_1\right) &=& \delta^{ac} \sigma^b d_{\sigma}^{\,1\,d} d_{\sigma}^{\,1\,i} \;,
        \label{coefficient_W21_Final}
        \\
        W^{abcd\,i}_{\left(22\right)}\left(\ve{x}_1\right) &=& \delta^{ac} d_{\sigma}^{\,1\,b} d_{\sigma}^{\,1\,d} d_{\sigma}^{\,1\,i} \;,
        \label{coefficient_W22_Final}
        \\
        W^{abcd\,i}_{\left(23\right)}\left(\ve{x}_1\right) &=& \sigma^a \sigma^b \sigma^c \sigma^d d_{\sigma}^{\,1\,i} \;,
        \label{coefficient_W23_Final}
        \\
        W^{abcd\,i}_{\left(24\right)}\left(\ve{x}_1\right) &=& \sigma^a \sigma^b \sigma^c d_{\sigma}^{\,1\,d} d_{\sigma}^{\,1\,i}\;,
        \label{coefficient_W24_Final}
        \\
        W^{abcd\,i}_{\left(25\right)}\left(\ve{x}_1\right) &=& \sigma^a \sigma^b d_{\sigma}^{\,1\,c} d_{\sigma}^{\,1\,d} d_{\sigma}^{\,1\,i}\;,
        \label{coefficient_W25_Final}
        \\
        W^{abcd\,i}_{\left(26\right)}\left(\ve{x}_1\right) &=& \sigma^a d_{\sigma}^{\,1\,b} \sigma^c d_{\sigma}^{\,1\,d} d_{\sigma}^{\,1\,i} \;,
        \label{coefficient_W26_Final}
        \\
        W^{abcd\,i}_{\left(27\right)}\left(\ve{x}_1\right) &=& \sigma^a d_{\sigma}^{\,1\,b} d_{\sigma}^{\,1\,c} d_{\sigma}^{\,1\,d} d_{\sigma}^{\,1\,i} \;,
        \label{coefficient_W27_Final}
        \\
        W^{abcd\,i}_{\left(28\right)}\left(\ve{x}_1\right) &=& d_{\sigma}^{\,1\,a} d_{\sigma}^{\,1\,b} d_{\sigma}^{\,1\,c} d_{\sigma}^{\,1\,d} d_{\sigma}^{\,1\,i} \;, 
        \label{coefficient_W28_Final}
\end{eqnarray}

\noindent
where the impact vector $\ve{d}^{\,1}_{\sigma}$ is defined by Eq.~(\ref{impact_vector_1}).

\section{Scalar functions of 1PN solution in (\ref{First_Integration_1PN_Final}) and (\ref{Second_Integration_1PN_Final})}\label{Appendix5}

In order to simplify the notation, it is useful to introduce the following abbreviations:
\begin{eqnarray}
	{a}_{\left(n\right)} &=& \left(x_1 + \ve{\sigma} \cdot \ve{x}_1\right)^n\,,
        \label{new_a_n}
        \\
        {b}_{\left(n\right)} &=& \frac{1}{\left(x_1\right)^n}\,,
        \label{new_b_n}
        \\
        {c}_{\left(n\right)} &=& \frac{\ve{\sigma} \cdot \ve{x}_1}{\left(x_1\right)^n}\,,
        \label{new_c_n}
        \\
        {d}_{\left(1\right)} &=& \ln \left(x_1 - \ve{\sigma} \cdot \ve{x}_1\right),
        \label{new_d_1}
        \\
	{d}_{\left(2\right)} &=& \arctan \frac{\ve{\sigma} \cdot \ve{x}_1}{d^{\,1}_{\sigma}} + \frac{\pi}{2}\,,
        \label{new_d_2}
        \\
	{d}_{\left(3\right)} &=& \arctan \frac{\ve{\sigma} \cdot \ve{x}_1}{d^{\,1}_{\sigma}}\,,
        \label{new_d_3}
        \\
	{d}_{\left(4\right)} &=& \frac{\ve{\sigma} \cdot \ve{x}_1}{d^{\,1}_{\sigma}}
	\left(\arctan \frac{\ve{\sigma} \cdot \ve{x}_1}{d^{\,1}_{\sigma}} + \frac{\pi}{2}\right), 
        \label{new_d_4}
\end{eqnarray}

\noindent 
where $n =1, 2, 3, \dots$ in Eqs.~(\ref{new_a_n}) - (\ref{new_c_n}) is a natural number. 

\subsection{Scalar functions of the 1PN monopole term in (\ref{First_Integration_1PN_Final}) and (\ref{Second_Integration_1PN_Final})}

\noindent
The scalar functions of the 1PN monopole term in Eqs.~(\ref{First_Integration_1PN_Final}) and (\ref{Second_Integration_1PN_Final}) are given by
\begin{eqnarray}
	\dot{F}_{\left(1\right)}\left(\ve{x}_1\right) &=& - 2\,{b}_{\left(1\right)}\,,
        \label{new_scalar_dot_F1}
        \\
	\dot{F}_{\left(2\right)}\left(\ve{x}_1\right) &=& - \frac{2}{({d^{\,1}_{\sigma}})^2}\left(1 + {c}_{\left(1\right)}\right),
        \label{new_scalar_dot_F2}
        \\
	F_{\left(1\right)}\left(\ve{x}_1\right) &=& + 2\,{d}_{\left(1\right)}\,,
        \label{new_scalar_F1}
        \\
	F_{\left(2\right)}\left(\ve{x}_1\right) &=& - \frac{2}{({d^{\,1}_{\sigma}})^2}\,{a}_{\left(1\right)}\,.  
        \label{new_scalar_F2}
\end{eqnarray}

\subsection{Scalar functions of the 1PN quadrupole term in (\ref{First_Integration_1PN_Final}) and (\ref{Second_Integration_1PN_Final})}

\noindent
The scalar functions of the 1PN quadrupole term in Eqs.~(\ref{First_Integration_1PN_Final}) and (\ref{Second_Integration_1PN_Final}) are given by
\begin{eqnarray}
	\dot{G}_{\left(1\right)}\left(\ve{x}_1\right) &=& - 2\,{b}_{\left(3\right)}\,,
        \label{new_scalar_dot_G1}
        \\
	\dot{G}_{\left(2\right)}\left(\ve{x}_1\right) &=& + \frac{4}{({d^{\,1}_{\sigma}})^4}\left(1 + {c}_{\left(1\right)}\right) 
	+ \frac{2}{({d^{\,1}_{\sigma}})^2}\,{c}_{\left(3\right)}\,,
        \label{new_scalar_dot_G2}
        \\
        \dot{G}_{\left(3\right)}\left(\ve{x}_1\right) &=& 
	- {b}_{\left(3\right)} + 3 \,({d^{\,1}_{\sigma}})^2 \,{b}_{\left(5\right)}\,,
        \label{new_scalar_dot_G3}
        \\
	\dot{G}_{\left(4\right)}\left(\ve{x}_1\right) &=& - \frac{4}{({d^{\,1}_{\sigma}})^4}\left(1 + {c}_{\left(1\right)}\right) 
	- \frac{2}{({d^{\,1}_{\sigma}})^2}\,{c}_{\left(3\right)} - 6\,{c}_{\left(5\right)}\,,
        \label{new_scalar_dot_G4}
        \\
	\dot{G}_{\left(5\right)}\left(\ve{x}_1\right) &=& - 3\,{b}_{\left(5\right)}\,,
        \label{new_scalar_dot_G5}
        \\
	\dot{G}_{\left(6\right)}\left(\ve{x}_1\right) &=& - \frac{8}{({d^{\,1}_{\sigma}})^6}\left(1 + {c}_{\left(1\right)}\right) 
	- \frac{4}{({d^{\,1}_{\sigma}})^4}\,{c}_{\left(3\right)} - \frac{3}{({d^{\,1}_{\sigma}})^2}\,{c}_{\left(5\right)}\,,
	\nonumber\\ 
       \label{new_scalar_dot_G6}
        \\
	\dot{G}_{\left(7\right)}\left(\ve{x}_1\right) &=& - \frac{2}{({d^{\,1}_{\sigma}})^4}\left(1 + {c}_{\left(1\right)}\right) 
	- \frac{1}{({d^{\,1}_{\sigma}})^2}\,{c}_{\left(3\right)} + 3\,{c}_{\left(5\right)}\,,
        \label{new_scalar_dot_G7}
        \\
	\dot{G}_{\left(8\right)}\left(\ve{x}_1\right) &=& + 6\,{b}_{\left(5\right)}\,,  
        \label{new_scalar_dot_G8}
\end{eqnarray}

\noindent 
and 
\begin{eqnarray}
	G_{\left(1\right)}\left(\ve{x}_1\right) &=& - \frac{2}{({d^{\,1}_{\sigma}})^2}\,{c}_{\left(1\right)}\,,
        \label{new_scalar_G1}
        \\
	G_{\left(2\right)}\left(\ve{x}_1\right) &=& + \frac{4}{({d^{\,1}_{\sigma}})^4}\,{a}_{\left(1\right)} 
	- \frac{2}{({d^{\,1}_{\sigma}})^2}\,{b}_{\left(1\right)}\,,
        \label{new_scalar_G2}
        \\
	G_{\left(3\right)}\left(\ve{x}_1\right) &=& + \frac{{c}_{\left(1\right)}}{({d^{\,1}_{\sigma}})^2} 
	+ {c}_{\left(3\right)}\,,
        \label{new_scalar_G3}
        \\
	G_{\left(4\right)}\left(\ve{x}_1\right) &=& - \frac{4}{({d^{\,1}_{\sigma}})^4}\,{a}_{\left(1\right)}
	+ \frac{2}{({d^{\,1}_{\sigma}})^2}\,{b}_{\left(1\right)} + 2\,{b}_{\left(3\right)}\,,
        \label{new_scalar_G4}
        \\
	G_{\left(5\right)}\left(\ve{x}_1\right) &=& - \frac{2}{({d^{\,1}_{\sigma}})^4}\,{c}_{\left(1\right)}
	- \frac{{c}_{\left(3\right)}}{({d^{\,1}_{\sigma}})^2}\,,
        \label{new_scalar_G5}
        \\
	G_{\left(6\right)}\left(\ve{x}_1\right) &=& - \frac{8}{({d^{\,1}_{\sigma}})^6}\,{a}_{\left(1\right)}
	+ \frac{4}{({d^{\,1}_{\sigma}})^4}\,{b}_{\left(1\right)}
	+ \frac{{b}_{\left(3\right)}}{({d^{\,1}_{\sigma}})^2}\,, 
        \label{new_scalar_G6}
        \\
	G_{\left(7\right)}\left(\ve{x}_1\right) &=& - \frac{2}{({d^{\,1}_{\sigma}})^4}\,{a}_{\left(1\right)} 
	+ \frac{{b}_{\left(1\right)}}{({d^{\,1}_{\sigma}})^2} - {b}_{\left(3\right)}\,,
        \label{new_scalar_G7}
        \\
	G_{\left(8\right)}\left(\ve{x}_1\right) &=& + \frac{4}{({d^{\,1}_{\sigma}})^4}\,{c}_{\left(1\right)}
	+ \frac{2}{({d^{\,1}_{\sigma}})^2}\,{c}_{\left(3\right)}\,. 
        \label{new_scalar_G8}
\end{eqnarray}

\section{The scalar functions of the 2PN solution in (\ref{First_Integration_2PN_Final}) and (\ref{Second_Integration_2PN_Final})}\label{Appendix6}

The scalar functions of the monopole-monopole term in (\ref{First_Integration_2PN_Final}) and (\ref{Second_Integration_2PN_Final}) are given by 
\begin{eqnarray}
	\dot{X}_{\left(1\right)}\left(\ve{x}_1\right) &=& - \frac{4}{\left(d^{\,1}_{\sigma}\right)^2} \left(1 + c_{(1)}\right) 
	+ 4\,b_{(2)} + \frac{b_{(4)}}{2} \left(d^{\,1}_{\sigma}\right)^2\,,
        \label{scalar_dot_X1}
        \\
	\dot{X}_{\left(2\right)}\left(\ve{x}_1\right) &=& + \frac{8}{\left(d^{\,1}_{\sigma}\right)^4}\,a_{(1)} + \frac{4}{\left(d^{\,1}_{\sigma}\right)^2}\,b_{(1)} 
	+ \frac{17}{4}\,\frac{c_{(2)}}{\left(d^{\,1}_{\sigma}\right)^2} - \frac{c_{(4)}}{2}
	\nonumber\\ 
	&& - \frac{15}{4}\, \frac{d_{(2)}}{\left(d^{\,1}_{\sigma}\right)^3}\,,
        \label{scalar_dot_X2}
\end{eqnarray}

\noindent 
and 
\begin{eqnarray}
	X_{\left(1\right)}\left(\ve{x}_1\right) &=& + \frac{4}{\left(d^{\,1}_{\sigma}\right)^2}\,a_{(1)} + \frac{c_{(2)}}{4} - \frac{15}{4}\,\frac{d_{(3)}}{d^{\,1}_{\sigma}}\,,
        \label{scalar_X1}
        \\
	X_{\left(2\right)}\left(\ve{x}_1\right) &=& + \frac{4}{\left(d^{\,1}_{\sigma}\right)^4}\,a_{(2)} + \frac{b_{(2)}}{4} 
	- \frac{15}{4}\,\frac{d_{(4)}}{\left(d^{\,1}_{\sigma}\right)^2}\,.
        \label{scalar_X2}
\end{eqnarray}

\noindent
These functions in Eqs.~(\ref{scalar_dot_X1}) and (\ref{scalar_dot_X2}) as well as in Eqs.~(\ref{scalar_X1}) and (\ref{scalar_X2}) are in coincidence with the results 
given by Eqs.~(48) and (51) in \cite{Klioner_Zschocke} (for GR values of the PPN parameter in \cite{Klioner_Zschocke}: $\alpha = \beta = \gamma = \epsilon = 1$).


\begin{widetext}

The scalar functions of the monopole-quadrupole term in (\ref{First_Integration_2PN_Final}) and (\ref{Second_Integration_2PN_Final}) are given by
\begin{eqnarray}
	\dot{Y}_{\left(1\right)}\left(\ve{x}_1\right) &=& + \frac{4}{\left(d^{\,1}_{\sigma}\right)^4} \left(1 + c_{(1)}\right) 
	+ \frac{2}{\left(d^{\,1}_{\sigma}\right)^2}\,b_{(2)} + \frac{7}{2}\,b_{(4)} - \frac{7}{4} \left(d^{\,1}_{\sigma}\right)^2 b_{(6)} 
	+ \frac{4}{\left(d^{\,1}_{\sigma}\right)^2}\,c_{(3)}\,,  
        \label{scalar_dot_Y1}
	\\
	\dot{Y}_{\left(2\right)}\left(\ve{x}_1\right) &=& - \frac{32}{\left(d^{\,1}_{\sigma}\right)^6}\,a_{(1)} 
	- \frac{8}{\left(d^{\,1}_{\sigma}\right)^4}\,b_{(1)} + \frac{4}{\left(d^{\,1}_{\sigma}\right)^2}\,b_{(3)} 
	- \frac{303}{32}\,\frac{c_{(2)}}{\left(d^{\,1}_{\sigma}\right)^4} 
	- \frac{37}{16}\,\frac{c_{(4)}}{\left(d^{\,1}_{\sigma}\right)^2} + \frac{7}{4}\,c_{(6)}  
	+ \frac{465}{32}\,\frac{d_{(2)}}{\left(d^{\,1}_{\sigma}\right)^5}\,,
        \label{scalar_dot_Y2}
        \\
	\dot{Y}_{\left(3\right)}\left(\ve{x}_1\right) &=& - \frac{12}{\left(d^{\,1}_{\sigma}\right)^4} \left(1 + c_{(1)}\right) 
        + \frac{21}{2}\,b_{(4)} - 7 \left(d^{\,1}_{\sigma}\right)^2 b_{(6)} - \frac{15}{4} \left(d^{\,1}_{\sigma}\right)^4 b_{(8)} 
	- \frac{6}{\left(d^{\,1}_{\sigma}\right)^2}\,c_{(3)} + 6\,c_{(5)}\,,
        \label{scalar_dot_Y3}
        \\
        \dot{Y}_{\left(4\right)}\left(\ve{x}_1\right) &=& + \frac{32}{\left(d^{\,1}_{\sigma}\right)^6}\,a_{(1)} 
	+ \frac{8}{\left(d^{\,1}_{\sigma}\right)^4}\,b_{(1)} - \frac{8}{\left(d^{\,1}_{\sigma}\right)^2}\,b_{(3)} + 12\,b_{(5)} 
	+ \frac{303}{32}\,\frac{c_{(2)}}{\left(d^{\,1}_{\sigma}\right)^4} 
	- \frac{27}{16}\,\frac{c_{(4)}}{\left(d^{\,1}_{\sigma}\right)^2} + \frac{81}{4}\,c_{(6)} + \frac{15}{2} \left(d^{\,1}_{\sigma}\right)^2 c_{(8)} 
	\nonumber\\ 
	&& - \frac{465}{32}\,\frac{d_{(2)}}{\left(d^{\,1}_{\sigma}\right)^5}\,,
        \label{scalar_dot_Y4}
        \\
        \dot{Y}_{\left(5\right)}\left(\ve{x}_1\right) &=& - \frac{16}{\left(d^{\,1}_{\sigma}\right)^6} \left(1 + c_{(1)}\right)
	+ \frac{4}{\left(d^{\,1}_{\sigma}\right)^4}\,b_{(2)} - \frac{2}{\left(d^{\,1}_{\sigma}\right)^2}\,b_{(4)}
        + \frac{19}{2}\,b_{(6)} + \frac{15}{4} \left(d^{\,1}_{\sigma}\right)^2 b_{(8)} 
	- \frac{4}{\left(d^{\,1}_{\sigma}\right)^4}\,c_{(3)} - \frac{6}{\left(d^{\,1}_{\sigma}\right)^2}\,c_{(5)}\,,
        \label{scalar_dot_Y5}
        \\
	\dot{Y}_{\left(6\right)}\left(\ve{x}_1\right) &=& + \frac{96}{\left(d^{\,1}_{\sigma}\right)^8}\,a_{(1)} 
        - \frac{8}{\left(d^{\,1}_{\sigma}\right)^4}\,b_{(3)} + \frac{6}{\left(d^{\,1}_{\sigma}\right)^2}\,b_{(5)} 
	+ \frac{747}{64}\,\frac{c_{(2)}}{\left(d^{\,1}_{\sigma}\right)^6} 
	+ \frac{121}{32}\,\frac{c_{(4)}}{\left(d^{\,1}_{\sigma}\right)^4} 
	+ \frac{101}{8}\,\frac{c_{(6)}}{\left(d^{\,1}_{\sigma}\right)^2} 
	- \frac{15}{4}\,c_{(8)}  
	- \frac{2325}{64}\,\frac{d_{(2)}}{\left(d^{\,1}_{\sigma}\right)^7}\,,
	\nonumber\\ 
        \label{scalar_dot_Y6}
        \\
        \dot{Y}_{\left(7\right)}\left(\ve{x}_1\right) &=& + \frac{32}{\left(d^{\,1}_{\sigma}\right)^6}\,a_{(1)} 
	- \frac{4}{\left(d^{\,1}_{\sigma}\right)^4}\,b_{(1)} + \frac{10}{\left(d^{\,1}_{\sigma}\right)^2}\,b_{(3)} - 6\,b_{(5)} 
	- \frac{87}{64}\,\frac{c_{(2)}}{\left(d^{\,1}_{\sigma}\right)^4} + \frac{355}{32}\,\frac{c_{(4)}}{\left(d^{\,1}_{\sigma}\right)^2} 
	- \frac{121}{8}\,c_{(6)} + \frac{15}{4} \left(d^{\,1}_{\sigma}\right)^2 c_{(8)}
	\nonumber\\ 
	&& - \frac{855}{64}\,\frac{d_{(2)}}{\left(d^{\,1}_{\sigma}\right)^5} \,, 
        \label{scalar_dot_Y7}
        \\
	\dot{Y}_{\left(8\right)}\left(\ve{x}_1\right) &=& - \frac{16}{\left(d^{\,1}_{\sigma}\right)^6} \left(1 + c_{(1)}\right) 
	- \frac{4}{\left(d^{\,1}_{\sigma}\right)^4}\,b_{(2)} + \frac{20}{\left(d^{\,1}_{\sigma}\right)^2}\,b_{(4)} - \frac{63}{2}\,b_{(6)} 
	+ \frac{15}{2} \left(d^{\,1}_{\sigma}\right)^2 b_{(8)} - \frac{12}{\left(d^{\,1}_{\sigma}\right)^4}\,c_{(3)} + \frac{12}{\left(d^{\,1}_{\sigma}\right)^2}\,c_{(5)}\,, 
        \label{scalar_dot_Y8}
	\\
	\nonumber\\
	\nonumber\\
	{Y}_{\left(1\right)}\left(\ve{x}_1\right) &=& + 12\,\frac{a_{\left(1\right)}}{\left(d^{\,1}_{\sigma}\right)^4}
        - 4\,\frac{b_{\left(1\right)}}{\left(d^{\,1}_{\sigma}\right)^2} 
        - \frac{93}{32}\,\frac{c_{\left(2\right)}}{\left(d^{\,1}_{\sigma}\right)^2} 
        - \frac{7}{16}\,c_{\left(4\right)}
        - \frac{285}{32}\,\frac{d_{\left(3\right)}}{\left(d^{\,1}_{\sigma}\right)^3}\,, 
        \label{scalar_Y1}
        \\
        {Y}_{\left(2\right)}\left(\ve{x}_1\right) &=& - 16\,\frac{a_{\left(2\right)}}{\left(d^{\,1}_{\sigma}\right)^6} 
        - \frac{91}{32}\,\frac{b_{\left(2\right)}}{\left(d^{\,1}_{\sigma}\right)^2} 
        - \frac{7}{16}\,b_{\left(4\right)}
        + 4\,\frac{c_{\left(1\right)}}{\left(d^{\,1}_{\sigma}\right)^4}  
        + \frac{465}{32}\,\frac{d_{\left(4\right)}}{\left(d^{\,1}_{\sigma}\right)^4}\,,  
        \label{scalar_Y2}
        \\
        {Y}_{\left(3\right)}\left(\ve{x}_1\right) &=& - 8\,\frac{a_{\left(1\right)}}{\left(d^{\,1}_{\sigma}\right)^4} 
        + 2\,\frac{b_{\left(1\right)}}{\left(d^{\,1}_{\sigma}\right)^2} 
	+2\,b_{\left(3\right)}
        + \frac{29}{64}\,\frac{c_{\left(2\right)}}{\left(d^{\,1}_{\sigma}\right)^2} 
        + \frac{111}{32}\,c_{\left(4\right)} 
        - \frac{5}{8} \left(d^{\,1}_{\sigma}\right)^2 c_{\left(6\right)}  
        + \frac{285}{64}\,\frac{d_{\left(3\right)}}{\left(d^{\,1}_{\sigma}\right)^3}\,,
        \label{scalar_Y3}
        \\ 
        {Y}_{\left(4\right)}\left(\ve{x}_1\right) &=& + 16\,\frac{a_{\left(2\right)}}{\left(d^{\,1}_{\sigma}\right)^6} 
        + \frac{27}{32}\,\frac{b_{\left(2\right)}}{\left(d^{\,1}_{\sigma}\right)^2} 
        + \frac{111}{16}\,b_{\left(4\right)} 
        - \frac{5}{4}\,\left(d^{\,1}_{\sigma}\right)^2 b_{\left(6\right)} 
        - 8\,\frac{c_{\left(1\right)}}{\left(d^{\,1}_{\sigma}\right)^4} 
	- 4\,\frac{c_{\left(3\right)}}{\left(d^{\,1}_{\sigma}\right)^2} 
        - \frac{465}{32}\,\frac{d_{\left(4\right)}}{\left(d^{\,1}_{\sigma}\right)^4}\,,  
        \label{scalar_Y4}
        \\  
        {Y}_{\left(5\right)}\left(\ve{x}_1\right) &=& + 8\,\frac{a_{\left(1\right)}}{\left(d^{\,1}_{\sigma}\right)^6} 
        - 4\,\frac{b_{\left(1\right)}}{\left(d^{\,1}_{\sigma}\right)^4} 
        - 2\,\frac{b_{\left(3\right)}}{\left(d^{\,1}_{\sigma}\right)^2} 
        - \frac{209}{64}\,\frac{c_{\left(2\right)}}{\left(d^{\,1}_{\sigma}\right)^4} 
        - \frac{91}{32}\,\frac{c_{\left(4\right)}}{\left(d^{\,1}_{\sigma}\right)^2} 
        + \frac{5}{8}\,c_{\left(6\right)} 
        - \frac{465}{64}\,\frac{d_{\left(3\right)}}{\left(d^{\,1}_{\sigma}\right)^5}\,, 
        \label{scalar_Y5}
        \\
        {Y}_{\left(6\right)}\left(\ve{x}_1\right) &=& + 48\,\frac{a_{\left(2\right)}}{\left(d^{\,1}_{\sigma}\right)^8} 
        + \frac{263}{64}\,\frac{b_{\left(2\right)}}{\left(d^{\,1}_{\sigma}\right)^4} 
        + \frac{91}{32}\,\frac{b_{\left(4\right)}}{\left(d^{\,1}_{\sigma}\right)^2} 
        + \frac{5}{8}\,b_{\left(6\right)} 
        - 16\,\frac{c_{\left(1\right)}}{\left(d^{\,1}_{\sigma}\right)^6}\,
        - 4\,\frac{c_{\left(3\right)}}{\left(d^{\,1}_{\sigma}\right)^4} 
        - \frac{2325}{64}\,\frac{d_{\left(4\right)}}{\left(d^{\,1}_{\sigma}\right)^6}\,, 
        \label{scalar_Y6}
        \\
        {Y}_{\left(7\right)}\left(\ve{x}_1\right) &=& + 16\,\frac{a_{\left(2\right)}}{\left(d^{\,1}_{\sigma}\right)^6} 
        + \frac{285}{64}\,\frac{b_{\left(2\right)}}{\left(d^{\,1}_{\sigma}\right)^2} 
        - \frac{71}{32}\,b_{\left(4\right)} 
        - \frac{5}{8} \left(d^{\,1}_{\sigma}\right)^2 b_{\left(6\right)}  
        + 4\,\frac{c_{\left(3\right)}}{\left(d^{\,1}_{\sigma}\right)^2} 
        - \frac{855}{64}\,\frac{d_{\left(4\right)}}{\left(d^{\,1}_{\sigma}\right)^4}\,,  
        \label{scalar_Y7}
        \\
        {Y}_{\left(8\right)}\left(\ve{x}_1\right) &=& - 32\,\frac{a_{\left(1\right)}}{\left(d^{\,1}_{\sigma}\right)^6} 
        + 12\,\frac{b_{\left(1\right)}}{\left(d^{\,1}_{\sigma}\right)^4} 
        + 8\,\frac{b_{\left(3\right)}}{\left(d^{\,1}_{\sigma}\right)^2} 
        + \frac{81}{32}\,\frac{c_{\left(2\right)}}{\left(d^{\,1}_{\sigma}\right)^4} 
        + \frac{91}{16}\,\frac{c_{\left(4\right)}}{\left(d^{\,1}_{\sigma}\right)^2} 
        + \frac{5}{4}\,c_{\left(6\right)} 
        + \frac{465}{32}\,\frac{d_{\left(3\right)}}{\left(d^{\,1}_{\sigma}\right)^5}\,. 
        \label{scalar_Y8}
\end{eqnarray}
\end{widetext}

\clearpage 

\begin{widetext}

The scalar functions of the quadrupole-quadrupole term in (\ref{First_Integration_2PN_Final}) and (\ref{Second_Integration_2PN_Final}) are given by 
\begin{eqnarray}
	\dot{Z}_{\left(1\right)}\left(\ve{x}_1\right) &=& + \frac{8}{\left(d^{\,1}_{\sigma}\right)^6} \left(1 + c_{(1)}\right) 
	- \frac{4}{\left(d^{\,1}_{\sigma}\right)^4}\,b_{(2)} - \frac{b_{(4)}}{\left(d^{\,1}_{\sigma}\right)^2} - \frac{3}{2}\,b_{(6)} + \frac{13}{8}  \left(d^{\,1}_{\sigma}\right)^2 b_{(8)}\,,
        \label{scalar_dot_Z1}
	\\
	\dot{Z}_{\left(2\right)}\left(\ve{x}_1\right) &=& - \frac{32}{\left(d^{\,1}_{\sigma}\right)^8}\,a_{(1)} + \frac{16}{\left(d^{\,1}_{\sigma}\right)^6}\,b_{(1)} 
	+ \frac{985}{128}\,\frac{c_{(2)}}{\left(d^{\,1}_{\sigma}\right)^6}  + \frac{217}{192}\,\frac{c_{(4)}}{\left(d^{\,1}_{\sigma}\right)^4} 
	+ \frac{5}{48}\,\frac{c_{(6)}}{\left(d^{\,1}_{\sigma}\right)^2} - \frac{13}{8}\,c_{(8)} + \frac{985}{128}\,\frac{d_{(2)}}{\left(d^{\,1}_{\sigma}\right)^7}\,, 
        \label{scalar_dot_Z2}
        \\
        \dot{Z}_{\left(3\right)}\left(\ve{x}_1\right) &=& - \frac{4}{\left(d^{\,1}_{\sigma}\right)^6} \left(1 + c_{(1)}\right) 
	+ \frac{6}{\left(d^{\,1}_{\sigma}\right)^4}\,b_{(2)} - \frac{3}{2}\,\frac{b_{(4)}}{\left(d^{\,1}_{\sigma}\right)^2} 
	+ \frac{37}{4}\,b_{(6)} - \frac{189}{16} \left(d^{\,1}_{\sigma}\right)^2 b_{(8)} + \frac{9}{4} \left(d^{\,1}_{\sigma}\right)^4 b_{(10)} + \frac{4}{\left(d^{\,1}_{\sigma}\right)^4} \,c_{(3)}\,,
	\nonumber\\ 
        \label{scalar_dot_Z3}
        \\
        \dot{Z}_{\left(4\right)}\left(\ve{x}_1\right) &=& - \frac{32}{\left(d^{\,1}_{\sigma}\right)^8}\,a_{(1)} 
	+ \frac{16}{\left(d^{\,1}_{\sigma}\right)^6}\,b_{(1)} - \frac{28}{\left(d^{\,1}_{\sigma}\right)^4}\,b_{(3)}
	+ \frac{24}{\left(d^{\,1}_{\sigma}\right)^2}\,b_{(5)} + \frac{5515}{512}\,\frac{c_{(2)}}{\left(d^{\,1}_{\sigma}\right)^6}
	- \frac{19061}{768} \frac{c_{(4)}}{\left(d^{\,1}_{\sigma}\right)^4} + \frac{2255}{192}\,\frac{c_{(6)}}{\left(d^{\,1}_{\sigma}\right)^2}
	\nonumber\\ 
	&& + \frac{329}{32}\,c_{(8)} - \frac{9}{4} \left(d^{\,1}_{\sigma}\right)^2 c_{(10)} 
	+ \frac{5515}{512}\,\frac{d_{(2)}}{\left(d^{\,1}_{\sigma}\right)^7}\,, 
        \label{scalar_dot_Z4}
        \\
        \dot{Z}_{\left(5\right)}\left(\ve{x}_1\right) &=& + \frac{32}{\left(d^{\,1}_{\sigma}\right)^8}\,a_{(1)} 
	- \frac{16}{\left(d^{\,1}_{\sigma}\right)^6}\,b_{(1)} - \frac{2285}{256}\,\frac{c_{(2)}}{\left(d^{\,1}_{\sigma}\right)^6} 
	- \frac{749}{384}\,\frac{c_{(4)}}{\left(d^{\,1}_{\sigma}\right)^4} 
	- \frac{73}{96}\,\frac{c_{(6)}}{\left(d^{\,1}_{\sigma}\right)^2}
	+ \frac{305}{16}\,c_{(8)} - \frac{9}{2} \left(d^{\,1}_{\sigma}\right)^2 c_{(10)}
	\nonumber\\ 
	&& - \frac{2285}{256}\,\frac{d_{(2)}}{\left(d^{\,1}_{\sigma}\right)^7}\,,
	\label{scalar_dot_Z5}
        \\
        \dot{Z}_{\left(6\right)}\left(\ve{x}_1\right) &=& + \frac{16}{\left(d^{\,1}_{\sigma}\right)^6}\,b_{(2)} 
	- \frac{56}{\left(d^{\,1}_{\sigma}\right)^4}\,b_{(4)} 
	+ \frac{22}{\left(d^{\,1}_{\sigma}\right)^2}\,b_{(6)} 
	+ \frac{41}{2}\,b_{(8)} - \frac{9}{2} \left(d^{\,1}_{\sigma}\right)^2 b_{(10)}
	+ \frac{16}{\left(d^{\,1}_{\sigma}\right)^6}\,c_{(3)} 
	- \frac{48}{\left(d^{\,1}_{\sigma}\right)^4}\,c_{(5)}\,,
        \label{scalar_dot_Z6}
        \\
        \dot{Z}_{\left(7\right)}\left(\ve{x}_1\right) &=& + \frac{8}{\left(d^{\,1}_{\sigma}\right)^6}\,b_{(2)} 
	- \frac{4}{\left(d^{\,1}_{\sigma}\right)^4}\,b_{(4)} 
	- \frac{1}{\left(d^{\,1}_{\sigma}\right)^2}\,b_{(6)}
	+ \frac{19}{2}\,b_{(8)} 
	- \frac{9}{4} \left(d^{\,1}_{\sigma}\right)^2 b_{(10)} 
        + \frac{8}{\left(d^{\,1}_{\sigma}\right)^6}\,c_{(3)}\,,
        \label{scalar_dot_Z7}
        \\
        \dot{Z}_{\left(8\right)}\left(\ve{x}_1\right) &=& + \frac{16}{\left(d^{\,1}_{\sigma}\right)^6}\,b_{(3)} 
	- \frac{24}{\left(d^{\,1}_{\sigma}\right)^4}\,b_{(5)}
	- \frac{2205}{512}\,\frac{c_{(2)}}{\left(d^{\,1}_{\sigma}\right)^8} 
        + \frac{3361}{256}\,\frac{c_{(4)}}{\left(d^{\,1}_{\sigma}\right)^6} 
	- \frac{1171}{64}\,\frac{c_{(6)}}{\left(d^{\,1}_{\sigma}\right)^4} 
	- \frac{255}{32}\,\frac{c_{(8)}}{\left(d^{\,1}_{\sigma}\right)^2} 
	+ \frac{9}{4}\,c_{(10)} 
	\nonumber\\ 
	&& - \frac{2205}{512}\,\frac{d_{(2)}}{\left(d^{\,1}_{\sigma}\right)^9}\,,
        \label{scalar_dot_Z8}
        \\
	\dot{Z}_{\left(9\right)}\left(\ve{x}_1\right) &=& - \frac{b_{(6)}}{2} + \frac{5}{8} \left(d^{\,1}_{\sigma}\right)^2 b_{(8)}\,,
        \label{scalar_dot_Z9}
        \\
	\dot{Z}_{\left(10\right)}\left(\ve{x}_1\right) &=& - \frac{8}{\left(d^{\,1}_{\sigma}\right)^6} \left(1 + c_{(1)}\right) 
	+ \frac{4}{\left(d^{\,1}_{\sigma}\right)^4}\,b_{(2)} 
	+ \frac{b_{(4)}}{\left(d^{\,1}_{\sigma}\right)^2} 
	+ \frac{5}{2}\,b_{(6)} 
	- \frac{95}{8} \left(d^{\,1}_{\sigma}\right)^2 b_{(8)} 
	+ \frac{15}{2} \left(d^{\,1}_{\sigma}\right)^4 b_{(10)}\,, 
        \label{scalar_dot_Z10}
        \\
        \dot{Z}_{\left(11\right)}\left(\ve{x}_1\right) &=& + \frac{32}{\left(d^{\,1}_{\sigma}\right)^8}\,a_{(1)} 
	- \frac{16}{\left(d^{\,1}_{\sigma}\right)^6}\,b_{(1)}
	- \frac{8}{\left(d^{\,1}_{\sigma}\right)^4}\,b_{(3)}
	- \frac{985}{128}\,\frac{c_{(2)}}{\left(d^{\,1}_{\sigma}\right)^6}
	+ \frac{1319}{192}\,\frac{c_{(4)}}{\left(d^{\,1}_{\sigma}\right)^4}
	+ \frac{187}{48}\,\frac{c_{(6)}}{\left(d^{\,1}_{\sigma}\right)^2} 
	+ \frac{85}{8}\,c_{(8)}
	\nonumber\\ 
	&& - 15 \left(d^{\,1}_{\sigma}\right)^2 c_{(10)} 
	- \frac{985}{128}\,\frac{d_{(2)}}{\left(d^{\,1}_{\sigma}\right)^7}\,,
        \label{scalar_dot_Z11}
        \\
        \dot{Z}_{\left(12\right)}\left(\ve{x}_1\right) &=& - \frac{16}{\left(d^{\,1}_{\sigma}\right)^8} \left(1 + c_{(1)}\right) 
	+ \frac{6}{\left(d^{\,1}_{\sigma}\right)^4}\,b_{(4)}
	+ \frac{2}{\left(d^{\,1}_{\sigma}\right)^2}\,b_{(6)}
	+ \frac{25}{4}\,b_{(8)} 
	- \frac{15}{2} \left(d^{\,1}_{\sigma}\right)^2 b_{(10)} 
	- \frac{8}{\left(d^{\,1}_{\sigma}\right)^6}\,c_{(3)}\,,
        \label{scalar_dot_Z12}
        \\
        \dot{Z}_{\left(13\right)}\left(\ve{x}_1\right) &=& - \frac{6}{\left(d^{\,1}_{\sigma}\right)^4}\,b_{(2)} 
	- \frac{9}{\left(d^{\,1}_{\sigma}\right)^2}\,b_{(4)} 
	+ \frac{33}{2}\,b_{(6)} 
	+ \frac{39}{16} \left(d^{\,1}_{\sigma}\right)^2 b_{(8)} 
	- \frac{27}{2} \left(d^{\,1}_{\sigma}\right)^4 b_{(10)}
	+ \frac{75}{8} \left(d^{\,1}_{\sigma}\right)^6 b_{(12)}
	\nonumber\\ 
	&& - \frac{6}{\left(d^{\,1}_{\sigma}\right)^4}\,c_{(3)}
	+ \frac{6}{\left(d^{\,1}_{\sigma}\right)^2}\,c_{(5)}\,,
        \label{scalar_dot_Z13}
        \\
        \dot{Z}_{\left(14\right)}\left(\ve{x}_1\right) &=& + \frac{16}{\left(d^{\,1}_{\sigma}\right)^4}\,b_{(3)}
	- \frac{12}{\left(d^{\,1}_{\sigma}\right)^2}\,b_{(5)} 
        - \frac{945}{512}\,\frac{c_{(2)}}{\left(d^{\,1}_{\sigma}\right)^6}
        + \frac{3781}{256}\,\frac{c_{(4)}}{\left(d^{\,1}_{\sigma}\right)^4}
	- \frac{319}{64}\,\frac{c_{(6)}}{\left(d^{\,1}_{\sigma}\right)^2}
	- \frac{411}{32}\,c_{(8)}
	+ \frac{81}{4} \left(d^{\,1}_{\sigma}\right)^2 c_{(10)} 
	\nonumber\\ 
	&& - \frac{75}{2} \left(d^{\,1}_{\sigma}\right)^4 c_{(12)} 
	- \frac{945}{512}\,\frac{d_{(2)}}{\left(d^{\,1}_{\sigma}\right)^7}\,,
        \label{scalar_dot_Z14}
        \\
	\dot{Z}_{\left(15\right)}\left(\ve{x}_1\right) &=& - \frac{16}{\left(d^{\,1}_{\sigma}\right)^8} \left(1 + c_{(1)}\right) 
	- \frac{8}{\left(d^{\,1}_{\sigma}\right)^6}\,b_{(2)} 
	- \frac{16}{\left(d^{\,1}_{\sigma}\right)^4}\,b_{(4)} 
	- 4\,b_{(8)}
	+ 12 \left(d^{\,1}_{\sigma}\right)^2 b_{(10)} 
	- \frac{75}{4} \left(d^{\,1}_{\sigma}\right)^4 b_{(12)} 
	- \frac{16}{\left(d^{\,1}_{\sigma}\right)^6}\,c_{(3)}
	\nonumber\\ 
	&& + \frac{6}{\left(d^{\,1}_{\sigma}\right)^4}\,c_{(5)}\,, 
        \label{scalar_dot_Z15}
        \\
        \dot{Z}_{\left(16\right)}\left(\ve{x}_1\right) &=& + \frac{16}{\left(d^{\,1}_{\sigma}\right)^8} \left(1 + c_{(1)}\right) 
	- \frac{16}{\left(d^{\,1}_{\sigma}\right)^6}\,b_{(2)} 
	+ \frac{50}{\left(d^{\,1}_{\sigma}\right)^4}\,b_{(4)} 
	- \frac{24}{\left(d^{\,1}_{\sigma}\right)^2}\,b_{(6)}
	- \frac{35}{4}\,b_{(8)}
	+ 24 \left(d^{\,1}_{\sigma}\right)^2 b_{(10)} 
	- \frac{75}{2} \left(d^{\,1}_{\sigma}\right)^4 b_{(12)}
	\nonumber\\ 
	&& - \frac{8}{\left(d^{\,1}_{\sigma}\right)^6}\,c_{(3)} 
	+ \frac{48}{\left(d^{\,1}_{\sigma}\right)^4}\,c_{(5)}\,,
        \label{scalar_dot_Z16}
        \end{eqnarray}
\end{widetext}
        
\begin{widetext}
	\begin{eqnarray}
        \dot{Z}_{\left(17\right)}\left(\ve{x}_1\right) &=& - \frac{32}{\left(d^{\,1}_{\sigma}\right)^6}\,b_{(3)}
	+ \frac{96}{\left(d^{\,1}_{\sigma}\right)^4}\,b_{(5)} 
	+ \frac{2205}{512}\,\frac{c_{(2)}}{\left(d^{\,1}_{\sigma}\right)^8} 
	- \frac{7457}{256}\,\frac{c_{(4)}}{\left(d^{\,1}_{\sigma}\right)^6}
	+ \frac{2195}{64}\,\frac{c_{(6)}}{\left(d^{\,1}_{\sigma}\right)^4}
	+ \frac{447}{32}\,\frac{c_{(8)}}{\left(d^{\,1}_{\sigma}\right)^2}
	+ \frac{39}{4}\,c_{(10)}
	\nonumber\\ 
	&& + \frac{75}{2} \left(d^{\,1}_{\sigma}\right)^2 c_{(12)} 
	+ \frac{2205}{512}\,\frac{d_{(2)}}{\left(d^{\,1}_{\sigma}\right)^9}\,,
        \label{scalar_dot_Z17}
        \\
        \dot{Z}_{\left(18\right)}\left(\ve{x}_1\right) &=& - \frac{12}{\left(d^{\,1}_{\sigma}\right)^6}\,b_{(4)} 
	+ \frac{6}{\left(d^{\,1}_{\sigma}\right)^4}\,b_{(6)} 
	+ \frac{15}{2}\,\frac{b_{(8)}}{\left(d^{\,1}_{\sigma}\right)^2} 
	+ \frac{3}{2} \,b_{(10)} 
	+ \frac{75}{8} \left(d^{\,1}_{\sigma}\right)^2 b_{(12)} 
	- \frac{12}{\left(d^{\,1}_{\sigma}\right)^6}\,c_{(5)}\,, 
        \label{scalar_dot_Z18}
        \\
        \dot{Z}_{\left(19\right)}\left(\ve{x}_1\right) &=& + \frac{5}{128}\,\frac{c_{(2)}}{\left(d^{\,1}_{\sigma}\right)^6}
        + \frac{5}{192}\,\frac{c_{(4)}}{\left(d^{\,1}_{\sigma}\right)^4}
	+ \frac{1}{48}\,\frac{c_{(6)}}{\left(d^{\,1}_{\sigma}\right)^2} 
	- \frac{5}{8}\,c_{(8)} 
	+ \frac{5}{128}\,\frac{d_{(2)}}{\left(d^{\,1}_{\sigma}\right)^7}\,,
        \label{scalar_dot_Z19}
        \\
        \dot{Z}_{\left(20\right)}\left(\ve{x}_1\right) &=& + \frac{925}{256}\,\frac{c_{(2)}}{\left(d^{\,1}_{\sigma}\right)^6} 
	+ \frac{925}{384}\,\frac{c_{(4)}}{\left(d^{\,1}_{\sigma}\right)^4} 
	+ \frac{185}{96}\,\frac{c_{(6)}}{\left(d^{\,1}_{\sigma}\right)^2}
	+ \frac{95}{16}\,c_{(8)} 
	- \frac{15}{2} \left(d^{\,1}_{\sigma}\right)^2 c_{(10)}  
	+ \frac{925}{256}\,\frac{d_{(2)}}{\left(d^{\,1}_{\sigma}\right)^7}\,,
	\label{scalar_dot_Z20}
        \\
	\dot{Z}_{\left(21\right)}\left(\ve{x}_1\right) &=& - \frac{48}{\left(d^{\,1}_{\sigma}\right)^8} \left(1 + c_{(1)}\right) 
        + \frac{16}{\left(d^{\,1}_{\sigma}\right)^6}\,b_{(2)} 
	+ \frac{10}{\left(d^{\,1}_{\sigma}\right)^4}\,b_{(4)}
	+ \frac{4}{\left(d^{\,1}_{\sigma}\right)^2}\,b_{(6)}
	+ \frac{55}{4}\,b_{(8)} 
	- 15 \left(d^{\,1}_{\sigma}\right)^2 b_{(10)} 
	- \frac{8}{\left(d^{\,1}_{\sigma}\right)^6}\,c_{(3)}\,,
        \label{scalar_dot_Z21}
        \\
	\dot{Z}_{\left(22\right)}\left(\ve{x}_1\right) &=& + \frac{128}{\left(d^{\,1}_{\sigma}\right)^{10}}\,a_{(1)} 
	- \frac{64}{\left(d^{\,1}_{\sigma}\right)^8}\,b_{(1)} 
	- \frac{32}{\left(d^{\,1}_{\sigma}\right)^6}\,b_{(3)} 
	+ \frac{24}{\left(d^{\,1}_{\sigma}\right)^4}\,b_{(5)}
	- \frac{6895}{256}\,\frac{c_{(2)}}{\left(d^{\,1}_{\sigma}\right)^8} 
	- \frac{13039}{384}\,\frac{c_{(4)}}{\left(d^{\,1}_{\sigma}\right)^6}
	- \frac{227}{96}\,\frac{c_{(6)}}{\left(d^{\,1}_{\sigma}\right)^4}
	\nonumber\\ 
	&& - \frac{5}{16}\,\frac{c_{(8)}}{\left(d^{\,1}_{\sigma}\right)^2}
	+ \frac{15}{2}\,c_{(10)}
	- \frac{6895}{256}\,\frac{d_{(2)}}{\left(d^{\,1}_{\sigma}\right)^9}\,,
        \label{scalar_dot_Z22}
        \\
        \dot{Z}_{\left(23\right)}\left(\ve{x}_1\right) &=& + \frac{24}{\left(d^{\,1}_{\sigma}\right)^8}\,a_{(1)} 
	- \frac{12}{\left(d^{\,1}_{\sigma}\right)^6}\,b_{(1)}
	+ \frac{14}{\left(d^{\,1}_{\sigma}\right)^4}\,b_{(3)} 
	- \frac{6}{\left(d^{\,1}_{\sigma}\right)^2}\,b_{(5)}
	- \frac{25875}{2048}\,\frac{c_{(2)}}{\left(d^{\,1}_{\sigma}\right)^6}
	+ \frac{8783}{1024}\,\frac{c_{(4)}}{\left(d^{\,1}_{\sigma}\right)^4}
	- \frac{701}{256}\,\frac{c_{(6)}}{\left(d^{\,1}_{\sigma}\right)^2}
	\nonumber\\ 
	&& - \frac{2577}{128}\,c_{(8)}
	+ \frac{399}{16} \left(d^{\,1}_{\sigma}\right)^2 c_{(10)}
	- \frac{75}{8} \left(d^{\,1}_{\sigma}\right)^4 c_{(12)}
	- \frac{25875}{2048}\,\frac{d_{(2)}}{\left(d^{\,1}_{\sigma}\right)^7}\,,
        \label{scalar_dot_Z23}
        \\
	\dot{Z}_{\left(24\right)}\left(\ve{x}_1\right) &=& + \frac{24}{\left(d^{\,1}_{\sigma}\right)^8} \left(1 + c_{(1)}\right) 
	- \frac{24}{\left(d^{\,1}_{\sigma}\right)^6}\,b_{(2)} 
	+ \frac{15}{\left(d^{\,1}_{\sigma}\right)^4}\,b_{(4)} 
	- \frac{6}{\left(d^{\,1}_{\sigma}\right)^2}\,b_{(6)}
	- \frac{609}{8}\,b_{(8)} 
	+ \frac{207}{2} \left(d^{\,1}_{\sigma}\right)^2 b_{(10)}
	- \frac{75}{2} \left(d^{\,1}_{\sigma}\right)^4 b_{(12)}
	\nonumber\\ 
	&& - \frac{12}{\left(d^{\,1}_{\sigma}\right)^6}\,c_{(3)}
	+ \frac{12}{\left(d^{\,1}_{\sigma}\right)^4}\,c_{(5)}\,,
        \label{scalar_dot_Z24}
        \\
	\dot{Z}_{\left(25\right)}\left(\ve{x}_1\right) &=& + \frac{128}{\left(d^{\,1}_{\sigma}\right)^{10}}\,a_{(1)} 
	- \frac{64}{\left(d^{\,1}_{\sigma}\right)^8}\,b_{(1)} 
	+ \frac{48}{\left(d^{\,1}_{\sigma}\right)^6}\,b_{(3)}
	- \frac{30}{\left(d^{\,1}_{\sigma}\right)^4}\,b_{(5)}
	- \frac{19405}{1024}\,\frac{c_{(2)}}{\left(d^{\,1}_{\sigma}\right)^8} 
	+ \frac{78899}{1536}\,\frac{c_{(4)}}{\left(d^{\,1}_{\sigma}\right)^6} 
	- \frac{41}{384}\,\frac{c_{(6)}}{\left(d^{\,1}_{\sigma}\right)^4}
	\nonumber\\ 
	&& + \frac{49}{64}\,\frac{c_{(8)}}{\left(d^{\,1}_{\sigma}\right)^2}
	- \frac{279}{8}\,c_{(10)} 
	+ \frac{75}{4} \left(d^{\,1}_{\sigma}\right)^2 c_{(12)} 
	- \frac{19405}{1024}\,\frac{d_{(2)}}{\left(d^{\,1}_{\sigma}\right)^9}\,,
        \label{scalar_dot_Z25}
        \\
        \dot{Z}_{\left(26\right)}\left(\ve{x}_1\right) &=& - \frac{128}{\left(d^{\,1}_{\sigma}\right)^{10}}\,a_{(1)} 
	+ \frac{64}{\left(d^{\,1}_{\sigma}\right)^8}\,b_{(1)}
	+ \frac{8}{\left(d^{\,1}_{\sigma}\right)^6}\,b_{(3)} 
	+ \frac{6395}{512}\,\frac{c_{(2)}}{\left(d^{\,1}_{\sigma}\right)^8} 
	+ \frac{251}{768}\,\frac{c_{(4)}}{\left(d^{\,1}_{\sigma}\right)^6} 
	- \frac{1025}{192}\,\frac{c_{(6)}}{\left(d^{\,1}_{\sigma}\right)^4} 
	+ \frac{73}{32}\,\frac{c_{(8)}}{\left(d^{\,1}_{\sigma}\right)^2}
	\nonumber\\ 
	&& - \frac{279}{4}\,c_{(10)} 
	+ \frac{75}{2} \left(d^{\,1}_{\sigma}\right)^2 c_{(12)} 
	+ \frac{6395}{512}\,\frac{d_{(2)}}{\left(d^{\,1}_{\sigma}\right)^9}\,,
        \label{scalar_dot_Z26}
        \\
	\dot{Z}_{\left(27\right)}\left(\ve{x}_1\right) &=& - \frac{16}{\left(d^{\,1}_{\sigma}\right)^{10}} \left(1 + c_{(1)}\right) 
	- \frac{48}{\left(d^{\,1}_{\sigma}\right)^8}\,b_{(2)}
	+ \frac{96}{\left(d^{\,1}_{\sigma}\right)^6}\,b_{(4)} 
	- \frac{30}{\left(d^{\,1}_{\sigma}\right)^4}\,b_{(6)}
	- \frac{6}{\left(d^{\,1}_{\sigma}\right)^2}\,b_{(8)}
	- \frac{147}{2}\,b_{(10)} 
	+ \frac{75}{2} \left(d^{\,1}_{\sigma}\right)^2 b_{(12)}
	\nonumber\\ 
	&& - \frac{56}{\left(d^{\,1}_{\sigma}\right)^8}\,c_{(3)}
	+ \frac{72}{\left(d^{\,1}_{\sigma}\right)^6}\,c_{(5)}\,,
        \label{scalar_dot_Z27}
        \\
	\dot{Z}_{\left(28\right)}\left(\ve{x}_1\right) &=& - \frac{24}{\left(d^{\,1}_{\sigma}\right)^8}\,b_{(3)} 
	+ \frac{36}{\left(d^{\,1}_{\sigma}\right)^6}\,b_{(5)}
	+ \frac{19845}{2048}\,\frac{c_{(2)}}{\left(d^{\,1}_{\sigma}\right)^{10}}
	- \frac{17961}{1024}\,\frac{c_{(4)}}{\left(d^{\,1}_{\sigma}\right)^8} 
	+ \frac{7467}{256}\,\frac{c_{(6)}}{\left(d^{\,1}_{\sigma}\right)^6} 
	+ \frac{1719}{128}\,\frac{c_{(8)}}{\left(d^{\,1}_{\sigma}\right)^4}
	+ \frac{159}{16}\,\frac{c_{(10)}}{\left(d^{\,1}_{\sigma}\right)^2}
	\nonumber\\ 
	&& - \frac{75}{8}\,c_{(12)} 
	+ \frac{19845}{2048}\,\frac{d_{(2)}}{\left(d^{\,1}_{\sigma}\right)^{11}}\,, 
        \label{scalar_dot_Z28}
\end{eqnarray}

\begin{eqnarray}
	{Z}_{\left(1\right)}\left(\ve{x}_1\right) &=& + 8\,\frac{a_{\left(1\right)}}{\left(d^{\,1}_{\sigma}\right)^6} 
        - 8\,\frac{b_{\left(1\right)}}{\left(d^{\,1}_{\sigma}\right)^4}  
        - \frac{327}{128}\,\frac{c_{\left(2\right)}}{\left(d^{\,1}_{\sigma}\right)^4}
        - \frac{7}{192}\,\frac{c_{\left(4\right)}}{\left(d^{\,1}_{\sigma}\right)^2} 
        + \frac{13}{48}\,c_{\left(6\right)} 
        + \frac{185}{128} \frac{d_{\left(3\right)}}{\left(d^{\,1}_{\sigma}\right)^5} \,, 
        \label{scalar_Z_1}
        \\
        {Z}_{\left(2\right)}\left(\ve{x}_1\right) &=& - 16\,\frac{a_{\left(2\right)}}{\left(d^{\,1}_{\sigma}\right)^8} 
        - \frac{985}{384}\, \frac{b_{\left(2\right)}}{\left(d^{\,1}_{\sigma}\right)^4} 
        - \frac{5}{192}\, \frac{b_{\left(4\right)}}{\left(d^{\,1}_{\sigma}\right)^2} 
        + \frac{13}{48}\,b_{\left(6\right)} 
        + 8\,\frac{c_{\left(1\right)}}{\left(d^{\,1}_{\sigma}\right)^6} 
        + \frac{985}{128}\,\frac{d_{\left(4\right)}}{\left(d^{\,1}_{\sigma}\right)^6} \,,  
        \label{scalar_Z_2}
        \\
        {Z}_{\left(3\right)}\left(\ve{x}_1\right) &=& + 4\,\frac{a_{\left(1\right)}}{\left(d^{\,1}_{\sigma}\right)^6} 
        + 4\,\frac{b_{\left(1\right)}}{\left(d^{\,1}_{\sigma}\right)^4} 
        - \frac{2103}{512} \,\frac{c_{\left(2\right)}}{\left(d^{\,1}_{\sigma}\right)^4} 
        + \frac{451}{256} \,\frac{c_{\left(4\right)}}{\left(d^{\,1}_{\sigma}\right)^2} 
        + \frac{23}{64}\, c_{\left(6\right)} 
        + \frac{9}{32} \left(d^{\,1}_{\sigma}\right)^2 c_{\left(8\right)}  
        - \frac{5175}{512}\,\frac{d_{\left(3\right)}}{\left(d^{\,1}_{\sigma}\right)^5} \,, 
        \label{scalar_Z_3}
        \\
        {Z}_{\left(4\right)}\left(\ve{x}_1\right) &=& - 16\,\frac{a_{\left(2\right)}}{\left(d^{\,1}_{\sigma}\right)^8} 
        - \frac{27019}{1536}\,\frac{b_{\left(2\right)}}{\left(d^{\,1}_{\sigma}\right)^4}
        + \frac{1585}{768}\,\frac{b_{\left(4\right)}}{\left(d^{\,1}_{\sigma}\right)^2} 
        + \frac{55}{192}\,b_{\left(6\right)} 
        + \frac{9}{32} \left(d^{\,1}_{\sigma}\right)^2 b_{\left(8\right)} 
        + 20\,\frac{c_{\left(1\right)}}{\left(d^{\,1}_{\sigma}\right)^6} 
        - 8\,\frac{c_{\left(3\right)}}{\left(d^{\,1}_{\sigma}\right)^4} 
	\nonumber\\ 
	&& + \frac{5515}{512}\,\frac{d_{\left(4\right)}}{\left(d^{\,1}_{\sigma}\right)^6} \,,  
        \label{scalar_Z_4}
        \\
        {Z}_{\left(5\right)}\left(\ve{x}_1\right) &=& + 16\,\frac{a_{\left(2\right)}}{\left(d^{\,1}_{\sigma}\right)^8} 
        - \frac{3859}{768}\,\frac{b_{\left(2\right)}}{\left(d^{\,1}_{\sigma}\right)^4} 
        + \frac{1609}{384}\,\frac{b_{\left(4\right)}}{\left(d^{\,1}_{\sigma}\right)^2} 
        + \frac{79}{96}\,b_{\left(6\right)} 
        + \frac{9}{16}\,\left(d^{\,1}_{\sigma}\right)^2 b_{\left(8\right)} 
        - 8\,\frac{c_{\left(1\right)}}{\left(d^{\,1}_{\sigma}\right)^6} 
        - \frac{2285}{256}\,\frac{d_{\left(4\right)}}{\left(d^{\,1}_{\sigma}\right)^6}\,, 
        \label{scalar_Z_5}
        \\
        {Z}_{\left(6\right)}\left(\ve{x}_1\right) &=& - 16\,\frac{a_{\left(1\right)}}{\left(d^{\,1}_{\sigma}\right)^8} 
        + 24\,\frac{b_{\left(1\right)}}{\left(d^{\,1}_{\sigma}\right)^6} 
        - 16\,\frac{b_{\left(3\right)}}{\left(d^{\,1}_{\sigma}\right)^4} 
        + \frac{6381}{256}\,\frac{c_{\left(2\right)}}{\left(d^{\,1}_{\sigma}\right)^6} 
        - \frac{2323}{384}\,\frac{c_{\left(4\right)}}{\left(d^{\,1}_{\sigma}\right)^4} 
        - \frac{119}{96}\,\frac{c_{\left(6\right)}}{\left(d^{\,1}_{\sigma}\right)^2} 
        - \frac{9}{16}\,c_{\left(8\right)}  
	\nonumber\\ 
	&& + \frac{2285}{256}\,\frac{d_{\left(3\right)}}{\left(d^{\,1}_{\sigma}\right)^7}\,,
        \label{scalar_Z_6}
        \\
        {Z}_{\left(7\right)}\left(\ve{x}_1\right) &=& + 16\,\frac{a_{\left(1\right)}}{\left(d^{\,1}_{\sigma}\right)^8} 
        - \frac{1419}{512}\,\frac{c_{\left(2\right)}}{\left(d^{\,1}_{\sigma}\right)^6} 
        - \frac{2443}{768}\,\frac{c_{\left(4\right)}}{\left(d^{\,1}_{\sigma}\right)^4} 
        - \frac{143}{192}\,\frac{c_{\left(6\right)}}{\left(d^{\,1}_{\sigma}\right)^2}  
        - \frac{9}{32}\,c_{\left(8\right)} 
        - \frac{5515}{512}\,\frac{d_{\left(3\right)}}{\left(d^{\,1}_{\sigma}\right)^7}\,,
        \label{scalar_Z_7}
        \\
        {Z}_{\left(8\right)}\left(\ve{x}_1\right) &=& + \frac{4831}{512}\,\frac{b_{\left(2\right)}}{\left(d^{\,1}_{\sigma}\right)^6} 
        -  \frac{877}{256}\,\frac{b_{\left(4\right)}}{\left(d^{\,1}_{\sigma}\right)^4} 
        -  \frac{43}{64}\,\frac{b_{\left(6\right)}}{\left(d^{\,1}_{\sigma}\right)^2} 
        -  \frac{9}{32}\,b_{\left(8\right)} 
        + 8\,\frac{c_{\left(3\right)}}{\left(d^{\,1}_{\sigma}\right)^6}
        - \frac{2205}{512}\,\frac{d_{\left(4\right)}}{\left(d^{\,1}_{\sigma}\right)^8}\,,  
        \label{scalar_Z_8}
        \\
        {Z}_{\left(9\right)}\left(\ve{x}_1\right) &=& + \frac{1}{128}\,\frac{c_{\left(2\right)}}{\left(d^{\,1}_{\sigma}\right)^4} 
        + \frac{1}{192}\,\frac{c_{\left(4\right)}}{\left(d^{\,1}_{\sigma}\right)^2} 
        + \frac{5}{48}\,c_{\left(6\right)} 
        + \frac{1}{128}\,\frac{d_{\left(3\right)}}{\left(d^{\,1}_{\sigma}\right)^5}\,,
        \label{scalar_Z_9}
        \\
        {Z}_{\left(10\right)}\left(\ve{x}_1\right) &=& - 8\,\frac{a_{\left(1\right)}}{\left(d^{\,1}_{\sigma}\right)^6} 
        + 8\,\frac{b_{\left(1\right)}}{\left(d^{\,1}_{\sigma}\right)^4} 
        + \frac{839}{256}\,\frac{c_{\left(2\right)}}{\left(d^{\,1}_{\sigma}\right)^4} 
        + \frac{199}{384}\,\frac{c_{\left(4\right)}}{\left(d^{\,1}_{\sigma}\right)^2}
        - \frac{85}{96}\,c_{\left(6\right)} 
        + \frac{15}{16} \left(d^{\,1}_{\sigma}\right)^2 c_{\left(8\right)} 
        - \frac{185}{256}\,\frac{d_{\left(3\right)}}{\left(d^{\,1}_{\sigma}\right)^5}\,, 
        \label{scalar_Z_10} 
        \\
        {Z}_{\left(11\right)}\left(\ve{x}_1\right) &=& + 16\,\frac{a_{\left(2\right)}}{\left(d^{\,1}_{\sigma}\right)^8} 
        + \frac{2521}{384}\,\frac{b_{\left(2\right)}}{\left(d^{\,1}_{\sigma}\right)^4} 
        + \frac{197}{192}\,\frac{b_{\left(4\right)}}{\left(d^{\,1}_{\sigma}\right)^2} 
        - \frac{85}{48}\,b_{\left(6\right)}
        + \frac{15}{8} \left(d^{\,1}_{\sigma}\right)^2 b_{\left(8\right)} 
        - 8\,\frac{c_{\left(1\right)}}{\left(d^{\,1}_{\sigma}\right)^6} 
        - \frac{985}{128}\,\frac{d_{\left(4\right)}}{\left(d^{\,1}_{\sigma}\right)^6}\,,  
        \label{scalar_Z_11}
        \\
        {Z}_{\left(12\right)}\left(\ve{x}_1\right) &=& - \frac{985}{256}\,\frac{c_{\left(2\right)}}{\left(d^{\,1}_{\sigma}\right)^6}  
        - \frac{217}{384}\,\frac{c_{\left(4\right)}}{\left(d^{\,1}_{\sigma}\right)^4} 
        - \frac{5}{96}\,\frac{c_{\left(6\right)}}{\left(d^{\,1}_{\sigma}\right)^2}
        - \frac{15}{16}\,c_{\left(8\right)} 
        - \frac{985}{256}\,\frac{d_{\left(3\right)}}{\left(d^{\,1}_{\sigma}\right)^7}\,, 
        \label{scalar_Z_12}
        \\
        {Z}_{\left(13\right)}\left(\ve{x}_1\right) &=& - 4\,\frac{a_{\left(1\right)}}{\left(d^{\,1}_{\sigma}\right)^6}  
        - 4\,\frac{b_{\left(1\right)}}{\left(d^{\,1}_{\sigma}\right)^4} 
        + \frac{3237}{2048}\, \frac{c_{\left(2\right)}}{\left(d^{\,1}_{\sigma}\right)^4} 
        - \frac{969}{1024}\, \frac{c_{\left(4\right)}}{\left(d^{\,1}_{\sigma}\right)^2} 
        + \frac{395}{256}\,c_{\left(6\right)} 
        - \frac{369}{128} \left(d^{\,1}_{\sigma}\right)^2 c_{\left(8\right)} 
        + \frac{15}{16}\left(d^{\,1}_{\sigma}\right)^4 c_{\left(10\right)}  
        \nonumber\\ 
	&& + \frac{15525}{2048}\,\frac{d_{\left(3\right)}}{\left(d^{\,1}_{\sigma}\right)^5}\,, 
        \label{scalar_Z_13}
        \\
        {Z}_{\left(14\right)}\left(\ve{x}_1\right) &=& + \frac{8507}{512}\,\frac{b_{\left(2\right)}}{\left(d^{\,1}_{\sigma}\right)^4}  
        -  \frac{1217}{256}\,\frac{b_{\left(4\right)}}{\left(d^{\,1}_{\sigma}\right)^2} 
        +  \frac{393}{64}\,b_{\left(6\right)} 
        -  \frac{369}{32} \left(d^{\,1}_{\sigma}\right)^2 b_{\left(8\right)} 
        + \frac{15}{4} \left(d^{\,1}_{\sigma}\right)^4 b_{\left(10\right)} 
        - 12\,\frac{c_{\left(1\right)}}{\left(d^{\,1}_{\sigma}\right)^6} 
        + 8\,\frac{c_{\left(3\right)}}{\left(d^{\,1}_{\sigma}\right)^4}
	\nonumber\\ 
	&& - \frac{945}{512}\,\frac{d_{\left(4\right)}}{\left(d^{\,1}_{\sigma}\right)^6}\,, 
        \label{scalar_Z_14}
        \\
        {Z}_{\left(15\right)}\left(\ve{x}_1\right) &=& - 16\,\frac{a_{\left(1\right)}}{\left(d^{\,1}_{\sigma}\right)^8}  
        - \frac{2677}{1024}\, \frac{c_{\left(2\right)}}{\left(d^{\,1}_{\sigma}\right)^6} 
        + \frac{5515}{1536}\, \frac{c_{\left(4\right)}}{\left(d^{\,1}_{\sigma}\right)^4} 
        + \frac{335}{384}\, \frac{c_{\left(6\right)}}{\left(d^{\,1}_{\sigma}\right)^2}  
        + \frac{249}{64}\,c_{\left(8\right)}
        - \frac{15}{8} \left(d^{\,1}_{\sigma}\right)^2 c_{\left(10\right)}
        + \frac{5515}{1024}\,\frac{d_{\left(3\right)}}{\left(d^{\,1}_{\sigma}\right)^7}\,, 
        \label{scalar_Z_15}
        \\
        {Z}_{\left(16\right)}\left(\ve{x}_1\right) &=& + 16\,\frac{a_{\left(1\right)}}{\left(d^{\,1}_{\sigma}\right)^8} 
        - 24\,\frac{b_{\left(1\right)}}{\left(d^{\,1}_{\sigma}\right)^6} 
        + 16\,\frac{b_{\left(3\right)}}{\left(d^{\,1}_{\sigma}\right)^4} 
        - \frac{10477}{512}\,\frac{c_{\left(2\right)}}{\left(d^{\,1}_{\sigma}\right)^6} 
        + \frac{5395}{768}\,\frac{c_{\left(4\right)}}{\left(d^{\,1}_{\sigma}\right)^4} 
        + \frac{311}{192}\,\frac{c_{\left(6\right)}}{\left(d^{\,1}_{\sigma}\right)^2} 
        + \frac{249}{32}\,c_{\left(8\right)} 
        - \frac{15}{4} \left(d^{\,1}_{\sigma}\right)^2 c_{\left(10\right)}
        \nonumber\\ 
	&& - \frac{2285}{512}\,\frac{d_{\left(3\right)}}{\left(d^{\,1}_{\sigma}\right)^7}\,, 
        \label{scalar_Z_16}
        \end{eqnarray}
\end{widetext}

\begin{widetext}
        \begin{eqnarray}
        {Z}_{\left(17\right)}\left(\ve{x}_1\right) &=& - \frac{8927}{512}\,\frac{b_{\left(2\right)}}{\left(d^{\,1}_{\sigma}\right)^6} 
        + \frac{1901}{256}\,\frac{b_{\left(4\right)}}{\left(d^{\,1}_{\sigma}\right)^4}
        + \frac{107}{64}\,\frac{b_{\left(6\right)}}{\left(d^{\,1}_{\sigma}\right)^2} 
        + \frac{249}{32}\,b_{\left(8\right)} 
        - \frac{15}{4} \left(d^{\,1}_{\sigma}\right)^2 b_{\left(10\right)} 
	+ 32\,\frac{c_{\left(1\right)}}{\left(d^{\,1}_{\sigma}\right)^8} 
        - 8\,\frac{c_{\left(3\right)}}{\left(d^{\,1}_{\sigma}\right)^6} 
        + \frac{2205}{512}\,\frac{d_{\left(4\right)}}{\left(d^{\,1}_{\sigma}\right)^8}\,,
	\nonumber\\
	\label{scalar_Z_17}
        \\
        {Z}_{\left(18\right)}\left(\ve{x}_1\right) &=& + \frac{2205}{2048}\,\frac{c_{\left(2\right)}}{\left(d^{\,1}_{\sigma}\right)^8} 
        - \frac{3361}{1024}\,\frac{c_{\left(4\right)}}{\left(d^{\,1}_{\sigma}\right)^6}  
        - \frac{365}{256}\,\frac{c_{\left(6\right)}}{\left(d^{\,1}_{\sigma}\right)^4} 
        - \frac{129}{128}\,\frac{c_{\left(8\right)}}{\left(d^{\,1}_{\sigma}\right)^2}  
        + \frac{15}{16}\,c_{\left(10\right)} 
        + \frac{2205}{2048}\,\frac{d_{\left(3\right)}}{\left(d^{\,1}_{\sigma}\right)^9}\,, 
        \label{scalar_Z_18}
        \\ 
        {Z}_{\left(19\right)}\left(\ve{x}_1\right) &=& - \frac{5}{384}\,\frac{b_{\left(2\right)}}{\left(d^{\,1}_{\sigma}\right)^4} 
        - \frac{1}{192}\,\frac{b_{\left(4\right)}}{\left(d^{\,1}_{\sigma}\right)^2}
        + \frac{5}{48}\,b_{\left(6\right)} 
        + \frac{5}{128}\,\frac{d_{\left(4\right)}}{\left(d^{\,1}_{\sigma}\right)^6}\,, 
        \label{scalar_Z_19}
        \\
        {Z}_{\left(20\right)}\left(\ve{x}_1\right) &=& - \frac{3997}{768}\,\frac{b_{\left(2\right)}}{\left(d^{\,1}_{\sigma}\right)^4} 
        - \frac{569}{384}\,\frac{b_{\left(4\right)}}{\left(d^{\,1}_{\sigma}\right)^2} 
        - \frac{95}{96}\,b_{\left(6\right)} 
        + \frac{15}{16} \left(d^{\,1}_{\sigma}\right)^2 b_{\left(8\right)} 
        + \frac{925}{256}\,\frac{d_{\left(4\right)}}{\left(d^{\,1}_{\sigma}\right)^6}\,, 
        \label{scalar_Z_20}
        \\
        {Z}_{\left(21\right)}\left(\ve{x}_1\right) &=& - 64\,\frac{a_{\left(1\right)}}{\left(d^{\,1}_{\sigma}\right)^8} 
        + 48\,\frac{b_{\left(1\right)}}{\left(d^{\,1}_{\sigma}\right)^6} 
        + \frac{3033}{128}\,\frac{c_{\left(2\right)}}{\left(d^{\,1}_{\sigma}\right)^6} 
        + \frac{601}{192}\,\frac{c_{\left(4\right)}}{\left(d^{\,1}_{\sigma}\right)^4}  
        + \frac{5}{48}\,\frac{c_{\left(6\right)}}{\left(d^{\,1}_{\sigma}\right)^2}  
        - \frac{15}{8} c_{\left(8\right)} 
        + \frac{985}{128}\,\frac{d_{\left(3\right)}}{\left(d^{\,1}_{\sigma}\right)^7}\,, 
        \label{scalar_Z_21}
        \\
        {Z}_{\left(22\right)}\left(\ve{x}_1\right) &=& + 64\,\frac{a_{\left(2\right)}}{\left(d^{\,1}_{\sigma}\right)^{10}} 
        + \frac{13039}{768}\,\frac{b_{\left(2\right)}}{\left(d^{\,1}_{\sigma}\right)^6} 
        + \frac{611}{384}\,\frac{b_{\left(4\right)}}{\left(d^{\,1}_{\sigma}\right)^4} 
        + \frac{5}{96}\,\frac{b_{\left(6\right)}}{\left(d^{\,1}_{\sigma}\right)^2} 
        - \frac{15}{16}\,b_{\left(8\right)} 
        - \frac{48}{\left(d^{\,1}_{\sigma}\right)^8}\,c_{\left(1\right)} 
        - \frac{6895}{256}\,\frac{d_{\left(4\right)}}{\left(d^{\,1}_{\sigma}\right)^8}\,,  
        \label{scalar_Z_22}
        \\ 
        {Z}_{\left(23\right)}\left(\ve{x}_1\right) &=& + 12 \frac{a_{\left(2\right)}}{\left(d^{\,1}_{\sigma}\right)^8} 
        + \frac{31153}{2048} \frac{b_{\left(2\right)}}{\left(d^{\,1}_{\sigma}\right)^4}
        - \frac{2371}{1024} \frac{b_{\left(4\right)}}{\left(d^{\,1}_{\sigma}\right)^2}  
        - \frac{37}{256} b_{\left(6\right)} 
        - \frac{111}{128} \left(d^{\,1}_{\sigma}\right)^2 b_{\left(8\right)}   
        + \frac{15}{16} \left(d^{\,1}_{\sigma}\right)^4 b_{\left(10\right)} 
        - 4 \frac{c_{\left(1\right)}}{\left(d^{\,1}_{\sigma}\right)^6} 
        + 8 \frac{c_{\left(3\right)}}{\left(d^{\,1}_{\sigma}\right)^4} 
        \nonumber\\  
	&& - \frac{25875}{2048} \frac{d_{\left(4\right)}}{\left(d^{\,1}_{\sigma}\right)^6} \,,
        \label{scalar_Z_23}
        \\
        {Z}_{\left(24\right)}\left(\ve{x}_1\right) &=& + 24\,\frac{a_{\left(1\right)}}{\left(d^{\,1}_{\sigma}\right)^8} 
        - 36\,\frac{b_{\left(1\right)}}{\left(d^{\,1}_{\sigma}\right)^6} 
        + 16\,\frac{b_{\left(3\right)}}{\left(d^{\,1}_{\sigma}\right)^4}
        - \frac{11343}{512}\,\frac{c_{\left(2\right)}}{\left(d^{\,1}_{\sigma}\right)^6}  
        + \frac{59}{256}\,\frac{c_{\left(4\right)}}{\left(d^{\,1}_{\sigma}\right)^4} 
        - \frac{65}{64}\,\frac{c_{\left(6\right)}}{\left(d^{\,1}_{\sigma}\right)^2} 
        - \frac{9}{32}\,c_{\left(8\right)} 
        - \frac{15}{4} \left(d^{\,1}_{\sigma}\right)^2 c_{\left(10\right)} 
        \nonumber\\  
	&& + \frac{945}{512}\,\frac{d_{\left(3\right)}}{\left(d^{\,1}_{\sigma}\right)^7}\,, 
        \label{scalar_Z_24}
        \\
        {Z}_{\left(25\right)}\left(\ve{x}_1\right) &=& + 64 \frac{a_{\left(2\right)}}{\left(d^{\,1}_{\sigma}\right)^{10}} 
        + \frac{93133}{3072} \frac{b_{\left(2\right)}}{\left(d^{\,1}_{\sigma}\right)^6} 
        - \frac{11479}{1536} \frac{b_{\left(4\right)}}{\left(d^{\,1}_{\sigma}\right)^4} 
        - \frac{433}{384} \frac{b_{\left(6\right)}}{\left(d^{\,1}_{\sigma}\right)^2} 
        - \frac{9}{64} b_{\left(8\right)} 
        - \frac{15}{8} \left(d^{\,1}_{\sigma}\right)^2 b_{\left(10\right)} 
        - 48 \frac{c_{\left(1\right)}}{\left(d^{\,1}_{\sigma}\right)^8}
        + 16 \frac{c_{\left(3\right)}}{\left(d^{\,1}_{\sigma}\right)^6} 
        \nonumber\\
	&& - \frac{19405}{1024} \frac{d_{\left(4\right)}}{\left(d^{\,1}_{\sigma}\right)^8} \,,  
        \label{scalar_Z_25}
        \\
        {Z}_{\left(26\right)}\left(\ve{x}_1\right) &=& - 64\,\frac{a_{\left(2\right)}}{\left(d^{\,1}_{\sigma}\right)^{10}} 
        + \frac{5893}{1536}\,\frac{b_{\left(2\right)}}{\left(d^{\,1}_{\sigma}\right)^6} 
        - \frac{5887}{768}\,\frac{b_{\left(4\right)}}{\left(d^{\,1}_{\sigma}\right)^4} 
        - \frac{457}{192}\,\frac{b_{\left(6\right)}}{\left(d^{\,1}_{\sigma}\right)^2} 
        - \frac{9}{32}\,b_{\left(8\right)} 
        - \frac{15}{4} \left(d^{\,1}_{\sigma}\right)^2 b_{\left(10\right)} 
        + 48\,\frac{c_{\left(1\right)}}{\left(d^{\,1}_{\sigma}\right)^8}
	\nonumber\\ 
	&& + \frac{6395}{512}\,\frac{d_{\left(4\right)}}{\left(d^{\,1}_{\sigma}\right)^8} \,, 
        \label{scalar_Z_26}
        \\
        {Z}_{\left(27\right)}\left(\ve{x}_1\right) &=& - 48\,\frac{b_{\left(1\right)}}{\left(d^{\,1}_{\sigma}\right)^8} 
        + \frac{48}{\left(d^{\,1}_{\sigma}\right)^6}\,b_{\left(3\right)} 
        - \frac{26781}{512}\,\frac{c_{\left(2\right)}}{\left(d^{\,1}_{\sigma}\right)^8}  
        + \frac{7457}{256}\,\frac{c_{\left(4\right)}}{\left(d^{\,1}_{\sigma}\right)^6}  
        + \frac{493}{64}\,\frac{c_{\left(6\right)}}{\left(d^{\,1}_{\sigma}\right)^4}  
        + \frac{129}{32}\,\frac{c_{\left(8\right)}}{\left(d^{\,1}_{\sigma}\right)^2}  
        + \frac{15}{4}\,c_{\left(10\right)}  
	\nonumber\\ 
	&& - \frac{2205}{512}\,\frac{d_{\left(3\right)}}{\left(d^{\,1}_{\sigma}\right)^9} \,,
        \label{scalar_Z_27}
        \\
        {Z}_{\left(28\right)}\left(\ve{x}_1\right) &=& - \frac{55767}{2048}\,\frac{b_{\left(2\right)}}{\left(d^{\,1}_{\sigma}\right)^8} 
        + \frac{10965}{1024}\,\frac{b_{\left(4\right)}}{\left(d^{\,1}_{\sigma}\right)^6} 
        + \frac{579}{256}\,\frac{b_{\left(6\right)}}{\left(d^{\,1}_{\sigma}\right)^4} 
        + \frac{129}{128}\,\frac{b_{\left(8\right)}}{\left(d^{\,1}_{\sigma}\right)^2} 
        + \frac{15}{16}\,b_{\left(10\right)}
        - 24\,\frac{c_{\left(3\right)}}{\left(d^{\,1}_{\sigma}\right)^8} 
        + \frac{19845}{2048}\,\frac{d_{\left(4\right)}}{\left(d^{\,1}_{\sigma}\right)^{10}}\,.  
        \label{scalar_Z_28}
\end{eqnarray}

\clearpage




\clearpage 

\end{widetext} 

\end{document}


\title{Light propagation in 2PN approximation in the monopole and quadrupole field of a body at rest: Supplement}

\author{Sven Zschocke}

\affiliation{Lohrmann Observatory, TUD Dresden University of Technology, Helmholtzstrasse 10, D-01069 Dresden, Germany}


\begin{abstract}
In a recent investigation \cite{Zschocke_Quadrupole_1}, the propagation of a light signal in the gravitational field of a body at rest has been determined 
in the second post-Newtonian (2PN) approximation. In order to integrate the geodesic equations for the light rays, the individual terms have been separated 
into time-dependent scalar functions and time-independent tensorial coefficients. In this way one can reduce the integration of these scalar functions to only 
eight master integrals, which can be solved in closed form. On the other side, this procedure leads to tensorial coefficients, which contain linearly dependent 
tensors. For subsequent investigations it is more appropriate, to rearrange these tensorial terms into a new form, where these tensorial coefficients contain 
only tensors which are linearly independent of each other. This rearranged form of the solution will be represented in this investigation. This new form for the 
light trajectory is the basis for our recent investigations in astrometry on the sub-microarcsecond level of accuracy in \cite{Zschocke_Quadrupole_2}. 
\end{abstract}

\pacs{95.10.Jk, 95.10.Ce, 95.30.Sf, 04.25.Nx, 04.80.Cc}

\maketitle

\begin{center}


\end{center}

\section{Introduction}\label{Introduction}\label{Section0}

The success of the astrometry missions Hipparcos \cite{Hipparcos} and Gaia \cite{Gaia} 
of the of European Space Agency (ESA) has shown in an impressive way, that the precision 
of angular observations of celestial objects can considerably be improved by space based 
telescopes: the Hipparcos has achieved an accuracy on the milli-arcsecond (mas) level 
and the Gaia mission has achieved an accuracy on the micro-arcsecond (\muas) level 
in angular measurements of stellar objects. It is clear, that future astrometry is aiming 
at the sub-\muas{} scale of precision in angular observations. In fact, there are several 
concrete proposals for future space astrometry missions of European Space Agency (ESA), like GaiaNIR \cite{Gaia_NIR} 
and Theia \cite{Theia}, which are aiming at the sub-\muas{} scale of precision. These efforts are mainly triggered by a set of 
several highly impressive science themes, which come in the focus on such scales of accuracy \cite{Sub_Micro_1}. Among 
them there are science cases like detection of gravitational waves by astrometric measurements, discoveries of 
Earth-like exoplanets, new insights about dark matter distributions, or new tests of general relativity (GR). 

Astrometry on the sub-\muas{} scale requires a general-relativistic modeling of light propagation from the celestial 
light sources through the curved space-time of the solar system towards the observer. Already on the \muas{} scale 
of accuracy, it is necessary to determine the light trajectories in the gravitational field of monopoles in the second 
post-Newtonian (2PN) approximation \cite{Klioner_Zschocke,Ashby_Bertotti,Teyssandier}. This problem has been solved long time 
ago \cite{Brumberg1987,Brumberg1991,KlionerKopeikin1992}. However, on the sub-\muas{} level of accuracy it is necessary to determine 
the light trajectories in the quadrupole field of solar system bodies in the 2PN approximation. Such a solution has only recently 
been achieved in our investigation in \cite{Zschocke_Quadrupole_1}. In order to integrate the geodesic equation, each individual term 
has been separated into time-dependent scalar functions and time-independent tensorial coefficients. The integration of these 
scalar functions can be reduced to only eight master integrals, which can be solved in closed form \cite{Zschocke_Quadrupole_1}. 
However, this approach leads to tensorial coefficients, which contain linearly dependent tensors. For subsequent investigations, 
it is much more appropriate, to rearrange these tensorial terms into a new form, where these tensorial coefficients contain only 
tensors which are linearly independent of each other. This rearranged form of the solution will be represented in this investigation 
as supplementary material of our recent article \cite{Zschocke_Quadrupole_2}.

\section{The geodesic equation in 2PN approximation}\label{Section1}

In \cite{Zschocke_Quadrupole_1} we have considered the propagation of a light signal in the gravitational field of a body at rest,
and have taken into account the monopole and quadrupole structure of the body. Then, the geodesic equation in 2PN approximation 
for light rays is given by Eq.~(74) in \cite{Zschocke_Quadrupole_1}, which reads 
\begin{eqnarray}
        \frac{\ddot{\ve{x}}\left(t\right)}{c^2} &=& \frac{\ddot{\ve{x}}^{\rm M}_{\rm 1PN}\left(t\right)}{c^2} 
        + \frac{\ddot{\ve{x}}^{\rm Q}_{\rm 1PN}\left(t\right)}{c^2}
        + \frac{\ddot{\ve{x}}^{{\rm M} \times {\rm M}}_{\rm 2PN}\left(t\right)}{c^2}    
        + \frac{\ddot{\ve{x}}^{{\rm M} \times {\rm Q}}_{\rm 2PN}\left(t\right)}{c^2}    
        + \frac{\ddot{\ve{x}}^{{\rm Q} \times {\rm Q}}_{\rm 2PN}\left(t\right)}{c^2}\,.
\label{Geodesic_Equation_2PN_15}
\end{eqnarray}

\noindent
The individual terms of this equation were presented by Eqs.~(50), (51), (75), (78), and (79) in \cite{Zschocke_Quadrupole_1}. 
A solution of the geodesic equation is uniquely defined by imposing initial value conditions, 
\begin{eqnarray}
	\ve{\sigma} &=& \frac{\dot{\ve{x}}\left(t\right)}{c}\bigg|_{t = - \infty}\,,
\label{Initial_Condition_1}
\\
	\ve{x}_0 &=& \ve{x}\left(t\right)\bigg|_{t =t_0}\,,
\label{Initial_Condition_2}
\end{eqnarray}

\noindent
where $\ve{\sigma}$ is the unit-direction, $\ve{\sigma} \cdot \ve{\sigma} = 1$  of the light ray at past infinity and $\ve{x}_0$ 
is the spatial position of the light source at the moment of emission of the light signal.

\subsection{The first integration of geodesic equation}

The first integration of geodesic equation (\ref{Geodesic_Equation_2PN_15}) yields the coordinate velocity of the light signal in 2PN approximation 
(cf. Eq.~(80) in \cite{Zschocke_Quadrupole_1}): 
\begin{eqnarray}
        \frac{\dot{\ve{x}}\left(t\right)}{c} &=&  \ve{\sigma} + \int\limits_{- \infty}^{t} d c {\rm t} \,\frac{\ddot{\ve{x}}\left({\rm t}\right)}{c^2}
\label{First_Integration_2PN_A}
        \\
	&=& \ve{\sigma} + \frac{\Delta \dot{\ve{x}}_{\rm 1PN}^{\rm M}\left(t\right)}{c} + \frac{\Delta \dot{\ve{x}}_{\rm 1PN}^{\rm Q}\left(t\right)}{c} 
	+ \frac{\Delta \dot{\ve{x}}^{{\rm M} \times {\rm M}}_{\rm 2PN}\left(t\right)}{c} + \frac{\Delta \dot{\ve{x}}^{{\rm M} \times {\rm Q}}_{\rm 2PN}\left(t\right)}{c} 
	+ \frac{\Delta \dot{\ve{x}}^{{\rm Q} \times {\rm Q}}_{\rm 2PN}\left(t\right)}{c}\,.
\label{First_Integration_2PN}
\end{eqnarray}

\noindent
From (\ref{First_Integration_2PN_A}) follows that the perturbations in (\ref{First_Integration_2PN}) vanish in the limit $t \rightarrow - \infty$. 
These perturbations in Eq.~(\ref{First_Integration_2PN}) were given in their explicit form by Eqs.~(62), (63), (83), (84), and (85) in \cite{Zschocke_Quadrupole_1}.

\subsection{The second integration of geodesic equation}

The second integration of geodesic equation (\ref{Geodesic_Equation_2PN_15}) yields the trajectory of the light signal in 2PN approximation
(cf. Eq.~(86) in \cite{Zschocke_Quadrupole_1}:
\begin{eqnarray}
        \ve{x}\left(t\right) &=& \ve{x}_{\rm N} + \int\limits_{t_0}^{t} d c {\rm t} \,\frac{\dot{\ve{x}}\left({\rm t}\right)}{c}
\label{Second_Integration_2PN_A}
	\\
	 &=& \ve{x}_{\rm N} + \Delta \ve{x}_{\rm 1PN}^{\rm M}\left(t,t_0\right) + \Delta \ve{x}_{\rm 1PN}^{\rm Q}\left(t,t_0\right)
	 + \Delta \ve{x}_{\rm 2PN}^{{\rm M} \times {\rm M}}\left(t,t_0\right) + \Delta \ve{x}_{\rm 2PN}^{{\rm M} \times {\rm Q}}\left(t,t_0\right)
	 + \Delta \ve{x}_{\rm 2PN}^{{\rm Q} \times {\rm Q}}\left(t,t_0\right),  
\label{Second_Integration_2PN}
\end{eqnarray}

\noindent
where 
\begin{eqnarray}
\ve{x}_{\rm N} &=& \ve{x}_0 + \left(t - t_0\right) \ve{\sigma}
\label{unperturbed_lightray}
\end{eqnarray}

\noindent
is the unperturbed light ray. From (\ref{Second_Integration_2PN_A}) follows that the perturbations in (\ref{Second_Integration_2PN}) vanish in the limit $t \rightarrow t_0$. 
These perturbations in Eq.~(\ref{Second_Integration_2PN}) were given in their explicit form by Eqs.~(66), (67), (89), (90), and (91) in \cite{Zschocke_Quadrupole_1}.

\section{A new representation of the solution of geodesic equation}\label{Section2}

In our approach in \cite{Zschocke_Quadrupole_1}, the perturbations of geodesics equation in (\ref{Geodesic_Equation_2PN_15}) were written in such a form, 
that the time-dependent scalar functions were separated from the time-independent tensorial coefficients, given in Appendix~E in \cite{Zschocke_Quadrupole_1}. 
The integration of these scalar functions can be reduced to finally only eight master integrals, which can be solved in closed form, given by Eqs.~(D6) - (D9) 
and Eqs.~(D25) - (D28) in Appendix~D in \cite{Zschocke_Quadrupole_1}. Therefore, by this separation of time-dependent scalar functions and time-independent tensorial 
coefficients, the integration procedure has considerably been simplified in \cite{Zschocke_Quadrupole_1}. However, this approach leads to tensorial coefficients, 
which contain linearly dependent tensors. For subsequent investigations it is much more convenient, to rearrange these tensorial terms into a new form, where the 
tensorial coefficients contain only tensors which are linearly independent of each other. In the subsequent Section, this new form for the tensorial coefficients and 
the corresponding scalar functions will be represented. These expressions are the basis for our investigations in \cite{Zschocke_Quadrupole_2}.

\subsection{The first integration of geodesic equation}

\subsubsection{The 1PN terms of first integration of geodesic equation}

The spatial components of 1PN terms in Eq.~(\ref{First_Integration_2PN}) were given by Eqs.~(62) and (63) in \cite{Zschocke_Quadrupole_1}. The tensorial coefficients of the 
1PN monopole terms are given by Eqs.~(52) and (53) in \cite{Zschocke_Quadrupole_1}, which are linearly independent of each other. On the other side, the tensorial coefficients 
of the 1PN quadrupole terms are given by Eqs.~(54) - (57) in \cite{Zschocke_Quadrupole_1}, which contain tensors, which are linearly dependent on each other. A set of tensorial 
coefficients, which contain only tensors, which are linearly independent of each other, is given by Eqs.~(\ref{coefficient_U1_N}) - (\ref{coefficient_V8_N}) in Appendix~\ref{Appendix1}.
If these 1PN terms in Eq.~(\ref{First_Integration_2PN}) are expanded in terms of these new tensorial coefficients, then these spatial components ($i=1,2,3$) are given as follows:
\begin{eqnarray}  
	\frac{\Delta \dot{x}^{i\,{\rm M}}_{\rm 1PN}\left(t\right)}{c} &=& 
        \frac{G M}{c^2} \sum\limits_{n=1}^{2} U^i_{(n)}\left(\ve{x}_{\rm N}\right)\, \dot{F}_{(n)}\left(\ve{x}_{\rm N}\right), 
	\label{First_Integration_1PN_Terms_M}
	\\
        \frac{\Delta \dot{x}^{i\,{\rm Q}}_{\rm 1PN}\left(t\right)}{c} &=& 
	\frac{G M_{ab}}{c^2} \sum\limits_{n=1}^{8} V^{ab\,i}_{(n)}\left(\ve{x}_{\rm N}\right)\, \dot{G}_{(n)}\left(\ve{x}_{\rm N}\right). 
        \label{First_Integration_1PN_Terms_Q} 
\end{eqnarray}

\noindent 
The scalar functions in (\ref{First_Integration_1PN_Terms_M}) and (\ref{First_Integration_1PN_Terms_Q}) are obtained by equating the tensorial coefficients 
of the 1PN solution in \cite{Zschocke_Quadrupole_1} with the tensorial coefficients in Appendix~\ref{Appendix1}.  
These scalar function are given by Eqs.~(\ref{scalar_dot_F1}) - (\ref{scalar_dot_G8}) in Appendix~\ref{Appendix2}.

\subsubsection{The 2PN terms of first integration of geodesic equation}

The spatial components of 2PN terms in Eq.~(\ref{First_Integration_2PN}) were given by Eqs.~(83) - (85) in \cite{Zschocke_Quadrupole_1}. The tensorial coefficients are given by 
Eqs.~(E28) - (E39) in \cite{Zschocke_Quadrupole_1} for the 2PN monopole-monopole terms, by Eqs.~(E42) - (E65) in \cite{Zschocke_Quadrupole_1} for the 2PN monopole-quadrupole terms 
and by Eqs.~(E67) - (E87) in \cite{Zschocke_Quadrupole_1} for the 2PN quadrupole -monopole-quadrupole terms. These tensorial coefficients contain tensors, which are linearly 
dependent on each other. A set of tensorial coefficients, which contain only tensors, which are linearly independent of each other, is given by 
Eqs.~(\ref{coefficient_U1_N}) - (\ref{coefficient_W28}) in Appendix~\ref{Appendix1}.
If these 2PN terms in Eq.~(\ref{First_Integration_2PN}) are expanded in terms of these new tensorial coefficients, then these spatial components ($i=1,2,3$) are given as follows:  
\begin{eqnarray}
	\frac{\Delta \dot{x}^{i\,{\rm M} \times {\rm M}}_{\rm 2PN}\left(t\right)}{c} &=& 
        \frac{G M}{c^2}\,\frac{G M}{c^2} \sum\limits_{n=1}^{2} U^i_{(n)}\left(\ve{x}_{\rm N}\right)\, \dot{A}_{(n)}\left(\ve{x}_{\rm N}\right),
	\label{First_Integration_2PN_Terms_MM}
	\\
        \frac{\Delta \dot{x}^{i\,{\rm M} \times {\rm Q}}_{\rm 2PN}\left(t\right)}{c} &=& 
	\frac{G M}{c^2}\,\frac{G M_{ab}}{c^2} \sum\limits_{n=1}^{8} V^{ab\,i}_{(n)}\left(\ve{x}_{\rm N}\right)\, \dot{B}_{(n)}\left(\ve{x}_{\rm N}\right),
	\label{First_Integration_2PN_Terms_MQ}
        \\ 
        \frac{\Delta \dot{x}^{i\,{\rm Q} \times {\rm Q}}_{\rm 2PN}\left(t\right)}{c} &=& 
	\frac{G M_{ab}}{c^2}\,\frac{G M_{cd}}{c^2} \sum\limits_{n=1}^{28} W^{abcd\,i}_{(n)}\left(\ve{x}_{\rm N}\right)\, \dot{C}_{(n)}\left(\ve{x}_{\rm N}\right). 
	\label{First_Integration_2PN_Terms_QQ}
\end{eqnarray}

\noindent
The scalar functions in (\ref{First_Integration_2PN_Terms_MM}) - (\ref{First_Integration_2PN_Terms_QQ}) are obtained by equating the tensorial coefficients
of the 2PN solution in \cite{Zschocke_Quadrupole_1} with the tensorial coefficients in Appendix~\ref{Appendix1}. These scalar functions are given by
Eqs.~(\ref{scalar_function_dot_A1_2PN}) - (\ref{scalar_function_dot_C28_2PN}) in Appendix~\ref{Appendix2}.

\subsection{The second integration of geodesic equation} 

\subsubsection{The 1PN terms of second integration of geodesic equation}

The spatial components of 1PN terms in Eq.~(\ref{Second_Integration_2PN}) were given by Eqs.~(66) and (67) in \cite{Zschocke_Quadrupole_1}. The tensorial coefficients of the
1PN monopole terms are given by Eqs.~(52) and (53) in \cite{Zschocke_Quadrupole_1}, which are linearly independent of each other. The tensorial coefficients of the 1PN quadrupole 
terms are given by Eqs.~(54) - (57) in \cite{Zschocke_Quadrupole_1}, which, as mentioned above, contain tensors, which are linearly dependent on each other. A set of tensorial
coefficients, which contain only tensors, which are linearly independent of each other, is given by Eqs.~(\ref{coefficient_U1_N}) - (\ref{coefficient_V8_N}) in Appendix~\ref{Appendix1}.
If these 1PN terms in Eq.~(\ref{Second_Integration_2PN}) are expanded in terms of these new tensorial coefficients, then they are given as follows: 
\begin{eqnarray}  
	\Delta \ve{x}^{{\rm M}}_{\rm 1PN}\left(t,t_0\right) &=& \Delta \ve{x}^{{\rm M}}_{\rm 1PN}\left(\ve{x}_{\rm N}\right) - \Delta \ve{x}^{{\rm M}}_{\rm 1PN}\left(\ve{x}_0\right),
	\label{Delta_M}
	\\
	\Delta \ve{x}^{{\rm Q}}_{\rm 1PN}\left(t,t_0\right) &=& \Delta \ve{x}^{{\rm Q}}_{\rm 1PN}\left(\ve{x}_{\rm N}\right) - \Delta \ve{x}^{{\rm Q}}_{\rm 1PN}\left(\ve{x}_0\right),
	\label{Delta_Q}
\end{eqnarray}

\noindent
with 
\begin{eqnarray}  
	\Delta x^{i\,{\rm M}}_{\rm 1PN}\left(\ve{x}_{\rm N}\right) &=& 
        \frac{G M}{c^2} \sum\limits_{n=1}^{2} U^i_{(n)}\left(\ve{x}_{\rm N}\right)\,F_{(n)}\left(\ve{x}_{\rm N}\right),
\label{Second_Integration_1PN_Terms_M}
	\\
        \Delta x^{i\,{\rm Q}}_{\rm 1PN}\left(\ve{x}_{\rm N}\right) &=& 
	+ \frac{G M_{ab}}{c^2} \sum\limits_{n=1}^{8} V^{ab\,i}_{(n)}\left(\ve{x}_{\rm N}\right)\,G_{(n)}\left(\ve{x}_{\rm N}\right). 
\label{Second_Integration_1PN_Terms_Q}
\end{eqnarray}

\noindent
The scalar functions in (\ref{Second_Integration_1PN_Terms_M}) and (\ref{Second_Integration_1PN_Terms_Q}) are obtained by equating the tensorial coefficients
of the 1PN solution in \cite{Zschocke_Quadrupole_1} with the tensorial coefficients in Appendix~\ref{Appendix1}. These scalar function are given by 
Eqs.~(\ref{scalar_F1}) - (\ref{scalar_G8}) in Appendix~\ref{Appendix3}. In order to obtain the terms $\Delta \ve{x}^{{\rm M}}_{\rm 1PN}\left(\ve{x}_0\right)$ and 
$\Delta \ve{x}^{{\rm Q}}_{\rm 1PN}\left(\ve{x}_0\right)$ in Eqs.~(\ref{Delta_M}) and (\ref{Delta_Q}), respectively, one has to replace the arguments 
$\ve{x}_{\rm N}$ by $\ve{x}_0$ in Eqs.~(\ref{Second_Integration_1PN_Terms_M}) and (\ref{Second_Integration_1PN_Terms_Q}).

\subsubsection{The 2PN terms of second integration of geodesic equation}

The spatial components of 2PN terms in Eq.~(\ref{Second_Integration_2PN}) were given by Eqs.~(89) - (91) in \cite{Zschocke_Quadrupole_1}. The tensorial coefficients are given by 
Eqs.~(E28) - (E39) in \cite{Zschocke_Quadrupole_1} for the 2PN monopole-monopole terms, by Eqs.~(E42) - (E65) in \cite{Zschocke_Quadrupole_1} for the 2PN monopole-quadrupole terms
and by Eqs.~(E67) - (E87) in \cite{Zschocke_Quadrupole_1} for the 2PN quadrupole -monopole-quadrupole terms. As mentioned above, these tensorial coefficients contain tensors, 
which are linearly dependent on each other. A set of tensorial coefficients, which contain only tensors, which are linearly independent of each other, is given by
Eqs.~(\ref{coefficient_U1_N}) - (\ref{coefficient_W28}) in Appendix~\ref{Appendix1}.
If these 2PN terms in Eq.~(\ref{Second_Integration_2PN}) are expanded in terms of these new tensorial coefficients, then they are given as follows: 
\begin{eqnarray}
	\Delta \ve{x}^{{\rm M} \times {\rm M}}_{\rm 1PN}\left(t,t_0\right) &=& 
	\Delta \ve{x}^{{\rm M} \times {\rm M}}_{\rm 1PN}\left(\ve{x}_{\rm N}\right) - \Delta \ve{x}^{{\rm M} \times {\rm M}}_{\rm 1PN}\left(\ve{x}_0\right),
        \label{Delta_MM}
        \\
        \Delta \ve{x}^{{\rm M} \times {\rm Q}}_{\rm 1PN}\left(t,t_0\right) &=& 
	\Delta \ve{x}^{{\rm M} \times {\rm Q}}_{\rm 1PN}\left(\ve{x}_{\rm N}\right) - \Delta \ve{x}^{{\rm M} \times {\rm Q}}_{\rm 1PN}\left(\ve{x}_0\right),
        \label{Delta_MQ}
	\\
	\Delta \ve{x}^{{\rm Q} \times {\rm Q}}_{\rm 1PN}\left(t,t_0\right) &=&
        \Delta \ve{x}^{{\rm Q} \times {\rm Q}}_{\rm 1PN}\left(\ve{x}_{\rm N}\right) - \Delta \ve{x}^{{\rm Q} \times {\rm Q}}_{\rm 1PN}\left(\ve{x}_0\right),
        \label{Delta_QQ}
\end{eqnarray}

\noindent
with 
\begin{eqnarray}
       \Delta x^{i\,{\rm M} \times {\rm M}}_{\rm 2PN}\left(\ve{x}_{\rm N}\right) &=&  
       \frac{G M}{c^2}\,\frac{G M}{c^2} \sum\limits_{n=1}^{2} U^i_{(n)}\left(\ve{x}_{\rm N}\right)\, A_{(n)}\left(\ve{x}_{\rm N}\right), 
       \label{Second_Integration_2PN_Terms_MM}
	\\
       \Delta x^{i\,{\rm M} \times {\rm Q}}_{\rm 2PN}\left(\ve{x}_{\rm N}\right) &=& 
       \frac{G M}{c^2}\,\frac{G M_{ab}}{c^2} \sum\limits_{n=1}^{8} V^{ab\,i}_{(n)}\left(\ve{x}_{\rm N}\right)\,B_{(n)}\left(\ve{x}_{\rm N}\right),
       \label{Second_Integration_2PN_Terms_MQ}
       \\ 
       \Delta x^{i\,{\rm Q} \times {\rm Q}}_{\rm 2PN}\left(\ve{x}_{\rm N}\right) &=& 
       \frac{G M_{ab}}{c^2}\,\frac{G M_{cd}}{c^2} \sum\limits_{n=1}^{28} W^{abcd\,i}_{(n)}\left(\ve{x}_{\rm N}\right)\,C_{(n)}\left(\ve{x}_{\rm N}\right).
       \label{Second_Integration_2PN_Terms_QQ}
\end{eqnarray}

\noindent
The scalar functions in (\ref{Second_Integration_2PN_Terms_MM}) - (\ref{Second_Integration_2PN_Terms_QQ}) are obtained by equating the tensorial coefficients
of the 2PN solution in \cite{Zschocke_Quadrupole_1} with the tensorial coefficients in Appendix~\ref{Appendix1}. These scalar functions are given by
Eqs.~(\ref{scalar_function_A1_2PN}) - (\ref{scalar_function_C28_2PN}) in Appendix~\ref{Appendix3}. In order to obtain the terms 
$\Delta \ve{x}^{{\rm M} \times {\rm M}}_{\rm 2PN}\left(\ve{x}_0\right)$, $\Delta \ve{x}^{{\rm M} \times {\rm Q}}_{\rm 2PN}\left(\ve{x}_0\right)$, 
and $\Delta \ve{x}^{{\rm Q} \times {\rm Q}}_{\rm 2PN}\left(\ve{x}_0\right)$ in Eqs.~(\ref{Delta_MM}), (\ref{Delta_MQ}), and (\ref{Delta_QQ}), 
one has to replace the arguments $\ve{x}_{\rm N}$ by $\ve{x}_0$ in Eqs.~(\ref{Second_Integration_2PN_Terms_MM}) - (\ref{Second_Integration_2PN_Terms_QQ}).

\section{Summary}

Astrometry on the sub-\muas{} scale requires a general-relativistic modeling of light trajectories in the solar system 
at the second post-Newtonian (2PN) approximation. In particular, it is not sufficient to account only for the monopole 
term of the body, but also the impact of the quadrupole structure of solar system bodies on light propagation 
needs to be considered. Such a solution of the geodesic equation for light rays has recently been achieved in our
investigation in \cite{Zschocke_Quadrupole_1}. The first and second integration of geodesic equation has been
determined by iteration in the scheme of the initial value problem. In order to determine the light trajectory, 
the individual terms of the geodesic equation have been separated into time-dependent scalar functions and time-independent 
tensorial coefficients. This approach has lead to eight master integrals, which have been solved in closed form in \cite{Zschocke_Quadrupole_1}. 
This procedure, however, leads to tensorial coefficients, which contain linearly dependent tensors. For subsequent investigations, it is much 
more effective, to rearrange these tensorial terms into such a form, that these tensorial coefficients contain only tensors which are linearly 
independent of each other. This rearranged new form of the solution of geodesic equation has been given by 
Eqs.~(\ref{First_Integration_1PN_Terms_M}) - (\ref{First_Integration_2PN_Terms_QQ}) for the first integration of geodesic equation, and by 
Eqs.~(\ref{Delta_M}) - (\ref{Second_Integration_2PN_Terms_QQ}) for the second integration of geodesic equation. 
This rearranged new form allows for our subsequent investigations of light propagation in the monopole and quadrupole field of solar system bodies 
on the sub-\muas{} level of astrometric angular observations \cite{Zschocke_Quadrupole_2}.

\section*{Acknowledgment}

This work was funded by the German Research Foundation (Deutsche Forschungsgemeinschaft DFG) under grant number 447922800.

\appendix

\section{Tensorial coefficients}\label{Appendix1}

The tensorial coefficients in Eqs.~(\ref{First_Integration_1PN_Terms_M}) - (\ref{First_Integration_2PN_Terms_QQ}) and 
Eqs.~(\ref{Second_Integration_1PN_Terms_M}) - (\ref{Second_Integration_1PN_Terms_Q}) as well as Eqs.~(\ref{Second_Integration_2PN_Terms_MM}) - (\ref{Second_Integration_2PN_Terms_QQ}) 
are given by
\begin{eqnarray}
        U^{i}_{\left(1\right)}\left(\ve{x}_{\rm N}\right) &=& \sigma^i\;,
        \label{coefficient_U1_N}
        \\
        U^{i}_{\left(2\right)}\left(\ve{x}_{\rm N}\right) &=& d_{\sigma}^i \;,
        \label{coefficient_U2_N}
        \\
        V^{ab\,i}_{\left(1\right)}\left(\ve{x}_{\rm N}\right) &=& \sigma^a \delta^{bi}\;,
        \label{coefficient_V1_N}
        \\
        V^{ab\,i}_{\left(2\right)}\left(\ve{x}_{\rm N}\right) &=& d_{\sigma}^a \delta^{bi}\;,
        \label{coefficient_V2_N}
        \\
        V^{ab\,i}_{\left(3\right)}\left(\ve{x}_{\rm N}\right) &=& \sigma^a \sigma^b \sigma^i\;,
        \label{coefficient_V3_N}
        \\
        V^{ab\,i}_{\left(4\right)}\left(\ve{x}_{\rm N}\right) &=& \sigma^a d_{\sigma}^b \sigma^i\;,
        \label{coefficient_V4_N}
        \\
        V^{ab\,i}_{\left(5\right)}\left(\ve{x}_{\rm N}\right) &=& d_{\sigma}^a d_{\sigma}^b \sigma^i\;,
        \label{coefficient_V5_N}
        \\
        V^{ab\,i}_{\left(6\right)}\left(\ve{x}_{\rm N}\right) &=& d_{\sigma}^a d_{\sigma}^b d_{\sigma}^i\;,
        \label{coefficient_V6_N}
        \\
        V^{ab\,i}_{\left(7\right)}\left(\ve{x}_{\rm N}\right) &=& \sigma^a \sigma^b d_{\sigma}^i\;,
        \label{coefficient_V7_N}
        \\
        V^{ab\,i}_{\left(8\right)}\left(\ve{x}_{\rm N}\right) &=& \sigma^a d_{\sigma}^b d_{\sigma}^i\;.
        \label{coefficient_V8_N}
\end{eqnarray}

\noindent
and 
\begin{eqnarray}
        W^{abcd\,i}_{\left(1\right)}\left(\ve{x}_{\rm N}\right) &=& \delta^{ac} \sigma^b \delta^{di} \;, 
        \label{coefficient_W1}
        \\
        W^{abcd\,i}_{\left(2\right)}\left(\ve{x}_{\rm N}\right) &=& \delta^{ac} d_{\sigma}^b \delta^{di} \;,
        \label{coefficient_W2}
        \\
        W^{abcd\,i}_{\left(3\right)}\left(\ve{x}_{\rm N}\right) &=& \sigma^a \sigma^b \sigma^c \delta^{di} \;, 
        \label{coefficient_W3}
        \\
        W^{abcd\,i}_{\left(4\right)}\left(\ve{x}_{\rm N}\right) &=& \sigma^a \sigma^b d_{\sigma}^c \delta^{di} \;,
        \label{coefficient_W4}
        \\
        W^{abcd\,i}_{\left(5\right)}\left(\ve{x}_{\rm N}\right) &=& \sigma^a d_{\sigma}^b \sigma^c \delta^{di} \;, 
        \label{coefficient_W5}
        \\
        W^{abcd\,i}_{\left(6\right)}\left(\ve{x}_{\rm N}\right) &=& \sigma^a d_{\sigma}^b d_{\sigma}^c \delta^{di} \;, 
        \label{coefficient_W6}
        \\
        W^{abcd\,i}_{\left(7\right)}\left(\ve{x}_{\rm N}\right) &=& d_{\sigma}^a d_{\sigma}^b \sigma^c \delta^{di} \;,
        \label{coefficient_W7}
        \\
        W^{abcd\,i}_{\left(8\right)}\left(\ve{x}_{\rm N}\right) &=& d_{\sigma}^a d_{\sigma}^b d_{\sigma}^c \delta^{di} \;, 
        \label{coefficient_W8}
	\\ 
        W^{abcd\,i}_{\left(9\right)}\left(\ve{x}_{\rm N}\right) &=& \delta^{ac} \delta^{bd} \sigma^i \;, 
        \label{coefficient_W9} 
        \\
        W^{abcd\,i}_{\left(10\right)}\left(\ve{x}_{\rm N}\right) &=& \delta^{ac} \sigma^b \sigma^d \sigma^i \;,
        \label{coefficient_W10} 
        \\
        W^{abcd\,i}_{\left(11\right)}\left(\ve{x}_{\rm N}\right) &=& \delta^{ac} \sigma^b d_{\sigma}^d \sigma^i \;,
        \label{coefficient_W11}
        \\
        W^{abcd\,i}_{\left(12\right)}\left(\ve{x}_{\rm N}\right) &=& \delta^{ac} d_{\sigma}^b d_{\sigma}^d \sigma^i \;,
        \label{coefficient_W12}
        \\
        W^{abcd\,i}_{\left(13\right)}\left(\ve{x}_{\rm N}\right) &=& \sigma^a \sigma^b \sigma^c \sigma^d \sigma^i \;, 
        \label{coefficient_W13}
        \\
        W^{abcd\,i}_{\left(14\right)}\left(\ve{x}_{\rm N}\right) &=& \sigma^a \sigma^b \sigma^c d_{\sigma}^d \sigma^i \;,
        \label{coefficient_W14}
        \\ 
        W^{abcd\,i}_{\left(15\right)}\left(\ve{x}_{\rm N}\right) &=& \sigma^a \sigma^b d_{\sigma}^c d_{\sigma}^d \sigma^i \;,
        \label{coefficient_W15}
        \\
        W^{abcd\,i}_{\left(16\right)}\left(\ve{x}_{\rm N}\right) &=& \sigma^a d_{\sigma}^b \sigma^c d_{\sigma}^d \sigma^i \;,
        \label{coefficient_W16}
        \\ 
        W^{abcd\,i}_{\left(17\right)}\left(\ve{x}_{\rm N}\right) &=& \sigma^a d_{\sigma}^b d_{\sigma}^c d_{\sigma}^d \sigma^i \;,
        \label{coefficient_W17}
        \end{eqnarray}

        \begin{eqnarray}  
        W^{abcd\,i}_{\left(18\right)}\left(\ve{x}_{\rm N}\right) &=& d_{\sigma}^a d_{\sigma}^b d_{\sigma}^c d_{\sigma}^d \sigma^i \;,
        \label{coefficient_W18}
        \\ 
        W^{abcd\,i}_{\left(19\right)}\left(\ve{x}_{\rm N}\right) &=& \delta^{ac} \delta^{bd} d_{\sigma}^i \;,
        \label{coefficient_W19} 
        \\
        W^{abcd\,i}_{\left(20\right)}\left(\ve{x}_{\rm N}\right) &=& \delta^{ac} \sigma^b \sigma^d d_{\sigma}^i \;,
        \label{coefficient_W20} 
        \\ 
        W^{abcd\,i}_{\left(21\right)}\left(\ve{x}_{\rm N}\right) &=& \delta^{ac} \sigma^b d_{\sigma}^d d_{\sigma}^i \;,
        \label{coefficient_W21}
        \\
        W^{abcd\,i}_{\left(22\right)}\left(\ve{x}_{\rm N}\right) &=& \delta^{ac} d_{\sigma}^b d_{\sigma}^d d_{\sigma}^i \;,
        \label{coefficient_W22}
        \\
        W^{abcd\,i}_{\left(23\right)}\left(\ve{x}_{\rm N}\right) &=& \sigma^a \sigma^b \sigma^c \sigma^d d_{\sigma}^i \;,
        \label{coefficient_W23}
        \\
        W^{abcd\,i}_{\left(24\right)}\left(\ve{x}_{\rm N}\right) &=& \sigma^a \sigma^b \sigma^c d_{\sigma}^d d_{\sigma}^i \;,
        \label{coefficient_W24}
        \\
        W^{abcd\,i}_{\left(25\right)}\left(\ve{x}_{\rm N}\right) &=& \sigma^a \sigma^b d_{\sigma}^c d_{\sigma}^d d_{\sigma}^i \;,
        \label{coefficient_W25}
        \\
        W^{abcd\,i}_{\left(26\right)}\left(\ve{x}_{\rm N}\right) &=& \sigma^a d_{\sigma}^b \sigma^c d_{\sigma}^d d_{\sigma}^i \;,
        \label{coefficient_W26}
        \\
        W^{abcd\,i}_{\left(27\right)}\left(\ve{x}_{\rm N}\right) &=& \sigma^a d_{\sigma}^b d_{\sigma}^c d_{\sigma}^d d_{\sigma}^i \;,
        \label{coefficient_W27}
        \\
        W^{abcd\,i}_{\left(28\right)}\left(\ve{x}_{\rm N}\right) &=& d_{\sigma}^a d_{\sigma}^b d_{\sigma}^c d_{\sigma}^d d_{\sigma}^i \;.
        \label{coefficient_W28}
\end{eqnarray}

\section{Scalar functions of first integration}\label{Appendix2}

The scalar functions of the first integration are based on the set of $4$ scalar functions: 
$\dot{\cal W}_{\left(n\right)}$, $\dot{\cal X}_{\left(n\right)}$, $\dot{\cal Y}_{\left(n\right)}$, $\dot{\cal Z}_{\left(n\right)}$. 
These $4$ types of scalar functions are defined as $4$ types of master integrals, given by Eqs.~(D1) - (D4) in \cite{Zschocke_Quadrupole_1}, 
which have been solved in closed form by Eqs.~(D6) - (D9) in \cite{Zschocke_Quadrupole_1}. 

\subsection{Scalar functions of the 1PN Monopole term}

The scalar functions of the 1PN monopole term in Eq.~(\ref{First_Integration_1PN_Terms_M}) are given by 
\begin{eqnarray}
        \dot{F}_{\left(1\right)}\left(\ve{x}_{\rm N}\right) &=& + 2\,\dot{\cal W}_{\left(3\right)}\left(\ve{x}_{\rm N}\right)\,,
        \label{scalar_dot_F1}
        \\
        \dot{F}_{\left(2\right)}\left(\ve{x}_{\rm N}\right) &=& - 2\,\dot{\cal X}_{\left(3\right)}\left(\ve{x}_{\rm N}\right)\,. 
        \label{scalar_dot_F2}
\end{eqnarray}

\subsection{Scalar functions of the 1PN Quadrupole term}

The scalar functions of the 1PN quadrupole term in Eq.~(\ref{First_Integration_1PN_Terms_Q}) are given by
\begin{eqnarray}
        \dot{G}_{\left(1\right)}\left(\ve{x}_{\rm N}\right) &=& + 6\,\dot{\cal W}_{\left(5\right)}\left(\ve{x}_{\rm N}\right),
        \label{scalar_dot_G1}
        \\
        \dot{G}_{\left(2\right)}\left(\ve{x}_{\rm N}\right) &=& + 6\,\dot{\cal X}_{\left(5\right)}\left(\ve{x}_{\rm N}\right),
        \label{scalar_dot_G2}
        \\
        \dot{G}_{\left(3\right)}\left(\ve{x}_{\rm N}\right) &=& + 3\,\dot{\cal W}_{\left(5\right)}\left(\ve{x}_{\rm N}\right)
        - 15 \left(d_{\sigma}\right)^2 \dot{\cal W}_{\left(7\right)}\left(\ve{x}_{\rm N}\right),
        \label{scalar_dot_G3}
        \\
        \dot{G}_{\left(4\right)}\left(\ve{x}_{\rm N}\right) &=& + 18\,\dot{\cal X}_{\left(5\right)}\left(\ve{x}_{\rm N}\right)
        - 30 \left(d_{\sigma}\right)^2 \dot{\cal X}_{\left(7\right)}\left(\ve{x}_{\rm N}\right),
        \label{scalar_dot_G4}
        \\
        \dot{G}_{\left(5\right)}\left(\ve{x}_{\rm N}\right) &=& + 15\,\dot{\cal W}_{\left(7\right)}\left(\ve{x}_{\rm N}\right),
        \label{scalar_dot_G5}
        \\
        \dot{G}_{\left(6\right)}\left(\ve{x}_{\rm N}\right) &=& - 15\,\dot{\cal X}_{\left(7\right)}\left(\ve{x}_{\rm N}\right),
       \label{scalar_dot_G6}
        \\
        \dot{G}_{\left(7\right)}\left(\ve{x}_{\rm N}\right) &=& - 15\,\dot{\cal X}_{\left(5\right)}\left(\ve{x}_{\rm N}\right)
        + 15 \left(d_{\sigma}\right)^2 \dot{\cal X}_{\left(7\right)}\left(\ve{x}_{\rm N}\right),
        \label{scalar_dot_G7}
        \\
        \dot{G}_{\left(8\right)}\left(\ve{x}_{\rm N}\right) &=& - 30\,\dot{\cal W}_{\left(7\right)}\left(\ve{x}_{\rm N}\right). 
        \label{scalar_dot_G8}
\end{eqnarray}

\subsection{Scalar functions of the 2PN Monopole-Monopole term}

The scalar functions of the 2PN monopole-monopole term in Eq.~(\ref{First_Integration_2PN_Terms_MM}) are given by
\begin{eqnarray}
        \dot{A}_{\left(1\right)}\left(\ve{x}_{\rm N}\right) &=& - 4\,\dot{\cal W}_{\left(4\right)}
        - 12 \left(x_0 + \ve{\sigma} \cdot \ve{x}_0\right) \dot{\cal W}_{\left(5\right)}
        - 2 \left(d_{\sigma}\right)^2 \dot{\cal W}_{\left(6\right)}
        + 4\,\dot{\cal X}_{\left(3\right)}
        - 12 \left(d_{\sigma}\right)^2 \dot{\cal X}_{\left(5\right)}
        - 8\,\dot{\cal Z}_{\left(3\right)}
        + 12 \left(d_{\sigma}\right)^2 \dot{\cal Z}_{\left(5\right)},
        \label{scalar_function_dot_A1_2PN}
        \\
	\dot{A}_{\left(2\right)}\left(\ve{x}_{\rm N}\right) &=& - \frac{4}{\left(d_{\sigma}\right)^2}\,\dot{\cal W}_{\left(3\right)}
        - 12\,\dot{\cal W}_{\left(5\right)} - \frac{4}{\left(d_{\sigma}\right)^2}\,\dot{\cal X}_{\left(2\right)}
        - \frac{4}{\left(d_{\sigma}\right)^2} \left(x_0 + \ve{\sigma} \cdot \ve{x}_0\right) \dot{\cal X}_{\left(3\right)}
        + 6\,\dot{\cal X}_{\left(4\right)} + 12 \left(x_0 + \ve{\sigma} \cdot \ve{x}_0\right) \dot{\cal X}_{\left(5\right)}
        \nonumber\\
        && - 2 \left(d_{\sigma}\right)^2 \dot{\cal X}_{\left(6\right)}
        + 12 \,\dot{\cal Y}_{\left(5\right)}\,. 
        \label{scalar_function_dot_A2_2PN}
\end{eqnarray}

\subsection{Scalar functions of the 2PN Monopole-Quadrupole term}

The scalar functions of the 2PN monopole-quadrupole term in Eq.~(\ref{First_Integration_2PN_Terms_MQ}) are given by
\begin{eqnarray}
        \dot{B}_{\left(1\right)}\left(\ve{x}_{\rm N}\right) &=& + \frac{4}{\left(d_{\sigma}\right)^2}\,\dot{\cal W}_{\left(4\right)}
        + 22\,\dot{\cal W}_{\left(6\right)} - 60 \left(x_0 + \ve{\sigma} \cdot \ve{x}_0\right) \dot{\cal W}_{\left(7\right)}
        + \frac{21}{2} \left(d_{\sigma}\right)^2 \dot{\cal W}_{\left(8\right)}
        - \frac{4}{\left(d_{\sigma}\right)^2}\,\frac{\ve{\sigma} \cdot \ve{x}_0}{x_0}\,\dot{\cal X}_{\left(3\right)}
        + 60\,\dot{\cal X}_{\left(5\right)}
        \nonumber\\
        && - 60 \left(d_{\sigma}\right)^2 \dot{\cal X}_{\left(7\right)}
        - 48\,\dot{\cal Z}_{\left(5\right)} + 60 \left(d_{\sigma}\right)^2 \dot{\cal Z}_{\left(7\right)} \,,
        \label{scalar_function_dot_B1_2PN}
        \\
        \dot{B}_{\left(2\right)}\left(\ve{x}_{\rm N}\right) &=& + \frac{8}{\left(d_{\sigma}\right)^4}\,\dot{\cal W}_{\left(3\right)} 
	- \frac{12}{\left(d_{\sigma}\right)^2}\,\dot{\cal W}_{\left(5\right)} + 60\,\dot{\cal W}_{\left(7\right)} 
        + \frac{8}{\left(d_{\sigma}\right)^4}\,\dot{\cal X}_{\left(2\right)} - \frac{4}{\left(d_{\sigma}\right)^2}\,\frac{1}{x_0}\,\dot{\cal X}_{\left(3\right)} 
	+ \frac{8}{\left(d_{\sigma}\right)^4}\left(x_0 + \ve{\sigma} \cdot \ve{x}_0 \right) \dot{\cal X}_{\left(3\right)}
	\nonumber\\
	&& - \frac{16}{\left(d_{\sigma}\right)^2}\,\dot{\cal X}_{\left(4\right)}
	+ \frac{12}{\left(d_{\sigma}\right)^2}\left(x_0 + \ve{\sigma} \cdot \ve{x}_0 \right) \dot{\cal X}_{\left(5\right)} 
	+ 18\,\dot{\cal X}_{\left(6\right)} - 60 \left(x_0 + \ve{\sigma} \cdot \ve{x}_0 \right) \dot{\cal X}_{\left(7\right)} 
	+ \frac{21}{2} \left(d_{\sigma}\right)^2 \dot{\cal X}_{\left(8\right)} - 60\,\dot{\cal Y}_{\left(7\right)}\,,
	\label{scalar_function_dot_B2_2PN}
        \\
        \dot{B}_{\left(3\right)}\left(\ve{x}_{\rm N}\right) &=& + \frac{4}{\left(d_{\sigma}\right)^2}\,\dot{\cal W}_{\left(4\right)} 
	+ \frac{6}{x_0}\,\dot{\cal W}_{\left(5\right)} - \frac{6}{\left(x_0\right)^3} \left(d_{\sigma}\right)^2 \dot{\cal W}_{\left(5\right)}
	- \frac{12}{\left(d_{\sigma}\right)^2} \left(x_0 + \ve{\sigma} \cdot \ve{x}_0 \right) \dot{\cal W}_{\left(5\right)} 
	- 90 \left(x_0 + \ve{\sigma} \cdot \ve{x}_0 \right) \dot{\cal W}_{\left(7\right)} 
	\nonumber\\ 
	&& - 84 \left(d_{\sigma}\right)^2  \dot{\cal W}_{\left(8\right)}
	+ 210 \left(d_{\sigma}\right)^2 \left(x_0 + \ve{\sigma} \cdot \ve{x}_0 \right) \dot{\cal W}_{\left(9\right)} 
	+ 30 \left(d_{\sigma}\right)^4 \dot{\cal W}_{\left(10\right)} + \frac{4}{\left(d_{\sigma}\right)^2}\,\dot{\cal X}_{\left(3\right)} 
	+ 4\,\frac{\ve{\sigma} \cdot \ve{x}_0}{\left(x_0\right)^3}\,\dot{\cal X}_{\left(3\right)} + 18\,\dot{\cal X}_{\left(5\right)} 
	\nonumber\\ 
	&& + 6\,\frac{\ve{\sigma} \cdot \ve{x}_0}{x_0}\,\dot{\cal X}_{\left(5\right)}
	- 6 \left(d_{\sigma}\right)^2 \frac{\ve{\sigma} \cdot \ve{x}_0}{\left(x_0\right)^3}\,\dot{\cal X}_{\left(5\right)}
	- 240 \left(d_{\sigma}\right)^2 \dot{\cal X}_{\left(7\right)} + 210 \left(d_{\sigma}\right)^4 \dot{\cal X}_{\left(9\right)} 
	- 24\,\dot{\cal Z}_{\left(5\right)} + 210 \left(d_{\sigma}\right)^2 \dot{\cal Z}_{\left(7\right)} 
	\nonumber\\ 
	&& - 210 \left(d_{\sigma}\right)^4 \dot{\cal Z}_{\left(9\right)}\,,
        \label{scalar_function_dot_B3_2PN}
        \\
        \dot{B}_{\left(4\right)}\left(\ve{x}_{\rm N}\right) &=& - \frac{8}{\left(d_{\sigma}\right)^4}\,\dot{\cal W}_{\left(3\right)} 
	- \frac{12}{\left(d_{\sigma}\right)^2}\,\dot{\cal W}_{\left(5\right)} + \frac{12}{\left(d_{\sigma}\right)^2}\,\frac{\ve{\sigma} \cdot \ve{x}_0}{x_0}\,\dot{\cal W}_{\left(5\right)} 
	+ 12 \,\frac{\ve{\sigma} \cdot \ve{x}_0}{\left(x_0\right)^3}\,\dot{\cal W}_{\left(5\right)} 
	+ 240\,\dot{\cal W}_{\left(7\right)} - 420 \left(d_{\sigma}\right)^2 \dot{\cal W}_{\left(9\right)} 
	- \frac{8}{\left(d_{\sigma}\right)^4}\,\dot{\cal X}_{\left(2\right)} 
        \nonumber\\ 
	&& + \frac{4}{\left(d_{\sigma}\right)^2}\,\frac{1}{x_0}\,\dot{\cal X}_{\left(3\right)} 
	+ \frac{8}{\left(x_0\right)^3}\,\dot{\cal X}_{\left(3\right)} - \frac{8}{\left(d_{\sigma}\right)^4} \left(x_0 + \ve{\sigma} \cdot \ve{x}_0\right) \dot{\cal X}_{\left(3\right)} 
	- \frac{20}{\left(d_{\sigma}\right)^2}\,\dot{\cal X}_{\left(4\right)} - \frac{12}{\left(x_0\right)^3} \left(d_{\sigma}\right)^2 \dot{\cal X}_{\left(5\right)}
	\nonumber\\ 
	&& + \frac{36}{\left(d_{\sigma}\right)^2}\left(x_0 + \ve{\sigma} \cdot \ve{x}_0\right) \dot{\cal X}_{\left(5\right)} + 132\,\dot{\cal X}_{\left(6\right)} 
	- 360 \left(x_0 + \ve{\sigma} \cdot \ve{x}_0\right) \dot{\cal X}_{\left(7\right)} - 183 \left(d_{\sigma}\right)^2 \dot{\cal X}_{\left(8\right)} 
	+ 420 \left(d_{\sigma}\right)^2 \left(x_0 + \ve{\sigma} \cdot \ve{x}_0\right) \dot{\cal X}_{\left(9\right)}
	\nonumber\\
	&& + 60 \left(d_{\sigma}\right)^4 \dot{\cal X}_{\left(10\right)} - 180\,\dot{\cal Y}_{\left(7\right)} + 420 \left(d_{\sigma}\right)^2 \dot{\cal Y}_{\left(9\right)}\,,
        \label{scalar_function_dot_B4_2PN}
        \\
        \dot{B}_{\left(5\right)}\left(\ve{x}_{\rm N}\right) &=& + \frac{16}{\left(d_{\sigma}\right)^4}\,\dot{\cal W}_{\left(4\right)} 
	+ \frac{12}{\left(d_{\sigma}\right)^2}\,\frac{1}{x_0}\,\dot{\cal W}_{\left(5\right)} + \frac{6}{\left(x_0\right)^3}\,\dot{\cal W}_{\left(5\right)} 
	- \frac{24}{\left(d_{\sigma}\right)^4} \left(x_0 + \ve{\sigma} \cdot \ve{x}_0\right) \dot{\cal W}_{\left(5\right)} 
	- \frac{64}{\left(d_{\sigma}\right)^2}\,\dot{\cal W}_{\left(6\right)}
	\nonumber\\ 
	&& + \frac{60}{\left(d_{\sigma}\right)^2}\left(x_0 + \ve{\sigma} \cdot \ve{x}_0\right) \dot{\cal W}_{\left(7\right)} + 69\,\dot{\cal W}_{\left(8\right)}
	- 210 \left(x_0 + \ve{\sigma} \cdot \ve{x}_0\right) \dot{\cal W}_{\left(9\right)} - 30 \left(d_{\sigma}\right)^2 \dot{\cal W}_{\left(10\right)} 
	+ \frac{8}{\left(d_{\sigma}\right)^4}\,\dot{\cal X}_{\left(3\right)}
	\nonumber\\ 
	&& - \frac{8}{\left(d_{\sigma}\right)^4}\,\frac{\ve{\sigma} \cdot \ve{x}_0}{x_0}\,\dot{\cal X}_{\left(3\right)} 
	- \frac{4}{\left(d_{\sigma}\right)^2}\,\frac{\ve{\sigma} \cdot \ve{x}_0}{\left(x_0\right)^3}\,\dot{\cal X}_{\left(3\right)} 
	- \frac{60}{\left(d_{\sigma}\right)^2}\,\dot{\cal X}_{\left(5\right)} 
	+ \frac{12}{\left(d_{\sigma}\right)^2}\,\frac{\ve{\sigma} \cdot \ve{x}_0}{x_0}\,\dot{\cal X}_{\left(5\right)} 
	+ 6 \, \frac{\ve{\sigma} \cdot \ve{x}_0}{\left(x_0\right)^3}\,\dot{\cal X}_{\left(5\right)}
	\nonumber\\ 
	&& + 210\,\dot{\cal X}_{\left(7\right)} - 210 \left(d_{\sigma}\right)^2 \dot{\cal X}_{\left(9\right)} 
	- 180\,\dot{\cal Z}_{\left(7\right)} 
	+ 210 \left(d_{\sigma}\right)^2 \dot{\cal Z}_{\left(9\right)} \,,
        \label{scalar_function_dot_B5_2PN}
        \\
        \dot{B}_{\left(6\right)}\left(\ve{x}_{\rm N}\right) &=& - \frac{16}{\left(d_{\sigma}\right)^6}\,\dot{\cal W}_{\left(3\right)} 
	- \frac{24}{\left(d_{\sigma}\right)^4}\,\dot{\cal W}_{\left(5\right)} 
	+ \frac{12}{\left(d_{\sigma}\right)^4}\,\frac{\ve{\sigma} \cdot \ve{x}_0}{x_0}\,\dot{\cal W}_{\left(5\right)} 
	+ \frac{6}{\left(d_{\sigma}\right)^2}\,\frac{\ve{\sigma} \cdot \ve{x}_0}{\left(x_0\right)^3}\,\dot{\cal W}_{\left(5\right)} 
        + \frac{30}{\left(d_{\sigma}\right)^2}\,\dot{\cal W}_{\left(7\right)} - 210\,\dot{\cal W}_{\left(9\right)}
	\nonumber\\ 
	&& - \frac{16}{\left(d_{\sigma}\right)^6}\,\dot{\cal X}_{\left(2\right)} 
	+ \frac{8}{\left(d_{\sigma}\right)^4}\,\frac{1}{x_0}\,\dot{\cal X}_{\left(3\right)}
	+ \frac{2}{\left(d_{\sigma}\right)^2}\,\frac{1}{\left(x_0\right)^3}\,\dot{\cal X}_{\left(3\right)} 
	- \frac{16}{\left(d_{\sigma}\right)^6}\left(x_0 + \ve{\sigma} \cdot \ve{x}_0\right)\dot{\cal X}_{\left(3\right)} 
	- \frac{28}{\left(d_{\sigma}\right)^4}\,\dot{\cal X}_{\left(4\right)}
	\nonumber\\ 
	&& - \frac{12}{\left(d_{\sigma}\right)^2}\,\frac{1}{x_0}\,\dot{\cal X}_{\left(5\right)}
	- \frac{6}{\left(x_0\right)^3}\,\dot{\cal X}_{\left(5\right)} 
	+ \frac{24}{\left(d_{\sigma}\right)^4}\left(x_0 + \ve{\sigma} \cdot \ve{x}_0\right) \dot{\cal X}_{\left(5\right)} 
	+ \frac{50}{\left(d_{\sigma}\right)^2}\,\dot{\cal X}_{\left(6\right)} 
	- \frac{90}{\left(d_{\sigma}\right)^2}\left(x_0 + \ve{\sigma} \cdot \ve{x}_0\right) \dot{\cal X}_{\left(7\right)}
	- 24\,\dot{\cal X}_{\left(8\right)} 
	\nonumber\\ 
	&& + 210 \left(x_0 + \ve{\sigma} \cdot \ve{x}_0\right) \dot{\cal X}_{\left(9\right)} 
	- 30 \left(d_{\sigma}\right)^2 \dot{\cal X}_{\left(10\right)} + 210\,\dot{\cal Y}_{\left(9\right)}\,,
        \label{scalar_function_dot_B6_2PN}
        \end{eqnarray}

        \begin{eqnarray} 
        \dot{B}_{\left(7\right)}\left(\ve{x}_{\rm N}\right) &=& - \frac{4}{\left(d_{\sigma}\right)^4}\,\dot{\cal W}_{\left(3\right)} 
	- \frac{18}{\left(d_{\sigma}\right)^2}\,\dot{\cal W}_{\left(5\right)} 
	+ \frac{6}{\left(d_{\sigma}\right)^2}\,\frac{\ve{\sigma} \cdot \ve{x}_0}{x_0}\,\dot{\cal W}_{\left(5\right)}
	- 6\,\frac{\ve{\sigma} \cdot \ve{x}_0}{\left(x_0\right)^3}\,\dot{\cal W}_{\left(5\right)} - 180\,\dot{\cal W}_{\left(7\right)} 
	+ 210 \left(d_{\sigma}\right)^2 \dot{\cal W}_{\left(9\right)} 
        \nonumber\\ 
	&& - \frac{4}{\left(d_{\sigma}\right)^4}\,\dot{\cal X}_{\left(2\right)}
	+ \frac{2}{\left(d_{\sigma}\right)^2}\,\frac{1}{x_0}\,\dot{\cal X}_{\left(3\right)} 
	- \frac{2}{\left(x_0\right)^3}\,\dot{\cal X}_{\left(3\right)} 
	- \frac{4}{\left(d_{\sigma}\right)^4} \left(x_0 + \ve{\sigma} \cdot \ve{x}_0\right) \dot{\cal X}_{\left(3\right)} 
	- \frac{22}{\left(d_{\sigma}\right)^2}\,\dot{\cal X}_{\left(4\right)} - \frac{6}{x_0}\,\dot{\cal X}_{\left(5\right)}
	\nonumber\\ 
	&& + \frac{6}{\left(x_0\right)^3} \left(d_{\sigma}\right)^2 \dot{\cal X}_{\left(5\right)} 
	- \frac{18}{\left(d_{\sigma}\right)^2} \left(x_0 + \ve{\sigma} \cdot \ve{x}_0\right) \dot{\cal X}_{\left(5\right)} 
	- 20\,\dot{\cal X}_{\left(6\right)} + 240 \left(x_0 + \ve{\sigma} \cdot \ve{x}_0\right) \dot{\cal X}_{\left(7\right)} 
	+ 9 \left(d_{\sigma}\right)^2 \dot{\cal X}_{\left(8\right)} 
	\nonumber\\ 
	&& - 210 \left(d_{\sigma}\right)^2 \left(x_0 + \ve{\sigma} \cdot \ve{x}_0\right) \dot{\cal X}_{\left(9\right)} 
	+ 30 \left(d_{\sigma}\right)^4 \dot{\cal X}_{\left(10\right)} + 150\,\dot{\cal Y}_{\left(7\right)} 
	- 210 \left(d_{\sigma}\right)^2 \dot{\cal Y}_{\left(9\right)}\,,
        \label{scalar_function_dot_B7_2PN}
        \\ 
        \dot{B}_{\left(8\right)}\left(\ve{x}_{\rm N}\right) &=& - \frac{8}{\left(d_{\sigma}\right)^4}\,\dot{\cal W}_{\left(4\right)}
        - \frac{12}{\left(x_0\right)^3}\,\dot{\cal W}_{\left(5\right)}
        + \frac{32}{\left(d_{\sigma}\right)^2}\,\dot{\cal W}_{\left(6\right)} 
        - \frac{120}{\left(d_{\sigma}\right)^2} \left(x_0 + \ve{\sigma} \cdot \ve{x}_0\right) \dot{\cal W}_{\left(7\right)}
	- 63\,\dot{\cal W}_{\left(8\right)}
	\nonumber\\ 
	&& + 420 \left(x_0 + \ve{\sigma} \cdot \ve{x}_0\right) \dot{\cal W}_{\left(9\right)}
        - 60 \left(d_{\sigma}\right)^2 \dot{\cal W}_{\left(10\right)}
        + \frac{8}{\left(d_{\sigma}\right)^4}\,\frac{\ve{\sigma} \cdot \ve{x}_0}{x_0}\,\dot{\cal X}_{\left(3\right)}
        + \frac{4}{\left(d_{\sigma}\right)^2}\,\frac{\ve{\sigma} \cdot \ve{x}_0}{\left(x_0\right)^3}\,\dot{\cal X}_{\left(3\right)}
        + \frac{24}{\left(d_{\sigma}\right)^2}\,\dot{\cal X}_{\left(5\right)}
        \nonumber\\
	&& - \frac{12}{\left(d_{\sigma}\right)^2}\,\frac{\ve{\sigma} \cdot \ve{x}_0}{x_0}\,\dot{\cal X}_{\left(5\right)}
	- 12\,\frac{\ve{\sigma} \cdot \ve{x}_0}{\left(x_0\right)^3}\,\dot{\cal X}_{\left(5\right)}
        - 420\,\dot{\cal X}_{\left(7\right)} + 420 \left(d_{\sigma}\right)^2 \dot{\cal X}_{\left(9\right)}
        + 360 \,\dot{\cal Z}_{\left(7\right)} - 420 \left(d_{\sigma}\right)^2 \dot{\cal Z}_{\left(9\right)}\,.  
        \label{scalar_function_dot_B8_2PN}
\end{eqnarray}

The scalar functions of the 2PN quadrupole-quadrupole term in Eq.~(\ref{First_Integration_2PN_Terms_QQ}) are given by
        \begin{eqnarray} 
        \dot{C}_{\left(1\right)}\left(\ve{x}_{\rm N}\right) &=& - \frac{12}{\left(d_{\sigma}\right)^2}\,\dot{\cal W}_{\left(6\right)}
        + 9\,\dot{\cal W}_{\left(8\right)} - 13 \left(d_{\sigma}\right)^2 \dot{\cal W}_{\left(10\right)}
        + \frac{12}{\left(d_{\sigma}\right)^2}\,\frac{\ve{\sigma} \cdot \ve{x}_0}{x_0}\,\dot{\cal X}_{\left(5\right)}\,,
        \label{scalar_function_dot_C1_2PN}
        \\
	\dot{C}_{\left(2\right)}\left(\ve{x}_{\rm N}\right) &=& + \frac{24}{\left(d_{\sigma}\right)^4}\,\dot{\cal W}_{\left(5\right)} 
	+ \frac{24}{\left(d_{\sigma}\right)^4}\,\dot{\cal X}_{\left(4\right)} 
	+ \frac{12}{\left(d_{\sigma}\right)^2}\,\frac{1}{x_0}\,\dot{\cal X}_{\left(5\right)} 
	- \frac{24}{\left(d_{\sigma}\right)^4}\left(x_0 + \ve{\sigma} \cdot \ve{x}_0\right)\dot{\cal X}_{\left(5\right)} 
	- \frac{12}{\left(d_{\sigma}\right)^2}\,\dot{\cal X}_{\left(6\right)}
	\nonumber\\ 
	&& + 12\,\dot{\cal X}_{\left(8\right)} - 13 \left(d_{\sigma} \right)^2 \dot{\cal X}_{\left(10\right)}\,,
	\label{scalar_function_dot_C2_2PN}
        \\
        \dot{C}_{\left(3\right)}\left(\ve{x}_{\rm N}\right) &=& + \frac{126}{\left(d_{\sigma}\right)^2}\,\dot{\cal W}_{\left(6\right)} 
	+ \frac{30}{x_0}\,\dot{\cal W}_{\left(7\right)} - \frac{30}{\left(x_0\right)^3} \left(d_{\sigma}\right)^2 \dot{\cal W}_{\left(7\right)} 
	- \frac{60}{\left(d_{\sigma}\right)^2} \left(x_0 + \ve{\sigma} \cdot \ve{x}_0\right) \dot{\cal W}_{\left(7\right)}
        - \frac{399}{2}\,\dot{\cal W}_{\left(8\right)} + \frac{381}{2} \left(d_{\sigma}\right)^2 \dot{\cal W}_{\left(10\right)}
	\nonumber\\ 
	&& - \frac{45}{2} \left(d_{\sigma}\right)^4 \dot{\cal W}_{\left(12\right)} + \frac{60}{\left(d_{\sigma}\right)^2}\,\dot{\cal X}_{\left(5\right)} 
	- \frac{66}{\left(d_{\sigma}\right)^2}\,\frac{\ve{\sigma} \cdot \ve{x}_0}{x_0}\,\dot{\cal X}_{\left(5\right)} 
	+ 24\,\frac{\ve{\sigma} \cdot \ve{x}_0}{\left(x_0\right)^3}\,\dot{\cal X}_{\left(5\right)} 
	- 60\,\dot{\cal X}_{\left(7\right)} + 60\,\frac{\ve{\sigma} \cdot \ve{x}_0}{x_0}\,\dot{\cal X}_{\left(7\right)}
	\nonumber\\ 
	&& - 30 \left(d_{\sigma}\right)^2 \frac{\ve{\sigma} \cdot \ve{x}_0}{\left(x_0\right)^3}\,\dot{\cal X}_{\left(7\right)}\,, 
        \label{scalar_function_dot_C3_2PN}
        \\
        \dot{C}_{\left(4\right)}\left(\ve{x}_{\rm N}\right) &=& - \frac{30}{\left(d_{\sigma}\right)^2}\,\frac{\ve{\sigma} \cdot \ve{x}_0}{x_0}\,\dot{\cal W}_{\left(7\right)} 
	+ 30\,\frac{\ve{\sigma} \cdot \ve{x}_0}{\left(x_0\right)^3}\,\dot{\cal W}_{\left(7\right)} 
	- \frac{36}{\left(d_{\sigma}\right)^2}\,\frac{1}{x_0}\,\dot{\cal X}_{\left(5\right)} 
	+ \frac{6}{\left(x_0\right)^3}\,\dot{\cal X}_{\left(5\right)} 
	+ \frac{72}{\left(d_{\sigma}\right)^4} \left(x_0 + \ve{\sigma} \cdot \ve{x}_0\right) \dot{\cal X}_{\left(5\right)}
	\nonumber\\ 
	&& + \frac{30}{\left(d_{\sigma}\right)^2}\,\dot{\cal X}_{\left(6\right)} 
	+ \frac{60}{x_0}\,\dot{\cal X}_{\left(7\right)} - \frac{30}{\left(x_0\right)^3} \left(d_{\sigma}\right)^2 \dot{\cal X}_{\left(7\right)} 
	- \frac{120}{\left(d_{\sigma}\right)^2} \left(x_0 + \ve{\sigma} \cdot \ve{x}_0\right) \dot{\cal X}_{\left(7\right)} 
	- \frac{315}{2}\,\dot{\cal X}_{\left(8\right)} + \frac{397}{2} \left(d_{\sigma}\right)^2 \dot{\cal X}_{\left(10\right)}
	\nonumber\\ 
	&& -  \frac{45}{2} \left(d_{\sigma}\right)^4 \dot{\cal X}_{\left(12\right)} \,, 
        \label{scalar_function_dot_C4_2PN}
        \\
        \dot{C}_{\left(5\right)}\left(\ve{x}_{\rm N}\right) &=& - \frac{24}{\left(d_{\sigma}\right)^4}\,\dot{\cal W}_{\left(5\right)} 
	+ 60\,\frac{\ve{\sigma} \cdot \ve{x}_0}{\left(x_0\right)^3}\,\dot{\cal W}_{\left(7\right)} 
	- \frac{24}{\left(d_{\sigma}\right)^4}\,\dot{\cal X}_{\left(4\right)} 
	- \frac{12}{\left(d_{\sigma}\right)^2}\,\frac{1}{x_0}\,\dot{\cal X}_{\left(5\right)} 
	+ \frac{48}{\left(x_0\right)^3}\,\dot{\cal X}_{\left(5\right)} 
	+ \frac{24}{\left(d_{\sigma}\right)^4} \left(x_0 + \ve{\sigma} \cdot \ve{x}_0\right) \dot{\cal X}_{\left(5\right)}
	\nonumber\\ 
	&& + \frac{12}{\left(d_{\sigma}\right)^2}\,\dot{\cal X}_{\left(6\right)}
	- \frac{60}{\left(x_0\right)^3} \left(d_{\sigma}\right)^2 \dot{\cal X}_{\left(7\right)} - 306\,\dot{\cal X}_{\left(8\right)}
	+ 385 \left(d_{\sigma}\right)^2 \dot{\cal X}_{\left(10\right)} - 45 \left(d_{\sigma}\right)^4 \dot{\cal X}_{\left(12\right)}\,,  
        \label{scalar_function_dot_C5_2PN}
        \\
        \dot{C}_{\left(6\right)}\left(\ve{x}_{\rm N}\right) &=& + \frac{96}{\left(d_{\sigma}\right)^4}\,\dot{\cal W}_{\left(6\right)} 
	- \frac{60}{\left(d_{\sigma}\right)^2}\,\frac{1}{x_0}\,\dot{\cal W}_{\left(7\right)} 
        + \frac{60}{\left(x_0\right)^3}\,\dot{\cal W}_{\left(7\right)} 
	+  \frac{120}{\left(d_{\sigma}\right)^4} \left(x_0 + \ve{\sigma} \cdot \ve{x}_0\right) \dot{\cal W}_{\left(7\right)} 
	- \frac{132}{\left(d_{\sigma}\right)^2}\,\dot{\cal W}_{\left(8\right)} - 356\,\dot{\cal W}_{\left(10\right)}
	\nonumber\\
	&& + 45 \left(d_{\sigma}\right)^2 \dot{\cal W}_{\left(12\right)} 
	+ \frac{72}{\left(d_{\sigma}\right)^4}\,\dot{\cal X}_{\left(5\right)}
	- \frac{24}{\left(d_{\sigma}\right)^4}\,\frac{\ve{\sigma} \cdot \ve{x}_0}{x_0}\,\dot{\cal X}_{\left(5\right)}
	- \frac{12}{\left(d_{\sigma}\right)^2}\,\frac{\ve{\sigma} \cdot \ve{x}_0}{\left(x_0\right)^3}\,\dot{\cal X}_{\left(5\right)}
	- \frac{120}{\left(d_{\sigma}\right)^2}\,\dot{\cal X}_{\left(7\right)}
	\nonumber\\ 
	&& + \frac{60}{\left(d_{\sigma}\right)^2}\,\frac{\ve{\sigma} \cdot \ve{x}_0}{x_0}\,\dot{\cal X}_{\left(7\right)} 
	+ 60\,\frac{\ve{\sigma} \cdot \ve{x}_0}{\left(x_0\right)^3}\,\dot{\cal X}_{\left(7\right)}\,,
        \label{scalar_function_dot_C6_2PN}
        \\
        \dot{C}_{\left(7\right)}\left(\ve{x}_{\rm N}\right) &=& + \frac{168}{\left(d_{\sigma}\right)^4}\,\dot{\cal W}_{\left(6\right)} 
        + \frac{60}{\left(d_{\sigma}\right)^2}\,\frac{1}{x_0}\,\dot{\cal W}_{\left(7\right)} 
        + \frac{30}{\left(x_0\right)^3}\,\dot{\cal W}_{\left(7\right)} 
        -  \frac{120}{\left(d_{\sigma}\right)^4} \left(x_0 + \ve{\sigma} \cdot \ve{x}_0\right) \dot{\cal W}_{\left(7\right)} 
        - \frac{66}{\left(d_{\sigma}\right)^2}\,\dot{\cal W}_{\left(8\right)} - 172\,\dot{\cal W}_{\left(10\right)}
        \nonumber\\
	&& + \frac{45}{2} \left(d_{\sigma}\right)^2 \dot{\cal W}_{\left(12\right)} 
        + \frac{120}{\left(d_{\sigma}\right)^4}\,\dot{\cal X}_{\left(5\right)}
        - \frac{48}{\left(d_{\sigma}\right)^4}\,\frac{\ve{\sigma} \cdot \ve{x}_0}{x_0}\,\dot{\cal X}_{\left(5\right)}
        - \frac{24}{\left(d_{\sigma}\right)^2}\,\frac{\ve{\sigma} \cdot \ve{x}_0}{\left(x_0\right)^3}\,\dot{\cal X}_{\left(5\right)}
        - \frac{120}{\left(d_{\sigma}\right)^2}\,\dot{\cal X}_{\left(7\right)}
        \nonumber\\ 
        && + \frac{30}{\left(d_{\sigma}\right)^2}\,\frac{\ve{\sigma} \cdot \ve{x}_0}{x_0}\,\dot{\cal X}_{\left(7\right)} 
        + 30\,\frac{\ve{\sigma} \cdot \ve{x}_0}{\left(x_0\right)^3}\,\dot{\cal X}_{\left(7\right)}\,,
        \label{scalar_function_dot_C7_2PN}
        \\
        \dot{C}_{\left(8\right)}\left(\ve{x}_{\rm N}\right) &=& - \frac{48}{\left(d_{\sigma}\right)^6}\,\dot{\cal W}_{\left(5\right)}
	+ \frac{180}{\left(d_{\sigma}\right)^4}\,\dot{\cal W}_{\left(7\right)} 
	- \frac{60}{\left(d_{\sigma}\right)^4}\,\frac{\ve{\sigma} \cdot \ve{x}_0}{x_0}\,\dot{\cal W}_{\left(7\right)} 
	- \frac{30}{\left(d_{\sigma}\right)^2}\,\frac{\ve{\sigma} \cdot \ve{x}_0}{\left(x_0\right)^3}\,\dot{\cal W}_{\left(7\right)} 
	- \frac{48}{\left(d_{\sigma}\right)^6}\,\dot{\cal X}_{\left(4\right)}
	- \frac{24}{\left(d_{\sigma}\right)^4}\,\frac{1}{x_0}\,\dot{\cal X}_{\left(5\right)}
	\nonumber\\ 
	&& - \frac{6}{\left(d_{\sigma}\right)^2}\,\frac{1}{\left(x_0\right)^3}\,\dot{\cal X}_{\left(5\right)}
	+ \frac{48}{\left(d_{\sigma}\right)^6}\left(x_0 + \ve{\sigma} \cdot \ve{x}_0\right)\dot{\cal X}_{\left(5\right)} 
	+ \frac{264}{\left(d_{\sigma}\right)^4}\,\dot{\cal X}_{\left(6\right)} 
	+ \frac{30}{\left(d_{\sigma}\right)^2}\,\frac{1}{x_0}\,\dot{\cal X}_{\left(7\right)}
	+ \frac{30}{\left(x_0\right)^3}\,\dot{\cal X}_{\left(7\right)}
        \nonumber\\ 
	&& - \frac{60}{\left(d_{\sigma}\right)^4}\left(x_0 + \ve{\sigma} \cdot \ve{x}_0\right) \dot{\cal X}_{\left(7\right)}
	- \frac{114}{\left(d_{\sigma}\right)^2}\,\dot{\cal X}_{\left(8\right)} 
	- 180\,\dot{\cal X}_{\left(10\right)}
	+ \frac{45}{2} \left(d_{\sigma}\right)^2 \dot{\cal X}_{\left(12\right)}\,,
        \label{scalar_function_dot_C8_2PN}
        \\
        \dot{C}_{\left(9\right)}\left(\ve{x}_{\rm N}\right) &=& + 3\,\dot{\cal W}_{\left(8\right)} - 5 \left(d_{\sigma}\right)^2 \dot{\cal W}_{\left(10\right)}\,,
        \label{scalar_function_dot_C9_2PN}
        \\
        \dot{C}_{\left(10\right)}\left(\ve{x}_{\rm N}\right) &=& - \frac{36}{\left(d_{\sigma}\right)^2}\,\dot{\cal W}_{\left(6\right)} 
	+ 57\,\dot{\cal W}_{\left(8\right)} + 95 \left(d_{\sigma}\right)^2 \dot{\cal W}_{\left(10\right)} 
	- 75 \left(d_{\sigma}\right)^4 \dot{\cal W}_{\left(12\right)} 
	+ \frac{36}{\left(d_{\sigma}\right)^2}\,\frac{\ve{\sigma} \cdot \ve{x}_0}{x_0}\,\dot{\cal X}_{\left(5\right)} 
	- 60\,\frac{\ve{\sigma} \cdot \ve{x}_0}{x_0}\,\dot{\cal X}_{\left(7\right)}\,, 
        \label{scalar_function_dot_C10_2PN}
        \\
        \dot{C}_{\left(11\right)}\left(\ve{x}_{\rm N}\right) &=& + \frac{24}{\left(d_{\sigma}\right)^4}\,\dot{\cal W}_{\left(5\right)} 
	- \frac{120}{\left(d_{\sigma}\right)^2}\,\dot{\cal W}_{\left(7\right)} 
	+ \frac{60}{\left(d_{\sigma}\right)^2}\,\frac{\ve{\sigma} \cdot \ve{x}_0}{x_0}\,\dot{\cal W}_{\left(7\right)} 
	+ \frac{24}{\left(d_{\sigma}\right)^4}\,\dot{\cal X}_{\left(4\right)} 
	+ \frac{36}{\left(d_{\sigma}\right)^2}\,\frac{1}{x_0}\,\dot{\cal X}_{\left(5\right)} 
	- \frac{72}{\left(d_{\sigma}\right)^4} \left(x_0 + \ve{\sigma} \cdot \ve{x}_0 \right) \dot{\cal X}_{\left(5\right)}
	\nonumber\\ 
	&& - \frac{192}{\left(d_{\sigma}\right)^2}\,\dot{\cal X}_{\left(6\right)}
	- \frac{60}{x_0}\,\dot{\cal X}_{\left(7\right)} 
	+ \frac{120}{\left(d_{\sigma}\right)^2} \left(x_0 + \ve{\sigma} \cdot \ve{x}_0 \right) \dot{\cal X}_{\left(7\right)} 
	+ 93\,\dot{\cal X}_{\left(8\right)} + 220 \left(d_{\sigma}\right)^2 \dot{\cal X}_{\left(10\right)} 
	- 150 \left(d_{\sigma}\right)^4 \dot{\cal X}_{\left(12\right)} \,, 
        \label{scalar_function_dot_C11_2PN}
        \\
        \dot{C}_{\left(12\right)}\left(\ve{x}_{\rm N}\right) &=& + \frac{72}{\left(d_{\sigma}\right)^4}\,\dot{\cal W}_{\left(6\right)} 
	+ \frac{60}{\left(d_{\sigma}\right)^2}\,\frac{1}{x_0}\,\dot{\cal W}_{\left(7\right)}
	- \frac{120}{\left(d_{\sigma}\right)^4} \left(x_0 + \ve{\sigma} \cdot \ve{x}_0 \right) \dot{\cal W}_{\left(7\right)} 
	- \frac{84}{\left(d_{\sigma}\right)^2}\,\dot{\cal W}_{\left(8\right)} - 50\,\dot{\cal W}_{\left(10\right)}
	\nonumber\\ 
	&& + 75 \left(d_{\sigma}\right)^2 \dot{\cal W}_{\left(12\right)} + \frac{72}{\left(d_{\sigma}\right)^4}\,\dot{\cal X}_{\left(5\right)} 
	- \frac{120}{\left(d_{\sigma}\right)^2}\,\dot{\cal X}_{\left(7\right)}\,, 
        \label{scalar_function_dot_C12_2PN}
\end{eqnarray}

        \begin{eqnarray} 
        \dot{C}_{\left(13\right)}\left(\ve{x}_{\rm N}\right) &=& + \frac{48}{\left(d_{\sigma}\right)^2}\,\dot{\cal W}_{\left(6\right)} 
	+ \frac{45}{x_0}\,\dot{\cal W}_{\left(7\right)} - \frac{45}{\left(x_0\right)^3} \left(d_{\sigma}\right)^2 \dot{\cal W}_{\left(7\right)} 
	- \frac{90}{\left(d_{\sigma}\right)^2} \left(x_0 + \ve{\sigma} \cdot \ve{x}_0\right) \dot{\cal W}_{\left(7\right)} - 351\,\dot{\cal W}_{\left(8\right)}
	\nonumber\\ 
	&& - \frac{105}{x_0} \left(d_{\sigma}\right)^2 \dot{\cal W}_{\left(9\right)}  
	+ \frac{105}{\left(x_0\right)^3} \left(d_{\sigma}\right)^4 \dot{\cal W}_{\left(9\right)}  
        + 210 \left(x_0 + \ve{\sigma} \cdot \ve{x}_0\right) \dot{\cal W}_{\left(9\right)} 
	+ \frac{777}{2} \left(d_{\sigma}\right)^2 \dot{\cal W}_{\left(10\right)} - 165  \left(d_{\sigma}\right)^4 \dot{\cal W}_{\left(12\right)}
	\nonumber\\ 
	&& - \frac{225}{2} \left(d_{\sigma}\right)^6 \dot{\cal W}_{\left(14\right)} + \frac{30}{\left(d_{\sigma}\right)^2}\,\dot{\cal X}_{\left(5\right)} 
	- \frac{18}{\left(d_{\sigma}\right)^2}\,\frac{\ve{\sigma} \cdot \ve{x}_0}{x_0}\,\dot{\cal X}_{\left(5\right)} 
	+ 12\,\frac{\ve{\sigma} \cdot \ve{x}_0}{\left(x_0\right)^3}\,\dot{\cal X}_{\left(5\right)} - 240\,\dot{\cal X}_{\left(7\right)} 
	+ 135\,\frac{\ve{\sigma} \cdot \ve{x}_0}{x_0}\,\dot{\cal X}_{\left(7\right)}
	\nonumber\\
	&& - 105 \left(d_{\sigma}\right)^2 \frac{\ve{\sigma} \cdot \ve{x}_0}{\left(x_0\right)^3}\,\dot{\cal X}_{\left(7\right)}
	+ 210 \left(d_{\sigma}\right)^2 \dot{\cal X}_{\left(9\right)} 
	- 105 \left(d_{\sigma}\right)^2 \frac{\ve{\sigma} \cdot \ve{x}_0}{x_0}\,\dot{\cal X}_{\left(9\right)} 
	+ 105 \left(d_{\sigma}\right)^4 \frac{\ve{\sigma} \cdot \ve{x}_0}{\left(x_0\right)^3}\,\dot{\cal X}_{\left(9\right)} \,, 
        \label{scalar_function_dot_C13_2PN}
        \\
        \dot{C}_{\left(14\right)}\left(\ve{x}_{\rm N}\right) &=& - \frac{48}{\left(d_{\sigma}\right)^4}\,\dot{\cal W}_{\left(5\right)} 
	+ \frac{420}{\left(d_{\sigma}\right)^2}\,\dot{\cal W}_{\left(7\right)} 
	+ 180\,\frac{\ve{\sigma} \cdot \ve{x}_0}{\left(x_0\right)^3}\,\dot{\cal W}_{\left(7\right)} 
	- 420\,\dot{\cal W}_{\left(9\right)} - 420 \left(d_{\sigma}\right)^2 \frac{\ve{\sigma} \cdot \ve{x}_0}{\left(x_0\right)^3}\,\dot{\cal W}_{\left(9\right)} 
        - \frac{48}{\left(d_{\sigma}\right)^4}\,\dot{\cal X}_{\left(4\right)}
	\nonumber\\ 
	&& - \frac{24}{\left(d_{\sigma}\right)^2}\,\frac{1}{x_0}\,\dot{\cal X}_{\left(5\right)} 
	+ \frac{42}{\left(x_0\right)^3}\,\dot{\cal X}_{\left(5\right)}
	+ \frac{48}{\left(d_{\sigma}\right)^4} \left(x_0 + \ve{\sigma} \cdot \ve{x}_0\right) \dot{\cal X}_{\left(5\right)} 
	+ \frac{444}{\left(d_{\sigma}\right)^2}\,\dot{\cal X}_{\left(6\right)} 
	+ \frac{210}{x_0}\,\dot{\cal X}_{\left(7\right)} - \frac{390}{\left(x_0\right)^3} \left(d_{\sigma}\right)^2 \dot{\cal X}_{\left(7\right)}
	\nonumber\\ 
	&& - \frac{420}{\left(d_{\sigma}\right)^2} \left(x_0 + \ve{\sigma} \cdot \ve{x}_0\right) \dot{\cal X}_{\left(7\right)} 
	- 876\,\dot{\cal X}_{\left(8\right)} - \frac{210}{x_0} \left(d_{\sigma}\right)^2 \dot{\cal X}_{\left(9\right)} 
	+ \frac{420}{\left(x_0\right)^3} \left(d_{\sigma}\right)^4 \dot{\cal X}_{\left(9\right)} 
	+ 420 \left(x_0 + \ve{\sigma} \cdot \ve{x}_0\right) \dot{\cal X}_{\left(9\right)}
	\nonumber\\ 
	&& + 1371 \left(d_{\sigma}\right)^2 \dot{\cal X}_{\left(10\right)} 
	- 585 \left(d_{\sigma}\right)^4 \dot{\cal X}_{\left(12\right)} - 450 \left(d_{\sigma}\right)^6 \dot{\cal X}_{\left(14\right)} \,,
        \label{scalar_function_dot_C14_2PN}
        \\
        \dot{C}_{\left(15\right)}\left(\ve{x}_{\rm N}\right) &=& + \frac{48}{\left(d_{\sigma}\right)^4}\,\dot{\cal W}_{\left(6\right)} 
	+ \frac{60}{\left(d_{\sigma}\right)^2}\,\frac{1}{x_0}\,\dot{\cal W}_{\left(7\right)}
	+ \frac{75}{\left(x_0\right)^3}\,\dot{\cal W}_{\left(7\right)} 
	- \frac{120}{\left(d_{\sigma}\right)^4} \left(x_0 + \ve{\sigma} \cdot \ve{x}_0 \right)\dot{\cal W}_{\left(7\right)} 
	- \frac{396}{\left(d_{\sigma}\right)^2}\,\dot{\cal W}_{\left(8\right)}
	- \frac{105}{x_0}\,\dot{\cal W}_{\left(9\right)}
	\nonumber\\
	&& - \frac{210}{\left(x_0\right)^3} \left(d_{\sigma}\right)^2 \dot{\cal W}_{\left(9\right)} 
	+ \frac{210}{\left(d_{\sigma}\right)^2} \left(x_0 + \ve{\sigma} \cdot \ve{x}_0 \right) \dot{\cal W}_{\left(9\right)}
	- 112\,\dot{\cal W}_{\left(10\right)} + 480 \left(d_{\sigma}\right)^2 \dot{\cal W}_{\left(12\right)} 
	+ 225 \left(d_{\sigma}\right)^4 \dot{\cal W}_{\left(14\right)} 
	\nonumber\\
	&& + \frac{24}{\left(d_{\sigma}\right)^4}\,\dot{\cal X}_{\left(5\right)} 
	- \frac{24}{\left(d_{\sigma}\right)^4}\,\frac{\ve{\sigma} \cdot \ve{x}_0}{x_0}\,\dot{\cal X}_{\left(5\right)} 
	- \frac{12}{\left(d_{\sigma}\right)^2}\,\frac{\ve{\sigma} \cdot \ve{x}_0}{\left(x_0\right)^3}\,\dot{\cal X}_{\left(5\right)} 
	- \frac{270}{\left(d_{\sigma}\right)^2}\,\dot{\cal X}_{\left(7\right)} 
	+ \frac{150}{\left(d_{\sigma}\right)^2}\,\frac{\ve{\sigma} \cdot \ve{x}_0}{x_0}\,\dot{\cal X}_{\left(7\right)}
	\nonumber\\
	&& + 195\,\frac{\ve{\sigma} \cdot \ve{x}_0}{\left(x_0\right)^3}\,\dot{\cal X}_{\left(7\right)}
	+ 210\,\dot{\cal X}_{\left(9\right)}
	- 105\,\frac{\ve{\sigma} \cdot \ve{x}_0}{x_0}\,\dot{\cal X}_{\left(9\right)}
	- 210 \left(d_{\sigma}\right)^2 \frac{\ve{\sigma} \cdot \ve{x}_0}{\left(x_0\right)^3}\,\dot{\cal X}_{\left(9\right)}\,,
        \label{scalar_function_dot_C15_2PN}
        \\
        \dot{C}_{\left(16\right)}\left(\ve{x}_{\rm N}\right) &=& - \frac{72}{\left(d_{\sigma}\right)^4}\,\dot{\cal W}_{\left(6\right)} 
	+ \frac{180}{\left(x_0\right)^3}\,\dot{\cal W}_{\left(7\right)} 
	- \frac{216}{\left(d_{\sigma}\right)^2}\,\dot{\cal W}_{\left(8\right)}
	- \frac{420}{\left(x_0\right)^3} \left(d_{\sigma}\right)^2 \dot{\cal W}_{\left(9\right)} 
	- 218\,\dot{\cal W}_{\left(10\right)} + 960 \left(d_{\sigma}\right)^2 \dot{\cal W}_{\left(12\right)}
	\nonumber\\ 
	&& + 450 \left(d_{\sigma}\right)^4 \dot{\cal W}_{\left(14\right)} 
	- \frac{144}{\left(d_{\sigma}\right)^4}\,\dot{\cal X}_{\left(5\right)} 
	- \frac{72}{\left(d_{\sigma}\right)^4}\,\frac{\ve{\sigma} \cdot \ve{x}_0}{x_0}\,\dot{\cal X}_{\left(5\right)} 
	- \frac{36}{\left(d_{\sigma}\right)^2}\,\frac{\ve{\sigma} \cdot \ve{x}_0}{\left(x_0\right)^3}\,\dot{\cal X}_{\left(5\right)} 
	+ \frac{240}{\left(d_{\sigma}\right)^2}\,\dot{\cal X}_{\left(7\right)}
	\nonumber\\ 
	&& + \frac{420}{\left(d_{\sigma}\right)^2}\,\frac{\ve{\sigma} \cdot \ve{x}_0}{x_0}\,\dot{\cal X}_{\left(7\right)}
	+ 360\,\frac{\ve{\sigma} \cdot \ve{x}_0}{\left(x_0\right)^3}\,\dot{\cal X}_{\left(7\right)}
        - 420\,\frac{\ve{\sigma} \cdot \ve{x}_0}{x_0}\,\dot{\cal X}_{\left(9\right)}
	- 420 \left(d_{\sigma}\right)^2 \frac{\ve{\sigma} \cdot \ve{x}_0}{\left(x_0\right)^3}\,\dot{\cal X}_{\left(9\right)}\,,
        \label{scalar_function_dot_C16_2PN}
        \\
        \dot{C}_{\left(17\right)}\left(\ve{x}_{\rm N}\right) &=& - \frac{48}{\left(d_{\sigma}\right)^6}\,\dot{\cal W}_{\left(5\right)} 
	+ \frac{300}{\left(d_{\sigma}\right)^4}\,\dot{\cal W}_{\left(7\right)}
	- \frac{300}{\left(d_{\sigma}\right)^4}\,\frac{\ve{\sigma} \cdot \ve{x}_0}{x_0}\,\dot{\cal W}_{\left(7\right)}
	- \frac{150}{\left(d_{\sigma}\right)^2}\,\frac{\ve{\sigma} \cdot \ve{x}_0}{\left(x_0\right)^3}\,\dot{\cal W}_{\left(7\right)}
	- \frac{840}{\left(d_{\sigma}\right)^2}\,\dot{\cal W}_{\left(9\right)}
	+ \frac{630}{\left(d_{\sigma}\right)^2}\,\frac{\ve{\sigma} \cdot \ve{x}_0}{x_0}\,\dot{\cal W}_{\left(9\right)}
	\nonumber\\ 
	&& + 420\,\frac{\ve{\sigma} \cdot \ve{x}_0}{\left(x_0\right)^3}\,\dot{\cal W}_{\left(9\right)}
	- \frac{48}{\left(d_{\sigma}\right)^6}\,\dot{\cal X}_{\left(4\right)}
	- \frac{72}{\left(d_{\sigma}\right)^4}\,\frac{1}{x_0}\,\dot{\cal X}_{\left(5\right)}
	- \frac{18}{\left(d_{\sigma}\right)^2}\,\frac{1}{\left(x_0\right)^3}\,\dot{\cal X}_{\left(5\right)}
	+ \frac{144}{\left(d_{\sigma}\right)^6} \left(x_0 + \ve{\sigma} \cdot \ve{x}_0\right) \dot{\cal X}_{\left(5\right)}
	\nonumber\\ 
	&& + \frac{864}{\left(d_{\sigma}\right)^4}\,\dot{\cal X}_{\left(6\right)}
	+ \frac{360}{\left(x_0\right)^3}\,\dot{\cal X}_{\left(7\right)}
	+  \frac{450}{\left(d_{\sigma}\right)^2}\,\frac{1}{x_0}\,\dot{\cal X}_{\left(7\right)} 
	- \frac{900}{\left(d_{\sigma}\right)^4} \left(x_0 + \ve{\sigma} \cdot \ve{x}_0\right) \dot{\cal X}_{\left(7\right)} 
	- \frac{1884}{\left(d_{\sigma}\right)^2}\,\dot{\cal X}_{\left(8\right)}
	- \frac{420}{x_0}\,\dot{\cal X}_{\left(9\right)} 
	\nonumber\\ 
	&& - \frac{420}{\left(x_0\right)^3} \left(d_{\sigma}\right)^2 \dot{\cal X}_{\left(9\right)} 
	+ \frac{840}{\left(d_{\sigma}\right)^2} \left(x_0 + \ve{\sigma} \cdot \ve{x}_0\right) \dot{\cal X}_{\left(9\right)}
	- 96\,\dot{\cal X}_{\left(10\right)} 
	+ 885 \left(d_{\sigma}\right)^2 \dot{\cal X}_{\left(12\right)} 
	+ 450 \left(d_{\sigma}\right)^4 \dot{\cal X}_{\left(14\right)}\,,
        \label{scalar_function_dot_C17_2PN}
        \\
        \dot{C}_{\left(18\right)}\left(\ve{x}_{\rm N}\right) &=& - \frac{144}{\left(d_{\sigma}\right)^6}\,\dot{\cal W}_{\left(6\right)} 
	- \frac{120}{\left(d_{\sigma}\right)^4}\,\frac{1}{x_0}\,\dot{\cal W}_{\left(7\right)}
	- \frac{30}{\left(d_{\sigma}\right)^2}\,\frac{1}{\left(x_0\right)^3}\,\dot{\cal W}_{\left(7\right)}
	+ \frac{240}{\left(d_{\sigma}\right)^6} \left(x_0 + \ve{\sigma} \cdot \ve{x}_0\right) \dot{\cal W}_{\left(7\right)}
	+ \frac{648}{\left(d_{\sigma}\right)^4}\,\dot{\cal W}_{\left(8\right)}
	\nonumber\\ 
	&& + \frac{210}{\left(d_{\sigma}\right)^2}\,\frac{1}{x_0}\,\dot{\cal W}_{\left(9\right)} 
	+ \frac{105}{\left(x_0\right)^3}\,\dot{\cal W}_{\left(9\right)} 
	- \frac{420}{\left(d_{\sigma}\right)^4} \left(x_0 + \ve{\sigma} \cdot \ve{x}_0\right) \dot{\cal W}_{\left(9\right)} 
	- \frac{324}{\left(d_{\sigma}\right)^2}\,\dot{\cal W}_{\left(10\right)} - 315\,\dot{\cal W}_{\left(12\right)}
	\nonumber\\ 
	&& - \frac{225}{2} \left(d_{\sigma}\right)^2 \dot{\cal W}_{\left(14\right)} 
	- \frac{144}{\left(d_{\sigma}\right)^6}\,\dot{\cal X}_{\left(5\right)} 
	+ \frac{540}{\left(d_{\sigma}\right)^4}\,\dot{\cal X}_{\left(7\right)} 
	- \frac{180}{\left(d_{\sigma}\right)^4}\,\frac{\ve{\sigma} \cdot \ve{x}_0}{x_0}\,\dot{\cal X}_{\left(7\right)} 
	- \frac{90}{\left(d_{\sigma}\right)^2}\,\frac{\ve{\sigma} \cdot \ve{x}_0}{\left(x_0\right)^3}\,\dot{\cal X}_{\left(7\right)} 
	- \frac{420}{\left(d_{\sigma}\right)^2}\,\dot{\cal X}_{\left(9\right)}
	\nonumber\\ 
	&& + \frac{210}{\left(d_{\sigma}\right)^2}\,\frac{\ve{\sigma} \cdot \ve{x}_0}{x_0}\,\dot{\cal X}_{\left(9\right)} 
	+ 105\,\frac{\ve{\sigma} \cdot \ve{x}_0}{\left(x_0\right)^3}\,\dot{\cal X}_{\left(9\right)}\,,  
        \label{scalar_function_dot_C18_2PN}
        \\
        \dot{C}_{\left(19\right)}\left(\ve{x}_{\rm N}\right) &=& 
        + \frac{9}{2}\,\dot{\cal X}_{\left(8\right)} - 5 \left(d_{\sigma}\right)^2 \dot{\cal X}_{\left(10\right)}\,, 
        \label{scalar_function_dot_C19_2PN}
        \\
        \dot{C}_{\left(20\right)}\left(\ve{x}_{\rm N}\right) &=& 
	- \frac{60}{\left(d_{\sigma}\right)^2}\,\frac{\ve{\sigma} \cdot \ve{x}_0}{x_0}\,\dot{\cal W}_{\left(7\right)} 
	+ \frac{60}{\left(d_{\sigma}\right)^2}\,\dot{\cal X}_{\left(6\right)} - 102\,\dot{\cal X}_{\left(8\right)} 
	+ 115 \left(d_{\sigma}\right)^2 \dot{\cal X}_{\left(10\right)}  
	- 75 \left(d_{\sigma}\right)^4 \dot{\cal X}_{\left(12\right)}\,, 
        \label{scalar_function_dot_C20_2PN}
        \end{eqnarray}

\begin{eqnarray}
        \dot{C}_{\left(21\right)}\left(\ve{x}_{\rm N}\right) &=& - \frac{120}{\left(d_{\sigma}\right)^4}\,\dot{\cal W}_{\left(6\right)} 
	- \frac{60}{\left(d_{\sigma}\right)^2}\,\frac{1}{x_0}\,\dot{\cal W}_{\left(7\right)} 
	+ \frac{120}{\left(d_{\sigma}\right)^4} \left(x_0 + \ve{\sigma} \cdot \ve{x}_0\right) \dot{\cal W}_{\left(7\right)} 
	+ \frac{120}{\left(d_{\sigma}\right)^2}\,\dot{\cal W}_{\left(8\right)} - 110\,\dot{\cal W}_{\left(10\right)} 
	\nonumber\\ 
	&& + 150 \left(d_{\sigma}\right)^2 \dot{\cal W}_{\left(12\right)}
	- \frac{120}{\left(d_{\sigma}\right)^4}\,\dot{\cal X}_{\left(5\right)}
	+ \frac{120}{\left(d_{\sigma}\right)^2}\,\dot{\cal X}_{\left(7\right)}
	- \frac{60}{\left(d_{\sigma}\right)^2}\,\frac{\ve{\sigma} \cdot \ve{x}_0}{x_0}\,\dot{\cal X}_{\left(7\right)}\,, 
        \label{scalar_function_dot_C21_2PN}
       \\
       \dot{C}_{\left(22\right)}\left(\ve{x}_{\rm N}\right) &=& - \frac{120}{\left(d_{\sigma}\right)^4}\,\dot{\cal W}_{\left(7\right)} 
        - \frac{120}{\left(d_{\sigma}\right)^4}\,\dot{\cal X}_{\left(6\right)}
	- \frac{60}{\left(d_{\sigma}\right)^2}\,\frac{1}{x_0}\,\dot{\cal X}_{\left(7\right)}
	+ \frac{120}{\left(d_{\sigma}\right)^4} \left(x_0 + \ve{\sigma} \cdot \ve{x}_0\right) \dot{\cal X}_{\left(7\right)}
	+ \frac{60}{\left(d_{\sigma}\right)^2}\,\dot{\cal X}_{\left(8\right)}
	\nonumber\\ 
	&& - 70\,\dot{\cal X}_{\left(10\right)} + 75 \left(d_{\sigma}\right)^2 \dot{\cal X}_{\left(12\right)}\,,
        \label{scalar_function_dot_C22_2PN}
        \\
        \dot{C}_{\left(23\right)}\left(\ve{x}_{\rm N}\right) &=& - \frac{6}{\left(d_{\sigma}\right)^4}\,\dot{\cal W}_{\left(5\right)} 
	- \frac{180}{\left(d_{\sigma}\right)^2}\,\dot{\cal W}_{\left(7\right)} 
	+ \frac{135}{\left(d_{\sigma}\right)^2}\,\frac{\ve{\sigma} \cdot \ve{x}_0}{x_0}\,\dot{\cal W}_{\left(7\right)}
	- 75\,\frac{\ve{\sigma} \cdot \ve{x}_0}{\left(x_0\right)^3}\,\dot{\cal W}_{\left(7\right)}
	+ 210\,\dot{\cal W}_{\left(9\right)} - 105\,\frac{\ve{\sigma} \cdot \ve{x}_0}{x_0}\,\dot{\cal W}_{\left(9\right)}
	\nonumber\\ 
	&& + 105 \left(d_{\sigma}\right)^2 \frac{\ve{\sigma} \cdot \ve{x}_0}{\left(x_0\right)^3}\,\dot{\cal W}_{\left(9\right)} 
	- \frac{6}{\left(d_{\sigma}\right)^4}\,\dot{\cal X}_{\left(4\right)}
	+ \frac{15}{\left(d_{\sigma}\right)^2}\,\frac{1}{x_0}\,\dot{\cal X}_{\left(5\right)}
	- \frac{15}{\left(x_0\right)^3}\,\dot{\cal X}_{\left(5\right)} 
	- \frac{30}{\left(d_{\sigma}\right)^4} \left(x_0 + \ve{\sigma} \cdot \ve{x}_0\right) \dot{\cal X}_{\left(5\right)}
	\nonumber\\ 
	&& - \frac{312}{\left(d_{\sigma}\right)^2}\,\dot{\cal X}_{\left(6\right)} 
	- \frac{120}{x_0}\,\dot{\cal X}_{\left(7\right)} 
	+ \frac{120}{\left(x_0\right)^3} \left(d_{\sigma}\right)^2 \dot{\cal X}_{\left(7\right)} 
	+ \frac{240}{\left(d_{\sigma}\right)^2} \left(x_0 + \ve{\sigma} \cdot \ve{x}_0\right) \dot{\cal X}_{\left(7\right)}
	+ \frac{1683}{2}\, \dot{\cal X}_{\left(8\right)} 
	+ \frac{105}{x_0} \left(d_{\sigma}\right)^2 \dot{\cal X}_{\left(9\right)}
	\nonumber\\ 
	&& - \frac{105}{\left(x_0\right)^3} \left(d_{\sigma}\right)^4 \dot{\cal X}_{\left(9\right)} 
	- 210 \left(x_0 + \ve{\sigma} \cdot \ve{x}_0\right) \dot{\cal X}_{\left(9\right)} 
	- \frac{2127}{2} \left(d_{\sigma}\right)^2 \dot{\cal X}_{\left(10\right)} 
	+ \frac{1305}{2} \left(d_{\sigma}\right)^4 \dot{\cal X}_{\left(12\right)} 
	- \frac{225}{2} \left(d_{\sigma}\right)^6 \dot{\cal X}_{\left(14\right)} \,,
	\nonumber\\ 
        \label{scalar_function_dot_C23_2PN}
        \\
        \dot{C}_{\left(24\right)}\left(\ve{x}_{\rm N}\right) &=& + \frac{84}{\left(d_{\sigma}\right)^4}\,\dot{\cal W}_{\left(6\right)} 
	+ \frac{120}{\left(d_{\sigma}\right)^2}\,\frac{1}{x_0}\,\dot{\cal W}_{\left(7\right)}
	- \frac{210}{\left(x_0\right)^3}\,\dot{\cal W}_{\left(7\right)}
	- \frac{240}{\left(d_{\sigma}\right)^4} \left(x_0 + \ve{\sigma} \cdot \ve{x}_0\right) \dot{\cal W}_{\left(7\right)}
	- \frac{468}{\left(d_{\sigma}\right)^2}\,\dot{\cal W}_{\left(8\right)}
	- \frac{210}{x_0}\,\dot{\cal W}_{\left(9\right)} 
	\nonumber\\ 
	&& + \frac{420}{\left(x_0\right)^3} \left(d_{\sigma}\right)^2 \dot{\cal W}_{\left(9\right)}
	+ \frac{420}{\left(d_{\sigma}\right)^2} \left(x_0 + \ve{\sigma} \cdot \ve{x}_0\right) \dot{\cal W}_{\left(9\right)} 
	+ 1761\, \dot{\cal W}_{\left(10\right)} 
	- 2235 \left(d_{\sigma}\right)^2 \dot{\cal W}_{\left(12\right)} 
	+ 450 \left(d_{\sigma}\right)^4 \dot{\cal W}_{\left(14\right)} 
	\nonumber\\ 
	&& + \frac{144}{\left(d_{\sigma}\right)^4}\,\dot{\cal X}_{\left(5\right)}
	+ \frac{60}{\left(d_{\sigma}\right)^4}\,\frac{\ve{\sigma} \cdot \ve{x}_0}{x_0}\,\dot{\cal X}_{\left(5\right)} 
	+ \frac{30}{\left(d_{\sigma}\right)^2}\,\frac{\ve{\sigma} \cdot \ve{x}_0}{\left(x_0\right)^3}\,\dot{\cal X}_{\left(5\right)} 
	- \frac{540}{\left(d_{\sigma}\right)^2}\,\dot{\cal X}_{\left(7\right)}
	- \frac{30}{\left(d_{\sigma}\right)^2}\,\frac{\ve{\sigma} \cdot \ve{x}_0}{x_0}\,\dot{\cal X}_{\left(7\right)} 
	\nonumber\\ 
	&& - 420\,\frac{\ve{\sigma} \cdot \ve{x}_0}{\left(x_0\right)^3}\,\dot{\cal X}_{\left(7\right)} 
	+ 420\,\dot{\cal X}_{\left(9\right)} + 420 \left(d_{\sigma}\right)^2 \frac{\ve{\sigma} \cdot \ve{x}_0}{\left(x_0\right)^3}\,\dot{\cal X}_{\left(9\right)}\,, 
        \label{scalar_function_dot_C24_2PN}
        \\
        \dot{C}_{\left(25\right)}\left(\ve{x}_{\rm N}\right) &=& - \frac{24}{\left(d_{\sigma}\right)^6}\,\dot{\cal W}_{\left(5\right)} 
	- \frac{270}{\left(d_{\sigma}\right)^4}\,\dot{\cal W}_{\left(7\right)} 
	+ \frac{150}{\left(d_{\sigma}\right)^4}\,\frac{\ve{\sigma} \cdot \ve{x}_0}{x_0}\,\dot{\cal W}_{\left(7\right)}
	+ \frac{75}{\left(d_{\sigma}\right)^2}\,\frac{\ve{\sigma} \cdot \ve{x}_0}{\left(x_0\right)^3}\,\dot{\cal W}_{\left(7\right)} 
	+ \frac{210}{\left(d_{\sigma}\right)^2}\,\dot{\cal W}_{\left(9\right)} 
	- \frac{105}{\left(d_{\sigma}\right)^2}\,\frac{\ve{\sigma} \cdot \ve{x}_0}{x_0}\,\dot{\cal W}_{\left(9\right)}
	\nonumber\\
	&& - 210\,\frac{\ve{\sigma} \cdot \ve{x}_0}{\left(x_0\right)^3}\,\dot{\cal W}_{\left(9\right)}
	- \frac{24}{\left(d_{\sigma}\right)^6}\,\dot{\cal X}_{\left(4\right)}
	+ \frac{60}{\left(d_{\sigma}\right)^4}\,\frac{1}{x_0}\,\dot{\cal X}_{\left(5\right)}
	+ \frac{15}{\left(d_{\sigma}\right)^2}\,\frac{1}{\left(x_0\right)^3}\,\dot{\cal X}_{\left(5\right)}
	- \frac{120}{\left(d_{\sigma}\right)^6} \left(x_0 + \ve{\sigma} \cdot \ve{x}_0\right) \dot{\cal X}_{\left(5\right)}
	\nonumber\\
	&& - \frac{408}{\left(d_{\sigma}\right)^4}\,\dot{\cal X}_{\left(6\right)} 
	- \frac{225}{\left(d_{\sigma}\right)^2}\,\frac{1}{x_0}\,\dot{\cal X}_{\left(7\right)}
	- \frac{165}{\left(x_0\right)^3}\,\dot{\cal X}_{\left(7\right)} 
	+ \frac{450}{\left(d_{\sigma}\right)^4} \left(x_0 + \ve{\sigma} \cdot \ve{x}_0\right) \dot{\cal X}_{\left(7\right)} 
	+ \frac{588}{\left(d_{\sigma}\right)^2}\,\dot{\cal X}_{\left(8\right)} 
	+ \frac{105}{x_0}\,\dot{\cal X}_{\left(9\right)} 
	\nonumber\\
	&& + \frac{210}{\left(x_0\right)^3} \left(d_{\sigma}\right)^2 \dot{\cal X}_{\left(9\right)} 
	- \frac{210}{\left(d_{\sigma}\right)^2} \left(x_0 + \ve{\sigma} \cdot \ve{x}_0 \right) \dot{\cal X}_{\left(9\right)} 
	+ 716\,\dot{\cal X}_{\left(10\right)} 
	- 1155 \left(d_{\sigma}\right)^2 \dot{\cal X}_{\left(12\right)} 
	+ 225 \left(d_{\sigma}\right)^4 \dot{\cal X}_{\left(14\right)}\,,
        \label{scalar_function_dot_C25_2PN}
        \\
	\dot{C}_{\left(26\right)}\left(\ve{x}_{\rm N}\right) &=& + \frac{120}{\left(d_{\sigma}\right)^4}\,\dot{\cal W}_{\left(7\right)} 
	+ \frac{240}{\left(d_{\sigma}\right)^4}\,\frac{\ve{\sigma} \cdot \ve{x}_0}{x_0}\,\dot{\cal W}_{\left(7\right)}
	+ \frac{120}{\left(d_{\sigma}\right)^2}\,\frac{\ve{\sigma} \cdot \ve{x}_0}{\left(x_0\right)^3}\,\dot{\cal W}_{\left(7\right)} 
	- \frac{420}{\left(d_{\sigma}\right)^2}\,\frac{\ve{\sigma} \cdot \ve{x}_0}{x_0}\,\dot{\cal W}_{\left(9\right)}
	- 420\,\frac{\ve{\sigma} \cdot \ve{x}_0}{\left(x_0\right)^3}\,\dot{\cal W}_{\left(9\right)}
	\nonumber\\
	&& - \frac{120}{\left(d_{\sigma}\right)^4}\,\dot{\cal X}_{\left(6\right)}
	+ \frac{60}{\left(d_{\sigma}\right)^2}\,\frac{1}{x_0}\,\dot{\cal X}_{\left(7\right)}
	- \frac{360}{\left(x_0\right)^3}\,\dot{\cal X}_{\left(7\right)} 
	- \frac{120}{\left(d_{\sigma}\right)^4} \left(x_0 + \ve{\sigma} \cdot \ve{x}_0\right) \dot{\cal X}_{\left(7\right)} 
	+ \frac{420}{\left(d_{\sigma}\right)^2}\,\dot{\cal X}_{\left(8\right)}
	+ \frac{420}{\left(x_0\right)^3} \left(d_{\sigma}\right)^2 \dot{\cal X}_{\left(9\right)}
	\nonumber\\ 
	&& + 1438\,\dot{\cal X}_{\left(10\right)} 
	- 2310 \left(d_{\sigma}\right)^2 \dot{\cal X}_{\left(12\right)} 
	+ 450 \left(d_{\sigma}\right)^4 \dot{\cal X}_{\left(14\right)}\,,
        \label{scalar_function_dot_C26_2PN}
        \\
        \dot{C}_{\left(27\right)}\left(\ve{x}_{\rm N}\right) &=& + \frac{96}{\left(d_{\sigma}\right)^6}\,\dot{\cal W}_{\left(6\right)}
	+ \frac{240}{\left(d_{\sigma}\right)^4}\,\frac{1}{x_0}\,\dot{\cal W}_{\left(7\right)} 
	+ \frac{60}{\left(d_{\sigma}\right)^2}\,\frac{1}{\left(x_0\right)^3}\,\dot{\cal W}_{\left(7\right)} 
	- \frac{480}{\left(d_{\sigma}\right)^6} \left(x_0 + \ve{\sigma} \cdot \ve{x}_0\right) \dot{\cal W}_{\left(7\right)}
	- \frac{1332}{\left(d_{\sigma}\right)^4}\,\dot{\cal W}_{\left(8\right)}
	- \frac{420}{\left(d_{\sigma}\right)^2}\,\frac{1}{x_0}\,\dot{\cal W}_{\left(9\right)}
	\nonumber\\ 
	&& - \frac{420}{\left(x_0\right)^3}\,\dot{\cal W}_{\left(9\right)}
	+ \frac{840}{\left(d_{\sigma}\right)^4} \left(x_0 + \ve{\sigma} \cdot \ve{x}_0\right) \dot{\cal W}_{\left(9\right)} 
	+ \frac{816}{\left(d_{\sigma}\right)^2}\,\dot{\cal W}_{\left(10\right)}
	+ 1935\, \dot{\cal W}_{\left(12\right)} 
	- 450 \left(d_{\sigma}\right)^2 \dot{\cal W}_{\left(14\right)} 
	+ \frac{96}{\left(d_{\sigma}\right)^6}\, \dot{\cal X}_{\left(5\right)}
	\nonumber\\ 
	&& - \frac{840}{\left(d_{\sigma}\right)^4}\, \dot{\cal X}_{\left(7\right)} 
	+ \frac{540}{\left(d_{\sigma}\right)^4}\, \frac{\ve{\sigma} \cdot \ve{x}_0}{x_0}\,\dot{\cal X}_{\left(7\right)} 
	+ \frac{270}{\left(d_{\sigma}\right)^2}\, \frac{\ve{\sigma} \cdot \ve{x}_0}{\left(x_0\right)^3}\,\dot{\cal X}_{\left(7\right)} 
	+ \frac{840}{\left(d_{\sigma}\right)^2}\, \dot{\cal X}_{\left(9\right)} 
	- \frac{630}{\left(d_{\sigma}\right)^2}\,\frac{\ve{\sigma} \cdot \ve{x}_0}{x_0} \dot{\cal X}_{\left(9\right)} 
	- 420\, \frac{\ve{\sigma} \cdot \ve{x}_0}{\left(x_0\right)^3}\,\dot{\cal X}_{\left(9\right)}\,,
	\nonumber\\ 
        \label{scalar_function_dot_C27_2PN}
        \\
        \dot{C}_{\left(28\right)}\left(\ve{x}_{\rm N}\right) &=& + \frac{120}{\left(d_{\sigma}\right)^6}\,\dot{\cal W}_{\left(7\right)}
        - \frac{420}{\left(d_{\sigma}\right)^4}\,\dot{\cal W}_{\left(9\right)}
        + \frac{210}{\left(d_{\sigma}\right)^4}\,\frac{\ve{\sigma} \cdot \ve{x}_0}{x_0}\,\dot{\cal W}_{\left(9\right)}
        + \frac{105}{\left(d_{\sigma}\right)^2}\,\frac{\ve{\sigma} \cdot \ve{x}_0}{\left(x_0\right)^3}\,\dot{\cal W}_{\left(9\right)}
        + \frac{120}{\left(d_{\sigma}\right)^6}\,\dot{\cal X}_{\left(6\right)}
        + \frac{180}{\left(d_{\sigma}\right)^4}\,\frac{\dot{\cal X}_{\left(7\right)}}{x_0}
        \nonumber\\
        && + \frac{45}{\left(d_{\sigma}\right)^2}\,\frac{\dot{\cal X}_{\left(7\right)}}{\left(x_0\right)^3}
        - \frac{360}{\left(d_{\sigma}\right)^6} \left(x_0 + \ve{\sigma} \cdot \ve{x}_0\right) \dot{\cal X}_{\left(7\right)}
        - \frac{690}{\left(d_{\sigma}\right)^4}\,\dot{\cal X}_{\left(8\right)}
	- \frac{210}{\left(d_{\sigma}\right)^2}\,\frac{1}{x_0}\,\dot{\cal X}_{\left(9\right)} 
	- \frac{105}{\left(x_0\right)^3}\,\dot{\cal X}_{\left(9\right)}
	\nonumber\\ 
	&& + \frac{420}{\left(d_{\sigma}\right)^4} \left(x_0 + \ve{\sigma} \cdot \ve{x}_0\right) \dot{\cal X}_{\left(9\right)} 
        + \frac{300}{\left(d_{\sigma}\right)^2}\,\dot{\cal X}_{\left(10\right)}
        + \frac{1005}{2}\,\dot{\cal X}_{\left(12\right)}
        - \frac{225}{2} \left(d_{\sigma}\right)^2 \dot{\cal X}_{\left(14\right)}\,. 
        \label{scalar_function_dot_C28_2PN}
\end{eqnarray}

\section{Scalar functions of second integration}\label{Appendix3}

The scalar functions of the second integration are based on the set of $4$ scalar functions: 
${\cal W}_{\left(n\right)}$, ${\cal X}_{\left(n\right)}$, ${\cal Y}_{\left(n\right)}$, ${\cal Z}_{\left(n\right)}$.
These $4$ types of scalar functions are defined as $4$ types of master integrals, given by Eqs.~(D20) - (D23) in \cite{Zschocke_Quadrupole_1}, 
which have been solved in closed form by Eqs.~(D25) - (D28) in \cite{Zschocke_Quadrupole_1}.

\subsection{Scalar functions of the 1PN Monopole term}

The scalar functions of the 1PN monopole term in Eq.~(\ref{Second_Integration_1PN_Terms_M}) are given by
\begin{eqnarray}
        F_{\left(1\right)}\left(\ve{x}_{\rm N}\right) &=&  + 2\,{\cal W}_{\left(3\right)}\left(\ve{x}_{\rm N}\right)\,,
        \label{scalar_F1}
        \\
        F_{\left(2\right)}\left(\ve{x}_{\rm N}\right) &=& - 2\,{\cal X}_{\left(3\right)}\left(\ve{x}_{\rm N}\right). 
        \label{scalar_F2}
\end{eqnarray}

\subsection{Scalar functions of the 1PN Quadrupole term} 

The scalar functions of the 1PN quadrupole term in Eq.~(\ref{Second_Integration_1PN_Terms_Q}) are given by
\begin{eqnarray}
        G_{\left(1\right)}\left(\ve{x}_{\rm N}\right) &=& + 6\,{\cal W}_{\left(5\right)}\left(\ve{x}_{\rm N}\right),
        \label{scalar_G1}
        \\
        G_{\left(2\right)}\left(\ve{x}_{\rm N}\right) &=& + 6\,{\cal X}_{\left(5\right)}\left(\ve{x}_{\rm N}\right),
        \label{scalar_G2}
        \\
        G_{\left(3\right)}\left(\ve{x}_{\rm N}\right) &=& + 3\,{\cal W}_{\left(5\right)}\left(\ve{x}_{\rm N}\right)
        - 15 \left(d_{\sigma}\right)^2 {\cal W}_{\left(7\right)}\left(\ve{x}_{\rm N}\right),
        \label{scalar_G3}
        \\
        G_{\left(4\right)}\left(\ve{x}_{\rm N}\right) &=& + 18\,{\cal X}_{\left(5\right)}\left(\ve{x}_{\rm N}\right)
        - 30 \left(d_{\sigma}\right)^2 {\cal X}_{\left(7\right)}\left(\ve{x}_{\rm N}\right),
        \label{scalar_G4}
        \\
        G_{\left(5\right)}\left(\ve{x}_{\rm N}\right) &=& + 15\,{\cal W}_{\left(7\right)}\left(\ve{x}_{\rm N}\right),
        \label{scalar_G5}
        \\
        G_{\left(6\right)}\left(\ve{x}_{\rm N}\right) &=& - 15\,{\cal X}_{\left(7\right)}\left(\ve{x}_{\rm N}\right),
        \label{scalar_G6}
        \\
        G_{\left(7\right)}\left(\ve{x}_{\rm N}\right) &=& - 15\,{\cal X}_{\left(5\right)}\left(\ve{x}_{\rm N}\right)
        + 15 \left(d_{\sigma}\right)^2 {\cal X}_{\left(7\right)}\left(\ve{x}_{\rm N}\right),
        \label{scalar_G7}
        \\
        G_{\left(8\right)}\left(\ve{x}_{\rm N}\right) &=& - 30\,{\cal W}_{\left(7\right)}\left(\ve{x}_{\rm N}\right).  
        \label{scalar_G8}
\end{eqnarray}

\subsection{Scalar functions of the 2PN Monopole-Monopole term} 

The scalar functions of the 2PN monopole-monopole term in Eq.~(\ref{Second_Integration_2PN_Terms_MM}) are given by
\begin{eqnarray}
        A_{\left(1\right)}\left(\ve{x}_{\rm N}\right) &=& - 4\,{\cal W}_{\left(4\right)}
        - 12 \left(x_0 + \ve{\sigma} \cdot \ve{x}_0\right) {\cal W}_{\left(5\right)}
        - 2 \left(d_{\sigma}\right)^2 {\cal W}_{\left(6\right)}
        + 4\,{\cal X}_{\left(3\right)}
        - 12 \left(d_{\sigma}\right)^2 {\cal X}_{\left(5\right)}
        - 8\,{\cal Z}_{\left(3\right)}
        + 12 \left(d_{\sigma}\right)^2 {\cal Z}_{\left(5\right)} ,
        \label{scalar_function_A1_2PN}
        \\
	A_{\left(2\right)}\left(\ve{x}_{\rm N}\right) &=& - \frac{4}{\left(d_{\sigma}\right)^2}\,{\cal W}_{\left(3\right)}
        - 12\,{\cal W}_{\left(5\right)} - \frac{4}{\left(d_{\sigma}\right)^2}\,{\cal X}_{\left(2\right)}
        - \frac{4}{\left(d_{\sigma}\right)^2} \left(x_0 + \ve{\sigma} \cdot \ve{x}_0\right) {\cal X}_{\left(3\right)}
        + 6\,{\cal X}_{\left(4\right)} + 12 \left(x_0 + \ve{\sigma} \cdot \ve{x}_0\right) {\cal X}_{\left(5\right)}
        \nonumber\\
        && - 2 \left(d_{\sigma}\right)^2 {\cal X}_{\left(6\right)}
        + 12 \,{\cal Y}_{\left(5\right)}\,. 
        \label{scalar_function_A2_2PN}
\end{eqnarray}

\subsection{Scalar functions of the 2PN Monopole-Quadrupole term} 

The scalar functions of the 2PN monopole-quadrupole term in Eq.~(\ref{Second_Integration_2PN_Terms_MQ}) are given by
\begin{eqnarray}
        B_{\left(1\right)}\left(\ve{x}_{\rm N}\right) &=& + \frac{4}{\left(d_{\sigma}\right)^2}\,{\cal W}_{\left(4\right)}
        + 22\,{\cal W}_{\left(6\right)} - 60 \left(x_0 + \ve{\sigma} \cdot \ve{x}_0\right) {\cal W}_{\left(7\right)}
        + \frac{21}{2} \left(d_{\sigma}\right)^2 {\cal W}_{\left(8\right)}
        - \frac{4}{\left(d_{\sigma}\right)^2}\,\frac{\ve{\sigma} \cdot \ve{x}_0}{x_0}\,{\cal X}_{\left(3\right)}
        + 60\,{\cal X}_{\left(5\right)}
        \nonumber\\
        && - 60 \left(d_{\sigma}\right)^2 {\cal X}_{\left(7\right)}
        - 48\,{\cal Z}_{\left(5\right)} + 60 \left(d_{\sigma}\right)^2 {\cal Z}_{\left(7\right)} \,,
        \label{scalar_function_B1_2PN}
        \end{eqnarray}

        \begin{eqnarray} 
        B_{\left(2\right)}\left(\ve{x}_{\rm N}\right) &=& + \frac{8}{\left(d_{\sigma}\right)^4}\,{\cal W}_{\left(3\right)}
        - \frac{12}{\left(d_{\sigma}\right)^2}\,{\cal W}_{\left(5\right)} + 60\,{\cal W}_{\left(7\right)}
        + \frac{8}{\left(d_{\sigma}\right)^4}\,{\cal X}_{\left(2\right)} - \frac{4}{\left(d_{\sigma}\right)^2}\,\frac{1}{x_0}\,{\cal X}_{\left(3\right)}
        + \frac{8}{\left(d_{\sigma}\right)^4}\left(x_0 + \ve{\sigma} \cdot \ve{x}_0 \right) {\cal X}_{\left(3\right)}
        \nonumber\\
	&& - \frac{16}{\left(d_{\sigma}\right)^2}\,{\cal X}_{\left(4\right)}
        + \frac{12}{\left(d_{\sigma}\right)^2}\left(x_0 + \ve{\sigma} \cdot \ve{x}_0 \right) {\cal X}_{\left(5\right)}
        + 18\,{\cal X}_{\left(6\right)} - 60 \left(x_0 + \ve{\sigma} \cdot \ve{x}_0 \right) {\cal X}_{\left(7\right)}
        + \frac{21}{2} \left(d_{\sigma}\right)^2 {\cal X}_{\left(8\right)} - 60\,{\cal Y}_{\left(7\right)}\,,
        \label{scalar_function_B2_2PN}
	\\ 
        B_{\left(3\right)}\left(\ve{x}_{\rm N}\right) &=& + \frac{4}{\left(d_{\sigma}\right)^2}\,{\cal W}_{\left(4\right)}
        + \frac{6}{x_0}\,{\cal W}_{\left(5\right)} - \frac{6}{\left(x_0\right)^3} \left(d_{\sigma}\right)^2 {\cal W}_{\left(5\right)}
        - \frac{12}{\left(d_{\sigma}\right)^2} \left(x_0 + \ve{\sigma} \cdot \ve{x}_0 \right) {\cal W}_{\left(5\right)}
        - 90 \left(x_0 + \ve{\sigma} \cdot \ve{x}_0 \right) {\cal W}_{\left(7\right)}
        \nonumber\\ 
	&& - 84 \left(d_{\sigma}\right)^2  {\cal W}_{\left(8\right)}
        + 210 \left(d_{\sigma}\right)^2 \left(x_0 + \ve{\sigma} \cdot \ve{x}_0 \right) {\cal W}_{\left(9\right)}
        + 30 \left(d_{\sigma}\right)^4 {\cal W}_{\left(10\right)} + \frac{4}{\left(d_{\sigma}\right)^2}\,{\cal X}_{\left(3\right)}
        + 4\,\frac{\ve{\sigma} \cdot\ve{x}_0}{\left(x_0\right)^3}\,{\cal X}_{\left(3\right)} + 18\,{\cal X}_{\left(5\right)}
        \nonumber\\ 
	&& + 6\,\frac{\ve{\sigma} \cdot\ve{x}_0}{x_0}\,{\cal X}_{\left(5\right)}
        - 6 \left(d_{\sigma}\right)^2 \frac{\ve{\sigma} \cdot\ve{x}_0}{\left(x_0\right)^3}\,{\cal X}_{\left(5\right)}
        - 240 \left(d_{\sigma}\right)^2 {\cal X}_{\left(7\right)} + 210 \left(d_{\sigma}\right)^4 {\cal X}_{\left(9\right)}
        - 24\,{\cal Z}_{\left(5\right)} + 210 \left(d_{\sigma}\right)^2 {\cal Z}_{\left(7\right)}
	\nonumber\\ 
	&& - 210 \left(d_{\sigma}\right)^4 {\cal Z}_{\left(9\right)}\,,
        \label{scalar_function_B3_2PN}
        \\
        B_{\left(4\right)}\left(\ve{x}_{\rm N}\right) &=& - \frac{8}{\left(d_{\sigma}\right)^4}\,{\cal W}_{\left(3\right)}
        - \frac{12}{\left(d_{\sigma}\right)^2}\,{\cal W}_{\left(5\right)} + \frac{12}{\left(d_{\sigma}\right)^2}\,\frac{\ve{\sigma} \cdot\ve{x}_0}{x_0}\,{\cal W}_{\left(5\right)}
        + 12 \,\frac{\ve{\sigma} \cdot\ve{x}_0}{\left(x_0\right)^3}\,{\cal W}_{\left(5\right)}
        + 240\,{\cal W}_{\left(7\right)} - 420 \left(d_{\sigma}\right)^2 {\cal W}_{\left(9\right)}
        - \frac{8}{\left(d_{\sigma}\right)^4}\,{\cal X}_{\left(2\right)} 
        \nonumber\\ 
        && + \frac{4}{\left(d_{\sigma}\right)^2}\,\frac{1}{x_0}\,{\cal X}_{\left(3\right)}  
        + \frac{8}{\left(x_0\right)^3}\,{\cal X}_{\left(3\right)} - \frac{8}{\left(d_{\sigma}\right)^4} \left(x_0 + \ve{\sigma} \cdot\ve{x}_0\right) {\cal X}_{\left(3\right)}
        - \frac{20}{\left(d_{\sigma}\right)^2}\,{\cal X}_{\left(4\right)} - \frac{12}{\left(x_0\right)^3} \left(d_{\sigma}\right)^2 {\cal X}_{\left(5\right)}
        \nonumber\\
        && + \frac{36}{\left(d_{\sigma}\right)^2}\left(x_0 + \ve{\sigma} \cdot\ve{x}_0\right) {\cal X}_{\left(5\right)} + 132\,{\cal X}_{\left(6\right)}
        - 360 \left(x_0 + \ve{\sigma} \cdot\ve{x}_0\right) {\cal X}_{\left(7\right)} - 183 \left(d_{\sigma}\right)^2 {\cal X}_{\left(8\right)}
        + 420 \left(d_{\sigma}\right)^2 \left(x_0 + \ve{\sigma} \cdot\ve{x}_0\right) {\cal X}_{\left(9\right)}
        \nonumber\\
        && + 60 \left(d_{\sigma}\right)^4 {\cal X}_{\left(10\right)} - 180\,{\cal Y}_{\left(7\right)} + 420 \left(d_{\sigma}\right)^2 {\cal Y}_{\left(9\right)}\,,
        \label{scalar_function_B4_2PN}
        \\ 
        B_{\left(5\right)}\left(\ve{x}_{\rm N}\right) &=& + \frac{16}{\left(d_{\sigma}\right)^4}\,{\cal W}_{\left(4\right)}
        + \frac{12}{\left(d_{\sigma}\right)^2}\,\frac{1}{x_0}\,{\cal W}_{\left(5\right)} + \frac{6}{\left(x_0\right)^3}\,{\cal W}_{\left(5\right)}
        - \frac{24}{\left(d_{\sigma}\right)^4} \left(x_0 + \ve{\sigma} \cdot\ve{x}_0\right) {\cal W}_{\left(5\right)}
        - \frac{64}{\left(d_{\sigma}\right)^2}\,{\cal W}_{\left(6\right)}
        \nonumber\\
        && + \frac{60}{\left(d_{\sigma}\right)^2}\left(x_0 + \ve{\sigma} \cdot\ve{x}_0\right) {\cal W}_{\left(7\right)} + 69\,{\cal W}_{\left(8\right)}
        - 210 \left(x_0 + \ve{\sigma} \cdot\ve{x}_0\right) {\cal W}_{\left(9\right)} - 30 \left(d_{\sigma}\right)^2 {\cal W}_{\left(10\right)}
        + \frac{8}{\left(d_{\sigma}\right)^4}\,{\cal X}_{\left(3\right)}
        \nonumber\\
        && - \frac{8}{\left(d_{\sigma}\right)^4}\,\frac{\ve{\sigma} \cdot\ve{x}_0}{x_0}\,{\cal X}_{\left(3\right)}
        - \frac{4}{\left(d_{\sigma}\right)^2}\,\frac{\ve{\sigma} \cdot\ve{x}_0}{\left(x_0\right)^3}\,{\cal X}_{\left(3\right)}
        - \frac{60}{\left(d_{\sigma}\right)^2}\,{\cal X}_{\left(5\right)}
        + \frac{12}{\left(d_{\sigma}\right)^2}\,\frac{\ve{\sigma} \cdot\ve{x}_0}{x_0}\,{\cal X}_{\left(5\right)}
        + 6 \, \frac{\ve{\sigma} \cdot\ve{x}_0}{\left(x_0\right)^3}\,{\cal X}_{\left(5\right)}
        \nonumber\\
        && + 210\,{\cal X}_{\left(7\right)} - 210 \left(d_{\sigma}\right)^2 {\cal X}_{\left(9\right)}
        - 180\,{\cal Z}_{\left(7\right)}
        + 210 \left(d_{\sigma}\right)^2 {\cal Z}_{\left(9\right)} \,,
        \label{scalar_function_B5_2PN}
        \\
        B_{\left(6\right)}\left(\ve{x}_{\rm N}\right) &=& - \frac{16}{\left(d_{\sigma}\right)^6}\,{\cal W}_{\left(3\right)}
        - \frac{24}{\left(d_{\sigma}\right)^4}\,{\cal W}_{\left(5\right)}
        + \frac{12}{\left(d_{\sigma}\right)^4}\,\frac{\ve{\sigma} \cdot\ve{x}_0}{x_0}\,{\cal W}_{\left(5\right)}
        + \frac{6}{\left(d_{\sigma}\right)^2}\,\frac{\ve{\sigma} \cdot\ve{x}_0}{\left(x_0\right)^3}\,{\cal W}_{\left(5\right)}
        + \frac{30}{\left(d_{\sigma}\right)^2}\,{\cal W}_{\left(7\right)} - 210\,{\cal W}_{\left(9\right)}
        \nonumber\\
        && - \frac{16}{\left(d_{\sigma}\right)^6}\,{\cal X}_{\left(2\right)}
        + \frac{8}{\left(d_{\sigma}\right)^4}\,\frac{1}{x_0}\,{\cal X}_{\left(3\right)}
        + \frac{2}{\left(d_{\sigma}\right)^2}\,\frac{1}{\left(x_0\right)^3}\,{\cal X}_{\left(3\right)}
        - \frac{16}{\left(d_{\sigma}\right)^6}\left(x_0 + \ve{\sigma} \cdot\ve{x}_0\right){\cal X}_{\left(3\right)}
        - \frac{28}{\left(d_{\sigma}\right)^4}\,{\cal X}_{\left(4\right)}
        \nonumber\\
        && - \frac{12}{\left(d_{\sigma}\right)^2}\,\frac{1}{x_0}\,{\cal X}_{\left(5\right)}
        - \frac{6}{\left(x_0\right)^3}\,{\cal X}_{\left(5\right)}
        + \frac{24}{\left(d_{\sigma}\right)^4}\left(x_0 + \ve{\sigma} \cdot\ve{x}_0\right) {\cal X}_{\left(5\right)}
        + \frac{50}{\left(d_{\sigma}\right)^2}\,{\cal X}_{\left(6\right)}
        - \frac{90}{\left(d_{\sigma}\right)^2}\left(x_0 + \ve{\sigma} \cdot\ve{x}_0\right) {\cal X}_{\left(7\right)}
        - 24\,{\cal X}_{\left(8\right)}
        \nonumber\\
        && + 210 \left(x_0 + \ve{\sigma} \cdot\ve{x}_0\right) {\cal X}_{\left(9\right)}
        - 30 \left(d_{\sigma}\right)^2 {\cal X}_{\left(10\right)} + 210\,{\cal Y}_{\left(9\right)}\,,
        \label{scalar_function_B6_2PN}
        \\
        B_{\left(7\right)}\left(\ve{x}_{\rm N}\right) &=& - \frac{4}{\left(d_{\sigma}\right)^4}\,{\cal W}_{\left(3\right)}
        - \frac{18}{\left(d_{\sigma}\right)^2}\,{\cal W}_{\left(5\right)}
        + \frac{6}{\left(d_{\sigma}\right)^2}\,\frac{\ve{\sigma} \cdot\ve{x}_0}{x_0}\,{\cal W}_{\left(5\right)}
        - 6\,\frac{\ve{\sigma} \cdot\ve{x}_0}{\left(x_0\right)^3}\,{\cal W}_{\left(5\right)} - 180\,{\cal W}_{\left(7\right)}
        + 210 \left(d_{\sigma}\right)^2 {\cal W}_{\left(9\right)}
        \nonumber\\
        && - \frac{4}{\left(d_{\sigma}\right)^4}\,{\cal X}_{\left(2\right)}
        + \frac{2}{\left(d_{\sigma}\right)^2}\,\frac{1}{x_0}\,{\cal X}_{\left(3\right)}
        - \frac{2}{\left(x_0\right)^3}\,{\cal X}_{\left(3\right)}
        - \frac{4}{\left(d_{\sigma}\right)^4} \left(x_0 + \ve{\sigma} \cdot\ve{x}_0\right) {\cal X}_{\left(3\right)}
        - \frac{22}{\left(d_{\sigma}\right)^2}\,{\cal X}_{\left(4\right)} - \frac{6}{x_0}\,{\cal X}_{\left(5\right)}
        \nonumber\\
        && + \frac{6}{\left(x_0\right)^3} \left(d_{\sigma}\right)^2 {\cal X}_{\left(5\right)}
        - \frac{18}{\left(d_{\sigma}\right)^2} \left(x_0 + \ve{\sigma} \cdot\ve{x}_0\right) {\cal X}_{\left(5\right)}
        - 20\,{\cal X}_{\left(6\right)} + 240 \left(x_0 + \ve{\sigma} \cdot\ve{x}_0\right) {\cal X}_{\left(7\right)}
        + 9 \left(d_{\sigma}\right)^2 {\cal X}_{\left(8\right)}
        \nonumber\\
        && - 210 \left(d_{\sigma}\right)^2 \left(x_0 + \ve{\sigma} \cdot\ve{x}_0\right) {\cal X}_{\left(9\right)}
        + 30 \left(d_{\sigma}\right)^4 {\cal X}_{\left(10\right)} + 150\,{\cal Y}_{\left(7\right)}
        - 210 \left(d_{\sigma}\right)^2 {\cal Y}_{\left(9\right)}\,,
        \label{scalar_function_B7_2PN}
        \\
        B_{\left(8\right)}\left(\ve{x}_{\rm N}\right) &=& - \frac{8}{\left(d_{\sigma}\right)^4}\,{\cal W}_{\left(4\right)}
        - \frac{12}{\left(x_0\right)^3}\,{\cal W}_{\left(5\right)}
        + \frac{32}{\left(d_{\sigma}\right)^2}\,{\cal W}_{\left(6\right)}
        - \frac{120}{\left(d_{\sigma}\right)^2} \left(x_0 + \ve{\sigma} \cdot\ve{x}_0\right) {\cal W}_{\left(7\right)}
        - 63\,{\cal W}_{\left(8\right)}
        \nonumber\\
        && + 420 \left(x_0 + \ve{\sigma} \cdot\ve{x}_0\right) {\cal W}_{\left(9\right)}
        - 60 \left(d_{\sigma}\right)^2 {\cal W}_{\left(10\right)}
        + \frac{8}{\left(d_{\sigma}\right)^4}\,\frac{\ve{\sigma} \cdot\ve{x}_0}{x_0}\,{\cal X}_{\left(3\right)}
        + \frac{4}{\left(d_{\sigma}\right)^2}\,\frac{\ve{\sigma} \cdot\ve{x}_0}{\left(x_0\right)^3}\,{\cal X}_{\left(3\right)}
        + \frac{24}{\left(d_{\sigma}\right)^2}\,{\cal X}_{\left(5\right)}
        \nonumber\\
        && - \frac{12}{\left(d_{\sigma}\right)^2}\,\frac{\ve{\sigma} \cdot\ve{x}_0}{x_0}\,{\cal X}_{\left(5\right)}
        - 12\,\frac{\ve{\sigma} \cdot\ve{x}_0}{\left(x_0\right)^3}\,{\cal X}_{\left(5\right)}
        - 420\,{\cal X}_{\left(7\right)} + 420 \left(d_{\sigma}\right)^2 {\cal X}_{\left(9\right)}
        + 360 \,{\cal Z}_{\left(7\right)} - 420 \left(d_{\sigma}\right)^2 {\cal Z}_{\left(9\right)}\,.  
        \label{scalar_function_B8_2PN}
\end{eqnarray}

\subsection{Scalar functions of the 2PN Quadrupole-Quadrupole term} 

The scalar functions of the 2PN quadrupole-quadrupole term in Eq.~(\ref{Second_Integration_2PN_Terms_QQ}) are given by
        \begin{eqnarray} 
        {C}_{\left(1\right)}\left(\ve{x}_{\rm N}\right) &=& - \frac{12}{\left(d_{\sigma}\right)^2}\,{\cal W}_{\left(6\right)}
        + 9\,{\cal W}_{\left(8\right)} - 13 \left(d_{\sigma}\right)^2 {\cal W}_{\left(10\right)}
        + \frac{12}{\left(d_{\sigma}\right)^2}\,\frac{\ve{\sigma} \cdot \ve{x}_0}{x_0}\,{\cal X}_{\left(5\right)}\,,
        \label{scalar_function_C1_2PN}
       \\ 
        {C}_{\left(2\right)}\left(\ve{x}_{\rm N}\right) &=& + \frac{24}{\left(d_{\sigma}\right)^4}\,{\cal W}_{\left(5\right)} 
        + \frac{24}{\left(d_{\sigma}\right)^4}\,{\cal X}_{\left(4\right)} 
        + \frac{12}{\left(d_{\sigma}\right)^2}\,\frac{1}{x_0}\,{\cal X}_{\left(5\right)} 
        - \frac{24}{\left(d_{\sigma}\right)^4}\left(x_0 + \ve{\sigma} \cdot \ve{x}_0\right){\cal X}_{\left(5\right)} 
        - \frac{12}{\left(d_{\sigma}\right)^2}\,{\cal X}_{\left(6\right)}
        \nonumber\\ 
        && + 12\,{\cal X}_{\left(8\right)} - 13 \left(d_{\sigma} \right)^2 {\cal X}_{\left(10\right)}\,,
        \label{scalar_function_C2_2PN}
       \\
        {C}_{\left(3\right)}\left(\ve{x}_{\rm N}\right) &=& + \frac{126}{\left(d_{\sigma}\right)^2}\,{\cal W}_{\left(6\right)} 
        + \frac{30}{x_0}\,{\cal W}_{\left(7\right)} - \frac{30}{\left(x_0\right)^3} \left(d_{\sigma}\right)^2 {\cal W}_{\left(7\right)} 
        - \frac{60}{\left(d_{\sigma}\right)^2} \left(x_0 + \ve{\sigma} \cdot \ve{x}_0\right) {\cal W}_{\left(7\right)}
        - \frac{399}{2}\,{\cal W}_{\left(8\right)} + \frac{381}{2} \left(d_{\sigma}\right)^2 {\cal W}_{\left(10\right)}
        \nonumber\\ 
        && - \frac{45}{2} \left(d_{\sigma}\right)^4 {\cal W}_{\left(12\right)} + \frac{60}{\left(d_{\sigma}\right)^2}\,{\cal X}_{\left(5\right)} 
        - \frac{66}{\left(d_{\sigma}\right)^2}\,\frac{\ve{\sigma} \cdot \ve{x}_0}{x_0}\,{\cal X}_{\left(5\right)} 
        + 24\,\frac{\ve{\sigma} \cdot \ve{x}_0}{\left(x_0\right)^3}\,{\cal X}_{\left(5\right)} 
        - 60\,{\cal X}_{\left(7\right)} + 60\,\frac{\ve{\sigma} \cdot \ve{x}_0}{x_0}\,{\cal X}_{\left(7\right)}
        \nonumber\\ 
        && - 30 \left(d_{\sigma}\right)^2 \frac{\ve{\sigma} \cdot \ve{x}_0}{\left(x_0\right)^3}\,{\cal X}_{\left(7\right)}\,, 
        \label{scalar_function_C3_2PN}
        \\
        {C}_{\left(4\right)}\left(\ve{x}_{\rm N}\right) &=& - \frac{30}{\left(d_{\sigma}\right)^2}\,\frac{\ve{\sigma} \cdot \ve{x}_0}{x_0}\,{\cal W}_{\left(7\right)} 
        + 30\,\frac{\ve{\sigma} \cdot \ve{x}_0}{\left(x_0\right)^3}\,{\cal W}_{\left(7\right)} 
        - \frac{36}{\left(d_{\sigma}\right)^2}\,\frac{1}{x_0}\,{\cal X}_{\left(5\right)} 
        + \frac{6}{\left(x_0\right)^3}\,{\cal X}_{\left(5\right)} 
        + \frac{72}{\left(d_{\sigma}\right)^4} \left(x_0 + \ve{\sigma} \cdot \ve{x}_0\right) {\cal X}_{\left(5\right)}
        \nonumber\\ 
        && + \frac{30}{\left(d_{\sigma}\right)^2}\,{\cal X}_{\left(6\right)} 
        + \frac{60}{x_0}\,{\cal X}_{\left(7\right)} - \frac{30}{\left(x_0\right)^3} \left(d_{\sigma}\right)^2 {\cal X}_{\left(7\right)} 
        - \frac{120}{\left(d_{\sigma}\right)^2} \left(x_0 + \ve{\sigma} \cdot \ve{x}_0\right) {\cal X}_{\left(7\right)} 
        - \frac{315}{2}\,{\cal X}_{\left(8\right)} + \frac{397}{2} \left(d_{\sigma}\right)^2 {\cal X}_{\left(10\right)}
        \nonumber\\ 
        && -  \frac{45}{2} \left(d_{\sigma}\right)^4 {\cal X}_{\left(12\right)} \,, 
        \label{scalar_function_C4_2PN}
        \\
        {C}_{\left(5\right)}\left(\ve{x}_{\rm N}\right) &=& - \frac{24}{\left(d_{\sigma}\right)^4}\,{\cal W}_{\left(5\right)} 
        + 60\,\frac{\ve{\sigma} \cdot \ve{x}_0}{\left(x_0\right)^3}\,{\cal W}_{\left(7\right)} 
        - \frac{24}{\left(d_{\sigma}\right)^4}\,{\cal X}_{\left(4\right)} 
        - \frac{12}{\left(d_{\sigma}\right)^2}\,\frac{1}{x_0}\,{\cal X}_{\left(5\right)} 
        + \frac{48}{\left(x_0\right)^3}\,{\cal X}_{\left(5\right)} 
        + \frac{24}{\left(d_{\sigma}\right)^4} \left(x_0 + \ve{\sigma} \cdot \ve{x}_0\right) {\cal X}_{\left(5\right)}
        \nonumber\\ 
        && + \frac{12}{\left(d_{\sigma}\right)^2}\,{\cal X}_{\left(6\right)}
        - \frac{60}{\left(x_0\right)^3} \left(d_{\sigma}\right)^2 {\cal X}_{\left(7\right)} - 306\,{\cal X}_{\left(8\right)}
        + 385 \left(d_{\sigma}\right)^2 {\cal X}_{\left(10\right)} - 45 \left(d_{\sigma}\right)^4 {\cal X}_{\left(12\right)}\,,  
        \label{scalar_function_C5_2PN}
        \\
        {C}_{\left(6\right)}\left(\ve{x}_{\rm N}\right) &=& + \frac{96}{\left(d_{\sigma}\right)^4}\,{\cal W}_{\left(6\right)} 
        - \frac{60}{\left(d_{\sigma}\right)^2}\,\frac{1}{x_0}\,{\cal W}_{\left(7\right)} 
        + \frac{60}{\left(x_0\right)^3}\,{\cal W}_{\left(7\right)} 
        +  \frac{120}{\left(d_{\sigma}\right)^4} \left(x_0 + \ve{\sigma} \cdot \ve{x}_0\right) {\cal W}_{\left(7\right)} 
        - \frac{132}{\left(d_{\sigma}\right)^2}\,{\cal W}_{\left(8\right)} - 356\,{\cal W}_{\left(10\right)}
        \nonumber\\
        && + 45 \left(d_{\sigma}\right)^2 {\cal W}_{\left(12\right)} 
        + \frac{72}{\left(d_{\sigma}\right)^4}\,{\cal X}_{\left(5\right)}
        - \frac{24}{\left(d_{\sigma}\right)^4}\,\frac{\ve{\sigma} \cdot \ve{x}_0}{x_0}\,{\cal X}_{\left(5\right)}
        - \frac{12}{\left(d_{\sigma}\right)^2}\,\frac{\ve{\sigma} \cdot \ve{x}_0}{\left(x_0\right)^3}\,{\cal X}_{\left(5\right)}
        - \frac{120}{\left(d_{\sigma}\right)^2}\,{\cal X}_{\left(7\right)}
        \nonumber\\ 
        && + \frac{60}{\left(d_{\sigma}\right)^2}\,\frac{\ve{\sigma} \cdot \ve{x}_0}{x_0}\,{\cal X}_{\left(7\right)} 
        + 60\,\frac{\ve{\sigma} \cdot \ve{x}_0}{\left(x_0\right)^3}\,{\cal X}_{\left(7\right)}\,,
        \label{scalar_function_C6_2PN}
        \\
        {C}_{\left(7\right)}\left(\ve{x}_{\rm N}\right) &=& + \frac{168}{\left(d_{\sigma}\right)^4}\,{\cal W}_{\left(6\right)} 
        + \frac{60}{\left(d_{\sigma}\right)^2}\,\frac{1}{x_0}\,{\cal W}_{\left(7\right)} 
        + \frac{30}{\left(x_0\right)^3}\,{\cal W}_{\left(7\right)} 
        -  \frac{120}{\left(d_{\sigma}\right)^4} \left(x_0 + \ve{\sigma} \cdot \ve{x}_0\right) {\cal W}_{\left(7\right)} 
        - \frac{66}{\left(d_{\sigma}\right)^2}\,{\cal W}_{\left(8\right)} - 172\,{\cal W}_{\left(10\right)}
        \nonumber\\
        && + \frac{45}{2} \left(d_{\sigma}\right)^2 {\cal W}_{\left(12\right)} 
        + \frac{120}{\left(d_{\sigma}\right)^4}\,{\cal X}_{\left(5\right)}
        - \frac{48}{\left(d_{\sigma}\right)^4}\,\frac{\ve{\sigma} \cdot \ve{x}_0}{x_0}\,{\cal X}_{\left(5\right)}
        - \frac{24}{\left(d_{\sigma}\right)^2}\,\frac{\ve{\sigma} \cdot \ve{x}_0}{\left(x_0\right)^3}\,{\cal X}_{\left(5\right)}
        - \frac{120}{\left(d_{\sigma}\right)^2}\,{\cal X}_{\left(7\right)}
        \nonumber\\ 
        && + \frac{30}{\left(d_{\sigma}\right)^2}\,\frac{\ve{\sigma} \cdot \ve{x}_0}{x_0}\,{\cal X}_{\left(7\right)} 
        + 30\,\frac{\ve{\sigma} \cdot \ve{x}_0}{\left(x_0\right)^3}\,{\cal X}_{\left(7\right)}\,,
        \label{scalar_function_C7_2PN}
        \\
        {C}_{\left(8\right)}\left(\ve{x}_{\rm N}\right) &=& - \frac{48}{\left(d_{\sigma}\right)^6}\,{\cal W}_{\left(5\right)}
        + \frac{180}{\left(d_{\sigma}\right)^4}\,{\cal W}_{\left(7\right)} 
        - \frac{60}{\left(d_{\sigma}\right)^4}\,\frac{\ve{\sigma} \cdot \ve{x}_0}{x_0}\,{\cal W}_{\left(7\right)} 
        - \frac{30}{\left(d_{\sigma}\right)^2}\,\frac{\ve{\sigma} \cdot \ve{x}_0}{\left(x_0\right)^3}\,{\cal W}_{\left(7\right)} 
        - \frac{48}{\left(d_{\sigma}\right)^6}\,{\cal X}_{\left(4\right)}
        - \frac{24}{\left(d_{\sigma}\right)^4}\,\frac{1}{x_0}\,{\cal X}_{\left(5\right)}
        \nonumber\\ 
        && - \frac{6}{\left(d_{\sigma}\right)^2}\,\frac{1}{\left(x_0\right)^3}\,{\cal X}_{\left(5\right)}
        + \frac{48}{\left(d_{\sigma}\right)^6}\left(x_0 + \ve{\sigma} \cdot \ve{x}_0\right){\cal X}_{\left(5\right)} 
        + \frac{264}{\left(d_{\sigma}\right)^4}\,{\cal X}_{\left(6\right)} 
        + \frac{30}{\left(d_{\sigma}\right)^2}\,\frac{1}{x_0}\,{\cal X}_{\left(7\right)}
        + \frac{30}{\left(x_0\right)^3}\,{\cal X}_{\left(7\right)}
        \nonumber\\ 
        && - \frac{60}{\left(d_{\sigma}\right)^4}\left(x_0 + \ve{\sigma} \cdot \ve{x}_0\right) {\cal X}_{\left(7\right)}
        - \frac{114}{\left(d_{\sigma}\right)^2}\,{\cal X}_{\left(8\right)} 
        - 180\,{\cal X}_{\left(10\right)}
        + \frac{45}{2} \left(d_{\sigma}\right)^2 {\cal X}_{\left(12\right)}\,,
        \label{scalar_function_C8_2PN}
        \\
        {C}_{\left(9\right)}\left(\ve{x}_{\rm N}\right) &=& + 3\,{\cal W}_{\left(8\right)} - 5 \left(d_{\sigma}\right)^2 {\cal W}_{\left(10\right)}\,,
        \label{scalar_function_C9_2PN}
        \\
        {C}_{\left(10\right)}\left(\ve{x}_{\rm N}\right) &=& - \frac{36}{\left(d_{\sigma}\right)^2}\,{\cal W}_{\left(6\right)} 
        + 57\,{\cal W}_{\left(8\right)} + 95 \left(d_{\sigma}\right)^2 {\cal W}_{\left(10\right)} 
        - 75 \left(d_{\sigma}\right)^4 {\cal W}_{\left(12\right)} 
        + \frac{36}{\left(d_{\sigma}\right)^2}\,\frac{\ve{\sigma} \cdot \ve{x}_0}{x_0}\,{\cal X}_{\left(5\right)} 
        - 60\,\frac{\ve{\sigma} \cdot \ve{x}_0}{x_0}\,{\cal X}_{\left(7\right)}\,, 
        \label{scalar_function_C10_2PN}
        \\
        {C}_{\left(11\right)}\left(\ve{x}_{\rm N}\right) &=& + \frac{24}{\left(d_{\sigma}\right)^4}\,{\cal W}_{\left(5\right)} 
        - \frac{120}{\left(d_{\sigma}\right)^2}\,{\cal W}_{\left(7\right)} 
        + \frac{60}{\left(d_{\sigma}\right)^2}\,\frac{\ve{\sigma} \cdot \ve{x}_0}{x_0}\,{\cal W}_{\left(7\right)} 
        + \frac{24}{\left(d_{\sigma}\right)^4}\,{\cal X}_{\left(4\right)} 
        + \frac{36}{\left(d_{\sigma}\right)^2}\,\frac{1}{x_0}\,{\cal X}_{\left(5\right)} 
        - \frac{72}{\left(d_{\sigma}\right)^4} \left(x_0 + \ve{\sigma} \cdot \ve{x}_0 \right) {\cal X}_{\left(5\right)}
        \nonumber\\ 
        && - \frac{192}{\left(d_{\sigma}\right)^2}\,{\cal X}_{\left(6\right)}
        - \frac{60}{x_0}\,{\cal X}_{\left(7\right)} 
        + \frac{120}{\left(d_{\sigma}\right)^2} \left(x_0 + \ve{\sigma} \cdot \ve{x}_0 \right) {\cal X}_{\left(7\right)} 
        + 93\,{\cal X}_{\left(8\right)} + 220 \left(d_{\sigma}\right)^2 {\cal X}_{\left(10\right)} 
        - 150 \left(d_{\sigma}\right)^4 {\cal X}_{\left(12\right)} \,, 
        \label{scalar_function_C11_2PN}
        \\
        {C}_{\left(12\right)}\left(\ve{x}_{\rm N}\right) &=& + \frac{72}{\left(d_{\sigma}\right)^4}\,{\cal W}_{\left(6\right)} 
        + \frac{60}{\left(d_{\sigma}\right)^2}\,\frac{1}{x_0}\,{\cal W}_{\left(7\right)}
        - \frac{120}{\left(d_{\sigma}\right)^4} \left(x_0 + \ve{\sigma} \cdot \ve{x}_0 \right) {\cal W}_{\left(7\right)} 
        - \frac{84}{\left(d_{\sigma}\right)^2}\,{\cal W}_{\left(8\right)} - 50\,{\cal W}_{\left(10\right)}
        \nonumber\\ 
        && + 75 \left(d_{\sigma}\right)^2 {\cal W}_{\left(12\right)} + \frac{72}{\left(d_{\sigma}\right)^4}\,{\cal X}_{\left(5\right)} 
        - \frac{120}{\left(d_{\sigma}\right)^2}\,{\cal X}_{\left(7\right)}\,, 
        \label{scalar_function_C12_2PN}
\end{eqnarray}

        \begin{eqnarray} 
        {C}_{\left(13\right)}\left(\ve{x}_{\rm N}\right) &=& + \frac{48}{\left(d_{\sigma}\right)^2}\,{\cal W}_{\left(6\right)} 
        + \frac{45}{x_0}\,{\cal W}_{\left(7\right)} - \frac{45}{\left(x_0\right)^3} \left(d_{\sigma}\right)^2 {\cal W}_{\left(7\right)} 
        - \frac{90}{\left(d_{\sigma}\right)^2} \left(x_0 + \ve{\sigma} \cdot \ve{x}_0\right) {\cal W}_{\left(7\right)} - 351\,{\cal W}_{\left(8\right)}
        \nonumber\\ 
        && - \frac{105}{x_0} \left(d_{\sigma}\right)^2 {\cal W}_{\left(9\right)}  
        + \frac{105}{\left(x_0\right)^3} \left(d_{\sigma}\right)^4 {\cal W}_{\left(9\right)}  
        + 210 \left(x_0 + \ve{\sigma} \cdot \ve{x}_0\right) {\cal W}_{\left(9\right)} 
        + \frac{777}{2} \left(d_{\sigma}\right)^2 {\cal W}_{\left(10\right)} - 165  \left(d_{\sigma}\right)^4 {\cal W}_{\left(12\right)}
        \nonumber\\ 
        && - \frac{225}{2} \left(d_{\sigma}\right)^6 {\cal W}_{\left(14\right)} + \frac{30}{\left(d_{\sigma}\right)^2}\,{\cal X}_{\left(5\right)} 
        - \frac{18}{\left(d_{\sigma}\right)^2}\,\frac{\ve{\sigma} \cdot \ve{x}_0}{x_0}\,{\cal X}_{\left(5\right)} 
        + 12\,\frac{\ve{\sigma} \cdot \ve{x}_0}{\left(x_0\right)^3}\,{\cal X}_{\left(5\right)} - 240\,{\cal X}_{\left(7\right)} 
        + 135\,\frac{\ve{\sigma} \cdot \ve{x}_0}{x_0}\,{\cal X}_{\left(7\right)}
        \nonumber\\
        && - 105 \left(d_{\sigma}\right)^2 \frac{\ve{\sigma} \cdot \ve{x}_0}{\left(x_0\right)^3}\,{\cal X}_{\left(7\right)}
        + 210 \left(d_{\sigma}\right)^2 {\cal X}_{\left(9\right)} 
        - 105 \left(d_{\sigma}\right)^2 \frac{\ve{\sigma} \cdot \ve{x}_0}{x_0}\,{\cal X}_{\left(9\right)} 
        + 105 \left(d_{\sigma}\right)^4 \frac{\ve{\sigma} \cdot \ve{x}_0}{\left(x_0\right)^3}\,{\cal X}_{\left(9\right)} \,, 
        \label{scalar_function_C13_2PN}
        \\
        {C}_{\left(14\right)}\left(\ve{x}_{\rm N}\right) &=& - \frac{48}{\left(d_{\sigma}\right)^4}\,{\cal W}_{\left(5\right)} 
        + \frac{420}{\left(d_{\sigma}\right)^2}\,{\cal W}_{\left(7\right)} 
        + 180\,\frac{\ve{\sigma} \cdot \ve{x}_0}{\left(x_0\right)^3}\,{\cal W}_{\left(7\right)} 
        - 420\,{\cal W}_{\left(9\right)} - 420 \left(d_{\sigma}\right)^2 \frac{\ve{\sigma} \cdot \ve{x}_0}{\left(x_0\right)^3}\,{\cal W}_{\left(9\right)} 
        - \frac{48}{\left(d_{\sigma}\right)^4}\,{\cal X}_{\left(4\right)}
        \nonumber\\ 
        && - \frac{24}{\left(d_{\sigma}\right)^2}\,\frac{1}{x_0}\,{\cal X}_{\left(5\right)} 
        + \frac{42}{\left(x_0\right)^3}\,{\cal X}_{\left(5\right)}
        + \frac{48}{\left(d_{\sigma}\right)^4} \left(x_0 + \ve{\sigma} \cdot \ve{x}_0\right) {\cal X}_{\left(5\right)} 
        + \frac{444}{\left(d_{\sigma}\right)^2}\,{\cal X}_{\left(6\right)} 
        + \frac{210}{x_0}\,{\cal X}_{\left(7\right)} - \frac{390}{\left(x_0\right)^3} \left(d_{\sigma}\right)^2 {\cal X}_{\left(7\right)}
        \nonumber\\ 
        && - \frac{420}{\left(d_{\sigma}\right)^2} \left(x_0 + \ve{\sigma} \cdot \ve{x}_0\right) {\cal X}_{\left(7\right)} 
        - 876\,{\cal X}_{\left(8\right)} - \frac{210}{x_0} \left(d_{\sigma}\right)^2 {\cal X}_{\left(9\right)} 
        + \frac{420}{\left(x_0\right)^3} \left(d_{\sigma}\right)^4 {\cal X}_{\left(9\right)} 
        + 420 \left(x_0 + \ve{\sigma} \cdot \ve{x}_0\right) {\cal X}_{\left(9\right)}
        \nonumber\\ 
        && + 1371 \left(d_{\sigma}\right)^2 {\cal X}_{\left(10\right)} 
        - 585 \left(d_{\sigma}\right)^4 {\cal X}_{\left(12\right)} - 450 \left(d_{\sigma}\right)^6 {\cal X}_{\left(14\right)} \,,
        \label{scalar_function_C14_2PN}
        \\
        {C}_{\left(15\right)}\left(\ve{x}_{\rm N}\right) &=& + \frac{48}{\left(d_{\sigma}\right)^4}\,{\cal W}_{\left(6\right)} 
        + \frac{60}{\left(d_{\sigma}\right)^2}\,\frac{1}{x_0}\,{\cal W}_{\left(7\right)}
        + \frac{75}{\left(x_0\right)^3}\,{\cal W}_{\left(7\right)} 
        - \frac{120}{\left(d_{\sigma}\right)^4} \left(x_0 + \ve{\sigma} \cdot \ve{x}_0 \right){\cal W}_{\left(7\right)} 
        - \frac{396}{\left(d_{\sigma}\right)^2}\,{\cal W}_{\left(8\right)}
        - \frac{105}{x_0}\,{\cal W}_{\left(9\right)}
        \nonumber\\
        && - \frac{210}{\left(x_0\right)^3} \left(d_{\sigma}\right)^2 {\cal W}_{\left(9\right)} 
        + \frac{210}{\left(d_{\sigma}\right)^2} \left(x_0 + \ve{\sigma} \cdot \ve{x}_0 \right) {\cal W}_{\left(9\right)}
        - 112\,{\cal W}_{\left(10\right)} + 480 \left(d_{\sigma}\right)^2 {\cal W}_{\left(12\right)} 
        + 225 \left(d_{\sigma}\right)^4 {\cal W}_{\left(14\right)} 
        \nonumber\\
        && + \frac{24}{\left(d_{\sigma}\right)^4}\,{\cal X}_{\left(5\right)} 
        - \frac{24}{\left(d_{\sigma}\right)^4}\,\frac{\ve{\sigma} \cdot \ve{x}_0}{x_0}\,{\cal X}_{\left(5\right)} 
        - \frac{12}{\left(d_{\sigma}\right)^2}\,\frac{\ve{\sigma} \cdot \ve{x}_0}{\left(x_0\right)^3}\,{\cal X}_{\left(5\right)} 
        - \frac{270}{\left(d_{\sigma}\right)^2}\,{\cal X}_{\left(7\right)} 
        + \frac{150}{\left(d_{\sigma}\right)^2}\,\frac{\ve{\sigma} \cdot \ve{x}_0}{x_0}\,{\cal X}_{\left(7\right)}
        \nonumber\\
        && + 195\,\frac{\ve{\sigma} \cdot \ve{x}_0}{\left(x_0\right)^3}\,{\cal X}_{\left(7\right)}
        + 210\,{\cal X}_{\left(9\right)}
        - 105\,\frac{\ve{\sigma} \cdot \ve{x}_0}{x_0}\,{\cal X}_{\left(9\right)}
        - 210 \left(d_{\sigma}\right)^2 \frac{\ve{\sigma} \cdot \ve{x}_0}{\left(x_0\right)^3}\,{\cal X}_{\left(9\right)}\,,
        \label{scalar_function_C15_2PN}
        \\
        {C}_{\left(16\right)}\left(\ve{x}_{\rm N}\right) &=& - \frac{72}{\left(d_{\sigma}\right)^4}\,{\cal W}_{\left(6\right)} 
        + \frac{180}{\left(x_0\right)^3}\,{\cal W}_{\left(7\right)} 
        - \frac{216}{\left(d_{\sigma}\right)^2}\,{\cal W}_{\left(8\right)}
        - \frac{420}{\left(x_0\right)^3} \left(d_{\sigma}\right)^2 {\cal W}_{\left(9\right)} 
        - 218\,{\cal W}_{\left(10\right)} + 960 \left(d_{\sigma}\right)^2 {\cal W}_{\left(12\right)}
        \nonumber\\ 
        && + 450 \left(d_{\sigma}\right)^4 {\cal W}_{\left(14\right)} 
        - \frac{144}{\left(d_{\sigma}\right)^4}\,{\cal X}_{\left(5\right)} 
        - \frac{72}{\left(d_{\sigma}\right)^4}\,\frac{\ve{\sigma} \cdot \ve{x}_0}{x_0}\,{\cal X}_{\left(5\right)} 
        - \frac{36}{\left(d_{\sigma}\right)^2}\,\frac{\ve{\sigma} \cdot \ve{x}_0}{\left(x_0\right)^3}\,{\cal X}_{\left(5\right)} 
        + \frac{240}{\left(d_{\sigma}\right)^2}\,{\cal X}_{\left(7\right)}
        \nonumber\\ 
        && + \frac{420}{\left(d_{\sigma}\right)^2}\,\frac{\ve{\sigma} \cdot \ve{x}_0}{x_0}\,{\cal X}_{\left(7\right)}
        + 360\,\frac{\ve{\sigma} \cdot \ve{x}_0}{\left(x_0\right)^3}\,{\cal X}_{\left(7\right)}
        - 420\,\frac{\ve{\sigma} \cdot \ve{x}_0}{x_0}\,{\cal X}_{\left(9\right)}
        - 420 \left(d_{\sigma}\right)^2 \frac{\ve{\sigma} \cdot \ve{x}_0}{\left(x_0\right)^3}\,{\cal X}_{\left(9\right)}\,,
        \label{scalar_function_C16_2PN}
        \\
        {C}_{\left(17\right)}\left(\ve{x}_{\rm N}\right) &=& - \frac{48}{\left(d_{\sigma}\right)^6}\,{\cal W}_{\left(5\right)} 
        + \frac{300}{\left(d_{\sigma}\right)^4}\,{\cal W}_{\left(7\right)}
        - \frac{300}{\left(d_{\sigma}\right)^4}\,\frac{\ve{\sigma} \cdot \ve{x}_0}{x_0}\,{\cal W}_{\left(7\right)}
        - \frac{150}{\left(d_{\sigma}\right)^2}\,\frac{\ve{\sigma} \cdot \ve{x}_0}{\left(x_0\right)^3}\,{\cal W}_{\left(7\right)}
        - \frac{840}{\left(d_{\sigma}\right)^2}\,{\cal W}_{\left(9\right)}
        + \frac{630}{\left(d_{\sigma}\right)^2}\,\frac{\ve{\sigma} \cdot \ve{x}_0}{x_0}\,{\cal W}_{\left(9\right)}
        \nonumber\\ 
        && + 420\,\frac{\ve{\sigma} \cdot \ve{x}_0}{\left(x_0\right)^3}\,{\cal W}_{\left(9\right)}
        - \frac{48}{\left(d_{\sigma}\right)^6}\,{\cal X}_{\left(4\right)}
        - \frac{72}{\left(d_{\sigma}\right)^4}\,\frac{1}{x_0}\,{\cal X}_{\left(5\right)}
        - \frac{18}{\left(d_{\sigma}\right)^2}\,\frac{1}{\left(x_0\right)^3}\,{\cal X}_{\left(5\right)}
        + \frac{144}{\left(d_{\sigma}\right)^6} \left(x_0 + \ve{\sigma} \cdot \ve{x}_0\right) {\cal X}_{\left(5\right)}
        \nonumber\\ 
        && + \frac{864}{\left(d_{\sigma}\right)^4}\,{\cal X}_{\left(6\right)}
        + \frac{360}{\left(x_0\right)^3}\,{\cal X}_{\left(7\right)}
        +  \frac{450}{\left(d_{\sigma}\right)^2}\,\frac{1}{x_0}\,{\cal X}_{\left(7\right)} 
        - \frac{900}{\left(d_{\sigma}\right)^4} \left(x_0 + \ve{\sigma} \cdot \ve{x}_0\right) {\cal X}_{\left(7\right)} 
        - \frac{1884}{\left(d_{\sigma}\right)^2}\,{\cal X}_{\left(8\right)}
        - \frac{420}{x_0}\,{\cal X}_{\left(9\right)} 
        \nonumber\\ 
        && - \frac{420}{\left(x_0\right)^3} \left(d_{\sigma}\right)^2 {\cal X}_{\left(9\right)} 
        + \frac{840}{\left(d_{\sigma}\right)^2} \left(x_0 + \ve{\sigma} \cdot \ve{x}_0\right) {\cal X}_{\left(9\right)}
        - 96\,{\cal X}_{\left(10\right)} 
        + 885 \left(d_{\sigma}\right)^2 {\cal X}_{\left(12\right)} 
        + 450 \left(d_{\sigma}\right)^4 {\cal X}_{\left(14\right)}\,,
        \label{scalar_function_C17_2PN}
        \\
        {C}_{\left(18\right)}\left(\ve{x}_{\rm N}\right) &=& - \frac{144}{\left(d_{\sigma}\right)^6}\,{\cal W}_{\left(6\right)} 
        - \frac{120}{\left(d_{\sigma}\right)^4}\,\frac{1}{x_0}\,{\cal W}_{\left(7\right)}
        - \frac{30}{\left(d_{\sigma}\right)^2}\,\frac{1}{\left(x_0\right)^3}\,{\cal W}_{\left(7\right)}
        + \frac{240}{\left(d_{\sigma}\right)^6} \left(x_0 + \ve{\sigma} \cdot \ve{x}_0\right) {\cal W}_{\left(7\right)}
        + \frac{648}{\left(d_{\sigma}\right)^4}\,{\cal W}_{\left(8\right)}
        \nonumber\\ 
        && + \frac{210}{\left(d_{\sigma}\right)^2}\,\frac{1}{x_0}\,{\cal W}_{\left(9\right)} 
        + \frac{105}{\left(x_0\right)^3}\,{\cal W}_{\left(9\right)} 
        - \frac{420}{\left(d_{\sigma}\right)^4} \left(x_0 + \ve{\sigma} \cdot \ve{x}_0\right) {\cal W}_{\left(9\right)} 
        - \frac{324}{\left(d_{\sigma}\right)^2}\,{\cal W}_{\left(10\right)} - 315\,{\cal W}_{\left(12\right)}
        \nonumber\\ 
        && - \frac{225}{2} \left(d_{\sigma}\right)^2 {\cal W}_{\left(14\right)} 
        - \frac{144}{\left(d_{\sigma}\right)^6}\,{\cal X}_{\left(5\right)} 
        + \frac{540}{\left(d_{\sigma}\right)^4}\,{\cal X}_{\left(7\right)} 
        - \frac{180}{\left(d_{\sigma}\right)^4}\,\frac{\ve{\sigma} \cdot \ve{x}_0}{x_0}\,{\cal X}_{\left(7\right)} 
        - \frac{90}{\left(d_{\sigma}\right)^2}\,\frac{\ve{\sigma} \cdot \ve{x}_0}{\left(x_0\right)^3}\,{\cal X}_{\left(7\right)} 
        - \frac{420}{\left(d_{\sigma}\right)^2}\,{\cal X}_{\left(9\right)}
        \nonumber\\ 
        && + \frac{210}{\left(d_{\sigma}\right)^2}\,\frac{\ve{\sigma} \cdot \ve{x}_0}{x_0}\,{\cal X}_{\left(9\right)} 
        + 105\,\frac{\ve{\sigma} \cdot \ve{x}_0}{\left(x_0\right)^3}\,{\cal X}_{\left(9\right)}\,,  
        \label{scalar_function_C18_2PN}
        \\
        {C}_{\left(19\right)}\left(\ve{x}_{\rm N}\right) &=& 
        + \frac{9}{2}\,{\cal X}_{\left(8\right)} - 5 \left(d_{\sigma}\right)^2 {\cal X}_{\left(10\right)}\,, 
        \label{scalar_function_C19_2PN}
        \\
        {C}_{\left(20\right)}\left(\ve{x}_{\rm N}\right) &=& 
        - \frac{60}{\left(d_{\sigma}\right)^2}\,\frac{\ve{\sigma} \cdot \ve{x}_0}{x_0}\,{\cal W}_{\left(7\right)} 
        + \frac{60}{\left(d_{\sigma}\right)^2}\,{\cal X}_{\left(6\right)} - 102\,{\cal X}_{\left(8\right)} 
        + 115 \left(d_{\sigma}\right)^2 {\cal X}_{\left(10\right)}  
        - 75 \left(d_{\sigma}\right)^4 {\cal X}_{\left(12\right)}\,, 
        \label{scalar_function_C20_2PN}
        \end{eqnarray}

\begin{eqnarray}
        {C}_{\left(21\right)}\left(\ve{x}_{\rm N}\right) &=& - \frac{120}{\left(d_{\sigma}\right)^4}\,{\cal W}_{\left(6\right)} 
        - \frac{60}{\left(d_{\sigma}\right)^2}\,\frac{1}{x_0}\,{\cal W}_{\left(7\right)} 
        + \frac{120}{\left(d_{\sigma}\right)^4} \left(x_0 + \ve{\sigma} \cdot \ve{x}_0\right) {\cal W}_{\left(7\right)} 
        + \frac{120}{\left(d_{\sigma}\right)^2}\,{\cal W}_{\left(8\right)} - 110\,{\cal W}_{\left(10\right)} 
        \nonumber\\ 
        && + 150 \left(d_{\sigma}\right)^2 {\cal W}_{\left(12\right)}
        - \frac{120}{\left(d_{\sigma}\right)^4}\,{\cal X}_{\left(5\right)}
        + \frac{120}{\left(d_{\sigma}\right)^2}\,{\cal X}_{\left(7\right)}
        - \frac{60}{\left(d_{\sigma}\right)^2}\,\frac{\ve{\sigma} \cdot \ve{x}_0}{x_0}\,{\cal X}_{\left(7\right)}\,, 
        \label{scalar_function_C21_2PN}
       \\
       {C}_{\left(22\right)}\left(\ve{x}_{\rm N}\right) &=& - \frac{120}{\left(d_{\sigma}\right)^4}\,{\cal W}_{\left(7\right)} 
        - \frac{120}{\left(d_{\sigma}\right)^4}\,{\cal X}_{\left(6\right)}
        - \frac{60}{\left(d_{\sigma}\right)^2}\,\frac{1}{x_0}\,{\cal X}_{\left(7\right)}
        + \frac{120}{\left(d_{\sigma}\right)^4} \left(x_0 + \ve{\sigma} \cdot \ve{x}_0\right) {\cal X}_{\left(7\right)}
        + \frac{60}{\left(d_{\sigma}\right)^2}\,{\cal X}_{\left(8\right)}
        \nonumber\\ 
        && - 70\,{\cal X}_{\left(10\right)} + 75 \left(d_{\sigma}\right)^2 {\cal X}_{\left(12\right)}\,,
        \label{scalar_function_C22_2PN}
        \\
        {C}_{\left(23\right)}\left(\ve{x}_{\rm N}\right) &=& - \frac{6}{\left(d_{\sigma}\right)^4}\,{\cal W}_{\left(5\right)} 
        - \frac{180}{\left(d_{\sigma}\right)^2}\,{\cal W}_{\left(7\right)} 
        + \frac{135}{\left(d_{\sigma}\right)^2}\,\frac{\ve{\sigma} \cdot \ve{x}_0}{x_0}\,{\cal W}_{\left(7\right)}
        - 75\,\frac{\ve{\sigma} \cdot \ve{x}_0}{\left(x_0\right)^3}\,{\cal W}_{\left(7\right)}
        + 210\,{\cal W}_{\left(9\right)} - 105\,\frac{\ve{\sigma} \cdot \ve{x}_0}{x_0}\,{\cal W}_{\left(9\right)}
        \nonumber\\ 
        && + 105 \left(d_{\sigma}\right)^2 \frac{\ve{\sigma} \cdot \ve{x}_0}{\left(x_0\right)^3}\,{\cal W}_{\left(9\right)} 
        - \frac{6}{\left(d_{\sigma}\right)^4}\,{\cal X}_{\left(4\right)}
        + \frac{15}{\left(d_{\sigma}\right)^2}\,\frac{1}{x_0}\,{\cal X}_{\left(5\right)}
        - \frac{15}{\left(x_0\right)^3}\,{\cal X}_{\left(5\right)} 
        - \frac{30}{\left(d_{\sigma}\right)^4} \left(x_0 + \ve{\sigma} \cdot \ve{x}_0\right) {\cal X}_{\left(5\right)}
        \nonumber\\ 
        && - \frac{312}{\left(d_{\sigma}\right)^2}\,{\cal X}_{\left(6\right)} 
        - \frac{120}{x_0}\,{\cal X}_{\left(7\right)} 
        + \frac{120}{\left(x_0\right)^3} \left(d_{\sigma}\right)^2 {\cal X}_{\left(7\right)} 
        + \frac{240}{\left(d_{\sigma}\right)^2} \left(x_0 + \ve{\sigma} \cdot \ve{x}_0\right) {\cal X}_{\left(7\right)}
        + \frac{1683}{2}\, {\cal X}_{\left(8\right)} 
        + \frac{105}{x_0} \left(d_{\sigma}\right)^2 {\cal X}_{\left(9\right)}
        \nonumber\\ 
        && - \frac{105}{\left(x_0\right)^3} \left(d_{\sigma}\right)^4 {\cal X}_{\left(9\right)} 
        - 210 \left(x_0 + \ve{\sigma} \cdot \ve{x}_0\right) {\cal X}_{\left(9\right)} 
        - \frac{2127}{2} \left(d_{\sigma}\right)^2 {\cal X}_{\left(10\right)} 
        + \frac{1305}{2} \left(d_{\sigma}\right)^4 {\cal X}_{\left(12\right)} 
        - \frac{225}{2} \left(d_{\sigma}\right)^6 {\cal X}_{\left(14\right)} \,,
        \nonumber\\ 
        \label{scalar_function_C23_2PN}
        \\
        {C}_{\left(24\right)}\left(\ve{x}_{\rm N}\right) &=& + \frac{84}{\left(d_{\sigma}\right)^4}\,{\cal W}_{\left(6\right)} 
        + \frac{120}{\left(d_{\sigma}\right)^2}\,\frac{1}{x_0}\,{\cal W}_{\left(7\right)}
        - \frac{210}{\left(x_0\right)^3}\,{\cal W}_{\left(7\right)}
        - \frac{240}{\left(d_{\sigma}\right)^4} \left(x_0 + \ve{\sigma} \cdot \ve{x}_0\right) {\cal W}_{\left(7\right)}
        - \frac{468}{\left(d_{\sigma}\right)^2}\,{\cal W}_{\left(8\right)}
        - \frac{210}{x_0}\,{\cal W}_{\left(9\right)} 
        \nonumber\\ 
        && + \frac{420}{\left(x_0\right)^3} \left(d_{\sigma}\right)^2 {\cal W}_{\left(9\right)}
        + \frac{420}{\left(d_{\sigma}\right)^2} \left(x_0 + \ve{\sigma} \cdot \ve{x}_0\right) {\cal W}_{\left(9\right)} 
        + 1761\, {\cal W}_{\left(10\right)} 
        - 2235 \left(d_{\sigma}\right)^2 {\cal W}_{\left(12\right)} 
        + 450 \left(d_{\sigma}\right)^4 {\cal W}_{\left(14\right)} 
        \nonumber\\ 
        && + \frac{144}{\left(d_{\sigma}\right)^4}\,{\cal X}_{\left(5\right)}
        + \frac{60}{\left(d_{\sigma}\right)^4}\,\frac{\ve{\sigma} \cdot \ve{x}_0}{x_0}\,{\cal X}_{\left(5\right)} 
        + \frac{30}{\left(d_{\sigma}\right)^2}\,\frac{\ve{\sigma} \cdot \ve{x}_0}{\left(x_0\right)^3}\,{\cal X}_{\left(5\right)} 
        - \frac{540}{\left(d_{\sigma}\right)^2}\,{\cal X}_{\left(7\right)}
        - \frac{30}{\left(d_{\sigma}\right)^2}\,\frac{\ve{\sigma} \cdot \ve{x}_0}{x_0}\,{\cal X}_{\left(7\right)} 
        \nonumber\\ 
        && - 420\,\frac{\ve{\sigma} \cdot \ve{x}_0}{\left(x_0\right)^3}\,{\cal X}_{\left(7\right)} 
        + 420\,{\cal X}_{\left(9\right)} + 420 \left(d_{\sigma}\right)^2 \frac{\ve{\sigma} \cdot \ve{x}_0}{\left(x_0\right)^3}\,{\cal X}_{\left(9\right)}\,, 
        \label{scalar_function_C24_2PN}
        \\
        {C}_{\left(25\right)}\left(\ve{x}_{\rm N}\right) &=& - \frac{24}{\left(d_{\sigma}\right)^6}\,{\cal W}_{\left(5\right)} 
        - \frac{270}{\left(d_{\sigma}\right)^4}\,{\cal W}_{\left(7\right)} 
        + \frac{150}{\left(d_{\sigma}\right)^4}\,\frac{\ve{\sigma} \cdot \ve{x}_0}{x_0}\,{\cal W}_{\left(7\right)}
        + \frac{75}{\left(d_{\sigma}\right)^2}\,\frac{\ve{\sigma} \cdot \ve{x}_0}{\left(x_0\right)^3}\,{\cal W}_{\left(7\right)} 
        + \frac{210}{\left(d_{\sigma}\right)^2}\,{\cal W}_{\left(9\right)} 
        - \frac{105}{\left(d_{\sigma}\right)^2}\,\frac{\ve{\sigma} \cdot \ve{x}_0}{x_0}\,{\cal W}_{\left(9\right)}
        \nonumber\\
        && - 210\,\frac{\ve{\sigma} \cdot \ve{x}_0}{\left(x_0\right)^3}\,{\cal W}_{\left(9\right)}
        - \frac{24}{\left(d_{\sigma}\right)^6}\,{\cal X}_{\left(4\right)}
        + \frac{60}{\left(d_{\sigma}\right)^4}\,\frac{1}{x_0}\,{\cal X}_{\left(5\right)}
        + \frac{15}{\left(d_{\sigma}\right)^2}\,\frac{1}{\left(x_0\right)^3}\,{\cal X}_{\left(5\right)}
        - \frac{120}{\left(d_{\sigma}\right)^6} \left(x_0 + \ve{\sigma} \cdot \ve{x}_0\right) {\cal X}_{\left(5\right)}
        \nonumber\\
        && - \frac{408}{\left(d_{\sigma}\right)^4}\,{\cal X}_{\left(6\right)} 
        - \frac{225}{\left(d_{\sigma}\right)^2}\,\frac{1}{x_0}\,{\cal X}_{\left(7\right)}
        - \frac{165}{\left(x_0\right)^3}\,{\cal X}_{\left(7\right)} 
        + \frac{450}{\left(d_{\sigma}\right)^4} \left(x_0 + \ve{\sigma} \cdot \ve{x}_0\right) {\cal X}_{\left(7\right)} 
        + \frac{588}{\left(d_{\sigma}\right)^2}\,{\cal X}_{\left(8\right)} 
        + \frac{105}{x_0}\,{\cal X}_{\left(9\right)} 
        \nonumber\\
        && + \frac{210}{\left(x_0\right)^3} \left(d_{\sigma}\right)^2 {\cal X}_{\left(9\right)} 
        - \frac{210}{\left(d_{\sigma}\right)^2} \left(x_0 + \ve{\sigma} \cdot \ve{x}_0 \right) {\cal X}_{\left(9\right)} 
        + 716\,{\cal X}_{\left(10\right)} 
        - 1155 \left(d_{\sigma}\right)^2 {\cal X}_{\left(12\right)} 
        + 225 \left(d_{\sigma}\right)^4 {\cal X}_{\left(14\right)}\,,
        \label{scalar_function_C25_2PN}
        \\
        {C}_{\left(26\right)}\left(\ve{x}_{\rm N}\right) &=& + \frac{120}{\left(d_{\sigma}\right)^4}\,{\cal W}_{\left(7\right)} 
        + \frac{240}{\left(d_{\sigma}\right)^4}\,\frac{\ve{\sigma} \cdot \ve{x}_0}{x_0}\,{\cal W}_{\left(7\right)}
        + \frac{120}{\left(d_{\sigma}\right)^2}\,\frac{\ve{\sigma} \cdot \ve{x}_0}{\left(x_0\right)^3}\,{\cal W}_{\left(7\right)} 
        - \frac{420}{\left(d_{\sigma}\right)^2}\,\frac{\ve{\sigma} \cdot \ve{x}_0}{x_0}\,{\cal W}_{\left(9\right)}
        - 420\,\frac{\ve{\sigma} \cdot \ve{x}_0}{\left(x_0\right)^3}\,{\cal W}_{\left(9\right)}
        \nonumber\\
        && - \frac{120}{\left(d_{\sigma}\right)^4}\,{\cal X}_{\left(6\right)}
        + \frac{60}{\left(d_{\sigma}\right)^2}\,\frac{1}{x_0}\,{\cal X}_{\left(7\right)}
        - \frac{360}{\left(x_0\right)^3}\,{\cal X}_{\left(7\right)} 
        - \frac{120}{\left(d_{\sigma}\right)^4} \left(x_0 + \ve{\sigma} \cdot \ve{x}_0\right) {\cal X}_{\left(7\right)} 
        + \frac{420}{\left(d_{\sigma}\right)^2}\,{\cal X}_{\left(8\right)}
        + \frac{420}{\left(x_0\right)^3} \left(d_{\sigma}\right)^2 {\cal X}_{\left(9\right)}
        \nonumber\\ 
        && + 1438\,{\cal X}_{\left(10\right)} 
        - 2310 \left(d_{\sigma}\right)^2 {\cal X}_{\left(12\right)} 
        + 450 \left(d_{\sigma}\right)^4 {\cal X}_{\left(14\right)}\,,
        \label{scalar_function_C26_2PN}
        \\
        {C}_{\left(27\right)}\left(\ve{x}_{\rm N}\right) &=& + \frac{96}{\left(d_{\sigma}\right)^6}\,{\cal W}_{\left(6\right)}
        + \frac{240}{\left(d_{\sigma}\right)^4}\,\frac{1}{x_0}\,{\cal W}_{\left(7\right)} 
        + \frac{60}{\left(d_{\sigma}\right)^2}\,\frac{1}{\left(x_0\right)^3}\,{\cal W}_{\left(7\right)} 
        - \frac{480}{\left(d_{\sigma}\right)^6} \left(x_0 + \ve{\sigma} \cdot \ve{x}_0\right) {\cal W}_{\left(7\right)}
        - \frac{1332}{\left(d_{\sigma}\right)^4}\,{\cal W}_{\left(8\right)}
        - \frac{420}{\left(d_{\sigma}\right)^2}\,\frac{1}{x_0}\,{\cal W}_{\left(9\right)}
        \nonumber\\ 
        && - \frac{420}{\left(x_0\right)^3}\,{\cal W}_{\left(9\right)}
        + \frac{840}{\left(d_{\sigma}\right)^4} \left(x_0 + \ve{\sigma} \cdot \ve{x}_0\right) {\cal W}_{\left(9\right)} 
        + \frac{816}{\left(d_{\sigma}\right)^2}\,{\cal W}_{\left(10\right)}
        + 1935\, {\cal W}_{\left(12\right)} 
        - 450 \left(d_{\sigma}\right)^2 {\cal W}_{\left(14\right)} 
        + \frac{96}{\left(d_{\sigma}\right)^6}\, {\cal X}_{\left(5\right)}
        \nonumber\\ 
        && - \frac{840}{\left(d_{\sigma}\right)^4}\, {\cal X}_{\left(7\right)} 
        + \frac{540}{\left(d_{\sigma}\right)^4}\, \frac{\ve{\sigma} \cdot \ve{x}_0}{x_0}\,{\cal X}_{\left(7\right)} 
        + \frac{270}{\left(d_{\sigma}\right)^2}\, \frac{\ve{\sigma} \cdot \ve{x}_0}{\left(x_0\right)^3}\,{\cal X}_{\left(7\right)} 
        + \frac{840}{\left(d_{\sigma}\right)^2}\, {\cal X}_{\left(9\right)} 
        - \frac{630}{\left(d_{\sigma}\right)^2}\,\frac{\ve{\sigma} \cdot \ve{x}_0}{x_0} {\cal X}_{\left(9\right)} 
        - 420\, \frac{\ve{\sigma} \cdot \ve{x}_0}{\left(x_0\right)^3}\,{\cal X}_{\left(9\right)}\,,
	\nonumber\\ 
        \label{scalar_function_C27_2PN}
        \\
        {C}_{\left(28\right)}\left(\ve{x}_{\rm N}\right) &=& + \frac{120}{\left(d_{\sigma}\right)^6}\,{\cal W}_{\left(7\right)}
        - \frac{420}{\left(d_{\sigma}\right)^4}\,{\cal W}_{\left(9\right)}
        + \frac{210}{\left(d_{\sigma}\right)^4}\,\frac{\ve{\sigma} \cdot \ve{x}_0}{x_0}\,{\cal W}_{\left(9\right)}
        + \frac{105}{\left(d_{\sigma}\right)^2}\,\frac{\ve{\sigma} \cdot \ve{x}_0}{\left(x_0\right)^3}\,{\cal W}_{\left(9\right)}
        + \frac{120}{\left(d_{\sigma}\right)^6}\,{\cal X}_{\left(6\right)}
        + \frac{180}{\left(d_{\sigma}\right)^4}\,\frac{{\cal X}_{\left(7\right)}}{x_0}
        \nonumber\\
        && + \frac{45}{\left(d_{\sigma}\right)^2}\,\frac{{\cal X}_{\left(7\right)}}{\left(x_0\right)^3}
        - \frac{360}{\left(d_{\sigma}\right)^6} \left(x_0 + \ve{\sigma} \cdot \ve{x}_0\right) {\cal X}_{\left(7\right)}
        - \frac{690}{\left(d_{\sigma}\right)^4}\,{\cal X}_{\left(8\right)}
        - \frac{210}{\left(d_{\sigma}\right)^2}\,\frac{1}{x_0}\,{\cal X}_{\left(9\right)} 
        - \frac{105}{\left(x_0\right)^3}\,{\cal X}_{\left(9\right)}
        \nonumber\\ 
        && + \frac{420}{\left(d_{\sigma}\right)^4} \left(x_0 + \ve{\sigma} \cdot \ve{x}_0\right) {\cal X}_{\left(9\right)} 
        + \frac{300}{\left(d_{\sigma}\right)^2}\,{\cal X}_{\left(10\right)}
        + \frac{1005}{2}\,{\cal X}_{\left(12\right)}
        - \frac{225}{2} \left(d_{\sigma}\right)^2 {\cal X}_{\left(14\right)}\,. 
        \label{scalar_function_C28_2PN}
\end{eqnarray}

\section{Corrigendum}

Each of the tensorial coefficients in our previous work \cite{Zschocke_Quadrupole_1}, in particular,   
the equations (E28) - (E39), (E41) - (E65), (E67) - (E87) in \cite{Zschocke_Quadrupole_1}, 
have been recalculated and crossed-checked and a very few typos have been found. In this appendix, 
we will present a compete list of all these typos in our article \cite{Zschocke_Quadrupole_1}. 
Actually, most of these misprints have already been mentioned in the footnote on p.~$9$ in \cite{Zschocke_Quadrupole_3}. 
\begin{enumerate}
\item[$\bullet$] In Eq.~(E67) in \cite{Zschocke_Quadrupole_1}:
$- \frac{24}{\left(d_{\sigma}\right)^4} \sigma^a \sigma^b d_{\sigma}^c \delta^{di} 
\rightarrow - \frac{24}{\left(d_{\sigma}\right)^4} \sigma^a d_{\sigma}^b \sigma^c \delta^{di}$.
\item[$\bullet$] In Eq.~(E69) in \cite{Zschocke_Quadrupole_1}: 
$+ \frac{540}{\left(d_{\sigma}\right)^4}\,\sigma^a d_{\sigma}^b d_{\sigma}^c d_{\sigma}^d \sigma^i 
\rightarrow + \frac{300}{\left(d_{\sigma}\right)^4}\,\sigma^a d_{\sigma}^b d_{\sigma}^c d_{\sigma}^d \sigma^i$. 
\item[$\bullet$] In Eq.~(E69) in \cite{Zschocke_Quadrupole_1}:
$+ \frac{90}{\left(d_{\sigma}\right)^2} \frac{1}{\left(x_0\right)^3}\,\sigma^a d_{\sigma}^b d_{\sigma}^c d_{\sigma}^d d_{\sigma}^i 
\rightarrow + \frac{60}{\left(d_{\sigma}\right)^2} \frac{1}{\left(x_0\right)^3}\,\sigma^a d_{\sigma}^b d_{\sigma}^c d_{\sigma}^d d_{\sigma}^i$.
\item[$\bullet$] In Eq.~(E69) in \cite{Zschocke_Quadrupole_1}:
$- \frac{60}{\left(d_{\sigma}\right)^2} \frac{1}{\left(x_0\right)^3}\,d_{\sigma}^a d_{\sigma}^b d_{\sigma}^c d_{\sigma}^d \sigma^i 
\rightarrow - \frac{30}{\left(d_{\sigma}\right)^2} \frac{1}{\left(x_0\right)^3}\,d_{\sigma}^a d_{\sigma}^b d_{\sigma}^c d_{\sigma}^d \sigma^i$.
\item[$\bullet$] In Eq.~(E71) in \cite{Zschocke_Quadrupole_1}: 
$+ \frac{210}{\left(d_{\sigma}\right)^2} \frac{\bm{\sigma} \cdot \bm{x}_0}{x_0}\,\sigma^a d_{\sigma}^b d_{\sigma}^c d_{\sigma}^d \sigma^i 
\rightarrow + \frac{630}{\left(d_{\sigma}\right)^2} \frac{\bm{\sigma} \cdot \bm{x}_0}{x_0}\,\sigma^a d_{\sigma}^b d_{\sigma}^c d_{\sigma}^d \sigma^i$.
\end{enumerate}

\noindent
By taking account of these few corrections one arrives at the correct coefficients presented in this Supplement.
